# Two-Dimensional Transition Metal Dichalcogenides-Based Counter Electrodes for Dye-Sensitized Solar Cells

*Eric Singh,*[†, ‡, §] *Ki Seok Kim,*[‡] *Geun Young Yeom,**[*, ‡, §] *and Hari Singh Nalwa**[*, ⊥]

[†]Department of Computer Science, Stanford University, Stanford, California 94305, United States

[‡]School of Advanced Materials Science and Engineering and [§]SKKU Advanced Institute of Nano Technology, Sungkyunkwan University, 2066 Seobu-ro, Jangan-gu, Suwon-si, Gyeonggi-do 16419, South Korea

[⊥]Advanced Technology Research, 26650 The Old Road, Suite 208, Valencia, California 91381, United States

*E-mail: gyyeom@skku.edu (G. Y. Yeom).
*E-mail: nalwa@mindspring.com (H. S. Nalwa).

**Abstract**

Dye-sensitized solar cells (DSSCs) are gaining considerable interest as alternatives to the semiconductor-based thin film solar cells. The noble metal platinum (Pt) is conventionally used as counter electrode (CE) material for fabricating DSSCs. Since Pt is expensive and scarce, new materials have been explored to develop cost-effective Pt-free counter electrodes for DSSCs. Two-dimensional (2D) graphene-based counter electrodes have achieved the highest known power conversion efficiency ($\eta$) of 13%, which has stimulated research activities in 2D layered transition metal dichalcogenides (TMDs) for developing Pt-free DSSCs. In this review, progress made on alternative counter electrodes for fabricating low-cost Pt-free DSSCs, based on earth-abundant 2D TMDs including $MoS_2$, $WS_2$, $TiS_2$, $FeS_2$, $CoS_2$, $NiS_2$, $SnS_2$, $MoSe_2$, $NbSe_2$, $TaSe_2$, $NiSe_2$, $FeSe_2$, $CoSe_2$, $Bi_2Se_3$ and their based composites, are discussed and summarized. Also, the considerable progress made on thin films of $MoS_2$ and $MoS_2$ based carbon, graphene, carbon nanotubes (CNTs), carbon nanofibers (CNFs), and poly(3,4-ethylenedioxythiophene):poly(styrenesulfonate) (PEDOT:PSS) composites as efficient counter electrodes (CEs) for DSSCs are discussed, in terms of their electrochemical and photovoltaic properties. At present, PCE values higher than that of standard Pt CE have been recorded for a number of TMD-based CEs, which include $MoS_2$ and $MoSe_2$/thin films deposited on Mo foil, $MoS_2$/CNTs, $MoS_2$/graphene, $MoS_2$/carbon, $MoSe_2$/PEDOT:PSS, $NbSe_2$, $FeS_2$, $FeSe_2$ nanosheets, $TiS_2$/graphene, and $NiS_2$/graphene hybrid systems in DSSCs, for the reduction of triiodide ($I_3^-$) to iodide ($I^-$). The highest PCE ($\eta$=10.46%) versus Pt CE ($\eta$=8.25%) at 1 Sun (100 mW/cm$^2$, AM 1.5G) was measured for DSSCs having a low cost and flexible $CoSe_2$/carbon-nanoclimbing wall counter electrode deposited on a nickel foam. Though TMD-based materials show great potential for solar cell devices, their long-term stability is equally important. The DSSCs with a

TiS$_2$/graphene hybrid, and TiS$_2$/PEDOT:PSS composite CEs, showed stability up to 20 to 30 days, respectively, without any measurable degradation in the photovoltaic performance. The long-term stability of TMDs-based DSSCs under different environmental conditions is also described in view of their commercial applications.

**Keywords:** Dye-sensitized solar cells (DSSCs), 2D Transition Metal Dichalcogenides, MoS$_2$, WS$_2$, TiS$_2$, FeS$_2$, CoS$_2$, NiS$_2$, SnS$_2$, MoSe$_2$, NbSe$_2$, TaSe$_2$, NiSe$_2$, FeSe$_2$, CoSe$_2$, Bi$_2$Se$_3$, counter electrodes, composites, Stability of TMDs-Based DSSCs.

**Contents**

## 1. Introduction: Dye-Sensitized Solar Cells (DSSCs)

Silicon has been traditionally used in fabricating solar cells. O'Regan and Grätzel[1] in 1991 demonstrated dye-sensitized solar cells (DSSC) as a low-cost approach and as an alternative to silicon solar cells. The concept of a DSSC system was hypothesized based on photosynthesis in nature. The first DSSC was fabricated using nanocrystalline porous TiO$_2$ as optically transparent thin-film photoelectrodes and a ruthenium(II)–bipyridyl complex as a photosensitizer (dye), which generated a power conversion efficiency of 7.9% for the triiodide/iodide (I$_3^-$/I$^-$) redox couple. This inspired a new field of research based on DSSCs where a broad range of novel materials are now being explored for electrodes (photoanode and counter electrode (CE; cathode)), photosensitizers (dyes), and reduction-oxidation (redox) electrolytes for fabricating DSSC devices.

The commonly used components in dye-sensitized solar cells are a photoanode, a CE, a photosensitizing dye, and an electrolyte. These include tris(2,2'-bipyridyl)cobalt(II/III) [Co(bpy)$_3$] (Co$^{2+}$/Co$^{3+}$), triiodide/iodide (I$_3^-$/I$^-$) and 5,5'-dithiobis(1-methyltetrazole)/1-methyltetrazole-5-thiolate (T$_2$/T$^-$) as redox couples, poly(3,4-ethylenedioxythiophene) (PEDOT), poly(styrenesulfonate) (PSS), poly(3-hexylthiophene) (P3HT), 2,2′,7,7′-tetrakis(*N*,*N*-*p*-dimethoxy-phenylamino)-9,9′-spirobifluorene (spiro-OMETAD) as solid-state hole-transport materials (HTM); *cis*-bis(isothiocyanato)bis(2,2'-bipyridyl-4,4'-dicarboxylato)ruthenium(II), (N3 dye), ditetrabutylammonium *cis*-bis(isothiocyanato)bis(2,2′-bipyridyl-4,4′-dicarboxylato)ruthenium(II), where tetrabutyl-ammonium cation: [C$_4$H$_9$)$_4$N$^+$] (N719 dye), and (tris(cyanato)-2,2',2''-terpyridyl-4,4',4''-tricarboxylate)ruthenium(II) (N749 or black dye) as photosensitizing dyes, are regularly used. The dye-sensitized mesoporous TiO$_2$ is used as a photoanode, while the Pt-coated fluorine-doped tin oxide (FTO) on a glass substrate is used as a CE, which facilitates the catalysis process. The counter electrode in a DSSC device acts as a catalyst. Pt counter electrodes yield the maximum electrocatalytic activity for the triiodide/iodide (I$_3^-$/I$^-$) redox couple, but are poorly effective for iodine-free redox couple such as Co$^{2+}$/Co$^{3+}$ and T$_2$/T$^-$. Transparent conducting oxides such as FTO or indium-tin oxide (ITO) on glass is the common substrate used in assembling DSSCs with different CE materials. The power conversion efficiency of DSSCs is governed by many factors including light absorbing capacity of photosensitizing dyes and catalytic materials.

The classical photosensitizing dyes were developed by Grätzel's research team,[2, 3] including N3 dye that absorbs up to 800 nm, and N749 dye (also known as black dye) which absorbs sunlight in the longer wavelength region up to 920 nm. In 2014, Mathew et al.[4] used graphene nanoplatelets as a CE, a new push-pull porphyrin with a donor–π-bridge acceptor (D–π–A) chemical structure as a sensitizing dye, and Co(II)/Co(III) redox mediator for developing a DSSC, which exhibited a $J_{SC}$ of 18.1 mA cm$^{-2}$, $V_{OC}$ of 0.91 V, FF of 0.78 and the highest known PCE of 13% under 100 mW/cm$^2$ (AM 1.5) illumination. The high PCE of DSSCs originated from a molecularly engineered porphyrin dye, 4-(7-{2-[(2Z,7Z,11E,16Z)-7,17-bis[2,6-bis(octyloxy)phenyl]-12-[bis({4-[2,4-bis(hexyloxy)phenyl]phenyl})amino]-21,23,24,25-tetraaza-22-zincahexacyclo[9.9.3.1$^{3,6}$.1$^{13,16}$.0$^8$,$^{23}$.0$^{18,21}$]pentacosa 1(20),2,4,6(25),7,9,11,13(24),14,16,18-undecaen-2-yl]ethynyl}-2,1,3-benzothiadiazol-4-yl)benzoic acid, coded as SM315 dye. The sensitizing dyes as a light harvester play a very significant role in achieving high PCE for DSSCs, therefore, dyes such as N3, N719, N749, Y123, Z907, JK-303 those absorb as much sunlight as possible is of a great interest, i.e., Y123 dye = 3-[6-[4-[bis(2',4'-dihexyloxybiphenyl-4-yl)amino-]phenyl[-4,4-dihexylcyclopenta-[2,1-b:3,4-b]dithiophene-2-yl]-2-cyanoacrylic acid; Z907 = *cis*-Bis(isothiocyanato)(2,2'-bipyridyl-4,4' -dicarboxylato)(4,4'-di-nonyl-2'-bipyridyl)ruthenium(II); and JK-306= (E)-3-{5'-{4-[bis(2',4'-dihexyloxybiphenyl-4-yl)amino]phenyl}-2,2'-bithiophene-5-yl}-2-cyanoacrylic acid. The many types of dye-sensitizers such as ruthenium dyes, metal-free organic dyes, porphyrin dyes, quantum dots, and perovskites have been explored for DSSCs. Over the past 30 years, significant progress has been made in exploring diverse aspects of DSSC components, including dye-sensitizers, CEs, electrolytes and photoanodes, and many outstanding reviews are available in the literature on this exciting research topic.[5-17] Chemical structures of photosensitizing dyes generally used in evaluating the photovoltaic performance of CEs in DSSCs are also discussed elsewhere. Many different types of Pt-free CE materials have been developed for DSSCs including carbon-based materials such as carbon black, mesoporous carbon, carbon nanotubes (CNTs), carbon fibers, fullerenes, graphene-based materials, metals, transition metal oxides, sulfides, carbides, nitrides, selenides, tellurides, chalcogenides, layered double hydroxides, phosphides and their alloys, conducting polymers such as PEDOT:PSS and other composites  materials.[18-22]

Platinum (Pt) and ITO are the traditional materials used in fabricating different types of solar cell devices. In DSSCs, the Pt deposited on a FTO transparent conductive glass substrate is used as a CE for the reduction of triiodide ($I_3^-$) ions to iodide ($I^-$) ions. In this case, Pt acts as a catalyst for the regeneration of redox couples while FTO acts as an electron collector. Pt is a highly expensive metal which has been identified as one of the most critically important metals for the U.S. economic growth.[23, 24] Among electrocatalysts for DSSCs, Pt shows the best photovoltaic performance due to its high conductivity, chemical stability, and electrochemica4l activity but it is scarce. The current idea is to explore new alternative catalytic materials to replace the conventional Pt CE in DSSCs using easily available, inexpensive, highly electrically conductive, and high electrocatalytic activity materials. Earth-abundant 2D materials that offer high optical transparency and high electrocatalytic acivitiy have been explored as potential candidate materials for CEs (cathode) and for replacing Pt in DSSC devices. During the last decade, graphene-based materials in the form of their thin films, nanosheets, fibers, multilayers, nanoplatelets, quantum dots, nanofoams and their nanocomposites with metals, CNTs, organic polymers, upconversion nanoparticles, titanium dioxide ($TiO_2$), ionic liquids, and halide perovskites, have been extensively investigated as Pt-free CEs for DSSCs[25] and heterojunction solar cells.[26, 27] The merit of graphene-based CEs in DSSC devices includes their high optical transparency, high electrical conductivity, large effective specific surface area, and the flexibility to fabricate them on different types of substrates.

Layered 2D TMDs are graphene analogs, therefore, research activities have been strongly diverted towards this new class of low-cost 2D materials. Among 2D TMDs, transition-metal disulfides and diselenides such as $MoS_2$, $MoSe_2$, $WS_2$, $TiS_2$, $NbSe_2$, $TaSe_2$, $NiSe_2$, $FeSe_2$, $CoSe_2$, $SnS_2$, $Bi_2Se_3$ and other TMD thin films have been investigated as CEs to fabricate Pt-free DSSCs. The CE acts as an electrocatalyst, and is one of the main components of a DSSC device which facilitates the reduction of triiodide ($I_3^-$) ions to iodide ($I^-$) ions in a redox electrolyte for dye generation. In this review, electrochemical and photovoltaic properties of low-cost catalytic CEs developed from earth-abundant TMDs and their composites with carbon, graphene, CNTs, carbon nanofibers, PEDOT:PSS and other materials are discussed. The impact of materials processing and morphology associated with PCEs of DSSCs has also been analyzed. Finally, this review discusses the electrochemical and environmental stability of TMDs-based CEs for DSSCs.

## 2. Two-Dimensional Transition Metal Dichalcogenides (2D TMDs)

Two-dimensional (2D) transition metal dichalcogenides (TMDs), i.e. $MX_2$ where M is a transition metal (Mo, W, Ti, Zr, Hf, V, Nb, Ta, Tc, Re, Pd, Pt) and X is a chalcogen (S, Se, Te), such as $MoS_2$, $WS_2$, $MoSe_2$, $WSe_2$, form 2D layered structures and are abundantly available in nature. In a $MX_2$ monolayer structure (X-M-X), an atomic layer of a transition metal (M) is sandwiched between two chalcogen (X) atomic layers, where transition metal atoms (M) such as Mo and W are covalently bonded with chalcogen atoms (X) such as S, Se, Te. Each layer is 6~7 Å thick in the TMD layered structures. Weak Van der Waals interactions occur between two adjacent $MX_2$ layers of TMDs, whereas the intra-layer M–X bonds are covalent.[28-31] From an electrical point of view, $MoS_2$, $MoSe_2$, $WS_2$ and $WSe_2$ are semiconductors, $WTe_2$ and $TiSe_2$ are semimetals, $VSe_2$ and $NbS_2$ are metals, while $NbSe_2$ and $TaS_2$ are superconductors in their bulk crystalline forms.[26, 30] Geim and Grigorieva[32] created a library of 2D materials (Table 1) where the 2D materials were classified into three different classes. Their graphene family includes graphene-based materials hBN, and BCN; the 2D chalcogenides include a large family of TMDs having insulator to semiconducting to metallic behavior; and the 2D oxides include layered oxides, perovskite-based materials, hydroxides, and others. The 2D TMDs include $MoS_2$, $WS_2$, $VS_2$, $NbS_2$, $ZrS_2$, $TiS_2$, $NbSe_2$, $WSe_2$, $MoSe_2$, $ZrSe_2$, $WTe_2$, and $MoTe_2$, and layered semiconductor chalcogenides such as GaSe, GaTe, InSe, $Bi_2Se_3$, and $Bi_2Te_3$. 2D materials represented by different colors denote their stability under ambient conditions. 2D materials among the graphene family have been extensively studied so far. 2D TMDs such as $MoS_2$, $WS_2$, $WSe_2$ and $MoSe_2$ are structurally similar. Among 2D TMDs, W based materials have stronger spin-orbit coupling while Se materials exhibit lower stability.

Research activities into 2D TMDs were intensified after the monolayers of $MoS_2$ showed high carrier mobility of 100 $cm^2$/Vs and on/off current ratios of $>10^8$ because of the interesting bandgap of $MoS_2$.[33] The monolayers of $MoS_2$ are direct semiconductors whereas $MoS_2$ bilayers, trilayers, few-layers are indirect semiconductors. Thickness dependent bandgap energies of 2D TMDs-based semiconductors, including $MoS_2$, $MoSe_2$, $MoTe_2$, $WS_2$, and $WSe_2$ (2H-MX2), were calculated by Yun et al.[34] using the first-principles calculations as shown in Figure 1.[34] As the number of layers in the TMDs is reduced to a single layer, the band curvatures lead to significant changes of effective masses. When a single layer of the TMDs is strained, the direct bandgap turns to an indirect bandgap. It was found that bandgap energy and effective masses are reduced by tensile strain, contrary to the compressive strain that increases both parameters. Also the applied larger tensile stress gives rise to a metallic character. This study emphasized the electronic structures of 2D TMDs.

The photoluminescence and absorption spectra of monolayers of $MoS_2$, $MoSe_2$, $WS_2$, and $WSe_2$ have been measured.[35] TMDs as bulk crystals are indirect-bandgap semiconductors, however they become direct-bandgap semiconductors as their thickness is reduced to monolayers. The bandgap increases as the number of layers decreases. Photoluminescence peaks at 1.84 eV for $MoS_2$, 1.65 eV for $WSe_2$ and 1.56 eV for $MoSe_2$ have been observed.[36] The indirect bandgap at 1.2 eV for bulk crystals of $WSe_2$ was reported,[37] whereas the monolayer showed a photoluminescence peak at 1.65 eV (752 nm). In the case of bulk crystals of $MoSe_2$, the indirect bandgap was measured as 1.1 eV (1.13 µm) while the monolayer exhibited a photoluminescence peak at 1.57 eV (792 nm). The photoluminescence emission peak for monolayer $WS_2$ appeared at 1.97 eV.[38]

| Graphene family | Graphene | hBN 'white graphene' | | BCN | Fluorographene | Graphene oxide |
|---|---|---|---|---|---|---|
| 2D chalcogenides | $MoS_2$, $WS_2$, $MoSe_2$, $WSe_2$ | Semiconducting dichalcogenides: $MoTe_2$, $WTe_2$, $ZrS_2$, $ZrSe_2$ and so on | | Metallic dichalcogenides: $NbSe_2$, $NbS_2$, $TaS_2$, $TiS_2$, $NiSe_2$ and so on | | |
| | | | | Layered semiconductors: GaSe, GaTe, InSe, $Bi_2Se_3$ and so on | | |
| 2D oxides | Micas, BSCCO | $MoO_3$, $WO_3$ | Perovskite-type: $LaNb_2O_7$, $(Ca,Sr)_2Nb_3O_{10}$, $Bi_4Ti_3O_{12}$, $Ca_2Ta_2TiO_{10}$ and so on | | Hydroxides: $Ni(OH)_2$, $Eu(OH)_2$ and so on | |
| | Layered Cu oxides | $TiO_2$, $MnO_2$, $V_2O_5$, $TaO_3$, $RuO_2$ and so on | | | Others | |

**Table 1.** Library of 2D materials. Blue cells represent monolayers of 2D materials showing stability under ambient conditions; green denotes probability of stability of 2D materials in air; pink denotes 2D unstable materials in air, but may be stable under inert atmosphere. Grey cells show 3D compounds easily exfoliated down to monolayers. Reprinted from ref. 32. Copyright 2013, Nature Publishing Group.

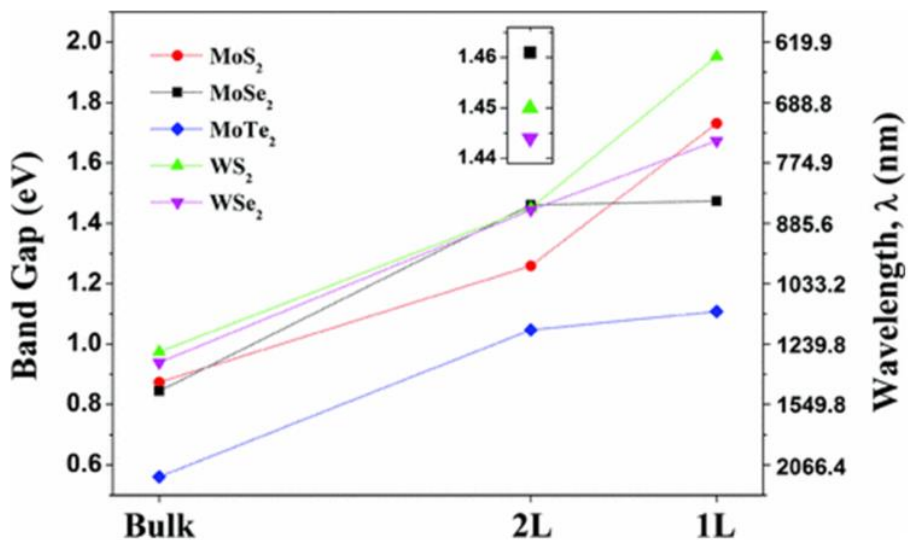

**Figure 1.** Colored lines representing thickness dependence of bandgap energies of 2D TMDs including $MoS_2$, $MoSe_2$, $MoTe_2$, $WS_2$, and $WSe_2$ attain (2H-$MX_2$ semiconductors). Reprinted with permission from ref. 34. Copyright 2012, American Physical Society.

Mechanical exfoliation is the most common approach for peeling off monolayers or a few-layers of TMD materials from their bulk crystals,[39-41] which is a top-down method. A second widely used approach is chemical vapor deposition (CVD), which is a bottom-up method to consecutively deposit desired thickness layers of TMDs.[42-45] Other approaches for preparing layers of TMDs include chemical,[46] lithium intercalation,[47-49] and ultrasonic-assisted liquid exfoliation in organic solvents,[50-54] and salt-assisted liquid-phase exfoliation,[55] all of which can prepare mono-layers to multi-layers of TMDs, such as $MoS_2$, $MoSe_2$, $WS_2$, and $WSe_2$.

Raman scattering methods for single-layer, multi-layer and bulk 2D TMDs, including $MoS_2$, $MoSe_2$, $WS_2$, and $WSe_2$, in terms of phonons with respect to the number of layers, has been reviewed and analyzed.[56, 57] Tonndorf et al.[37] studied photoluminescence and Raman characteristics of the monolayers of $MoS_2$, $MoSe_2$, and $WSe_2$. Figure 2a shows a schematic representation of the four Raman active modes and two Raman inactive modes of TMDs $MX_2$ (M = Mo, W and X = Se, Se) as predicted for the D6h point group.[58] The Raman active modes include three in-plane modes referred as $E_{1g}$, $E^1_{2g}$, and $E^2_{2g}$, and one out-of-plane mode referred as $A_{1g}$. However, experimentally only the two active Raman modes $E^1_{2g}$ and $A_{1g}$ were observed. The active Raman $E^2_{2g}$, mode appears at very low frequencies, whereas the $E_{1g}$ mode is forbidden.[59] A systematic study of the optical absorption of single-layer and few-layers of $MoS_2$ from the 385 nm (3.22 eV) to 1000 nm (1.24 eV) spectral range was conducted by Castellanos-Gomez et al.[60] using a hyperspectral imaging technique. Poly(dimethylsiloxane) (PDMS) was used as a transparent substrate for depositing mechanically exfoliated $MoS_2$ thin films. The optical absorbance spectra of $MoS_2$ flakes consisting of single-layer to six-layer were recorded as a function of different excitation wavelengths. The bandgap of a monolayer $MoS_2$ was measured as 1.85 eV, which decreased with increasing number of $MoS_2$ layers, reaching a bandgap value of 1.35 eV for the bulk $MoS_2$. Figure 2b and c shows Raman spectra for the $E^1_{2g}$ and $A_{1g}$ modes of the exfoliated $MoS_2$ flakes ranging from single-layer to six-layer, and the variation of peak frequency difference ($\Delta$) between the $E^1_{2g}$ and $A_{1g}$ modes as a function of the number of $MoS_2$ layers. The $\Delta$ value increases with the increasing number of layers from 19.8 cm$^{-1}$ for a single-layer to 25 cm$^{-1}$ for bulk $MoS_2$.

Raman spectroscopy for few-layer $MoSe_2$ nanosheets was recorded by Balasingam et al.[61] which showed Raman peaks at 239 cm$^{-1}$ and 287.11 cm$^{-1}$ (Figure 2d) corresponding to the $A_{1g}$ mode (out-of-plane) and $E^1_{2g}$ mode (in-plane) of $MoSe_2$, respectively. For bulk crystal $MoSe_2$, the $A_{1g}$ and $E^1_{2g}$ modes appear at 242 cm$^{-1}$ and 286 cm$^{-1}$, respectively. The red shift in the $A_{1g}$ mode and a blue shift in the $E^1_{2g}$ mode indicate the formation of a few-layered $MoSe_2$ nanosheet. Raman spectra of monolayers, few-layers and bulk $WS_2$ and $WSe_2$ were evaluated by Zhao et al.[62] Thickness dependent Raman $A_{1g}$ and $E^1_{2g}$ modes of 1 to 5 layers, as well as bulk $WS_2$ and WSe$_2$, are shown in (Figure 2e and f). As discussed above, the frequency difference ($\Delta$) can be used to distinguish the number of layers. McCreary et al.[38] reported the E$^1_{2g}$ mode at 357.5 cm$^{-1}$ (in-plane) and A$_{1g}$ mode at 419 cm$^{-1}$ (out-of-plane) with a frequency difference ($\Delta$) of 61.5 cm$^{-1}$ for WS$_2$ monolayer. Photoluminescence emission and Raman scattering of 2D TMDs have been extensively studied and analyzed by several research groups around the world.[63-71]

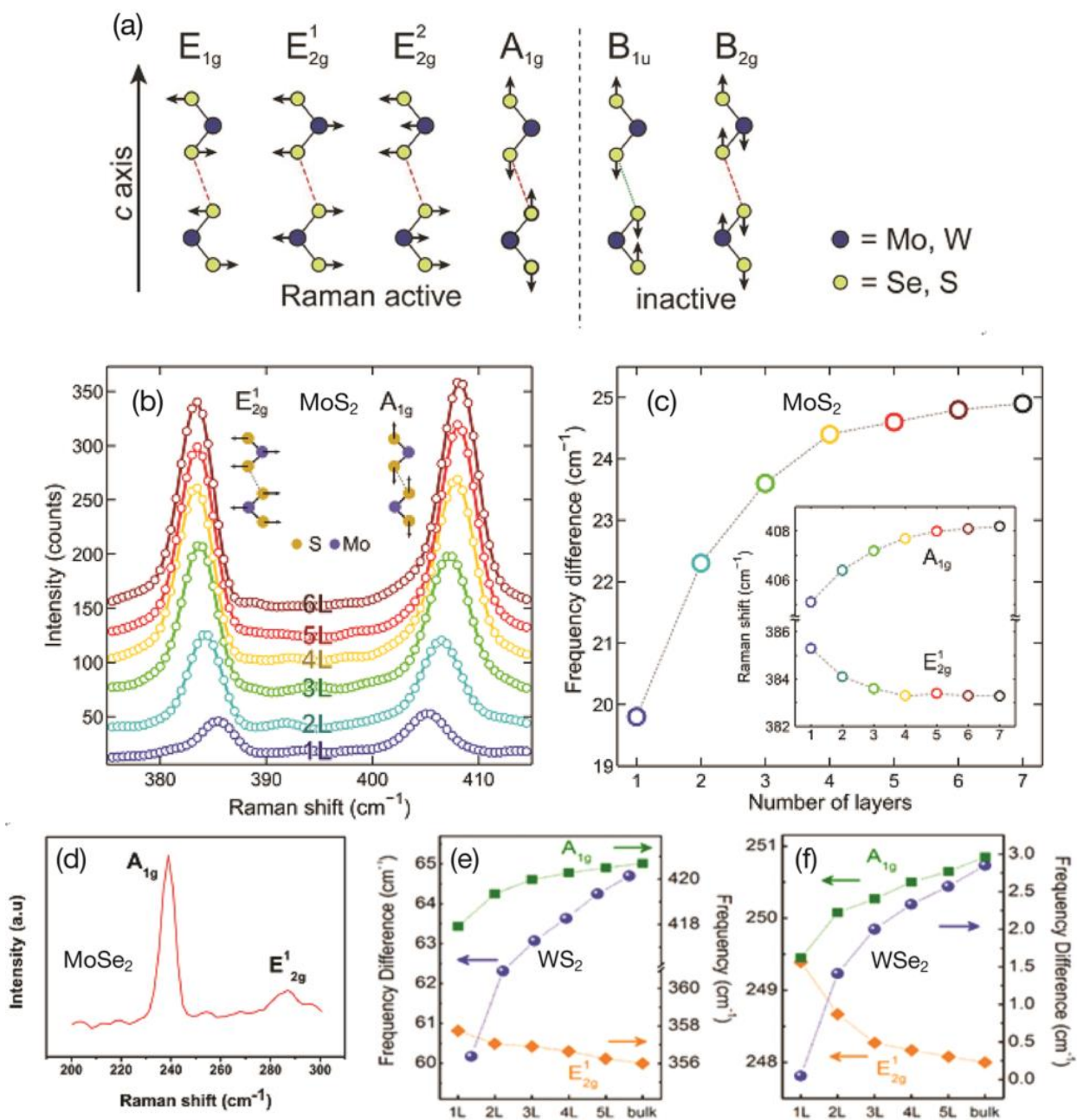

**Figure 2.** (a) Schematic representation of the four Raman active modes and two Raman inactive modes of TMDs $MX_2$ (M = Mo, W and X = Se, Se). Reprinted with permission from ref. 37. Copyright 2013, Optical Society of America. (b) Raman spectra of mechanically exfoliated $MoS_2$ flakes deposited onto a transparent poly(dimethylsiloxane) (PDMS) substrate with different number of layers, single-layer to six-layers of $MoS_2$. (c) Peak frequency difference ($\Delta$) between Raman modes as a function of the number of $MoS_2$ layers. Reprinted with permission from ref. 60. Copyright 2016, Institute of Physics (IOP). (d) Raman spectrum of few-layer $MoSe_2$ nanosheets showing two distinct Raman active modes; $A_{1g}$ and $E^1_{2g}$. Reprinted with permission from ref. 61. Copyright 2015, Dalton Transactuins. (e,f) Thickness dependence of Raman $A_{1g}$

and $E^1_{2g}$ modes of 1 to 5 layers and bulk $WS_2$ and $WSe_2$. Reprinted with permission from ref. 62. Copyright 2013, The Royal Society of Chemistry.

## 3. TMD-Based Counter Electrodes for DSSCs

The new materials used as CEs in DSSC devices are generally characterized by their molecular structure and surface morphology using a wide variety of spectroscopic techniques, including X-ray diffraction (XRD), Raman spectroscopy, X-ray photoelectron spectroscopy (XPS), UV-vis (Ultraviolet-visible) spectrometry, atomic force microscopy (AFM), field-emission scanning electron microscopy (FESEM), high-resolution transmission electron microscopy (HRTEM) and energy dispersive X-ray spectroscopy (EDX). The electrochemical catalytic activity and photovoltaic performance of CEs in DSSCs are evaluated using different methods. Electrocatalytic activity can be evaluated by cyclic voltammetry (CV), rotating disk electrode (RDE) for determining rate constant and effective catalytic surface area, interfacial charge transfer parameters (series resistance and charge-transfer resistance) measurements by electrochemical impedance spectroscopy (EIS) and Tafel polarization plots for electrocatalytic ability, incident photon-to-current conversion efficiency (IPCE) spectra measurements; these measurements of a DSSC device and their electro-catalytic activities are compared with a standard Pt CE. Photovoltaic tests of a DSSC device are performed by measuring Photocurrent density-voltage (*J-V*) characteristic curves of CEs under a simulated solar illumination at 100 mW/cm$^2$ (AM 1.5 G) corresponding to 1-Sun intensity unless specified for other sunlight intensity. The photovoltaic and electrochemical parameters of the DSSC system include short-circuit photocurrent density ($J_{SC}$), open-circuit voltage ($V_{OC}$), fill factor (FF), power conversion efficiency (PCE) which is the solar-to-electricity conversion efficiency ($\eta$), series resistance ($R_s$), charge-transfer resistance ($R_{CT}$) at the CE/electrolyte interface, capacitance, and Nernst diffusion impedance ($Z_N$). When using a new material as a CE catalyst, the electrochemical and photovoltaic properties of a DSSC device are compared with a conventional Pt CE due to its optimal electrocatalytic activity and chemical stability in the electrolyte. The dye-adsorbed $TiO_2$ mesoporous film deposited on a FTO glass substrate is used as a photoanode (working electrode) of the DSSC device. The ruthenium N719 dye has been commonly used for the reduction of triiodide ($I_3^-$) to iodide ($I^-$). DSSCs are assembled with different CEs to be studied, and filled with the same redox electrolyte solution such as triiodide/iodide ($I_3^-/I^-$) redox couple. The CE reduces the triiodide ($I_3^-$) ions to iodide ions ($I^-$). For example, a typical iodide electrolyte solution contains 0.1 M of LiI, 0.05 M of iodine ($I_2$), 0.6 M of 1,2-dimethyl-3-n-propylimidazolium iodide (DMPII), and 0.5 M of 4-*tert*-butylpyridine (TBP) in acetonitrile. All these spectroscopic techniques and electrochemical measurements were performed, which have been discussed throughout this review as needed in order to evaluate the potential of 2D TMDs (such as $MoS_2$, $WS_2$, $TiS_2$, $FeS_2$, $CoS_2$, $NiS_2$, $SnS_2$, $MoSe_2$, $NbSe_2$, $TaSe_2$, $NiSe_2$, $FeSe_2$, $CoSe_2$, $Bi_2Se_3$ and their based hybrids/composites) as CEs for fabricating low-cost Pt-free DSSCs.

### 3.1 $MoS_2$ Counter Electrodes

#### 3.1.1 **Pristine $MoS_2$**

TMDs have been traditionally used as solid lubricants[72-80] but recently have found the potential for applications in the fields of electronics and photonics. Like 2D graphene-based materials, layered transition-metal dichalcogenides $MX_2$ (M= W, Mo, Hf, Nb, Re, Ta; X = S, Se, Te) have become of great interest due to their unique electronic and photonic properties and a very wide range of applications in field-effect transistors (FETs),[81-99] photodetectors,[100-102] light-emitting diodes (LEDs),[103, 104] ferroelectric memories,[105] static random access memory (RAM),[83] high-frequency resonators,[106] supercapacitors,[107, 108] energy conversion and storage devices,[109-112] gas sensors,[113, 114] biosensors,[115, 116] biomedical applications,[117] organic solar cells,[118] and photonics /optoelectronics.[119, 120]

TMDs are one of the 2D graphene analogs which show great potential for both bulk-heterojunction (BHJ) and dye-sensitized solar cells. A review article on atomically thin $MoS_2$ layers used as the electron-transport layer (ETL), hole-transport layer (HTL), interfacial layer, and protective layer in fabricating bulk-heterojunction (BHJ) solar cells has been recently published by Singh et al.[118] Here, the applications of $MoS_2$ thin films as CEs for fabricating Pt-free dye- DSSCs are discussed.

Polytypism in $MoS_2$ has been studied by using Raman spectroscopy.[121] Layered $MoS_2$ exists in 2H, 3R and 1T phases where monolayers are stacked in a different sequence.[122-124] The semiconducting 2H-$MoS_2$ phase of the bulk crystal contains two-layer per unit cell stacked in a hexagonal symmetry, where each Mo atom is coordinated with six sulfur (S) atoms. The 3R-$MoS_2$ phase contains three-layer per unit cell stacked in a rhombohedral symmetry. The metallic 1T phase contains one $MoS_2$ layer per unit cell in tetragonal symmetry with octahedral coordination. A structural phase transition of 2H-$MoS_2$ to 1T-$MoS_2$ (2H→1T) has been reported due to a lithium ion intercalation process.[122-125] The metallic 1T phase of $MoS_2$ exhibits very interesting electronic properties.[126-129] Therefore, Wei et al.[130] used metastable 1T metallic phase $MoS_2$ in fabricating DSSCs. In that work, $MoS_2$ films were deposited on FTO glass substrate by a hydrothermal method, performing reactions at 180 or 200 °C for 24 hours. The $MoS_2$ films prepared at 200 °C showed lumps, while those grown at 180 °C had flower-like structures. High-angle annular dark-field scanning transmission electron microscopy (HAAD-FSTEM) images, Raman spectroscopy, and XPS also showed the formation of 2H and 1T metallic phases of $MoS_2$.

Figure 3 represents the J-V curves of 2H-MoS$_2$ and flower-structured 1T metallic phase MoS$_2$. The flower-structured 1T metallic MoS$_2$ film was grown onto an FTO substrate as a CE of DSSC, which exhibited PCE ($\eta$) of 7.08%, that is a three times higher PCE compared with 2H phase MoS$_2$ ($\eta = 1.72\%$) and comparable to a Pt CE ($\eta = 7.25\%$). Such a large difference in PCE values of 1T and 2H phases occurs because the electrical conductivity of 1T-MoS$_2$ is $10^7$ times larger compared with 2H-MoS$_2$, which gives rise to a higher electrocatalytic activity for the 1T phase than that of the 2H phase. The 1T-MoS$_2$ CE also demonstrated lower charge-transfer resistance ($R_{CT}$). The IPCE curves and CV measurements also showed a better electrocatalytic activity of 1T-MoS$_2$ CE in DSSC than that of 2H-MoS$_2$.

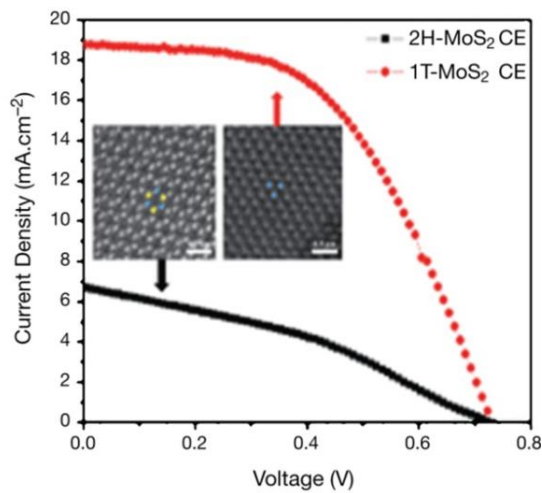

**Figure 3.** Photocurrent density-voltage (J-V) curves of 2H-type MoS$_2$ and flower-structured 1T metallic phase MoS$_2$. Insert shows high-angle annular dark-field scanning transmission electron microscopy (HAAD-FSTEM) confirming 2H and 1T metallic phase of MoS$_2$. Reprinted with permission from ref. 130. Copyright 2016, Royal Society of Chemistry.

A very interesting study was conducted by Infant et al.[131] who developed CE materials for DSSCs by vertically oriented MoS$_2$ on an FTO substrate, in order to increase the reflectivity of MoS$_2$ CE. Figure 4 shows high resolution SEM images of CVD-deposited MoS$_2$ thin films, the reflectivity of $MoS_2$ CE measured by UV–vis spectrophotometer, and CV measurements using standard hydrogen electrode (SHE) as a reference electrode. The high quality thin films of $MoS_2$ were obtained at the optimum conditions of 600 $^o$C with a 15 minute reaction time at a flow rate of 50 sccm. The FTO substrate is damaged if temperature is increased above 600 $^o$C and the reaction time is over 15 minutes, leading to excessive deposition of sulfur on the surface. The layered $MoS_2$ thin films are polycrystalline and have 0.18-0.27 nm spacing as evidenced by the TEM image and SEAD pattern. The CVD-prepared $MoS_2$ thin films are vertically oriented on the FTO substrate, which yields more active sites and eventually enhance the reflectivity so that more photons are absorbed, and also created active edge sites facilitating the generation of the $I^-/I_3^-$ redox couple. The reflectance of vertically inclined $MoS_2$ films on the FTO substrate was measured by a UV–vis spectrophotometer between 350-800 nm wavelength for reaction temperatures varying from 400 to 700 $^o$C, a reaction time ranging from 5 to 30 minutes, and different flow rates of Ar gas (Figure 4). The maximum reflectance of 38% was observed for a vertically inclined $MoS_2$ thin film under optimized conditions, due to high crystallinity, and the inclined angle of 26$^o$ was estimated that supports the reflectivity of photons to dye molecules. The reflectivity of 38% was observed for $MoS_2$ thin films prepared at 50 sccm due to its crystallinity and higher inclination angle, which contributes to more absorption photons, and hence leads to a higher PCE of 7.50% compared to PCE of 7.38% for 150 sccm with 25% reflectivity, due to smaller angle of $MoS_2$ inclination. This approach yielded a PCE of 7.50% for the $MoS_2$ CE, exceeding the PCE of Pt based CE ($\eta$ = 7.28%). The $MoS_2$ CE shows a higher *Jsc* value of 15.2 mA/cm$^2$, and is higher than Pt CE (*Jsc* = 14.6 mA/cm$^2$) due to high reflectivity. The $R_{CT}$ of $MoS_2$ CE was lower than Pt CE, indicating the higher electro-catalytic activity of vertically inclined $MoS_2$ film on the FTO substrate was due to more active edge sites, which gives rise to enhanced electrocatalytic activity of the $MoS_2$ CE. Two redox peaks are observed in the CV curves of Pt and $MoS_2$ CEs, one corresponding to the reduction of $I_3^-$, which is a cathodic peak ($V_{pc}$), and the other corresponding to the oxidation of $I^-$, which is an anodic peak ($V_{pa}$). The value of anodic peak to cathodic peak separation ($V_{pp}$) was found to be less for the $MoS_2$ CE (0.456 V) than that of Pt (0.484 V), which also indicates a better electrocatalytic activity of $MoS_2$ CE due to the presence of active edge sites, as evidenced in the SEM images.

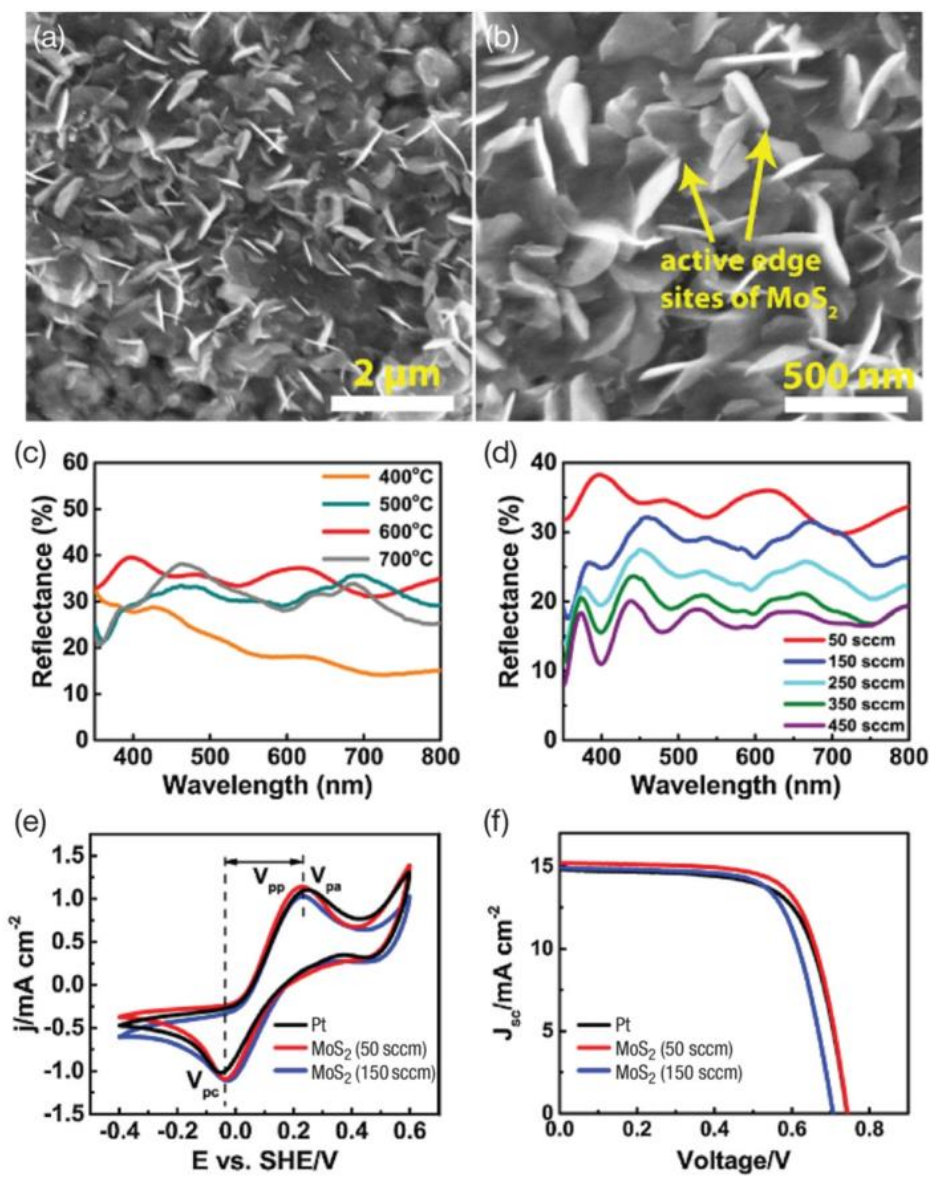

**Figure 4.** SEM (a) and high resolution SEM (b) of CVD-deposited $MoS_2$ thin films. The highest quality thin films of $MoS_2$ were prepared at 600 $^o$C with 15 minute reaction time at the Ar flow rate of 50 sccm. (c) Optical properties of $MoS_2$ thin films between 350-800 nm range prepared at different temperature. (d) Flow rates of Ar gas between 50 to 450 sccm for preparing $MoS_2$ thin films by CVD technique. (e) Cyclic voltammetry (CV) plots using standard hydrogen electrode (SHE) as a reference electrode and (f) Photocurrent density-voltage (J-V) curves of $MoS_2$ and Pt based CEs prepared at 50 and 150 sccm flow rates of Ar gas. Reprinted with permission from ref. 131. Copyright 2016, Elsevier.

Vertical $MoS_2$ nanosheets on different substrates using CVD and $CS_2$ as a sulfur precursor have been developed.[132] The DSSC CEs with vertical $MoS_2$ nanosheets showed a comparable electro-catalytic activity to Pt CE for the triiodide ($I_3^-$) reduction, resulting from large specific surface areas and more active edges. Li et al.[133] prepared molybdenum disulfide-based CEs for DSSCs with different morphologies (multilayers, few-layers and nanoparticles) using the thermal decomposition method. The X-ray diffraction and transmission electron microscopy showed edge area to basal-plane ratio in the following order: $MoS_2$ nanoparticles > multilayered $MoS_2$ > few-layered $MoS_2$. A similar order was observed for the PCE values with corresponding CEs-based DSSCs. The $MoS_2$ nanoparticles-based CE had the minimum $R_{CT}$, while the few-layered $MoS_2$ based CE had the maximum, as measured by EIS. The active sites of $MoS_2$ responsible for the reduction of triiodide lie on the edges of layered materials, instead of their basal planes. $MoS_2$ nanoparticle CE showed the highest PCE value of 5.41%, compared with 6.58% of Pt CE. A novel approach for improving PCE of $MoS_2$ CE based DSSCs has been developed,[134] where electrocatalytic activity was enhanced by artificially generating active edge sites on $MoS_2$ atomic layers by hole patterning. The PCE of the DSSC increased from 2% to 5.8% after applying the hole patterning approach. Al-Mamun et al.[135] deposited $MoS_2$ nanoscale thin films onto FTO substrates using a low temperature one-pot hydrazine assisted hydrothermal method. Both the hydrothermal reaction temperature as well as the different molar ratio of reaction precursors was found to impact the structure and performance of $MoS_2$ films used as CEs for DSSCs. The $MoS_2$ thin films having surface exposed layered nanosheets were obtained by the hydrothermal process with a molar ratio of reaction precursors as 1:28 of $(NH_4)_6Mo_7O_{24}\cdot 4H_2O$ and $NH_2CSNH_2$ (thiourea) at 150 ℃ for 24 hours, referred to as MS-150-28. The molar ratio of $(NH_4)_6Mo_7O_{24}\cdot 4H_2O$ and $NH_2CSNH_2$ was fixed at 1:7, 1:14, 1:28 and 1:42 and the hydrothermal temperature was maintained at either 120, 150, 180 or 210 ℃. The DSSCs having $MoS_2$ CEs fabricated using different molar ratios of reaction precursors at a temperature of 150 ℃ (referred to as MS-150-7, MS-150-14, MS-150-28 and MS-150-42) exhibited PCEs of 3.70, 4.97, 7.41 and 4.96%, respectively. The DSSCs with MS-120-28, MS-180-28, and MS-210-28 CEs showed PCEs of 5.52, 7.15 and 5.47%, respectively. The $MoS_2$ film based CEs showed a PCE of 7.41%, higher than the Pt electrode based DSSCs ($\eta = 7.13\%$) using $TiO_2$ photoanodes sensitized by N719 dye.

Pulse electrochemical deposited thin films of molybdenum sulfide ($MoS_x$) on indium tin oxide/poly(ethylene naphthalate) (ITO/PEN) substrates have been studied as flexible CEs for DSSC, and these showed a PCE of 4.39% for the triiodide ($I_3^-$) reduction.[136] The nanostructured $MoS_2$ thin films developed by a low-temperature thermally reduced technique on a FTO substrate have also been used for DSSCs.[137] $MoS_2$ thin film annealed at 300 ℃ were also used as CEs for DSSCs, which showed a PCE of 6.351 %, slightly lower than the Pt reference CE ($\eta = $ 6.929 %). The performance of DSSCs was impacted by the molar ratio of reaction precursors and the temperature of thermal reaction. Thermally reduced (TR) $MoS_2$ thin film annealed at 250 ℃ showed PCE of 1.917%, while those annealed at 350 ℃ showed PCE of 3.479%. The 300 ℃ annealed TR-$MoS_2$ CE also has larger exchange current density than those of 250 ℃ and 350 ℃ annealed TR-$MoS_2$ CEs and comparable with thermally deposited (TD) Pt CE. The $R_{CT}$ values that correspond to the charge-transfer resistance at the electrolyte–electrode interface were 14.98 $\Omega$ cm$^2$ for TD-Pt CE and 30.98, 141.41 $\Omega$.cm$^2$ for TR-MoS$_2$ CEs annealed at 300 °C and 350 °C, respectively. TR-MoS$_2$ CEs annealed at 250 °C has no $R_{CT}$ value being too large. The annealing temperature of 300 °C generates much larger active area, providing the highest electrocatalytic activity for the reduction of I$_3^-$, while 350 °C annealing decreases the active sites of the edge-planes in MoS$_2$. A PCE of 7.01% was achieved for DSSCs using pristine MoS$_2$ as a CE, chemically synthesized by low-temperature wet-chemical process, which has a comparable PCE of 7.31% for DSSCs with Pt CEs.[138] The R$_s$ and R$_{CT}$ values of 23.51 and 18.50 $\Omega$.cm$^2$, respectively, for the MoS$_2$ CE were lower than those of Pt CE (26.73 and 22.88 $\Omega$.cm$^2$), suggesting better electro-catalytic activity of MoS$_2$ for the reduction of triiodide (I$_3^-$).

   A correlation between the electrical conductivity of the CE, PCE, and the crystallinity of MoS$_2$ was also demonstrated.[139] The XRD, XPS, EIS and Hall measurements established a link between the PCE, carrier concentration, mobility, and *Jsc* value. The DSSCs having pristine (non-annealed), vacuum-annealed and N$_2$-annealed MoS$_2$ CEs showed PCE values of 1.0, 1.7 and 0.8%, respectively. Thermal annealing in vacuum was found to reduce the over-potential that leads to an increased *Jsc* value of 7.95 mA/cm$^2$ due to high MoS$_2$ crystallinity, whereas the N$_2$-annealing of MoS$_2$ CEs increases over-potential, which gives rise to lower *Jsc* value of 4.35 mA/cm$^2$ due to the poor crystallinity of MoS$_2$. Interestingly, the electrical conductivity of MoS$_2$ CEs follows the order:N$_2$-annealed > vacuum-annealed > non-annealed MoS$_2$. This indicates that the PCE of the DSSCs is influenced by the over-potential that involves an electron transferring from the MoS$_2$ CE to the electrolyte, instead of the electrical conductivity of the CE. Antonelou et al.[140] reported the growth of monolayer and few-layer MoS$_2$ films by the sulfurization of molybdenum (Mo) foils at 800 °C. The few-layer thick MoS$_2$ films were used as CEs for DSSCs for the reduction of $I_3^-$ to $I^-$. The electrocatalytic activity of $MoS_2$ CE on flexible Mo substrates depends upon the number of monolayers in the DSSC. DSSCs having the $MoS_2$/Mo CE yield a PCE of 8.4%, close to Pt/FTO-based DSSCs (PCE of 8.7%). Stability of a three-monolayer thick $MoS_2$ CE was studied for 100 consecutive cycles, where no degradation of the peak current density was noticed for 100 repeated cycles, confirming long term electrochemical stability in an electrolyte solution. $MoS_2$ layers with 1-2 nm thickness showed long term chemical stability of the DSSC device for the electrolyte solution comparable to Pt CE. The increased number of active sites due to a grainy surface texture of Mo foil leads to the higher electro-catalytic activity of $MoS_2$ films.

A very interesting comparison was made by Wu et al.[141] for chemically synthesized $MoS_2$ and $WS_2$ as CEs in the reduction of $I_3^-$ to $I^-$ and disulfide/thiolate ($T_2$/$T^-$) based DSSCs. The $R_{CT}$ values of $0.5\ \Omega\cdot cm^2$ for $MoS_2$ and $0.3\ \Omega\cdot cm^2$ for $WS_2$, respectively, compared with $R_{CT}$ of $3.0\ \Omega\cdot cm^2$ for Pt CE, indicates that both $MoS_2$ and $WS_2$ are as effective as CEs as standard Pt for triiodide ($I_3^-$) reduction in DSSCs. The high FFs of 73% for $MoS_2$ CEs and 70% for $WS_2$ CEs also confirm high electro-catalytic activities for the reduction of triiodide ($I_3^-$) to iodide ($I^-$). Therefore, high PCE values of 7.59% for $MoS_2$ and 7.73% for $WS_2$ were observed, which are comparable to the PCE of 7.64% for Pt CEs in DSSCs under simulated AM 1.5 illumination. The $Z_N$ of $>100\ \Omega$ for triiodide ($I_3^-$) reduction on the sulfide electrodes was found to be higher compared with a $Z_N$ of $9.5\ \Omega$ on the Pt CE. The photocurrent density voltage (J–V) curves of the DSSCs having $MoS_2$, $WS_2$, and Pt based CEs for the triiodide/iodide ($I_3^-$/$I^-$) redox couple and disulfide/thiolate ($T_2$/$T^-$) redox couple are shown in (Figure 5). The $TiO_2$ film photoanode was obtained after pre-heating at 80 °C and immersing in a $5 \times 10^{-4}$ M solution of N719 dye in acetonitrile/*tert*-butyl alcohol for 20 hours. The triiodide/iodide ($I_3^-$/$I^-$) electrolyte is made of 0.06 M of lithium iodide (LiI), 0.03 M of $I_2$, 0.5 M of 4-*tert*-butyl pyridine, 0.6 M of 1-butyl-3-methylimidazolium iodide, and 0.1 M of guanidiniumthiocyanate in acetonitrile. The 5-mercapto-1-methyltetrazole di-5-(1-methyltetrazole) disulfide/$N$-tetramethylammonium salt ($^+NMe_4T$-) ($T_2/T^-$) electrolyte consists of 0.4 M of $^+NMe_4T^-$, 0.05M of $LiClO_4$, 0.4 M of di-5-(1-methyltetrazole) disulfide ($T_2$,), and 0.5 M 4-*tert*-butylpyridine in acetonitrile and ethylene carbonate solution. For both DSSCs, a spray-coating technique was used for fabricating $MoS_2$ and $WS_2$ CEs. The J–V curves of the DSSCs having sulfide CEs and $T_2/T^-$ electrolyte show PCE values of 4.97% for $MoS_2$, 5.24% for $WS_2$, and 3.70% for Pt CE. The PCE values were increased 36% for $MoS_2$ and 41% for $WS_2$ compared to the Pt CE, showing that DSSCs having $MoS_2$ and $WS_2$ are better than that of the Pt CE for the $T_2/T^-$ redox couple.

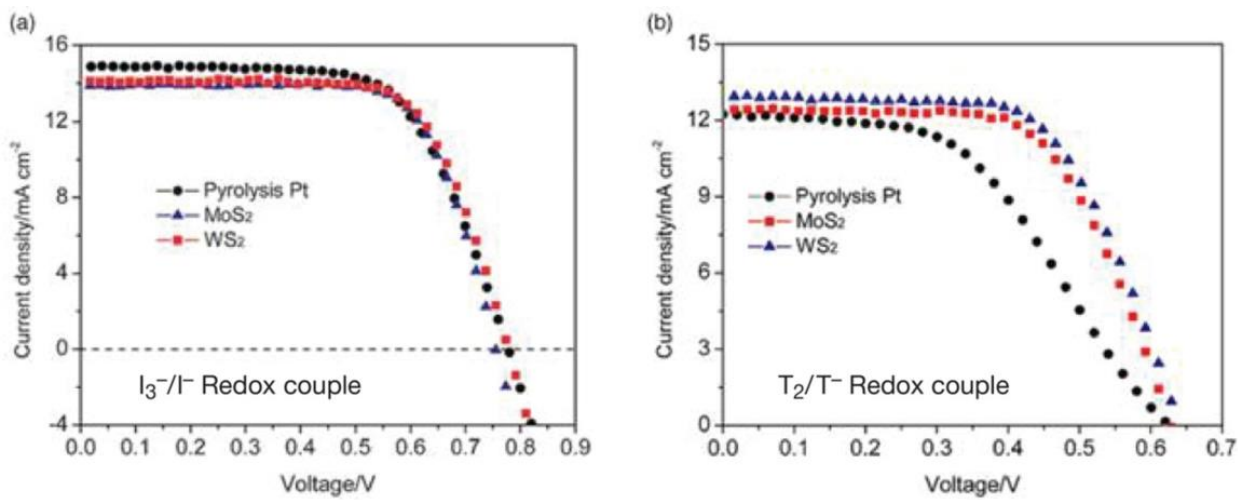

**Figure 5.** (a) Photocurrent density–voltage (J–V) curves of the DSSCs having $MoS_2$, $WS_2$, and Pt counter electrodes for triiodide to iodide ($I_3^-/I^-$) reduction. (b) *J-V* curves of the DSSCs using $MoS_2$, $WS_2$ and Pt CEs for redox couple of disulfide/thiolate ($T_2/T^-$). Reprinted with permission from ref. 141. Copyright 2011, The Royal Society of Chemistry/Owner Societies.

The interfaced exfoliated $MoS_2$ thin films with different porphyrin molecules, where the $MoS_2$ was functionalized with zinc(II) porphyrin (ZnPP), showed a 10-fold increase in photocurrent compared to $MoS_2$ films.[142] Exfoliated ultrathin porous $MoS_2$ sheets prepared by a tetraethylorthosilicate (TEOS)-assisted hydrothermal method were used as CEs in DSSCs.[143] The cyclic voltammograms and electrochemical impedance spectroscopy showed low $R_{CT}$ and high electro-catalytic activity for porous $MoS_2$ sheets CEs in DSSCs, with a PCE of 6.35%, slightly better than that of Pt CEs ($\eta$ = 6.19%) under similar experimental conditions. A CE created by spin-coating of $MoS_2$ nanosheets followed by thermal treatment was also prepared.[144] DSSCs having $MoS_2$ nanosheets thermally treated at 100 °C showed a PCE value of 7.35%, comparable to conventional a Pt CE (7.53%). When $MoS_2$ nanosheets were thermally treated at 300 °C, the PCE value decreased significantly due to the transformation of $MoS_2$ to $MoO_3$. $MoS_2$ films deposited on FTO glass using an RF sputtering method were also used as CEs for $TiO_2$-based DSSCs.[145] CV, EIS, and Tafel polarization curve measurements conducted on the $MoS_2$ CE showed high electrocatalytic activity, low charge-transfer resistance as well as the fast reaction kinetics for triiodide ($I_3^-$) reduction. The $MoS_2$ CE prepared after 5 minutes of sputtering showed a PCE of 6.0 %, comparable to Pt CEs ($\eta$ = 6.6 %) in DSCCs. The PCE of DSSCs having $MoS_2$ CEs sputtered for 1, 3, 5 and 7 minutes were 5.7, 5.8, 6.0, and 5.2%, respectively. $MoS_2$ CEs were also developed by synthesizing $MoS_2$ films at 70 °C using molybdenum(V) chloride and thioacetamide, followed by near-infrared laser-sintering for 1 minute to enhance crystallinity and interconnectivity between the $MoS_2$ particles.[146] The DSSC with laser-sintered $MoS_2$ CE exhibited a PCE of 7.19%, much higher than heat-sintered $MoS_2$ CE (5.96%) and comparable with a Pt CE ($\eta$ = 7.42%). The laser-sintered $MoS_2$ CE offers superior electrocatalytic activity for the triiodide ($I_3^-$) to iodide ($I^-$) redox couple. Also, a solution-phase process was used to grow $MoS_2$ nanofilms on FTO glass as a CE for a DSSC, which showed a PCE of 8.3%.[147] Finally, exfoliated and annealed $MoS_{2x}Se_2(1-x)$ as well as exfoliated-$MoS_2$ films were used as CEs.[148]

The thickness and size of exfoliated $MoS_2$ nanosheets were 0.9 by 1.2 nm and 0.2 by 2 μm, corresponding to a single-layer. The $MoS_2$ based CE showed a PCE of 6%, compared to 5.1% for the annealed $MoS_2$ films.

### 3.1.2 $MoS_2$/Graphene Composites

2D TMDs based composites have been extensively investigated.[149, 150] Like graphene, $MoS_2$ can be mechanically exfoliated into nanosheets from its bulk crystals and used for studying electronics and photonic properties. These 2D layered materials can be combined into a single hybrid structure, so that both $MoS_2$ and graphene components can synergistically enhance photovoltaic properties. The nanocomposites of $MoS_2$ and graphene have been investigated as CEs for DSSCs.

To fabricate a Pt-free DSCC, Lin et al.[151] used a $MoS_2$/graphene nanosheet composite as a CE and nanocrystalline $TiO_2$ as a photoanode. The redox electrolyte solution in the DSSC was made of 1 M 1,3-dimethylimidazolium iodide, 0.15 M iodine, 0.5 M 4-*tert*-butylpyridine and 0.1 M guanidine thiocyanate in 3-methoxypropionitrile. The $MoS_2$/graphene nanosheet based CE showed a PCE of 5.81%, in comparison to a PCE of 6.24% for the conventional sputtered Pt CE. Yue et al.[152] used a $MoS_2$/graphene flake composite film as the CEs for DSSCs with N719 dye for the reduction of triiodide ($I_3^-$) to iodide ($I^-$). The $MoS_2$ powder (ARCOS, 99%) and 8 nm flakes of multi-layer graphene nanopowder (Uni-Onward Corp., 99.5%) were mixed in specific ratios to prepare the composites. The electrocatalytic activity increased after adding graphene flakes to the $MoS_2$ film. Charge-transfer resistances ($R_{CT}$) of 3.98 $\Omega.cm^2$ for graphene, 2.71 $\Omega.cm^2$ for $MoS_2$, 2.09 $\Omega.cm^2$ for $MoS_2$/graphene, and 2.01 $\Omega.cm^2$ for Pt CEs were measured. The current density of the $MoS_2$/graphene composite CE was recorded to be higher than that of the MoS$_2$, graphene and Pt CEs. The J–V characteristics of the DSSCs having MoS$_2$/graphene as CEs and ranging in thickness from single-layer to six-layer were investigated and compared with conventional Pt CEs. The thickness of the MoS$_2$/graphene layers affected the PCE of the DSSCs, and *Jsc* and *Voc* increased with increasing number of plaster layers from single to 3-layer, and, thereafter, a decrease was noticed for 4-layers and 5-layers. The effect of graphene content was also studied, where, the R$_{CT}$ was found to decrease for the MoS$_2$/graphene CE, from 0.5 wt.% to 1.5 wt.% of graphene content. The PCE of the MoS$_2$/graphene composite CE having 1.5 wt.% graphene was 5.98%, compared to the PCE of 6.23% for Pt CE. Yu et al.[153] used MoS$_2$ nanosheets and graphene composites for fabricating CEs for the triiodide (I$_3^-$) reduction. Graphene thin films were prepared by the chemical vapor deposition (CVD) technique in combination with a hydrothermal process. MoS$_2$ nanosheets with 210 nm thickness were in-situ grown on FTO glass substrate, and uniformly dispersed on the surface of a graphene film. Figure 6 shows the morphology of the synthesized graphene-MoS$_2$ composites, using field-emission scanning electron microscopy (FESEM) and high-resolution transmission electron microscopy (HRTEM) with different magnifications. The thickness of the graphene film was found to be 2.48 nm by atomic force microscopy, which corresponds to seven layers of graphene. The top-view FESEM images of graphene-MoS$_2$ hybrids showed fully covered graphene film with 5 to 20 nm thick MoS$_2$ nanosheets. The nucleation and growth of MoS$_2$ nanosheets depends upon the graphene film. Graphene films play an active role for higher electrical conductivity by speeding up the charge transfer process and generating active sites for dispersion and integration of MoS$_2$ nanosheets. The presence of MoS$_2$ nanosheets increases the electrode-electrolyte contact area, and therefore helps in improving the electrocatalytic activity. The low R$_{CT}$ of 1.5 $\Omega$.cm$^2$ for MoS$_2$/graphene CE, 1.7 $\Omega$.cm$^2$ for MoS$_2$, 1.70 $\Omega$.cm$^2$ for graphene, and a high R$_{CT}$ of 2.1 $\Omega$.cm$^2$ for a Pt CE, indicates better interaction and contact formation of $MoS_2$ nanosheets and graphene film with fluorine doped tin oxide (FTO) substrates and a fast charge transfer. The FF of graphene is 24% compared to a FF of 65% for $MoS_2$, however when both materials are combined into a composite system, the FF raises to 68%. Therefore, $MoS_2$ nanosheet/graphene composite CEs resulted in a higher PCE of 7.1% because of the synergetic interactions between graphene and $MoS_2$, compared to the low PCE values of 2.8% for graphene and 5.6% for $MoS_2$, and a comparable PCE of 7.4% for Pt reference CEs. $MoS_2$/graphene based CEs in a DSSC offer higher electrocatalytic activity for triiodide ($I_3^-$) reduction induced by the synergetic interactions. Figure 7 represents the $J$–$V$ characteristics of DSSCs having graphene-$MoS_2$ as CEs with increasing number of layers[152] and graphene-$MoS_2$ nanosheet based CEs for the triiodide ($I_3^-$) reduction.[153]

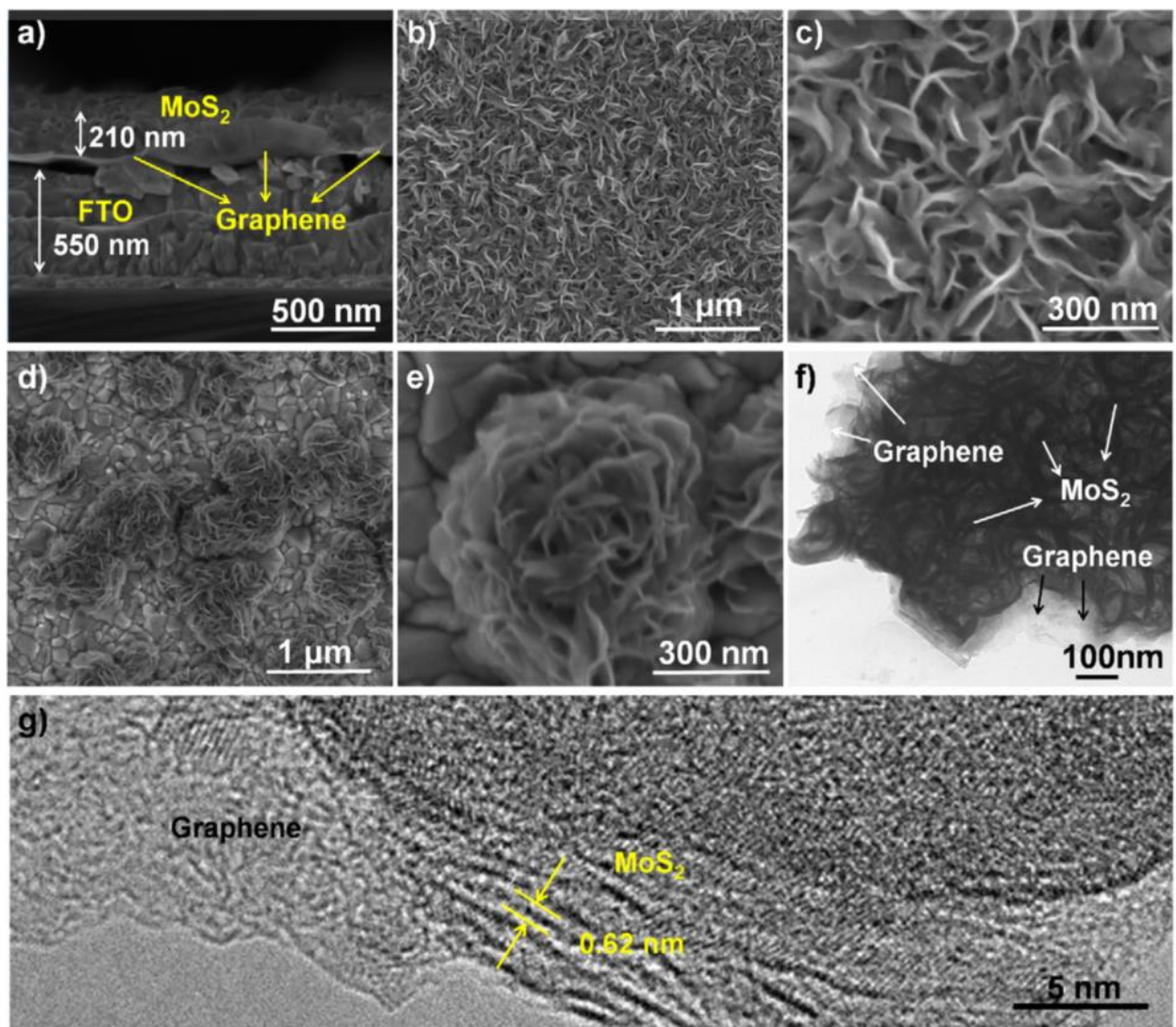

**Figure 6.** Field-emission scanning electron microscopy (FESEM) images of graphene-$MoS_2$ composites with different magnifications; (a) side-view, (b) and (c) top-view. (d) and (e) top-view FESEM images of flower-like $MoS_2$ clusters without graphene film. (f) and (g) high-resolution transmission electron microscopy (HRTEM) images of as-prepared graphene-$MoS_2$ hybrids. Reprinted with permission from ref. 153. Copyright 2016, Elsevier.

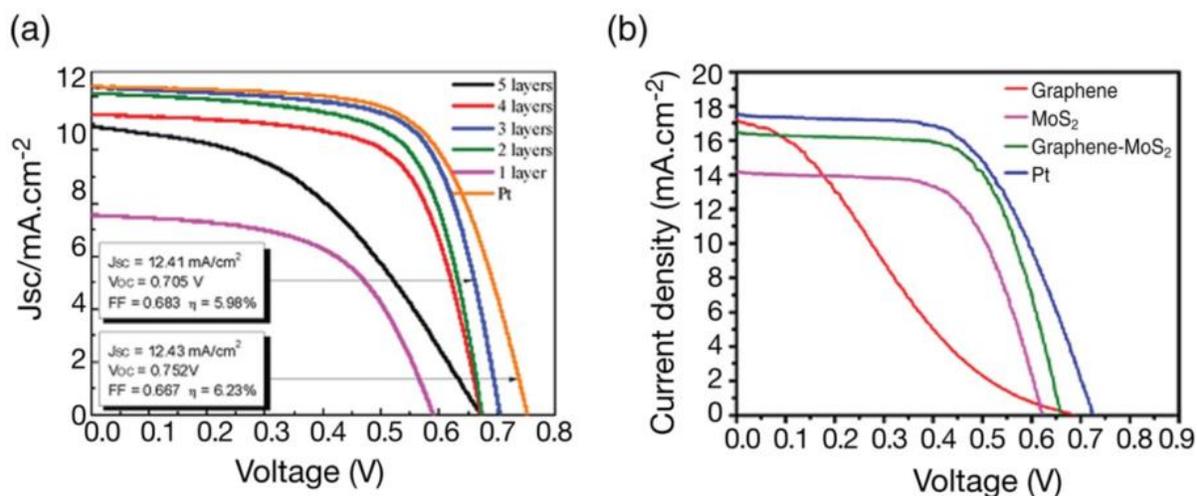

**Figure 7.** (a) J–V characteristics of DSSCs having MoS₂/graphene as counter electrodes with different thickness ranging from single-layer to six-layer and a comparison with sputtered Pt counter electrode (PEC of 6.23%). Reprinted with permission from ref. 152. Copyright 2012, Elsevier. (b) J-V curves of DSSCs based on Pt, graphene/MoS₂, MoS₂, and graphene CEs. Reprinted with permission from ref. 153. Copyright 2016, Elsevier.

MoS₂ nanosheets/graphene electrodes were also studied by Lynch et al.[154] The PCE of 95% of the Pt electrode was achieved after adding 10 wt.% MoS₂ nanosheets to a graphene film CE. This again confirms that the MoS₂ nanosheet/graphene composite ECs have higher catalytic activity than graphene CEs, though the graphene nanosheets contribute to higher electrical conductivity in the composite. An electrochemical strategy[155] was used for preparing MoS₂/graphene composite as CEs of DSSCs, which included electro-deposition and electro-reduction of graphene oxide, and thereafter electro-deposition of MoS₂ on reduced graphene oxide (GO) layers. The MoS₂/graphene composites were characterized by SEM, TEM and Raman spectroscopy. The MoS₂/graphene CEs based DSCCs exhibited a PCE of 8.01%, comparable to a PCE of 8.21% for the Pt CE. In another study, nanocomposites of MoS₂ and nitrogen-doped graphene oxide (N-GO) were used as a CE for DSSCs.[156] The MoS₂/N-GO nanocomposites were characterized by HRTEM, XPS, and Raman spectroscopy, and their electrochemical properties were evaluated by EIS, CV, and Tafel measurements. The MoS₂/N-GO nanocomposite formation offered high specific surface area of N-GO and many edge sites of $MoS_2$. The $MoS_2$/N-GO nanocomposite based CE exhibited a PCE of 5.95 %, lower than the standard Pt CE (PCE of 6.43 %).

Composite films of $MoS_2$ with nitrogen-doped graphene (N-graphene) were prepared using a drop-coating method and used as a CE of DSSCs.[157] The N-graphene supported an increase in electrical conductivity, whereas $MoS_2$ increased the electrocatalytic activity in the composite thin film. The N-graphene/$MoS_2$ composite film showed a PCE of 7.82%, lower than the Pt CE ($\eta$ = 8.25%) based DSSC. The electrocatalytic capability of N-graphene/$MoS_2$ composite films for the triiodide ($I_3^-$) reduction was much higher compared with pristine N-graphene and $MoS_2$ thin films, as studied using CV, RDE, the Tafel polarization curve, and EIS. The graphene flakes (GF) into a nanosheet-like $MoS_2$ matrix were dispersed using an *in-situ* hydrothermal method, and used a $MoS_2$/GF hybrid as a CE to fabricate Pt-free DSSCs.[158] The incorporation of GFs into the $MoS_2$ matrix was confirmed using scanning electron microscopy (SEM), transmission electron microscopy (TEM), XRD, and Raman spectroscopy. The electrochemical measurements showed improvement in the electrocatalytic activity after the GFs were integrated into the $MoS_2$ matrix, where the hybrid containing 1.5 wt.% of graphene flakes exhibited the highest electrocatalytic activity. The DSSC with the $MoS_2$/GF hybrid CE showed a PCE of 6.07%, which was 95% of the Pt CE ($\eta$ = 6.41%).

### 3.1.3 $MoS_2$/Carbon Nanotubes Composites

Carbon nanotubes (CNTs) are one of the most interesting materials because they offer high optical transparency, high electrical conductivity, high mechanical strength, and high thermal stability, and therefore CNTs have been studied as CEs for DSSCs. It is of great interest to combine the unique properties of both $MoS_2$ and CNTs into a single hybrid system which could synergistically enhance the electrocatalytic activity of a DSSC system. Yue et al.[159] used a flower-like structure of $MoS_2$/single-wall carbon nanotubes ($MoS_2$/SWCNTs) as CE catalyst for DSSCs, synthesized using a glucose and PEDOT:PSS assisted (G-P-A) hydrothermal process. The flower-like structure and surface morphology of the (G-P-A) $MoS_2$/SWCNTs was confirmed by SEM and TEM techniques. The DSSC having (G-P-A) $MoS_2$/SWCNTs CE exhibits a lower $R_{CT}$ of 1.46 $\Omega \cdot cm^2$ compared to the $R_{CT}$ of 2.44 $\Omega \cdot cm^2$ for the Pt electrode. The PCE reached 8.14%, better than that of the Pt-based DSSC (7.78%) with the iodide/triiodide ($I_3^-$) electrolyte. The speedy reduction of triiodide ($I_3^-$) to iodide ($I^-$) by the (G-P-A) $MoS_2$/SWCNTs CE is attributed to the fast transport of the electrolyte via the flower-like structure. The $MoS_2$ and multi-walled carbon nanotube (MWCNTs) nanocomposites were employed as a CE in DSSCs for the reduction of triiodide ($I_3^-$).[160] The microstructure of the MWCNTs@$MoS_2$ nanocomposite studied by transmission electron microscopy (TEM) confirmed the deposition of few-layers $MoS_2$ nanosheets on the MWCNTs surface. A MWCNTs@$MoS_2$ composite based CE resulted in higher cathodic current density than those of MWCNTs, $MoS_2$, and Pt CEs. The MWCNTs@$MoS_2$ CE showed a low charge-transfer resistance of 1.69 $\Omega.cm^2$ and no degradation up to 100 repeated cyclic voltammogram tests. The DSSC having a MWCNTs@$MoS_2$ composite CE exhibited a PCE of 6.45%, slightly better than that of the DSSC having sputtered Pt CE ($\eta$ = 6.41%). The MWCNT@$MoS_2$ based CE also showed improved catalytic activity for the reduction of $I_3^-$.

The $MoS_2$ nanosheets anchored onto the CNT surfaces were used in Pt-free CEs for DSSCs.[161] The $MoS_2$ nanosheets offer edge-plane electrocatalytically active sites for the reduction of $I_3^-$. The large surface area of CNTs supports the loading of $MoS_2$ nanosheets in order to increase the electrochemical activity. The CNTs deposited onto the FTO substrate promoted charge transport, leading to a higher exchange current density and also to the lower charge-transfer resistance. The $MoS_2$/CNT hybrid based DSSCs achieved a PCE of 7.83%, which is 9.5% higher than that of the Pt CE ($\eta$ = 7.15%) based DSSC.

Another research group[162] used flower-like $MoS_2$ and a multi-walled carbon nanotubes ($MoS_2$/MWCNTs) hybrid as a CE for dye-sensitized solar cells. The flower-like $MoS_2$/MWCNTs hybrid contains a large specific surface area and lamellar structure, as evidenced by field emission scanning electron microscopy (FESEM). The optimized $MoS_2$/MWCNTs has a $R_{CT}$ of 2.05 $\Omega.cm^2$ and series resistance of 1.13 $\Omega \cdot cm^2$ as measured by electrochemical impedance spectroscopy. Cyclic voltammogram measurements showed larger current density for $MoS_2$/MWCNTs based CEs than those of $MoS_2$, MWCNTs, and Pt CEs. $MoS_2$/MWCNTs CE based DSSCs exhibited a PCE of 7.50 %, comparable with the DSSC based on the Pt CE ($\eta$ = 7.49%). A carbon nanotubes-$MoS_2$-carbon (CNTs-$MoS_2$-carbon) hybrid was prepared using wet impregnation and calcination with polyethylene glycol as a CE for DSSCs.[163] Spectroscopic characterization by Raman spectra, XRD, TEM, XPS, BET and thermal methods indicated a homogenous coating of CNTs with thin layers of $MoS_2$, as a result of wetting and emulsification of polyethylene glycol 400. The CNTs-$MoS_2$-carbon heterostructure was used as CEs for DSSCs, and showed high stability and electrocatalytic activity in the reduction of $I_3^-$ to $I^-$ because of low $R_{CT}$. Interestingly, a PCE of 7.23% achieved for the CNTs-$MoS_2$-carbon CEs based DSSC was higher than Pt CEs ($\eta$ = 6.19%).

In another study, nanocomposites of $MoS_2$ and CNTs using a glucose aided (G-A) hydrothermal method were prepared by Yue et al.[164] The (G-A)$MoS_2$/CNTs nanocomposites obtained by adding 0.5, 1.0 , 1.5, 2.0 and 2.5 wt.% of acid-treated CNTs were deposited onto a FTO substrate and used as CEs in DSSCs for the reduction of triiodide ($I_3^-$) to iodide ($I^-$). The dye-sensitized $TiO_2$ anode was prepared by dipping the $TiO_2$ anode in 0.3 mM of dye N719 ethanol for 24 hours. Figure 8 shows the SEM images of $MoS_2$ nanopower and $MoS_2$-MWCNTs composites, J-V characteristics of the DSSCs fabricated with $MoS_2$, $MoS_2/C$, (G-A)$MoS_2$/CNTs and Pt CEs, and the effect of CNT contents on the PCE of the DSSCs using (G-A)$MoS_2$/CNTs. Scanning and transmission electron microscopy showed tentacle-like structures of the $MoS_2$/CNTs composites, having large active surface area and interconnected networks for fast transport for the electrolyte. The CNTs functionalized with a -COOH functional group were used and $MoS_2$ was anchored onto the functionalized CNTs. The (G-A)$MoS_2$/CNTs had specific surface area of 411.7 $m^2$/g and exhibited a small overpotential and better conductivity. The Nernst diffusion impedance ($Z_w$) values of 1.85 $\Omega\,cm^2$ for (G-A)$MoS_2$/CNTs CEs and 2.25 $\Omega\,cm^2$ for Pt CEs were measured, which indicates that the (G-A) $MoS_2$/CNTs catalyst accelerated the reduction of triiodide ions ($I_3^-$) to iodide ions ($I^-$). The (G-A)$MoS_2$/CNTs based CEs were found to exhibit enhanced electrocatalytic activity as evidenced by the CV, EIS, and Tafel polarization measurements. The photovoltaic performance of the DSSCs having $MoS_2$, $MoS_2/C$, (G-A)$MoS_2$/CNTs, and Pt CEs were compared. The photovoltaic performance of (G-A)$MoS_2$/CNTs CEs were also studied as a function of the CNT contents. The PCE of the DSSCs increased as the contents of the CNTs increased from 0 to 1.5 wt.%, however, further increase in CNT content leads to a decreased PCE. Likely, the maximum photovoltaic performance of the (G-A)$MoS_2$/CNTs CE in the DSSCs was achieved for a film thickness of 12 μm. The (G-A)$MoS_2$/CNTs composite CE has a lower $R_{CT}$ of 1.77 $\Omega\,cm^2$ at the electrolyte/electrode interface than those of $MoS_2$, $MoS_2$/carbon and conventional sputtered Pt CEs, and a PCE of 7.92% higher than the PCE of 7.11% for the Pt electrode in DSSCs for the triiodide/iodide ($I_3^-/I^-$) system. Lin et al.[165] fabricated nanocomposites of $MoS_2$/reduced graphene oxide (RGO) with CNTs using electrophoretic deposition. The $MoS_2$/RGO-CNTs hybrid was then used as a CE in DSSCs. In this hybrid, CNTs offer conductive networks for electron transport to increase the rate of charge-transfer at the CE/electrolyte interface. The $MoS_2$/RGO-CNTs hybrid CEs show improved electrocatalytic activity in comparison with the $MoS_2$/RGO alone. The DSSC having a $MoS_2$/RGO-CNTs hybrid CE achieved a PCE of 7.46 %, exceeding PCE values of DSSCs containing $MoS_2$/RGO CE ($\eta = 6.82$ %) and Pt CE ($\eta = 7.23$ %). Thin films of $MoS_2$/carbon nanotube composites have also been applied as electrodes for lithium ion batteries.[166]

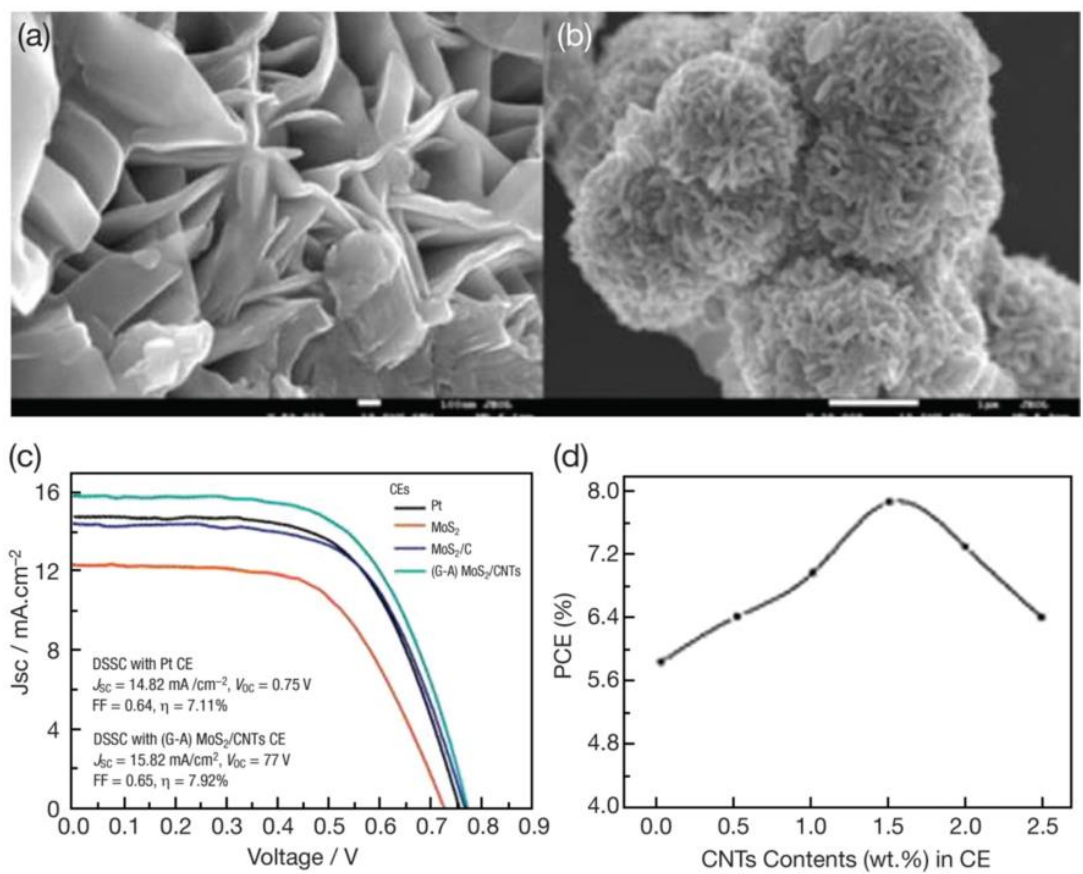

**Figure 8.** Scanning electron microscopic image of $MoS_2$ nanopower and $MoS_2$-MWCNTs composites prepared by glucose aided (G-A) hydrothermal method. Photocurrent–voltage (J-V) characteristics of the DSSCs fabricated with $MoS_2$, $MoS_2$/C, (G-A)$MoS_2$/CNTs and Pt CEs.

Relationship between the contents of CNTs in (G-A)MoS$_2$/CNTs CE and the PCE of DSSCs. Reprinted with permission from ref. 164. Copyright 2013, Elsevier.

### 3.1.4 MoS$_2$/TiO$_2$ Composites

MoS$_2$/TiO$_2$ heterostructures were developed by Du et al.[167] by depositing few-layer MoS$_2$ on mesoporous TiO$_2$ by a chemical-bath method. Raman spectrum and HRTEM images indicated a few-layer structure of the MoS$_2$. After few-layer MoS$_2$ deposition, the UV absorption spectra of the TiO$_2$ photoanode showed enhanced absorption in the visible wavelength region and a PCE of 1.08% for the MoS$_2$/TiO$_2$ based DSSC. Photovoltaic performance of the DSSC was found to be optimized by both thermal annealing and the thickness of the MoS$_2$ sensitized layer. Jhang et al.[168] modified the interface of a MoS$_2$ CE/electrolyte by incorporating TiO$_2$ nanoparticles, in order to control the overpotential loss. Their MoS$_2$/TiO$_2$ nanocomposite had a weight ratio of 5:1. Thermal annealing of the MoS$_2$ and MoS$_2$/TiO$_2$ CEs was performed at 400 $^o$C in vacuum for 2 hours. Figure 9 shows the J-V curves for MoS$_2$, MoS$_2$/TiO$_2$, and Pt CEs in the DSSCs. The addition of TiO$_2$ nanoparticles into the MoS$_2$ CE significantly increased the *Jsc* value from 7.24 mA/cm$^2$ for the MoS$_2$ CE to 13.76 mA/cm$^2$ for the MoS$_2$/TiO$_2$ CE. The MoS$_2$/TiO$_2$ CE based DSSC ($\eta$ = 5.08%) shows two times better photovoltaic performance compared with the MoS$_2$ CE (2.54%) and comparable to the Pt CE (5.27%), because of an increased active surface area from the incorporated TiO$_2$ nanoparticles and the low overpotential loss. Though the MoS$_2$/TiO$_2$ CE exhibited high PCE, its electrical conductivity was found to be lower compared to the MoS$_2$ CE, and, therefore, the high electrocatalytic activity was not related to its electrical conductivity.

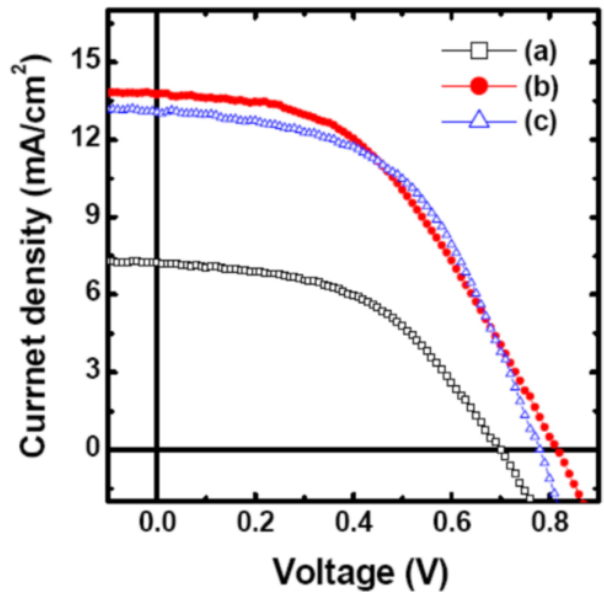

**Figure 9** Photocurrent density-voltage (J-V) characteristics curves of DSSCs with (a) MoS$_2$, (b) MoS$_2$/TiO$_2$ and (c) Pt counter electrodes measured under the solar light illumination of 100 mW/cm$^2$ (AM 1.5G). Reprinted with permission from ref. 168. Copyright 2015, Elsevier.

The MoS$_2$/TiO$_2$ based CE doped with different Co contents for the DSSCs were found to influence the PCE by improving electrocatalytic activity.[169] The Co content-optimized MoS$_2$/TiO$_2$ CE had enhanced catalytic activity at the electrolyte interfaces. The Co 3d orbit plays a role in increasing in the reduction of I$_3^-$ to I$^-$. The photoanodes were developed using MoS$_2$ and TiO$_2$ nanoparticles.[170] The DSSCs with MoS$_2$@TiO$_2$ photoanode showed a PCE of 6.02%, 1.5 times higher than that of the TiO$_2$ film photoanode ($\eta$ = 4.43%).

### 3.1.5 MoS$_2$/Carbon Composites

A very interesting study on MoS$_2$/carbon composites was conducted by Yue et al.[171] who developed DSSCs using MoS$_2$ and carbon composites as a CE with different contents of carbon. Figure 10 represents the SEM images of hydrothermally synthesized MoS$_2$ and a porous MoS$_2$–carbon hybrid, a HRTEM image of MoS$_2$–carbon hybrid, an IPCE spectra, and J-V curves of the DSSCs with MoS$_2$, MoS$_2$–carbon hybrid, and Pt CEs. The commercial MoS$_2$ particles have lamellar structure with large surface area. The hydrothermally prepared $MoS_2$ and $MoS_2$–carbon hybrid contains a large number of interlaced nanosheets. The intercalated nanosheets in the $MoS_2$–carbon hybrid are thinner compared with $MoS_2$, indicating an increase in the specific surface area, which is highly suitable for improving the electrocatalytic activity. The $MoS_2$ and $MoS_2$–carbon hybrid CEs were characterized by CV, EIS, and Tafel polarization curve measurements, and compared with the Pt CE. IPCE spectra of $MoS_2$, $MoS_2$–C, and Pt CEs in DSSCs measured in the 300–750 nm range show a similar photoelectric response. A strong photoelectric peak in the IPCE spectra appears at 340 nm. DSSCs with $MoS_2$ CE show the highest photoelectric response of 44.6% in the IPCE spectra at 520 nm, which is lower compared to a DSSC having a Pt CE (60.9%). Interestingly, in the IPCE spectra of the DSSC having porous $MoS_2$–carbon CE, the highest photoelectric response of 67.3% is observed at 520 nm, which is higher than that of the Pt CE based DSSC. If one compares the $R_{CT}$ values of 4.13 $\Omega.cm^2$ for $MoS_2$ and 2.29 $\Omega.cm^2$ for Pt electrodes, the $MoS_2$/carbon electrode had a low $R_{CT}$ value of 2.07 $\Omega.cm^2$, appearing as a result of the high conductivity and large surface area for the $MoS_2$/carbon composite, which eventually contributes to enhanced electrocatalytic activity of the CE. The $R_{CT}$ values of the $MoS_2$/carbon electrode are also influenced by the carbon content, which decreases with the increase in carbon content, from 3.60 $\Omega.cm^2$ for 1.13 wt.% carbon to 2.07 $\Omega.cm^2$ for 3.30 wt.% carbon. Thereafter, the $R_{CT}$ values increased from 2.40 $\Omega.cm^2$ for 4.35 wt.% carbon content to 3.13 $\Omega.cm^2$ for 5.38 wt.% carbon content. PCE values of 6.01, 7.03, 7.69, 7.33 and 5.76% were measured for $MoS_2$–carbon electrodes having 1.13, 2.23, 3.30, 4.35 and 5.38 wt.% of carbon content, respectively. The highest value of 7.69% was achieved at 3.30 wt.% carbon content in a $MoS_2$/carbon composite CE based DSSC for the $I^-/I_3^-$ redox reaction, and exceeded the PCE value of 6.74% for a Pt CE under similar experimental conditions. PCE values of the DSSCs were also found to vary as a function of the thickness of MoS$_2$/carbon CE. PCE values of 5.10, 6.89, 7.69, 7.01 and 4.85 % were measured for the CE thicknesses of 4, 8, 12, 16 and 20 μm for the MoS$_2$/carbon CE. EIS, CV, and Tafel curve analysis showed low R$_{CT}$, high electrocatalytic activity, and faster reduction of triiodide (I$_3^-$) to iodide (I$^-$) for the MoS$_2$/carbon CE compared to that of the Pt CE.

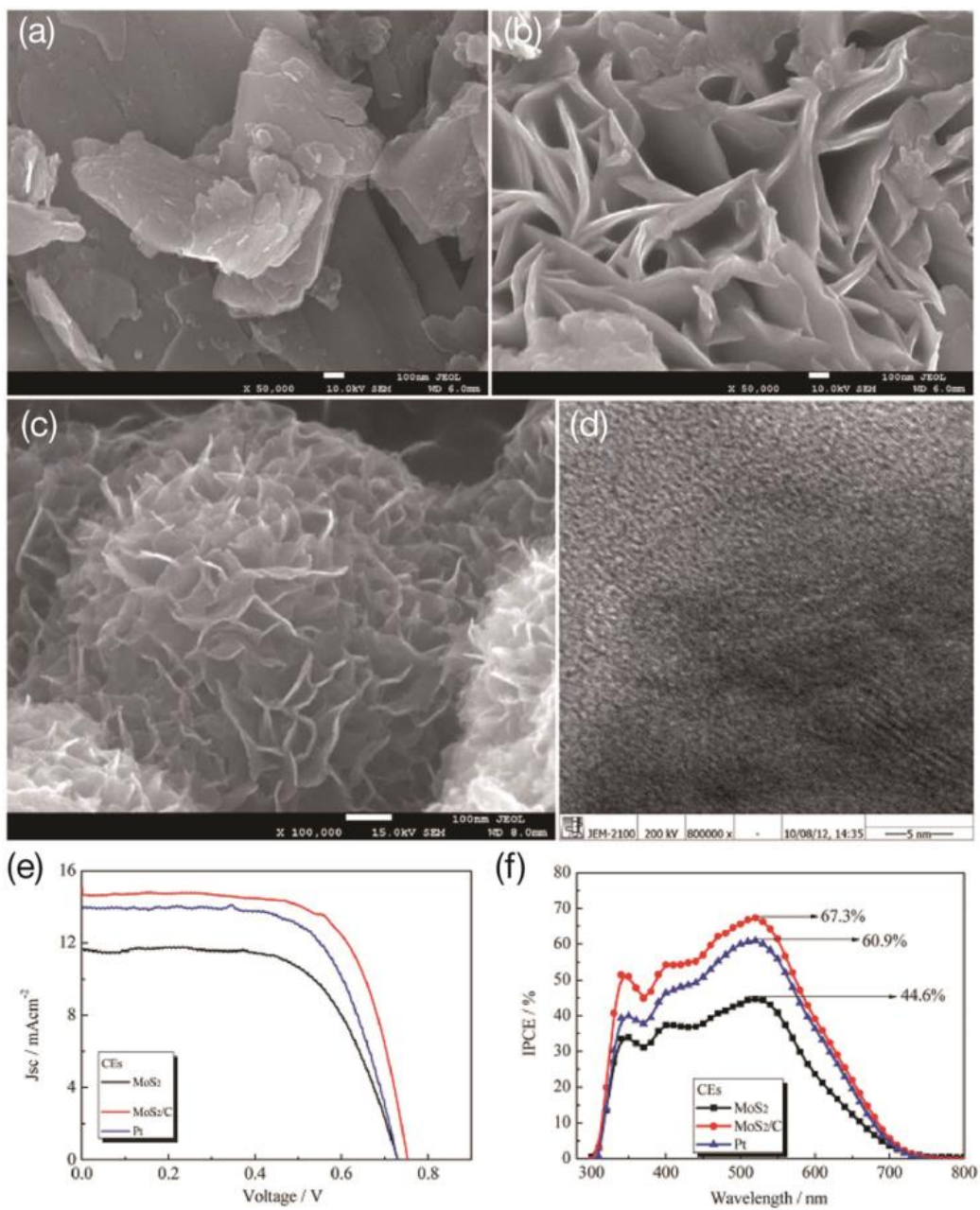

**Figure 10.** Scanning electron microscopy (SEM) images of (a) commercial $MoS_2$, (b) hydrothermal route synthesized $MoS_2$, (c) porous $MoS_2$–Carbon hybrid prepared by a hydrothermal route, (d) High-resolution TEM image of $MoS_2$–Carbon hybrid, (e) Photocurrent–voltage (J-V) curves of the DSSCs with $MoS_2$, $MoS_2$-C and Pt CEs and (f) incident photon-to-current efficiency (IPCE) spectra of the DSSCs with $MoS_2$, $MoS_2$–C and Pt CEs. Reprinted with permission from ref. 171. Copyright 2013, The Royal Society of Chemistry.

In yet another study, $MoS_2$/carbon fibers were used as CEs for DSSCs.[172] Both electrocatalytic activity and the PCE (3.26%) of the $MoS_2$/carbon fiber based CE was found be better than that of Pt/carbon fiber CE ($\eta$ = 2.93%). In another study, composites of flower-like $MoS_2$ microspheres and carbon materials such as vulcan carbon, acetylene black, MWCNTs, carbon nanofibers (CNFs), and rice husk ash were studied as cost-effective Pt-free CEs for DSSCs.[173] The electrolyte used in the DSSC was a phthaloylchitosan-based polymer. The carbon materials/$MoS_2$ CEs showed low $R_{CT}$ at the CE/electrolyte interface and high electro-catalytic activity for $I_3^-$ reduction. The DSSC with $MoS_2$/CNF CE showed a PCE of 3.17 %, compared to a PCE of 1.04% for the pure $MoS_2$ CE.

Another study used $MoS_2$ and PEDOT:PSS composites as CEs for DSSCs.[174] The $MoS_2$/PEDOT:PSS composite CE exhibits a PCE of 5.7% and FF of 58%, comparable to the Pt CE. The high PCE of the $MoS_2$/PEDOT:PSS CE originated from high electrocatalytic activity of the $MoS_2$ active sites for triiodide ($I_3^-$) reduction and high conductivity of PEDOT:PSS. The inorganic/organic $MoS_2$/PEDOT:PSS composite may be useful as a low-cost Pt-free CEs for DSSCs. Another study used $Bi_5FeTi_3O_{15}$ (BFTO) nanofibers of 40–100 nm diameter developed by a sol-gel aided electrospinning method.[175] The $MoS_2$/BFTO nanocomposite-based CE for DSSCs was prepared by uniformly dispersing $MoS_2$ nanoparticles into the BFTO matrix. The optical bandgap of the $MoS_2$/BFTO nanocomposites was found to decrease with increasing $MoS_2$ contents. The DSSC with a $MoS_2$/BFTO nanocomposite-based CE showed a PCE of 5.20%, 24 times higher than that of the pure BFTO nanofiber based CE. Table 2 summarizes the electrochemical and photovoltaic properties of all types of MoS$_2$ based CEs discussed in this section above, and a comparison is made with conventional Pt CEs for DSSCs.

**Table 2.** Photovoltaic parameters of MoS$_2$ based counter electrodes (CEs) used for DSSCs. FTO glass is the common substrate used in assembling DSSCs with different CE materials. The measurements were conducted at a simulated solar light intensity of 100 mW/cm$^2$ (AM 1.5G) unless specified. The photovoltaic parameters short-circuit photocurrent density ($J_{SC}$), open-circuit voltage ($V_{OC}$), fill factor (FF), power conversion efficiency ($\eta$), series resistance (R$_s$), charge-transfer resistance (R$_{CT}$), electrolyte and dye used of DSSCs of graphene/MoS$_2$ nanocomposites are summarized and compared with standard Pt counter electrode.

| Counter Electrodes | Redox Couple | Dye | $J_{SC}$ (mA/cm$^2$) | $V_{OC}$ (V) | FF (%) | PCE ($\eta$, %) | R$_s$ ($\Omega.cm^2$) | R$_{CT}$ ($\Omega.cm^2$) | Ref. |
|---|---|---|---|---|---|---|---|---|---|
| 2H-MoS$_2$ (hydrothermal, 200 $^o$C) | I$^-$/I$_3^-$ | N719 | 6.78 | 0.73 | 35 | 1.72 | 16 | 49 | 130 |
| 1T-MoS$_2$ (hydrothermal, 180 $^o$C) | I$^-$/I$_3^-$ | N719 | 8.76 | 0.73 | 52 | 7.08 | 16 | 19 | 130 |
| Pt reference | I$^-$/I$_3^-$ | N719 | 17.75 | 0.702 | 58 | 7.25 | - | - | 130 |
| MoS$_2$ (CVD) vertically inclined | I$^-$/I$_3^-$ | N719 | 15.2 | 0.707 | 69.7 | 7.50 | 9.5 | 3.10 | 131 |
| Pt reference | I$^-$/I$_3^-$ | N719 | 14.6 | 0.712 | 70.0 | 7.28 | 6.7 | 5.36 | 131 |
| MoS$_2$/graphite | I$^-$/I$_3^-$ | N719 | 15.64 | 0.685 | 67 | 7.18 | - | 8.05 | 132 |
| Graphite | I$^-$/I$_3^-$ | N719 | 11.62 | 0.445 | 63 | 3.26 | - | 15.70 | 132 |
| Pt reference | I$^-$/I$_3^-$ | N719 | 15.84 | 0.735 | 65 | 7.57 | - | 6.35 | 132 |
| MoS$_2$ (multi-layer) | I$^-$/I$_3^-$ | N719 | 15.81 | 0.745 | 25 | 2.92 | 27.3 | 186.2 | 133 |
| MoS$_2$ (few-layer) | I$^-$/I$_3^-$ | N719 | 14.90 | 0.744 | 16 | 1.74 | 35.8 | 281.2 | 133 |
| MoS$_2$ (nanoparticle) | I$^-$/I$_3^-$ | N719 | 14.72 | 0.745 | 49 | 5.41 | 26.9 | 93.0 | 133 |
| Pt reference | I$^-$/I$_3^-$ | N719 | 13.41 | 0.754 | 65 | 6.58 | 34.2 | 3.9 | 133 |
| MoS$_2$ (hydrothermal method) | I$^-$/I$_3^-$ | N719 | 18.37 | 0.698 | 57.8 | 7.41 | - | 0.619 | 135 |
| Pt reference | I$^-$/I$_3^-$ | N719 | 16.78 | 0.722 | 58.8 | 7.13 | - | 3.78 | 135 |
| MoS$_2$ (300 $^0$C annealed) | I$^-$/I$_3^-$ | N719 | 16.905 | 0.727 | 51.7 | 6.351 | 23.89 | 30.98 | 137 |
| Pt reference | I$^-$/I$_3^-$ | N719 | 17.056 | 0.724 | 55.7 | 6.929 | 27.17 | 14.98 | 137 |
| MoS$_2$ (chemical deposition) | I$^-$/I$_3^-$ | N719 | 18.46 | 0.68 | 58 | 7.01 | 23.51 | 18.50 | 138 |
| Pt reference | I$^-$/I$_3^-$ | N719 | 16.80 | 0.71 | 60 | 7.31 | 26.73 | 22.88 | 138 |
| MoS$_2$ (non-annealed) | I$^-$/I$_3^-$ | N719 | 5.24 | 0.74 | 27 | 1.0 | 28 | - | 139 |
| MoS$_2$ (vacuum-annealed) | I$^-$/I$_3^-$ | N719 | 7.95 | 0.74 | 29 | 1.7 | 22 | - | 139 |
| MoS$_2$ (N$_2$-annealed) | I$^-$/I$_3^-$ | N719 | 4.35 | 0.67 | 27 | 0.8 | 44 | - | 139 |
| MoS$_2$/Mo (*in-situ* sulfurization) | I$^-$/I$_3^-$ | N719 | 22.6 | 0.74 | 50 | 8.4 | - | - | 140 |
| Pt reference | $I^-/I_3^-$ | N719 | 21.9 | 0.735 | 53.4 | 8.7 | - | - | 140 |
| MoS$_2$ (hydrothermal method) | $I^-/I_3^-$ | N719 | 13.84 | 0.76 | 73 | 7.59 | 20.8 | 0.5 | 141 |
| WS$_2$ (hydrothermal method) | $I^-/I_3^-$ | N719 | 14.13 | 0.78 | 70 | 7.73 | 19.4 | 0.3 | 141 |
| Pt reference | $I^-/I_3^-$ | N719 | 14.89 | 0.78 | 66 | 7.64 | 12.7 | 3.0 | 141 |
| MoS$_2$ (hydrothermal method) | $T_2/T^-$ | N719 | 12.52 | 0.63 | 63 | 4.97 | - | - | 141 |
| WS$_2$ (hydrothermal method) | $T_2/T^-$ | N719 | 12.99 | 0.64 | 64 | 5.24 | - | - | 141 |
| Pt reference | $T_2/T^-$ | N719 | 12.23 | 0.63 | 48 | 3.70 | - | - | 141 |
| MoS$_2$ (porous sheets) | $I^-/I_3^-$ | N719 | 15.40 | 0.763 | 53 | 6.35 | 7.95 | 1.73 | 143 |
| MoS$_2$ (flower-shaped) | $I^-/I_3^-$ | N719 | 13.73 | 0.700 | 52 | 5.23 | 7.89 | 2.67 | 143 |
| Pt reference | $I^-/I_3^-$ | N719 | 16.34 | 0.745 | 51 | 6.19 | 8.06 | 1.82 | 143 |
| MoS$_2$ (sputtering, 5 min) | $I^-/I_3^-$ | N719 | 13.17 | 0.71 | 64 | 6.6 | 30.1 | 2.2 | 145 |
| Pt reference | $I^-/I_3^-$ | N719 | 14.70 | 0.71 | 66 | 6.0 | 3.1 | 1.5 | 145 |
| MoS$_2$ (as-prepared) | $I^-/I_3^-$ | N719 | 11.92 | 0.656 | 35 | 2.74 | - | $1.01 \times 10^4$ | 146 |
| MoS$_2$ (heat-sintered) | $I^-/I_3^-$ | N719 | 13.01 | 0.705 | 65 | 5.96 | - | 18.50 | 146 |
| MoS$_2$ (laser-sintered) | $I^-/I_3^-$ | N719 | 14.94 | 0.718 | 67 | 7.19 | - | 15.29 | 146 |
| Pt reference | $I^-/I_3^-$ | N719 | 14.30 | 0.741 | 70 | 7.42 | - | 3.99 | 146 |
| MoS$_2$ (growth time, 5 h) | $I^-/I_3^-$ | N719 | 15.15 | 0.76 | 52 | 5.96 | 50.9 | 118.8 | 147 |
| MoS$_2$ (growth time, 10 h) | $I^-/I_3^-$ | N719 | 15.94 | 0.71 | 63 | 7.14 | 50.4 | 21.2 | 147 |
| MoS$_2$ (growth time, 15 h) | $I^-/I_3^-$ | N719 | 16.96 | 0.74 | 66 | 8.28 | 38.8 | 12.9 | 147 |
| Pt reference | $I^-/I_3^-$ | N719 | 13.77 | 0.74 | 74 | 7.53 | 36.3 | 13.1 | 147 |
| MoS$_2$ (Exfoliated) | $I^-/I_3^-$ | N719 | 11.54 | 0.80 | 65 | 6.0 | 29.60 | 19.60 | 148 |
| MoS$_2$ (Annealed) | $I^-/I_3^-$ | N719 | 10.92 | 0.80 | 58 | 5.1 | 28.10 | 121.10 | 148 |
| Pt reference | $I^-/I_3^-$ | N719 | - | | | | | | 148 |
| MoS$_2$ | $I^-/I_3^-$ | N719 | 12.92 | 0.701 | 46 | 4.15 | 11.69 | 3.65 | 151 |
| Graphene nanosheet | $I^-/I_3^-$ | N719 | 11.99 | 0.754 | 30 | 2.68 | 9.31 | 6.24 | 151 |
| MoS$_2$–graphene nanosheet | $I^-/I_3^-$ | N719 | 12.79 | 0.773 | 59 | 5.81 | 9.52 | 2.34 | 151 |
| Pt reference | $I^-/I_3^-$ | N719 | 13.12 | 0.763 | 62 | 6.24 | 9.11 | 1.79 | 151 |
| MoS$_2$/Graphene | $I^-/I_3^-$ | N719 | 12.41 | 0.71 | 68 | 5.98 | 24.42 | 4.94 | 152 |
| Pt reference | $I^-/I_3^-$ | N719 | 12.43 | 0.73 | 67 | 6.23 | 24.72 | 4.74 | 152 |
| MoS$_2$ (CVD) | $I^-/I_3^-$ | N719 | 14.0 | 0.62 | 65 | 5.6 | 1.5 | 2.3 | 153 |
| MoS$_2$/Graphene (CVD) | $I^-/I_3^-$ | N719 | 16.1 | 0.66 | 67 | 7.1 | 1.7 | 1.6 | 153 |
| Graphene | $I^-/I_3^-$ | N719 | 16.9 | 0.68 | 24 | 2.8 | 1.7 | 2.8 | 153 |
| Pt reference | $I^-/I_3^-$ | N719 | 17.2 | 0.72 | 60 | 7.4 | 2.1 | 1.8 | 153 |
| MoS$_2$ | $I^-/I_3^-$ | N719 | 9.14 | 0.589 | 47 | 2.53 | - | - | 154 |
| Graphene | $I^-/I_3^-$ | N719 | 10.7 | 0.652 | 51.9 | 3.62 | - | - | 154 |
| MoS$_2$:Graphene (10:90) | $I^-/I_3^-$ | N719 | 11.91 | 0.646 | 56.5 | 4.35 | - | - | 154 |
| Pt reference | $I^-/I_3^-$ | N719 | 13.39 | 0.657 | 50 | 4.40 | - | - | 154 |
| MoS$_2$/Graphene Oxide (N$_2$ doped) | $I^-/I_3^-$ | N719 | 15.98 | 0.70 | 53 | 5.95 | 25.7 | 5.4 | 156 |
| Graphene Oxide (N$_2$ doped) | $I^-/I_3^-$ | N719 | 14.66 | 0.71 | 38 | 3.95 | 25.7 | 21.3 | 156 |
| MoS$_2$ | $I^-/I_3^-$ | N719 | 15.39 | 0.69 | 39 | 4.09 | 26.0 | 10.1 | 156 |
| Pt reference | $I^-/I_3^-$ | N719 | 16.14 | 0.70 | 57 | 6.43 | 26.3 | 4.3 | 156 |
| MoS$_2$ (drop-coating) | $I^-/I_3^-$ | N719 | 13.46 | 0.79 | 58 | 6.20 | 21.14 | 24.93 | 157 |
| Nitrogen-doped Graphene (NGr) | $I^-/I_3^-$ | N719 | 12.72 | 0.68 | 64 | 5.50 | 16.31 | 30.17 | 157 |
| MoS$_2$/NGr (8 wt.%) | $I^-/I_3^-$ | N719 | 15.36 | 0.77 | 66 | 7.82 | 15.60 | 16.73 | 157 |
| Pt reference | $I^-/I_3^-$ | N719 | 15.71 | 0.77 | 68 | 8.25 | 15.23 | 10.15 | 157 |
| MoS$_2$ | $I^-/I_3^-$ | N719 | 10.56 | 0.67 | 58 | 4.10 | - | - | 158 |
| Graphene flake (GF) | $I^-/I_3^-$ | N719 | 10.96 | 0.69 | 48 | 3.63 | - | - | 158 |
| MoS$_2$ (GF is 0.5 wt.%) | $I^-/I_3^-$ | N719 | 12.09 | 0.69 | 59 | 4.85 | - | - | 158 |
| MoS$_2$ (GF is 1 wt.%) | $I^-/I_3^-$ | N719 | 12.68 | 0.74 | 59 | 5.35 | - | - | 158 |
| MoS$_2$ (GF is 1.5 wt.%) | $I^-/I_3^-$ | N719 | 13.27 | 0.75 | 61 | 6.07 | - | - | 158 |
| MoS$_2$ (GF is 2 wt.%) | $I^-/I_3^-$ | N719 | 12.95 | 0.74 | 58 | 5.56 | - | - | 158 |
| MoS$_2$ (GF is 2.5 wt.%) | $I^-/I_3^-$ | N719 | 12.63 | 0.70 | 58 | 5.04 | - | - | 158 |
| Pt reference | $I^-/I_3^-$ | N719 | 12.95 | 0.75 | 66 | 6.41 | - | - | 158 |
| MoS$_2$ | $I^-/I_3^-$ | N719 | 11.25 | 0.72 | 61 | 4.99 | 11.37 | 2.43 | 160 |
| Multi-walled CNT (MWCNT) | $I^-/I_3^-$ | N719 | 9.11 | 0.65 | 58 | 3.53 | 9.91 | 8.59 | 160 |
| MoS$_2$/MWCNT | $I^-/I_3^-$ | N719 | 13.69 | 0.73 | 65 | 6.45 | 10.22 | 1.69 | 160 |
| Pt reference | $I^-/I_3^-$ | N719 | 13.24 | 0.74 | 66 | 6.41 | 9.06 | 1.91 | 160 |
| MoS$_2$ | $I^-/I_3^-$ | N719 | 14.44 | 0.74 | 64 | 6.81 | - | - | 161 |
| Carbon nanotubes | $I^-/I_3^-$ | N719 | 13.33 | 0.75 | 62 | 6.15 | - | - | 161 |
| MoS$_2$/Carbon nanotubes | $I^-/I_3^-$ | N719 | 16.65 | 0.74 | 66 | 7.83 | - | - | 161 |
| Pt reference | $I^-/I_3^-$ | N719 | 14.83 | 0.74 | 65 | 7.15 | - | - | 161 |
| MoS$_2$/CNTs | $I^-/I_3^-$ | N719 | 14.93 | 0.65 | 47 | 4.51 | 8.42 | 4.35 | 163 |
| CNTs/MoS$_2$/carbon | $I^-/I_3^-$ | N719 | 16.44 | 0.79 | 57 | 7.23 | 8.14 | 1.73 | 163 |
| Pt reference | $I^-/I_3^-$ | N719 | 15.40 | 0.75 | 55 | 6.19 | 8.31 | 1.95 | 163 |
| MoS$_2$/CNTs (G-A) | $I^-/I_3^-$ | N719 | 15.82 | 0.77 | 65 | 7.92 | 5.20 | 1.77 | 164 |
| MoS$_2$/Carbon | $I^-/I_3^-$ | N719 | 14.52 | 0.76 | 64 | 7.06 | 5.24 | 2.35 | 164 |
| MoS$_2$ | $I^-/I_3^-$ | N719 | 12.33 | 0.72 | 61 | 5.42 | 5.33 | 4.16 | 164 |
| Pt reference | $I^-/I_3^-$ | N719 | 14.82 | 0.75 | 64 | 7.11 | 5.06 | 2.22 | 164 |
| MoS$_2$/reduced graphene oxide (RGO) | $I^-/I_3^-$ | N719 | 14.31 | 0.76 | 63 | 6.82 | 20.58 | 4.42 | 165 |
| MoS$_2$/RGO-CNTs | $I^-/I_3^-$ | N719 | 14.59 | 0.76 | 67 | 7.46 | 20.37 | 3.31 | 165 |
| Pt reference | $I^-/I_3^-$ | N719 | 14.53 | 0.77 | 65 | 7.23 | 20.13 | 4.06 | 165 |
| MoS$_2$ (spin-coating) | $I^-/I_3^-$ | N719 | 7.24 | 0.70 | 49 | 2.54 | 59.5 | - | 168 |
| MoS$_2$/TiO$_2$ (5:1 wt. ratio) | $I^-/I_3^-$ | N719 | 13.76 | 0.82 | 45 | 5.08 | 56.5 | - | 168 |
| Pt reference | $I^-/I_3^-$ | N719 | 13.06 | 0.78 | 52 | 5.27 | - | - | 168 |
| MoS$_2$/TiO$_2$ | $I^-/I_3^-$ | N719 | 4.67 | 0.68 | 44 | 1.4 | - | - | 169 |
| MoS$_2$/TiO$_2$/Co | $I^-/I_3^-$ | N719 | 9.21 | 0.70 | 50 | 3.2 | - | - | 169 |
| MoS$_2$/Carbon (C is 2.23 wt.%) | $I^-/I_3^-$ | N719 | 13.98 | 0.74 | 68 | 7.03 | 5.87 | 2.67 | 171 |
| MoS$_2$/Carbon (C is 3.30 wt.%) | I$^-$/I$_3^-$ | N719 | 15.07 | 0.75 | 68 | 7.69 | 5.77 | 2.07 | 171 |
| MoS$_2$/Carbon (C is 4.35 wt.%) | I$^-$/I$_3^-$ | N719 | 14.37 | 0.75 | 68 | 7.33 | 5.83 | 2.40 | 171 |
| MoS$_2$ | I$^-$/I$_3^-$ | N719 | 11.66 | 0.73 | 63 | 5.36 | 5.87 | 4.13 | 171 |
| Pt reference | I$^-$/I$_3^-$ | N719 | 13.98 | 0.73 | 66 | 6.74 | 5.79 | 2.29 | 171 |
| | | | | | | | | | |
| MoS$_2$/PEDOT–PSS | I$^-$/I$_3^-$ | N719 | 14.55 | 0.68 | 58 | 5.7 | - | - | 174 |
| PEDOT–PSS | I$^-$/I$_3^-$ | N719 | 14.6 | 0.68 | 26 | 2.5 | - | - | 174 |
| Pt reference | I$^-$/I$_3^-$ | N719 | 15.26 | 0.73 | 59 | 6.6 | - | - | 174 |

_disulfide/thiolate (T$_2$/T$^-$) redox couple.

In the case of R$_s$ and R$_{CT}$: Some of the authors used $\Omega$ instead of $\Omega$.cm$^2$ for the resistances without mentioning the size of the electrode.

## 3.2 WS$_2$ Counter Electrodes

Tungsten disulfide (WS$_2$), traditionally used as a lubricant, is a semiconductor having van der Waals bonding which forms 2D layered-structures similar to other TMDs. WS$_2$ can form atomically thin nanosheets,[176, 177] nanorods,[178] and nanotubes,[179, 180] which have been actively studied for potential applications.

Carbon-coated WS$_2$ CEs have been fabricated for DSSCs at low temperature and characterized using FESEM, XRD, and Raman spectroscopy.[181] The electrocatalytic activity of the WS$_2$ CEs was studied using CV and EIS. The DSSCs with carbon-coated WS$_2$ CEs show a PCE of 5.5%, comparable to that of Pt CE based DSSCs ($\eta$ = 5.6%). A DSSC having plastic WS$_2$ CEs exhibited a PCE of 5.0%. Carbon-coated WS$_2$ seems promising to develop low cost Pt-free CEs for DSSCs. The WS$_2$ films were deposited by radio frequency (RF) sputtering and a sulfurization process as CE for DSSCs.[182] The WS$_2$ films were characterized using XRD, FESEM, Raman spectroscopy, and XPS techniques. The transparent WS$_2$ CEs demonstrated high electrocatalytic activity and fast reduction of triiodide, (I$_3^-$) as characterized using CV, EIS, and Tafel polarization curve. WS$_2$ CE sputtered for 10 minutes showed a PCE of 6.3%, slightly lower than the Pt-based CE ($\eta$ = 6.64%) used in the DSSC. The J-V characteristics as a function of sputtering time used to prepare $WS_2$ films as a CE were also studied. $WS_2$ film CEs prepared at sputtering time of 5, 10 and 15 minutes showed PCEs of 5.4%, 6.3% and 5.8%, respectively.

Another research team[183] used edge-oriented $WS_2$ based CEs for DSSCs. Edge-oriented $WS_2$ was obtained from mesoporous interconnected $WO_3$ structures using a high temperature sulfurization process. The DSSCs with edge-oriented $WS_2$ CEs show a PCE of 8.85%, higher compared to the Pt CE ($\eta = 7.20\%$). The large number of active edge sites in edge-oriented $WS_2$ is responsible for high electrocatalytic activity for the reduction of triiodide ($I_3^-$) in the DSSCs. The $WS_2$ films were fabricated by the doctor-blade method (or tape casting method; a method removing excessive liquid material using a moving blade for uniform coating) to use as CEs for DSSCs.[184] The $TiO_2$ (P25) and carbon nanoparticles were introduced into $WS_2$ films to increase electrical conductivity and bonding strength. The electrochemical catalytic activity of $WS_2$/P25/C CEs was compared with Pt for the triiodide ($I_3^-$) to iodide ($I^-$) electrolyte system using CV and EIS measurements. The DSSC developed with $WS_2$/P25/C CE was shown to yield a PCE of 4.56%.

Yue et al.[185] prepared $WS_2$ decorated multi-walled carbon nanotubes (MWCNTs) by applying a hydrothermal method, for use as a low-cost Pt-free CE for DSSCs. The contents of MWCNTs in MWCNTs-$WS_2$ CEs varied from 1 to 10 wt.%. PCE values of 5.20, 5.45, 6.41, 5.53 and 5.22% were measured in the DSSCs for CEs having 1, 3, 5, 7 and 10 wt.% contents of MWCNTs, respectively. CV and EIS showed high electrocatalytic activity for the MWCNTs-$WS_2$ CE for the triiodide ($I_3^-$) reduction. The $R_{CT}$ of MWCNTs-$WS_2$ CEs having 1, 3, 7, and 10 wt.% contents of MWCNTs were 4.54, 3.47, 3.24, and 4.59 $\Omega.cm^2$, respectively. The $R_{CT}$ of the MWCNTs-$WS_2$ CE with 5 wt.% contents of MWCNTs shows 2.53 $\Omega.cm^2$, comparable to the $R_{CT}$ of 2.74 $\Omega.cm^2$ for a Pt CE . The DSSCs based on $WS_2$/MWCNTs CEs showed a PCE of 6.41% for 5 wt.% MWCNTs, comparable to the PCE of 6.56% for Pt CE under a simulated AM 1.5 solar illumination (100 mW/cm$^2$). The low $R_{CT}$ of WS$_2$/MWCNTs CEs at the electrolyte/electrode interface contributed to the higher PCEs.

The WS$_2$ based CEs prepared by a hydrothermal method was also used for DSSC by Wu et al.[141], which exhibited a PCE of 7.73%. The same research team[186] also synthesized WS$_2$/MWCNTs hybrids by a glucose-aided (G-A) hydrothermal route, which is discussed here in detail. Figure 11 presents the SEM images of WS$_2$, MWCNTs, and (G-A)WS$_2$/MWCNTs composites, and *J-V* curves of the DSSCs with WS$_2$, MWCNTs, WS$_2$/MWCNTs* (prepared without the aid of glucose), WS$_2$/MWCNTs, and Pt CEs, under a simulated solar illumination of 100 mW/cm$^2$. The WS$_2$ exhibits graphene-like lamellar structure, whereas the MWCNTs have a fiber-like structure, indicating both materials have a large specific surface area. The specific surface area of the (G-A)WS$_2$/MWCNTs composite was estimated to be 230 m$^2$/g by the Brunauer−Emmett−Teller (BET) technique, indicating high electrochemical activity as well as photovoltaic efficiency for CEs. The cathodic peak potentials of the WS$_2$, WS$_2$/MWCNTs, and WS$_2$/MWCNTs* CEs showed cathodic peak potentials of −0.14, −0.13 and -01.7 V, respectively, which implies that the MWCNTs help in improving the electrocatalytic activity, and a lower cathodic peak potential observed for (G-A)WS$_2$/MWCNTs as opposed to WS$_2$/MWCNTs* results from the large specific surface area generated by glucose aided preparation. The EIS measurements yielded an $R_{CT}$ of 2.49 Ω·cm$^2$ and $R_s$ of 2.54 Ω·cm$^2$ for the WS$_2$/MWCNTs hybrid CE, which is smaller compared with WS$_2$ CE, and indicates the synergistic effect between WS$_2$ and MWCNTs that enhanced the electrical conductivity of the hybrid. The (G-A)WS$_2$/MWCNT CE based DSSC resulted in a PCE of 7.36%, comparable to WS$_2$ CE (5.32%), MWCNTs CE (4.34%), and the Pt CE (7.54%). The *J$_{SC}$* and PCE values increased with increasing content of MWCNTs, up to 5 wt.% in the $WS_2$/MWCNT hybrid CEs, and thereafter started decreasing with further increases in MWCNTs content. The (G-A)$WS_2$/MWCNT (5 wt.%) film also exhibits a smaller transmission between 320 to 800 nm than Pt film, therefore the $WS_2$/MWCNT film absorbs more incident light, which also further improves photovoltaic performance. The glucose aided (G-A)$WS_2$/MWCNTs (5 wt.%) CE in the DSSC had low $R_{CT}$ and high electrocatalytic activity for the reduction of triiodide ($I_3^-$), due to the synergistic effects induced by glucose.

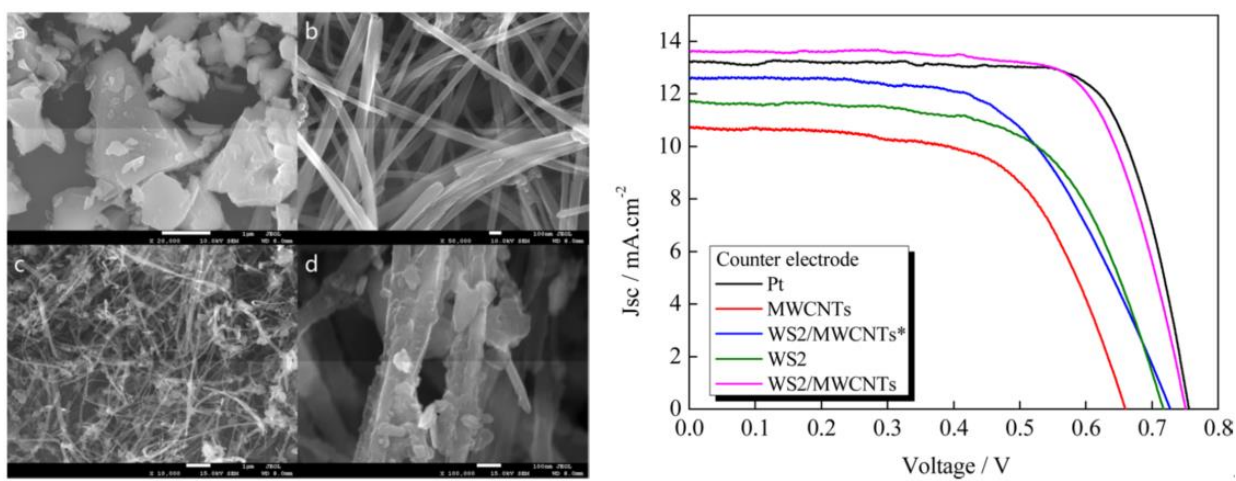

**Figure 11** (Left) SEM images of (a) $WS_2$, (b) MWCNTs, and (c and d) $WS_2$/MWCNTs composites. (Right) Photocurrent−voltage (J-V) curves of the DSSCs with Pt, $WS_2$, MWCNTs, $WS_2$/MWCNTs* (prepared without glucose aid), and (G-A)$WS_2$/MWCNTs counter electrodes under a simulated solar illumination of 100 mW/cm$^2$. Reprinted with permission from ref. 186. Copyright 2012, American Chemical Society.

### 3.3 TiS$_2$ Counter Electrodes

Titanium Disulfide (TiS$_2$) is a metallic thermoelectric material which attains unique morphologies such as nanotubes, nanoclusters and nanodisks, and exhibits interesting physical properties.[187-198] Meng et al.[199] has prepared 2D TiS$_2$ nanosheets decorated on graphene using a ball milling method and followed by high-temperature annealing. The electroactive surface areas of 1.70 cm$^2$ for TiS$_2$–graphene and 0.232 cm$^2$ for Pt electrodes, measured by CV, indicate more active sites for the hybrid. Figure 12 compares J-V curves of DSSCs having TiS$_2$-graphene and Pt/FTO CEs. The CEs based on TiS$_2$–graphene hybrids exhibited higher electrocatalytic activity for the reduction of triiodide (I$_3^-$) to iodide (I$^-$) in electrolyte with a PCE of 8.80%, which is higher than the Pt CE ($\eta$ = 8.00%). The R$_s$ of TiS$_2$–graphene CEs (2.32 $\Omega$.cm$^2$) was found to be smaller compared to the Pt CEs (6.90 $\Omega$.cm$^2$), showing better contact between the hybrid CE and FTO glass. Furthermore, the R$_{CT}$ of TiS$_2$–graphene CE (0.63 $\Omega$.cm$^2$) was also lower than the Pt CE (1.32 $\Omega$.cm$^2$), which again indicates a higher electrocatalytic activity for the reduction of triiodide(I$_3^-$). Also, the Z$_N$ of TiS$_2$–graphene CE (10.52 $\Omega$.cm$^2$) was found to be higher than the Pt CE (6.89 $\Omega$.cm$^2$). The high electrocatalytic activity of the TiS$_2$–graphene hybrids is attributed to the highly electroactivity of TiS$_2$ and enhanced transport facilitated by the graphene conductive network.

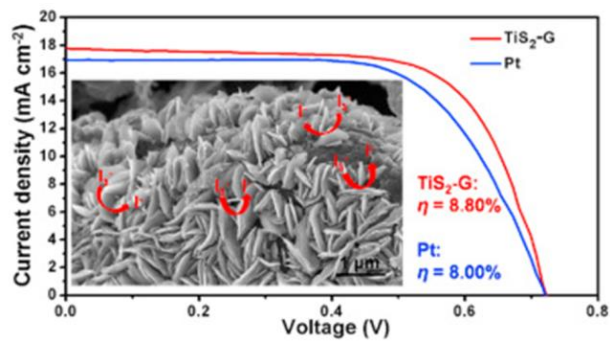

**Figure 12** A comparison of photocurrent density-voltage (*J-V*) curves of DSSCs having TiS$_2$-graphene hybrid and Pt counter electrodes. Reprinted with permission from ref. 199. Copyright 2016, Elsevier.

Li et al. [200] deposited composite films of TiS$_2$/ PEDOT:PSS on ITO substrates by drop coating, to study CEs of DSSCs. The wt.% of TiS$_2$ particles in TiS$_2$/PEDOT:PSS composite films varied from 5 to 15 wt.%. TiS$_2$ particles were dispersed in a PEDOT:PSS matrix to be used as an electrocatalyst for the $I^-/I_3^-$ redox reaction. In the composite, conducing polymer PEDOT:PSS plays the role of a binder for the $TiS_2$ nanoparticles, as well as a linking agent between $TiS_2$ particles and the ITO substrate, and also facilitates electron transfer. The $TiO_2$ photoanode for a DSSC was prepared by immersing it in N719 dye solution for 24 hours at room temperature. Figure 13 shows the photocurrent density-voltage curves and IPCE curves of the DSSCs with Pt, bare $TiS_2$, bare PEDOT:PSS, and 10 wt.% $TiS_2$/PEDOT:PSS composite CEs. The $TiS_2$/PEDOT:PSS composite CE offered a large surface area, yielding a high PCE of 7.04%. The CEs of bare $TiS_2$, bare PEDOT:PSS, the $TiS_2$/PEDOT:PSS composite, and Pt were characterized by AFM, SEM, and EDX. AFM i PSS and of 378 nm for $TiS_2$/PEDOT:PSS composite film; therefore, the higher roughness could lead to a larger active surface area and mages indicated a roughness of 45 nm for bare PEDOT: higher electrocatalytic activity for the $TiS_2$/PEDOT:PSS composite thin. The electrocatalytic properties of the DSSCs using the CEs of bare $TiS_2$, bare PEDOT:PSS, the $TiS_2$/PEDOT:PSS composite,  and Pt were evaluated by CV, RDE, EIS, and Tafel polarization measurements. The high PCE of the $TiS_2$/PEDOT:PSS composite CE based DSSC was also measured by IPCE curves. The maximum of the IPCE spectra at 520 nm increased from 52% to 63% as the content of $TiS_2$ particles increased from 0 to 10 wt.%, respectively, and similar characteristics were observed for the $J_{SC}$ values from the J-V curves of the DSSCs. The 10 wt.% $TiS_2$/PEDOT:PSS composite film based CE shows a higher redox current density compared with bare $TiS_2$ and PEDOT:PSS CEs, therefore, it possesses high electrocatalytic activity for triiodide ($I_3^-$) reduction, and it also exhibits high electrochemical stability after 100 consecutive cycles in the $I^-/I_3^-$ redox electrolyte.

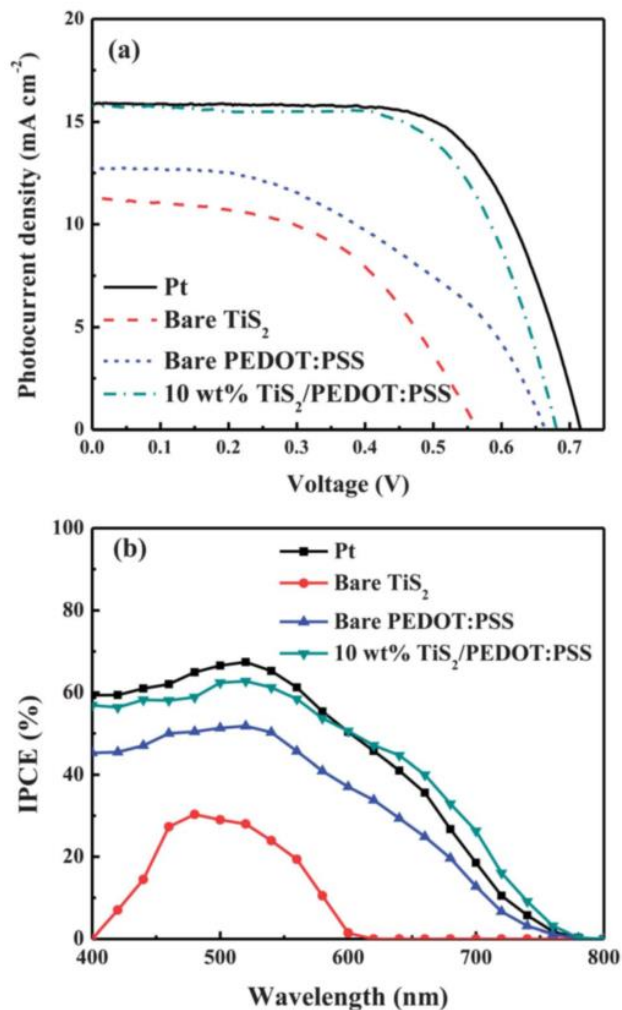

**Figure 13** (a) Photocurrent density-voltage curves of DSSCs with Pt, bare TiS$_2$, bare PEDOT:PSS, and 10 wt.% TiS$_2$/PEDOT:PSS composite based CEs recorded under light illumination of 100 mW/cm$^2$ (AM 1.5). (b) incident photon-to-current conversion efficiency (IPCE) curves of the DSSCs with similar CEs. Reprinted with permission from ref. 200. Copyright 2013, The Royal Society of Chemistry.

### 3.4 NiS$_2$ Counter Electrodes

Nickel Disulfide (NiS$_2$) is a semiconducting material with pyrite structure which acquires unique morphologies such as hollow prisms, nano/microspheres, nanocubes, nanosheets, nanoparticles, and also exhibits interesting electrical, optical, magnetic, and catalytic properties.[201-205] Depending upon the synthesis procedures, the stoichiometric composition of nickel sulfide varies to a great extent (NiS, NiS$_2$, Ni$_3$S$_2$, Ni$_3$S$_4$, Ni$_6$S$_5$, Ni$_7$S$_6$, Ni$_9$S$_8$, etc).

In one study, hierarchical $NiS_2$ hollow microspheres on a FTO substrate were prepared by a hydrothermal method to use as a CE for a DSSC.[206] The $NiS_2$ hollow microspheres were partially broken, offering more active sites for electrocatalysis and electrolyte adsorptions. The IPCE values of 81.3% for the $NiS_2$ microspheres CE and 76.6% for the Pt CEs at 500 nm were observed. The peak current density of the $NiS_2$ microspheres CE was found to be higher than the Pt CE, whereas the peak-to-peak separation ($E_{pp}$) value was lower by 10 mV compared to Pt CE, which suggests a high electrocatalytic activity for the $NiS_2$ microspheres CE in the reduction of triiodide ($I_3^-$) in the electrolyte. The $NiS_2$ hollow microspheres CE based DSSC showed a PCE of 7.84%, equal to the Pt CE (7.89%), indicating their potential as low-cost CEs for DSSCs. The $NiS/NiS_2$ composite hollow spheres prepared by a solvothermal method exhibited a low $R_{CT}$ of $0.34\ \Omega\,cm^{-2}$ at the CE/electrolyte interface, and a PCE of 7.66 %, outperforming the Pt CE (7.01 %), and showed high electrocatalytic activity for $I_3^-$ reduction, and also better electrochemical stability.[207] The NiS and $NiS_2$ hollow spheres were synthesized through a solvothermal process.[208] The Ni/S molar ratio controlled the different stoichiometric ratios of nickel sulfides. The $NiS_2$ CE based DSSC showed a higher electrocatalytic activity than that of the NiS CE for $I_3^-$ reduction. The DSSC with $NiS_2$ CE yielded a PCE value of 7.13% in comparison to 6.49% for NiS CE.

$NiS_2$ polyhedrons were studied as CEs for DSSCs by Zheng et al.[209] Figure 14 shows SEM, TEM, and selected area electron diffraction (SAED) images of $NiS_2$ octahedrons and $NiS_2$ cubes, and electrochemical characteristics of DSSCs having $NiS_2$ octahedrons, $NiS_2$ cubes and Pt CEs, under simulated AM1.5G solar light. The average size of $NiS_2$ octahedrons and cubes were about 250 nm. The electrochemical performance of $NiS_2$ octahedron and cube based CEs were evaluated using CV, J-V characteristics, EIS, and the Tafel polarization method. The $NiS_2$ octahedron CEs showed peak current density of 1.40 mA/cm$^2$, compared to 1.22 mA/cm$^2$ for the NiS$_2$ cubes, indicating better electrocatalytic activity. The NiS$_2$ octahedron CEs also exhibited a higher $J_{SC}$ of 13.55 mA/cm$^2$ and FF of 62%, higher than that of the NiS$_2$ cube CE ($J_{SC}$ of 12.62 mA/cm$^2$ and FF of 60%), giving rise to a higher PCE. Octahedral NiS$_2$ nanocrystals based CEs incorporated into DSSCs exhibited a PCE of 5.98%, slightly higher than that of the NiS$_2$ cube nanocrystals ($\eta$ = 5.43%). The NiS$_2$ octahedron CE had a PCE of up to 91% of the conventional Pt CE in DSSCs ($\eta$ = 6.55%). The R$_s$ of the NiS$_2$ octahedron based CE was 13.14 $\Omega \cdot$cm$^2$, somewhat lower than that of the NiS$_2$ cube CE (R$_s$ of 14.98 $\Omega \cdot$cm$^2$), indicating higher electrical conductivity of the NiS$_2$ octahedron based CE. The R$_{CT}$ of the NiS$_2$ octahedron CE was measured as 9.86 $\Omega \cdot$cm$^2$, also lower than that of the NiS$_2$ cube CE (R$_{CT}$ of 13.17 $\Omega \cdot$cm$^2$), which demonstrates that the NiS$_2$ octahedrons with {111} facets possesses better electrocatalytic activity than the NiS$_2$ cubes with {100} facets.

The electrocatalytic performance of NiS$_2$ nanoparticles and their nanocomposites with RGO were compared by Li et al.[210] In a hydrothermal process, graphene oxide was transformed to RGO, and then NiS$_2$@RGO nanocomposites were formed by depositing NiS$_2$ nanoparticles on the surface of RGO. CEs for DSSCs were fabricated by drop-casting solutions of NiS$_2$, NiS$_2$@RGO, and RGO nanocomposites on FTO-coated glass substrate. The surface areas measured by the BET method were 11.4, 9.4, and 8.6, and 5.8 m$^2$/g for NiS$_2$@RGO, NiS$_2$, and RGO, respectively. Figure 15 shows *J-V* curves of DSSCs with bare NiS$_2$, NiS$_2$@RGO nanocomposites, bare RGO, and Pt CEs. The NiS$_2$@RGO nanocomposites based CE showed a PCE of 8.55% (*Jsc* = 16.55 mA/cm$^2$, *Voc* = 0.749 V, and FF = 0.69), much higher than that of the NiS$_2$ CE ($\eta$ = 7.02%), RGO CE ($\eta$ = 3.14%), or standard Pt CE ($\eta$ = 8.15%) for the DSSCs under the same experimental conditions. The larger R$_{CT}$ values of 100.2 $\Omega \cdot$cm$^2$ for RGO and 8.8 $\Omega$.cm$^2$ for NiS$_2$ also suggest the low electrocatalytic activity. On the other hand, the smaller R$_{CT}$ value of 2.9 $\Omega$.cm$^2$ for the NiS$_2$@RGO nanocomposite indicates much higher electrocatalytic activity for the reduction of triiodide (I$_3{}^-$) in electrolyte due the cooperative synergetic effect and the increased conductivity from the RGO nanosheets. This study demonstrates that NiS$_2$@RGO nanocomposites are a promising alternate CE to conventional Pt CE for DSSC devices.

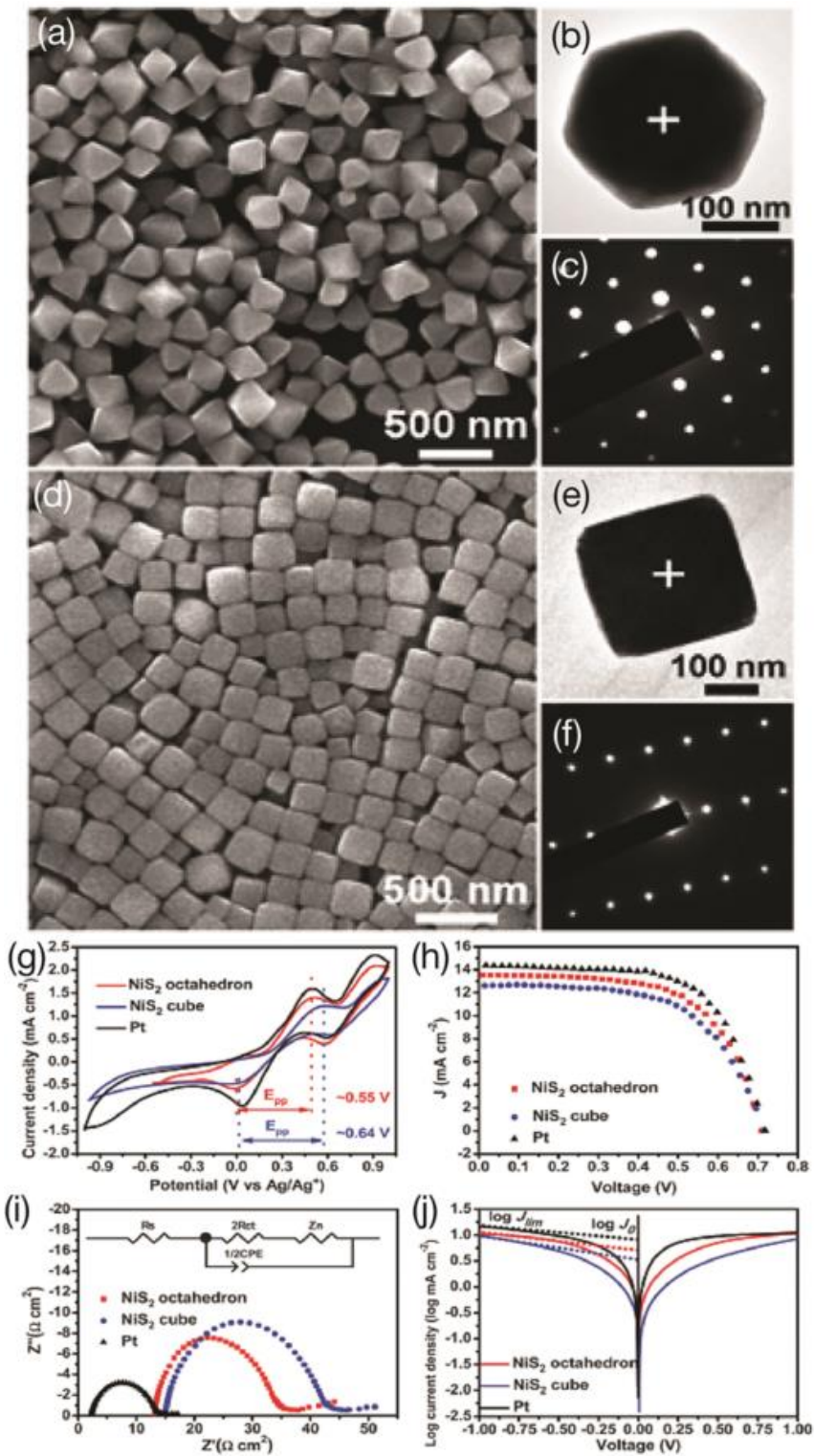

**Figure 14.** SEM, TEM, and SAED images of NiS$_2$ octahedrons (a–c) and NiS$_2$ cubes (d–f). SAED patterns of NiS$_2$ octahedrons (c) and NiS$_2$ cubes (f) were recorded from the corresponding particles depicted (b) and (e), respectively. (g) C-V curves of DSSCs having counter electrodes

of NiS$_2$ octahedrons, NiS$_2$ cubes and Pt for the reduction of tri-iodide (h) J–V curves of DSSCs with NiS$_2$ octahedrons, NiS$_2$ cubes and Pt CEs under simulated AM1.5G solar light. Nyquist plots (i) and Tafel polarization curves (j) of DSSCs having NiS$_2$ octahedrons, NiS$_2$ cubes and Pt CEs. Reprinted with permission from ref. 209. Copyright 2015, The Royal Society of Chemistry.

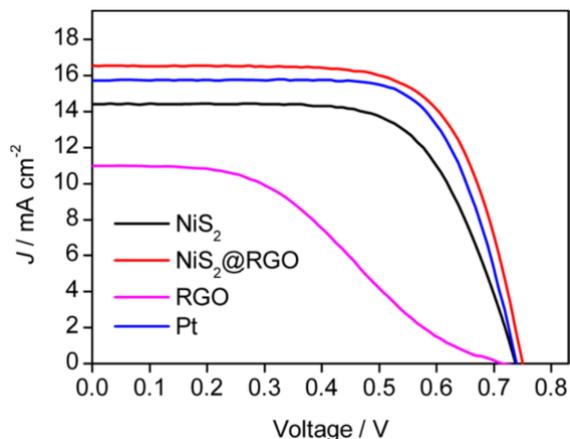

**Figure 15.** (a) Photocurrent density-voltage (*J-V*) curves of DSSCs with bare NiS$_2$, NiS$_2$@RGO nanocomposites, bare RGO, and Pt CEs measured under light illumination of 100 mW/cm$^2$ (AM 1.5). Reprinted with permission from ref. 210. Copyright 2013, American Chemical Society.

### 3.5 FeS$_2$ Based Counter Electrodes

Pyrite iron disulfide (NiS$_2$) acquires unique morphological structures including nanocrystals, nanowires, nanosheets, nanocubes, and exhibits interesting electrical, photovoltaic, and catalytic properties.[211-220] An interesting comparative study was done by Shukla et al.[221] on pyrite iron disulfide (FeS$_2$) as a CE material in comparison with Pt and PEDOT CEs in DSSCs. The FeS$_2$ film CEs were fabricated on a FTO glass substrate by a spray pyrolysis method and used in I$_3^-$/I$^-$ and Co(III)/Co(II) electrolyte-mediated DSSCs. N719 dye was used for the DSSC with I$_3^-$/I$^-$ redox electrolyte and C128 dye for the [Co(bpy)3]$^{2+/3+}$ redox electrolyte. The I$_3^-$/I$^-$ redox electrolyte contained 1.0 mM of 1,3-dimethylimidazolium iodide, 50 mM of LiI, 30 mM of I$_2$, 0.5 mM of *tert*-butylpyridine, and 0.1 mM of guanidinium thiocyanate in a acetonitrile and valeronitrile (v/v, 85/15) mixed solution. The cobalt electrolyte was made up of 0.22 M of Co(bpy)$_3$(TFSI)$_2$, (bpy=2,2'-bipyridine and TFSI = [*bis*(trifluoromethane)-sulfonimide], 0.05 M of Co(bpy)$_3$(TFSI)$_3$, 0.1 M of lithium *bis*(trifluoromethanesulfonyl)imide (LiTFSI), and 0.2 M of *tert*-butylpyridine (tBP) in acetonitrile. Figure 16 shows the *J-V* curves and IPCE of the DSSCs with FeS$_2$ and Pt CEs in I$_3^-$/I$^-$ electrolyte, and DSSCs with FeS$_2$ and PEDOT CEs in the Co(III)/Co(II) electrolyte. The catalytic activity of the FeS$_2$ film CEs was found to be comparable to the Pt and PEDOT CEs, which were both in I$_3^-$/I$^-$ and Co$^{2+}$/Co$^{3+}$ electrolytes, respectively. With the I$_3^-$/I$^-$ electrolyte, a PCE of 7.97% was observed for the FeS$_2$ film CE and a PCE of 7.54% for the Pt CE in the I$_3^-$/I$^-$ electrolyte, whereas the PCEs were almost the same (6.3%) for the FeS$_2$ film and PEDOT CEs in the [Co(bpy)$_3$]$^{2+/3+}$ redox-mediated DSSCs. The performance of the DSSCs with FeS$_2$ and Pt CEs was studied by varying solar light illumination intensities between 1 Sun and 0.1 Sun**.** The current density was observed to be higher for the FeS$_2$ CE (15.20 mA/cm$^2$) than for the Pt CE (14.77 mA/cm$^2$) at 1 Sun light intensity. When the solar light intensity was reduced to 0.5 and 0.1 Sun, the difference in current density between the FeS$_2$ and Pt CEs was not noticeable, with the *Jsc* values being 8.97 and 8.91 mA/cm$^2$ at 0.5 Sun, and 1.75 and 1.74 mA/cm$^2$ at 0.1 Sun for the Fe and Pt CEs, respectively. The PCE values for the FeS$_2$ and Pt CEs were 9.27% and 8.92% at 0.5 Sun, and 8.68% and 8.32% at 0.1 Sun, respectively. The excellent performance of FeS$_2$ film in both electrolyte systems makes it very interesting for applications in DSSCs.

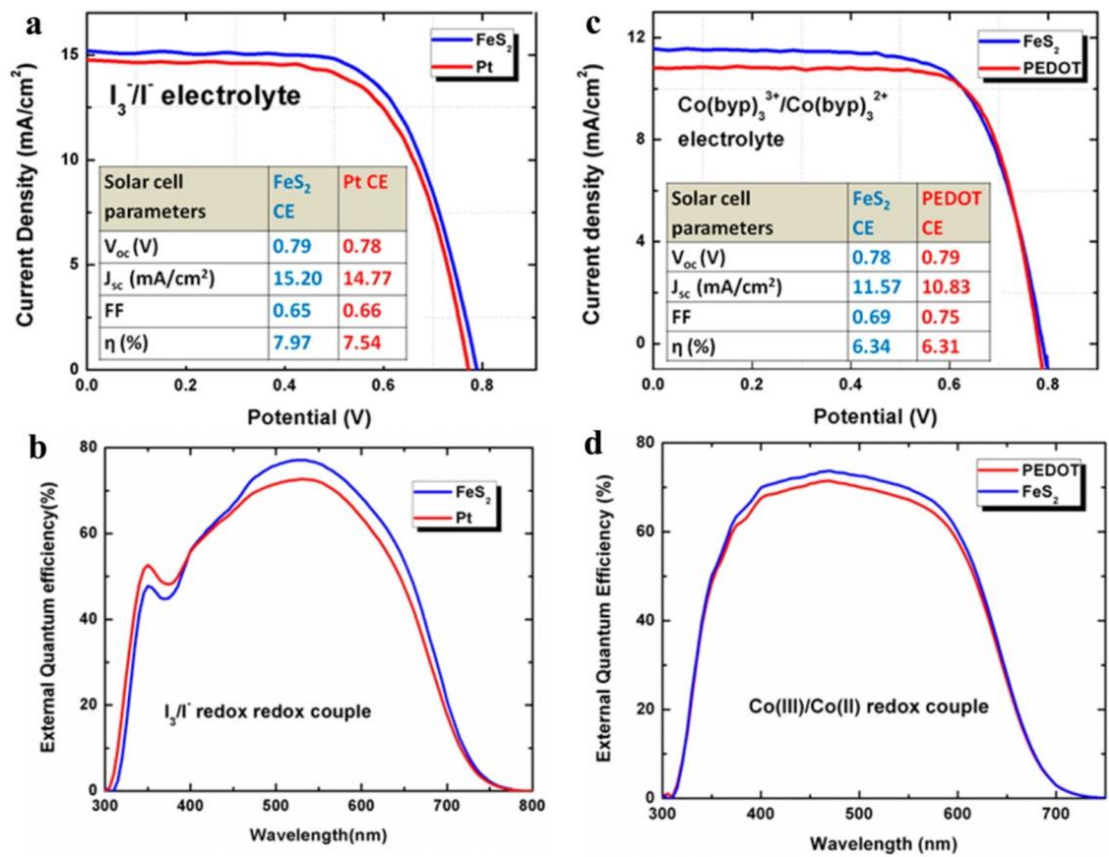

**Figure 16.** (a) Photocurrent density-voltage (*J-V*) curves and (b) IPCE of the DSSCs with FeS$_2$ and Pt counter electrodes using N719 dye in I$_3^-$/I$^-$ electrolyte. (c) J-V curve and (d) IPCE of the DSSCs with FeS$_2$ and PEDOT counter electrodes suing C128 dye in Co$^{3+}$/Co$^{2+}$ electrolyte. Reprinted with permission from ref. 221. Copyright 2014, American Chemical Society.

In an interesting study, FeS$_2$ nanorod arrays were fabricated on a FTO substrate after sulfurizing FeO(OH) nanorods, and used as a CE for DSSCs.[222] The FeS$_2$ nanorods exhibited better electrocatalytic activity than FeS$_2$ films and Pt–based CEs due to more active sites, which resulted in high *J$_{SC}$* value of the DSSCs. The FeS$_2$ nanorods-based CEs showed lower interface resistance compared with FeS$_2$ thin films, which leads to a higher FF and hence a higher PCE comparable to Pt CE based DSSCs. The electrochemical stability of the FeS$_2$ nanorod arrays-based CE measured in I$_3^-$/I$^-$ electrolyte showed a slight change in CV plots up to 10 consecutive days of aging time. In another study, FeS$_2$ powder prepared through a hydrothermal method was used as a CE for fabricating DSSCs.[223] The effect of NaOH addition on FeS$_2$ crystal size and electrocatalytic activities was then studied. It was observed that the size of FeS$_2$ nanoparticles decreased after adding NaOH, and the resulting photovoltaic performance and electrocatalytic activity of DSSC with FeS$_2$ powder significantly increased, achieving a PCE of 5.78% under simulated sunlight irradiation of 1 Sun.

A chemically prepared FeS$_2$ nanocrystal ink was used to fabricate a CE for a DSSC, which showed a PCE of 7.31% after ethanedithiol (EDT) treatment.[224] FeS$_2$ nanocrystal ink casted on a flexible ITO/PET substrate exhibited a $J_{SC}$ of 14.93 mA/cm$^2$, $V_{OC}$ of 0.71 V, FF of 0.60, and a PCE of 6.36%. The semi-transparent FeS$_2$ nanocrystal/ITO glass CE has an optical transmittance of 50–70% between 300 to 800 nm in comparison to 15% transmittance for the reference Pt/ITO glass CE. When the DSSC with the FeS$_2$ nanocrystal CE was illuminated from the rear side, it showed a PCE of 4.17%, which was 57% of the front illumination value, whereas opaque Pt CE had a PCE of 1.06% from the rear side. Semi-transparent FeS$_2$ nanocrystal CEs offer bifacial DSSCs utilizing incident light from both front and rear sides, and thus could be cost-effective for energy production. The FeS$_2$ nanocrystal ink also demonstrated high electrocatalytic activity and electrochemical stability. Additionally, MWCNT/TiO$_2$ hybrid and pure TiO$_2$ mesoporous photoanodes with FeS$_2$ thin films as CEs were studied for DSSCs by Kilic et al.[225] In the MWCNT/TiO$_2$ hybrid photoanode, CNT played an important role of increasing the optical absorption and shifting it toward a longer wavelength region, where the bandgap of 3.15 eV for mesoporous TiO$_2$ shifted to 2.5 eV for the MWCNT/TiO$_2$ hybrid. The DSSC with the MWCNT/TiO$_2$ hybrid photoanode and the Pt CE showed values of $J_{SC}$ of 15.96 mA/cm$^2$, $V_{OC}$ of 0.77 V, FF of 0.57 and a PCE of 7.0%. The DSSC with the pure mesoporous TiO$_2$ photoanode and the Pt CE both resulted in PCEs of 6.51%. The enhancement in PCE value of the hybrid photoanode is associated with MWCNTs, which offer an electrical conduction pathway for speedy electron transport. The MWCNT/TiO$_2$ hybrid photoanode also showed an increase in IPCE in the 350-600 nm wavelength range compared to the mesoporous TiO$_2$ photoanode. When FeS$_2$ thin films were used as a CE with a MWCNT/TiO$_2$ hybrid photoanode, the PCE of the DSSC increased to 7.27% under 1 Sun. The DSSC with a FeS$_2$ CE and a pureTiO$_2$ photoanode both yielded a PCE of 6.65%. The FeS$_2$ thin films showed an optical bandgap of 1.27 eV and large effective surface area, which contribute to more light absorption and increased electrocatalytic activity for the reduction of triiodide (I$_3^-$).

## 3.6 CoS$_2$ Counter Electrodes

Cobalt disulfide (CoS$_2$) is a semiconducting material that exhibits interesting magnetic, electrical, and catalytic properties for energy storage applications.[226-231] The properties of CoS$_2$ nanocrystalline thin films prepared by a hydrothermal method were reported by Jin et al.[232] CoS$_2$ powder was dispersed in ethanol in order to prepare a nanoink (40 mg m/L) for fabricating a CE. The CoS$_2$ nanoink (10 μL) was drop-cast on an FTO glass or flexible ITO/PET substrate, followed by ethanol evaporation. The self-assembled CoS$_2$ film CE has a thickness of 2.5 μm which can be controlled by the nanoink content. DSSCs were assembled by using a N719 dye-sensitized TiO$_2$ photoanode as the working electrode, a CoS$_2$ nanocrystal film or a Pt as the CE, and an electrolyte solution containing LiI (0.1 M), I2 (0.05 M) I2, 1,2-dimethyl-3-n-propylimidazolium iodide (DMPII, 0.6 M), and 4-*tert*-butylpyridine (TBP, 0.5 M) in acetonitrile. The morphology and structure of the CoS$_2$ self-assemblies were studied by SEM and TEM techniques. Figure 17 shows an SEM image of a self-assembled CoS$_2$ nanocrystal film, a photographic image of CoS$_2$ nanoink, a self-assembled CoS$_2$ CE, and *J-V* curves of the DSSCs, fabricated with the CoS$_2$ nanocrystal thin film and a Pt CE on the FTO and flexible ITO/polyethylene terephthalate (PET) substrate. The CoS$_2$ nanoparticles were stabilized with poly(vinylpyrrolidone) (PVP) to form an oriented structure. The DSSC with self-assembled CoS$_2$ CE shows a $J_{SC}$ of 14.62 mA/cm$^2$, $V_{OC}$ of 0.71 V, and a FF of 0.64, resulting in a PCE of 6.78%, which is comparable to a Pt CE with a value of 7.38%. The PCE was found to decrease when the thickness of the self-assembled CoS$_2$ film was either higher or lower than 2.5 μm. The CoS$_2$ shows a R$_s$ value of 34.20 Ω.cm$^2$, slightly higher compared to Pt (27.13 Ω.cm$^2$), indicating comparable electrical conductivity. The R$_{CT}$ value of 7.21 Ω.cm$^2$ for the CoS$_2$ CE indicates higher electrocatalytic activity for the reduction of triiodide (I$_3^-$). The CoS$_2$ CE has a high chemical capacitance (Cμ) of 27.57 μF, compared to 2.76 μF for the Pt CE, which also is evidence of a high surface area for CoS$_2$ CE which is beneficial to electrocatalysis. The DSSCs with CoS$_2$ deposited on a flexible ITO/PET substrate showed a $J_{SC}$ of 13.17 mA/cm$^2$, $V_{OC}$ of 0.70 V, FF of 0.68 and a PCE of 6.40%. The fabrication of a CoS$_2$ CE is low-cost and solution processable at room temperature, which could make it an alternative to Pt CEs for flexible DSSCs.

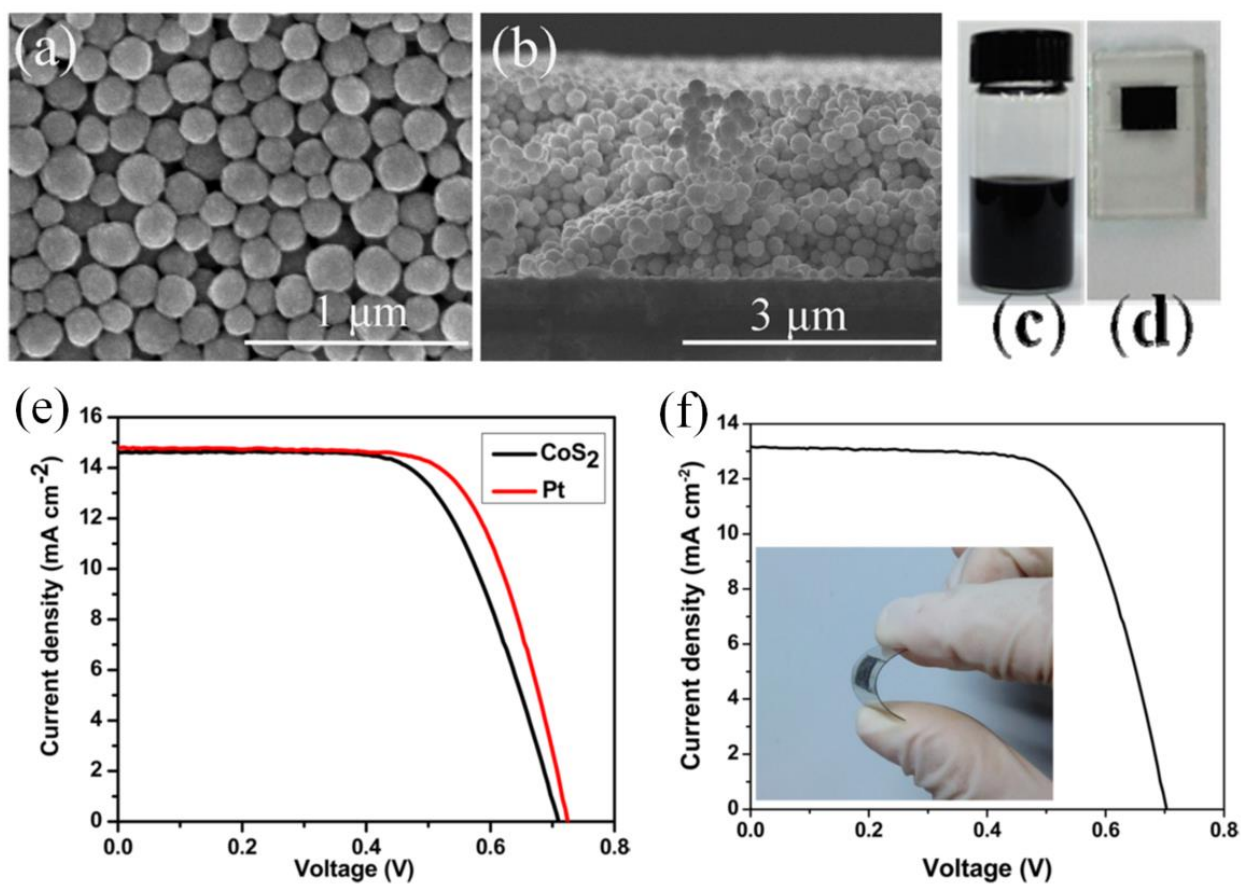

**Figure 17.** (a) SEM image of a self-assembled $CoS_2$ nanocrystal film on FTO glass substrate, (b) cross-sectional SEM image of the $CoS_2$ counter electrode (CE), (c) photographic image of $CoS_2$ nanoink, and (d) self-assembled $CoS_2$ CE. (e) Photocurrent density-voltage (*J-V*) curves of the DSSCs with $CoS_2$ nanocrystal thin film and Pt CEs. (f) *J−V* curves of the DSSCs with $CoS_2$ deposited on a flexible ITO/PET substrate. Reprinted with permission from ref. 232. Copyright 2014, American Chemical Society.

Another study used mesoporous $CoS_2$ nanotube arrays deposited on an FTO glass substrate and used as a CE for a DSSC.[233] The $CoS_2$ nanotube arrays were characterized by SEM, TEM, and XRD techniques for their morphology and crystal structures. The electrocatalytic properties of the $CoS_2$ nanotube arrays were measured using CV and Tafel polarization curve measurements. The DSSCs having $CoS_2$ CEs achieved a PCE of 6.13%, comparable to that of sputtered Pt CE (6.04 %). The $R_{CT}$ of the mesoporous $CoS_2$ nanotube array CE was found to be 3.51 $\Omega.cm^2$ and comparable to the Pt CE (5.78 $\Omega.cm^2$). The $CoS_2$ nanotube array based CEs have large active surface area due to the mesoporous nanotube structure. The $CoS_2$ nanotube array CE also exhibits electrocatalytic activities comparable to Pt CE. Tsai et al.[234] also prepared $CoS_2$ nanoflake arrays from $Co(OH)_2$ nanoflake arrays through an ion exchange reaction to develop CEs for DSSCs. The $CoS_2$ nanoflakes were found to be composed of $CoS_2$ single crystals as well as their aggregates. The DSSC with $CoS_2$ nanoflake arrays as a CE showed a PCE of 5.20%, comparable to a sputtered Pt CE (5.34%).

Different types of cobalt sulfide (CoS) have been used as CEs for DSSCs. The CoS nanoparticles deposited onto FTO glass substrates showed good transparency and high electrocatalytic activity for the I-/$I_3$- redox couple for a DSSC.[235] CoS nanoparticle CEs showed a low $R_{CT}$ value of 1.3 $\Omega.cm^2$, less than that of Pt on FTO glass ($R_{CT}$ of 2.3 $\Omega.cm^2$) and achieved a PCE of 6.6%. An optimized CoS nanoparticle CE was also studied for a ferrocene-based liquid electrolyte. The rose-petal like $CoS_2$ was deposited on an FTO as CE using a chemical bath deposition method.[236] The DSSC assembled with a $CoS_2$ CE achieved a PCE of 5.32%, higher than that of a Pt CE (5.02%). The PCEs of $CoS_2$ CEs depend on the deposition parameters including the concentrations of urea and thioacetamide, and the deposition time of the CEs. Also, $CoS_2$ embedded carbon nanocages were fabricated as CEs for DSSCs through a zeolitic imidazolate framework-67, Co(2-methylimidolate)$_2$ template.[237] The performance of the $CoS_2$ CE in a DSSC was optimized via a sulfurization process, where $CoS_2$ nanoparticles with embedded carbon nanocages were sulfurized for a period of 4 hours, and showed the highest PCE of 8.20%, even higher than Pt-based CE (7.88%). The synergic effect of $CoS_2$ nanoparticles and the carbon matrix resulted in the CE having high electrical conductivity and catalytic activity. Kim et al.[238] deposited $CoS_2$, nickel sulfide (NiS), and Ni-doped $CoS_2$ nanoparticles on a FTO substrate as CEs for DSSCs via a chemical bath deposition method. The surface morphology of the thin films was analyzed by SEM. Electrochemical properties of Ni-doped $CoS_2$ thin films evaluated by EIS, CV, and Tafel polarization curves indicated increased electrocatalytic activity for the reduction of $I_3^-$ in the DSSCs compared to Pt CEs. The Ni-doped $CoS_2$ CE (15% Ni) showed a PCE of 5.50% under 1 Sun illumination, exceeding the PCE of the Pt CE ($\eta = 5.21\%$). PCE and $R_{CT}$ values of the DSSCs were found to depend on the amount of Ni-doping of the $CoS_2$ nanoparticles. These yielded PCEs of 4.81, 5.17, 5.50, and 4.12%, for $R_{CT}$ values of 279.7, 36.63, 8.53 and 82.72 $\Omega.cm^2$, at 5, 10, 15 and 20% Ni contents in the $CoS_2$ CE, respectively. Comparatively, DSSCs with bare $CoS_2$ and NiS CEs showed poor electrocatalytic activity of $I_3^-$ reduction.

In another study, $CoS_2$/graphene composites were prepared *via* a hydrothermal method using Co ions with thiourea in the presence of graphene oxide (GO).[239] The distribution and size of the $CoS_2$ nanoparticles deposited onto a flexible graphene sheet was controlled in order to optimize the electrocatalytic activity for $I_3^-$ reduction. A $CoS_2$ nanoparticles/graphene sheet ($CoS_2/G_{50}$) CE was prepared by incorporating 50 mg graphene oxide, and exhibited the lowest electrolyte diffusion resistance and the highest electrocatalytic activity. The DSSC with $CoS_2/G_{50}$ CE achieved a PCE of 6.55%, higher than bare $CoS_2$ or graphene CEs or a conventional Pt CE ($\eta = 6.20\%$). Also, $CoS_2$/RGO composite films for a CE of a DSSC were prepared using the layer-by-layer (LbL) assembly method, followed by thermal annealing.[240] The photovoltaic parameters of the $CoS_2$/RGO CE based DSSCs were found to depend upon the deposition times of graphene oxide. PCE values of 2.6, 4.1, 5.4, 2.9 and 1.4% were measured for 2, 4, 6, 8 and 10 deposition times of graphene oxide, respectively. It appeared that the lowest $R_{CT}$ of 4.8 $\Omega \cdot cm^2$ was observed for a $CoS_2$/RGO CE prepared with 6 deposition times.

### 3.7 SnS$_2$ Counter Electrodes

Tin disulfide (SnS$_2$) attains morphological structures which include nanocrystals, nanosheets, nanowires, nanobelts, and these show potential for field-effect transistors, gas sensors, photocatalysts, and solar cells.[241-248] In one study, semitransparent SnS$_2$ nanosheets were prepared as a CE to develop a Pt-free DSSC for the reduction of I$_3^-$.[249] The SnS$_2$-based CE with 300 nm thickness showed high electrocatalytic activity, with a PCE of 7.64 % compared to a PCE of 7.71% for a Pt CE based DSSC. When SnS$_2$ nanosheets were functionalized with carbon nanoparticles, the CE exhibited a PCE of 8.06%, a better electrocatalytic performance than a Pt CE. SnS$_2$ nanosheets could therefore be used as a low cost electrocatalytic electrode material for DSSCs. Yang et al.[250] prepared SnS$_2$ nanoparticles and a RGO nanocomposite as a Pt-free CE for a DSSC. The SnS$_2$ nanoparticles dispersed onto RGO sheets exhibited improved electrocatalytic activity for reducing I$_3^-$, and also increased conductivity. Figure 18 illustrates the preparation of SnS$_2$@RGO nanocomposites and compares $J-V$ curves for DSSCs having Pt, RGO, SnS$_2$, and SnS$_2$@RGO composite CEs. The DSSC with SnS$_2$@RGO nanocomposite CE showed a PCE of 7.12%, much higher than that of the RGO sheet alone (3.73 %) and SnS$_2$ nanoparticles (5.58 %),and a comparable PCE value to Pt CE (6.79 %). The Z$_N$ values were 0.64, 4.36, 5.01 and 0.95 $\Omega$ for the SnS$_2$@RGO nanocomposite, SnS$_2$, RGO and Pt CEs, respectively. The RGO and SnS$_2$ CEs have larger R$_{CT}$ values, hence lower electrocatalytic ability among these CEs, whereas the SnS$_2$@RGO composite has a lower R$_{CT}$ value, leading to a higher electrocatalytic activity. The role of the RGO sheets is to facilitate the conduction pathway. The E$_{pp}$ of the SnS$_2$@RGO nanocomposite CE is 593 mV, similar to a Pt CE (E$_{pp}$ of 605 mV). This indicates that the electrocatalytic activity of the SnS$_2$@RGO nanocomposite CE is similar to that of a conventional

Pt CE. The electrochemical stability of the SnS$_2$@RGO nanocomposite CE was analyzed up to 25 cycles of CV curves, where no drastic change was noticed in peak current density, confirming that SnS$_2$@RGO nanocomposite CEs are stable for electrocatalysis of I$_3^-$ in the electrolyte. The SnS$_2$@RGO nanocomposite CEs are as promising as Pt CEs in DSSCs. Table 3 lists a summary of the photovoltaic parameters of WS$_2$, NiS$_2$, TiS$_2$, FeS$_2$, CoS$_2$, and SnS$_2$ based CEs for DSSCs and their comparison with a standard Pt CE.

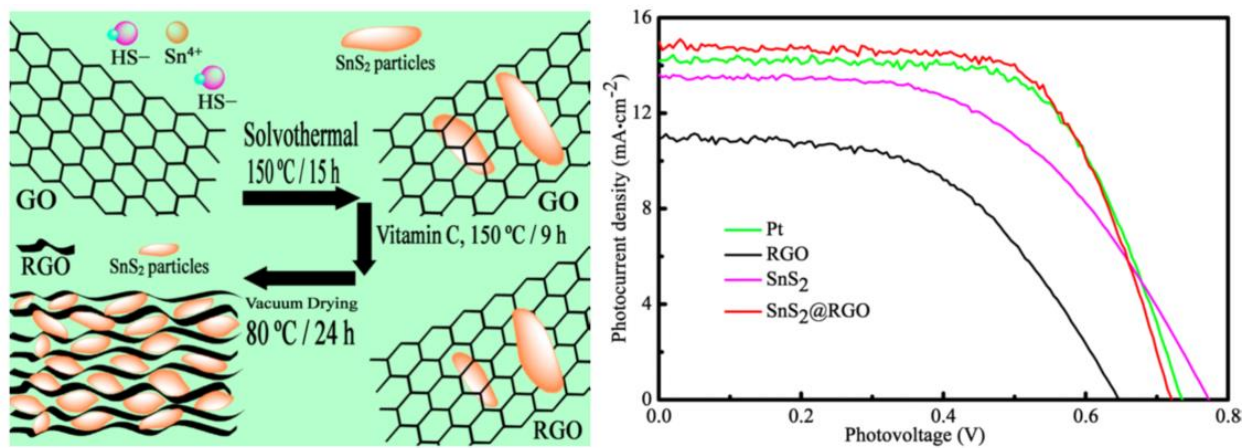

**Figure 18.** (Left) Illustration depicting preparation of tin sulfide nanoparticles/reduced graphene oxide (SnS$_2$@RGO) Nanocomposites. (Right) Photocurrent density−photovoltage *(J−V)* curves for DSSCs having Pt, RGO, SnS$_2$, and SnS$_2$@RGO composite CEs. Reprinted with permission from ref. 250. Copyright 2015, American Chemical Society.

**Table 3.** Photovoltaic parameters of WS$_2$, NiS$_2$, TiS$_2$, FeS$_2$, CoS$_2$, and SnS$_2$ based CEs used in DSSCs. FTO glass is the common substrate used in assembling DSSCs with different CE materials. The measurements were conducted at a simulated solar light intensity of 100 mW/cm$^2$ (AM 1.5G) unless specified. The photovoltaic parameters short-circuit photocurrent density ($J_{SC}$), open-circuit voltage ($V_{OC}$), fill factor (FF), and power conversion efficiency (η), series resistance (R$_s$), charge-transfer resistance (R$_{CT}$), electrolyte and dye used for DSSCs are summarized and compared with standard Pt counter electrode.

| Counter Electrodes | Redox Couples | Dye | $J_{SC}$ (mA/cm$^2$) | $V_{OC}$ (V) | FF (%) | PCE (η, %) | R$_s$ (Ω.cm$^2$) | R$_{CT}$ (Ω.cm$^2$) | Ref. |
|---|---|---|---|---|---|---|---|---|---|
| WS$_2$ | I$^-$/I$_3^-$ | N719 | 12.1 | 0.662 | 55 | 4.4 | - | - | 181 |
| WS$_2$ (glucose solution, 0.3 M) | I$^-$/I$_3^-$ | N719 | 12.8 | 0.658 | 63 | 5.3 | - | - | 181 |
| WS$_2$ (glucose solution, 0.6 M) | I$^-$/I$_3^-$ | N719 | 13.1 | 0.670 | 62 | 5.5 | - | - | 181 |
| WS$_2$ (glucose solution, 1.2 M) | I$^-$/I$_3^-$ | N719 | 12.4 | 0.675 | 63 | 5.3 | - | - | 181 |
| Pt reference | $I^-/I_3^-$ | N719 | 13.2 | 0.668 | 63 | 5.6 | - | - | 181 |
| | | | | | | | | | |
| WS$_2$ (sputtering time, 10 min) | $I^-/I_3^-$ | N719 | 13.43 | 0.71 | 66 | 6.3 | - | - | 182 |
| WS$_2$ (sputtering time, 15 min) | $I^-/I_3^-$ | N719 | 15.01 | 0.69 | 55 | 5.8 | - | - | 182 |
| Pt reference | $I^-/I_3^-$ | N719 | 16.50 | 0.66 | 61 | 6.8 | - | - | 182 |
| | | | | | | | | | |
| WS$_2$ (hydrothermal method) | $I^-/I_3^-$ | N719 | 11.28 | 0.72 | 59 | 4.79 | 4.86 | 5.13 | 185 |
| WS$_2$ /MWCNTs (3 wt.%) | $I^-/I_3^-$ | N719 | 12.65 | 0.73 | 59 | 5.45 | 3.75 | 3.47 | 185 |
| WS$_2$ /MWCNTs (5 wt.%) | $I^-/I_3^-$ | N719 | 13.51 | 0.73 | 65 | 6.41 | 3.01 | 2.53 | 185 |
| WS$_2$ /MWCNTs (10 wt.%) | $I^-/I_3^-$ | N719 | 12.09 | 0.72 | 60 | 5.22 | 4.17 | 4.59 | 185 |
| MWCNTs | $I^-/I_3^-$ | N719 | 10.77 | 0.66 | 61 | 4.34 | 6.52 | 6.60 | 185 |
| Pt reference | $I^-/I_3^-$ | N719 | 13.23 | 0.74 | 67 | 6.56 | 2.26 | 2.74 | 185 |
| | | | | | | | | | |
| WS$_2$ (hydrothermal method) | $I^-/I_3^-$ | N719 | 11.72 | 0.72 | 63 | 5.32 | 2.86 | 4.60 | 186 |
| WS$_2$ /MWCNTs (1 wt.%) | $I^-/I_3^-$ | N719 | 11.95 | 0.73 | 63 | 5.50 | 2.80 | 3.86 | 186 |
| WS$_2$ /MWCNTs (5 wt.%) | $I^-/I_3^-$ | N719 | 13.63 | 0.75 | 72 | 7.36 | 2.54 | 2.49 | 186 |
| WS$_2$ /MWCNTs (7 wt.%) | $I^-/I_3^-$ | N719 | 12.75 | 0.75 | 68 | 6.50 | 2.67 | 2.94 | 186 |
| WS$_2$ /MWCNTs (10 wt.%) | $I^-/I_3^-$ | N719 | 12.47 | 0.74 | 68 | 6.27 | 2.78 | 3.46 | 186 |
| WS$_2$ /MWCNTs* | $I^-/I_3^-$ | N719 | 12.65 | 0.73 | 59 | 5.45 | 2.85 | 3.47 | 186 |
| MWCNTs | $I^-/I_3^-$ | N719 | 10.77 | 0.66 | 61 | 4.34 | 2.95 | 6.60 | 186 |
| Pt reference | $I^-/I_3^-$ | N719 | 13.23 | 0.76 | 75 | 7.54 | 2.27 | 2.74 | 186 |
| | | | | | | | | | |
| TiS$_2$ nanosheets | $I^-/I_3^-$ | N719 | 17.48 | 0.73 | 60.3 | 7.66 | - | - | 199 |
| TiS$_2$/graphene hybrid | $I^-/I_3^-$ | N719 | 17.76 | 0.72 | 68.5 | 8.80 | 2.32 | 0.63 | 199 |
| Graphene | $I^-/I_3^-$ | N719 | 15.41 | 0.71 | 48.4 | 5.33 | - | - | 199 |
| Pt reference | $I^-/I_3^-$ | N719 | 16.93 | 0.72 | 65.6 | 8.00 | 6.90 | 1.32 | 199 |
| | | | | | | | | | |
| TiS$_2$ (drop coating method) | $I^-/I_3^-$ | N719 | 11.27 | 0.565 | 51 | 3.24 | 16.12 | - | 200 |
| TiS$_2$ /PEDOT:PSS (5 wt.%) | $I^-/I_3^-$ | N719 | 13.81 | 0.686 | 62 | 5.91 | - | - | 200 |
| TiS$_2$ /PEDOT:PSS (10 wt.%) | $I^-/I_3^-$ | N719 | 15.78 | 0.681 | 66 | 7.04 | 15.78 | 4.78 | 200 |
| PEDOT:PSS | $I^-/I_3^-$ | N719 | 12.74 | 0.664 | 46 | 3.91 | 14.92 | 7.27 | 200 |
| Pt reference | $I^-/I_3^-$ | N719 | 15.83 | 0.716 | 68 | 7.65 | 14.29 | 3.02 | 200 |
| | | | | | | | | | |
| NiS$_2$-Hollow microspheres | $I^-/I_3^-$ | N719 | 17.48 | 0.712 | 63 | 7.84 | 10.56 | 9.68 | 206 |
| Pt reference | $I^-/I_3^-$ | N719 | 17.04 | 0.747 | 62 | 7.89 | 12.78 | 8.76 | 206 |
| | | | | | | | | | |
| NiS$_2$-Octahedron | $I^-/I_3^-$ | N719 | 13.55 | 0.712 | 62 | 5.98 | 13.14 | 9.86 | 209 |
| NiS$_2$-Cube | $I^-/I_3^-$ | N719 | 12.62 | 0.715 | 60 | 5.43 | 14.98 | 13.17 | 209 |
| Pt reference | $I^-/I_3^-$ | N719 | 14.37 | 0.718 | 63 | 6.55 | 2.24 | 6.25 | 209 |
| | | | | | | | | | |
| NiS$_2$ (hydrothermal method) | $I^-/I_3^-$ | N719 | 14.42 | 0.738 | 66 | 7.02 | 5.1 | 8.8 | 210 |
| NiS$_2$/RGO | $I^-/I_3^-$ | N719 | 16.55 | 0.749 | 69 | 8.55 | 6.4 | 2.9 | 210 |
| Reduced Graphene Oxide (RGO) | $I^-/I_3^-$ | N719 | 10.98 | 0.716 | 40 | 3.14 | 14.2 | 100.2 | 210 |
| Pt reference | $I^-/I_3^-$ | N719 | 15.75 | 0.739 | 70 | 8.15 | 2.2 | 0.5 | 210 |
| | | | | | | | | | |
| FeS$_2$ (spray pyrolysis) | $I^-/I_3^-$ | N719 | 15.20 | 0.79 | 65 | 7.97 | - | - | 221 |
| Pt reference | $I^-/I_3^-$ | N719 | 14.77 | 0.78 | 66 | 7.54 | - | - | 221 |
| FeS$_2$ | $Co^{2+}/Co^{3+}$ | C128 | 11.57 | 0.78 | 69 | 6.34 | 4.9 | 7.2 | 221 |
| PEDOT | $Co^{2+}/Co^{3+}$ | C128 | 10.83 | 0.79 | 75 | 6.31 | 6.0 | 3.9 | 221 |
| FeS$_2$ films | I$^-$/I$_3^-$ | N719 | 12.56 | 0.658 | 57.8 | 4.78 | 9.60 | 213.1 | 222 |
| FeS$_2$ nanorods | I$^-$/I$_3^-$ | N719 | 13.68 | 0.653 | 65.7 | 5.88 | 9.61 | 11.0 | 222 |
| Pt reference | I$^-$/I$_3^-$ | N719 | 13.36 | 0.685 | 68.2 | 6.23 | 5.62 | 9.2 | 222 |
| FeS$_2$ (without NaOH) | I$^-$/I$_3^-$ | N719 | 10.20 | 0.70 | 66 | 4.76 | 3.73 | 13.6 | 223 |
| FeS$_2$ (with NaOH) | I$^-$/I$_3^-$ | N719 | 12.08 | 0.74 | 64 | 5.78 | 2.91 | 5.99 | 223 |
| Pt reference | I$^-$/I$_3^-$ | N719 | 11.58 | 0.74 | 69 | 5.93 | 3.09 | 1.16 | 223 |
| FeS$_2$ (with ethanedithiol) | I$^-$/I$_3^-$ | N719 | 15.14 | 0.71 | 68 | 7.31 | - | 1.60 | 224 |
| FeS$_2$ (without ethanedithiol) | I$^-$/I$_3^-$ | N719 | 12.63 | 0.71 | 64 | 5.74 | - | 4.45 | 224 |
| Pt reference | I$^-$/I$_3^-$ | N719 | 15.37 | 0.71 | 69 | 7.52 | - | 1.47 | 224 |
| FeS$_2$ (MWCNT/TiO$_2$ photoanode) | I$^-$/I$_3^-$ | N719 | 16.86 | 0.77 | 56 | 7.27 | - | - | 225 |
| FeS$_2$ (TiO$_2$ photoanode) | I$^-$/I$_3^-$ | N719 | 15.16 | 0.77 | 57 | 6.65 | - | - | 225 |
| Pt (MWCNT/TiO$_2$ photoanode) | I$^-$/I$_3^-$ | N719 | 15.96 | 0.77 | 57 | 7.00 | - | - | 225 |
| Pt (TiO$_2$ photoanode) | I$^-$/I$_3^-$ | N719 | 15.68 | 0.77 | 54 | 6.51 | - | - | 225 |
| CoS$_2$ nanocrystals | I$^-$/I$_3^-$ | N719 | 14.62 | 0.71 | 64 | 6.78 | 34.20 | 7.21 | 232 |
| Pt reference | I$^-$/I$_3^-$ | N719 | 14.78 | 0.72 | 68 | 7.38 | 27.13 | 4.57 | 232 |
| CoS$_2$ nanotube (NT1) | I$^-$/I$_3^-$ | N719 | 5.26 | 0.765 | 52.5 | 2.13 | - | - | 233 |
| CoS$_2$ nanotube (NT2) | I$^-$/I$_3^-$ | N719 | 10.68 | 0.794 | 64.6 | 5.48 | - | - | 233 |
| CoS$_2$ nanotube (NT3) | I$^-$/I$_3^-$ | N719 | 11.70 | 0.797 | 65.5 | 6.11 | - | - | 233 |
| CoS$_2$ nanotube (NT4) | I$^-$/I$_3^-$ | N719 | 11.58 | 0.804 | 65.8 | 6.13 | - | - | 233 |
| Pt reference | I$^-$/I$_3^-$ | N719 | 12.28 | 0.770 | 63.9 | 6.04 | - | - | 233 |
| CoS$_2$ nanoflakes | I$^-$/I$_3^-$ | N719 | 10.13 | 0.747 | 68.8 | 5.20 | - | - | 234 |
| Pt reference | I$^-$/I$_3^-$ | N719 | 10.04 | 0.767 | 69.4 | 5.34 | - | - | 234 |
| CoS(50) (chloroform, 50 mM) | I$^-$/I$_3^-$ | N3 | 8.48 | 0.703 | 53.1 | 3.4 | - | - | 235 |
| CoS(25) (chloroform, 25 mM) | I$^-$/I$_3^-$ | N3 | 9.23 | 0.700 | 51.3 | 3.5 | - | - | 235 |
| CoS(25)A (annealed for 240 min) | I$^-$/I$_3^-$ | N3 | 3.86 | 0.560 | 23.2 | 0.5 | - | - | 235 |
| CoS(12.5) (chloroform, 12.5 mM) | I$^-$/I$_3^-$ | N3 | 6.08 | 0.701 | 54.1 | 2.3 | - | - | 235 |
| CoS(2.5) (chloroform, 2.5 mM) | I$^-$/I$_3^-$ | N3 | 4.32 | 0.700 | 47.9 | 1.4 | - | - | 235 |
| Pt reference | I$^-$/I$_3^-$ | N3 | 9.05 | 0.702 | 56.4 | 3.6 | - | - | 235 |
| CoS(25) (chloroform, 25 mM) | I$^-$/I$_3^-$ | N719 | 14.15 | 0.703 | 66.7 | 6.6 | - | - | 235 |
| Pt reference | I$^-$/I$_3^-$ | N719 | 13.90 | 0.692 | 62.4 | 6.0 | - | - | 235 |
| CoS$_2$ rose-petal structure | I$^-$/I$_3^-$ | N719 | 12.14 | 0.64 | 68 | 5.32 | 9.88 | 3.52 | 236 |
| Pt reference | I$^-$/I$_3^-$ | N719 | 13.35 | 0.61 | 61 | 5.02 | 9.09 | 3.65 | 236 |
| CoS$_2$ (chemical bath deposition) | I$^-$/I$_3^-$ | N719 | 8.41 | 0.628 | 73.6 | 4.01 | 12.02 | 326.6 | 238 |
| CoS$_2$ (Ni-doped, 15%) | I$^-$/I$_3^-$ | N719 | 12.12 | 0.649 | 69.8 | 5.50 | 9.21 | 8.53 | 238 |
| NiS nanoparticles | I$^-$/I$_3^-$ | N719 | 9.54 | 0.592 | 67.8 | 3.83 | - | - | 238 |
| Pt reference | I$^-$/I$_3^-$ | N719 | 12.33 | 0.649 | 65.0 | 5.21 | 10.4 | 14.18 | 238 |
| CoS$_2$ | I$^-$/I$_3^-$ | N719 | 6.25 | 0.58 | 51 | 1.86 | 9.7 | 3.4 | 239 |
| CoS$_2$-G$_{20}$ (GO powder, 20 mg) | I$^-$/I$_3^-$ | N719 | 14.38 | 0.71 | 57 | 5.86 | 7.7 | 2.3 | 239 |
| CoS$_2$-G$_{50}$ (GO powder, 50 mg) | I$^-$/I$_3^-$ | N719 | 15.12 | 0.73 | 60 | 6.55 | 7.6 | 1.3 | 239 |
| CoS$_2$-G$_{80}$ (GO powder, 80 mg) | I$^-$/I$_3^-$ | N719 | 13.11 | 0.71 | 52 | 4.83 | 7.9 | 2.7 | 239 |

| | | | | | | | | |
|---|---|---|---|---|---|---|---|---|
| Graphene | $I^-/I_3^-$ | N719 | 3.68 | 0.65 | 58 | 1.37 | 7.6 | 4.0 | 239 |
| Pt reference | $I^-/I_3^-$ | N719 | 14.69 | 0.73 | 58 | 6.20 | 7.2 | 1.9 | 239 |
| CoS$_2$/Reduced graphene oxide | $I^-/I_3^-$ | Z907 | 12.87 | 0.67 | 63 | 5.4 | 22.5 | 4.8 | 240 |
| SnS$_2$ (350 $^\circ$C, 30 min) | $I^-/I_3^-$ | N719 | 15.63 | 0.725 | 55.3 | 6.27 | 21.5 | 8.3 | 249 |
| SnS$_2$ (400 $^\circ$C, 30 min) | $I^-/I_3^-$ | N719 | 16.96 | 0.743 | 60.7 | 7.64 | 18.6 | 5.6 | 249 |
| C/SnS$_2$ (400 $^\circ$C, 30 min) | $I^-/I_3^-$ | N719 | 17.47 | 0.745 | 61.9 | 8.06 | 17.4 | 5.2 | 249 |
| SnS$_2$ (450 $^\circ$C, 30min) | $I^-/I_3^-$ | N719 | 13.37 | 0.734 | 62.6 | 6.14 | 20.8 | 10.6 | 249 |
| Pt reference | $I^-/I_3^-$ | N719 | 16.53 | 0.730 | 63.9 | 7.71 | 16.2 | 6.7 | 249 |
| SnS$_2$@RGO hybrid | $I^-/I_3^-$ | N719 | 14.80 | 0.718 | 67.02 | 7.12 | 17.96 | 7.24 | 250 |
| SnS$_2$ | $I^-/I_3^-$ | N719 | 13.60 | 0.770 | 53.28 | 5.58 | 39.73 | 11.24 | 250 |
| RGO | $I^-/I_3^-$ | N719 | 10.08 | 0.661 | 52.29 | 3.73 | 34.20 | 50.25 | 250 |
| Pt reference | $I^-/I_3^-$ | N719 | 14.00 | 0.720 | 67.36 | 6.79 | 24.21 | 5.08 | 250 |

WS$_2$/MWCNTs* prepared without glucose aid.
In the case of R$_s$ and R$_{CT}$: Some of the authors used $\Omega$ instead of $\Omega.cm^2$ for the resistances without mentioning the size of the electrode.

## 4. Transition-Metal Diselenides Based Counter Electrodes

### 4.1 MoSe$_2$ Counter Electrodes

Molybdenum dichalcogenides, such as MoS$_2$, MoSe$_2$, and MoTe$_2$, are a very interesting class of materials that has attracted a great attention because of their unique properties.[251-257] Molybdenum diselenide (MoSe$_2$) is a semiconductor that has been studied for device applications.[92, 258, 259] The nanosheets of MoSe$_2$ have weak van der Waals forces, therefore, poorly adhere to the surface of substrates. To resolve this problem, Lee et al.[260] and Chen et al.[261] applied a CVD technique to grow catalytic MoSe$_2$ films on Mo foils. Lee et al.[260] used few-layer MoSe$_2$ in a DSSC as CE for the reduction of I$_3^-$ to I$^-$. Figure 19 shows a schematic representation for the preparation of few-layer MoSe$_2$, and their SEM and TEM images. The few-layer MoSe$_2$ on Mo film was developed by selenizing the Mo-coated soda-lime glass in a tube furnace. X-ray diffraction analysis exhibits body-centered cubic (*bcc*) crystal structures of the Mo film. The Mo film thickness was 1 μm and selenization was performed at 550 °C for 5 minutes, and produced the best performance for the DSSC. The $MoSe_2$ layer was about 70 nm as confirmed by the cross-sectional SEM image. The surface of the selenized Mo/glass contains nanoparticles, and the HRTEM image shows the few-layer structures. The interlayer spacing was measured as 0.63−0.64 nm. The few-layer $MoSe_2$ was prepared by surface selenization of Mo-coated soda-lime glass, which yielded a PCE of 9.0%, compared to a PCE of 8.68% for the Pt/FTO-based photoanode, thus showing the Pt/FTO free $MoSe_2$ CE outperformed the conventional CE. An $MoS_2$/Mo combination as a CE in a DSSC shows a PCE of 8.69%. Figure 20 shows a comparison of J-V curves of the DSSCs having $MoSe_2$/Mo and $MoS_2$/Mo CEs with different temperatures and times of selenization and sulfurization, and the Pt/FTO CE. The PCE of the DSSCs decreased from 7.14% for $MoSe_2$ CEs selenized at 580℃ for 60 minutes, to a PCE of 4.26% for 120 minutes. The edge sites were also found to be important for the high catalytic activity. The Mo substrate plays an important role in significantly reducing the sheet resistance of the CE, which eventually leads to a high DSSC performance. The low sheet resistance of 0.29 $\Omega$/sq for $MoSe_2$/Mo, compared to 12.60 $\Omega$/sq for Pt/FTO, indicates that $MoSe_2$/Mo is better as a CE in a DSSC than a Pt/FTO. The $R_{CT}$ of 0.87 $\Omega\cdot cm^2$ for $MoSe_2$ and 0.61 $\Omega\cdot cm^2$ for $MoS_2$ is lower compared to Pt (6.26 $\Omega\cdot cm^2$) due to high conductivity of the Mo substrate, therefore the Mo substrate seems superior to the FTO substrate. Chen et al.[261] also used a $MoSe_2$/Mo based CE in a DSSC to develop a Pt/FTO-free CE, which showed a PCE of 8.13%, slightly higher than the Pt/FTO CE (PCE of 8.06%). The $R_s$ and $R_{CT}$ values of the $MoSe_2$/Mo CEs were found to be 16.5% and 3.35% of the Pt/FTO CE, indicating a higher electrocatalytic activity of the $MoSe_2$/Mo.

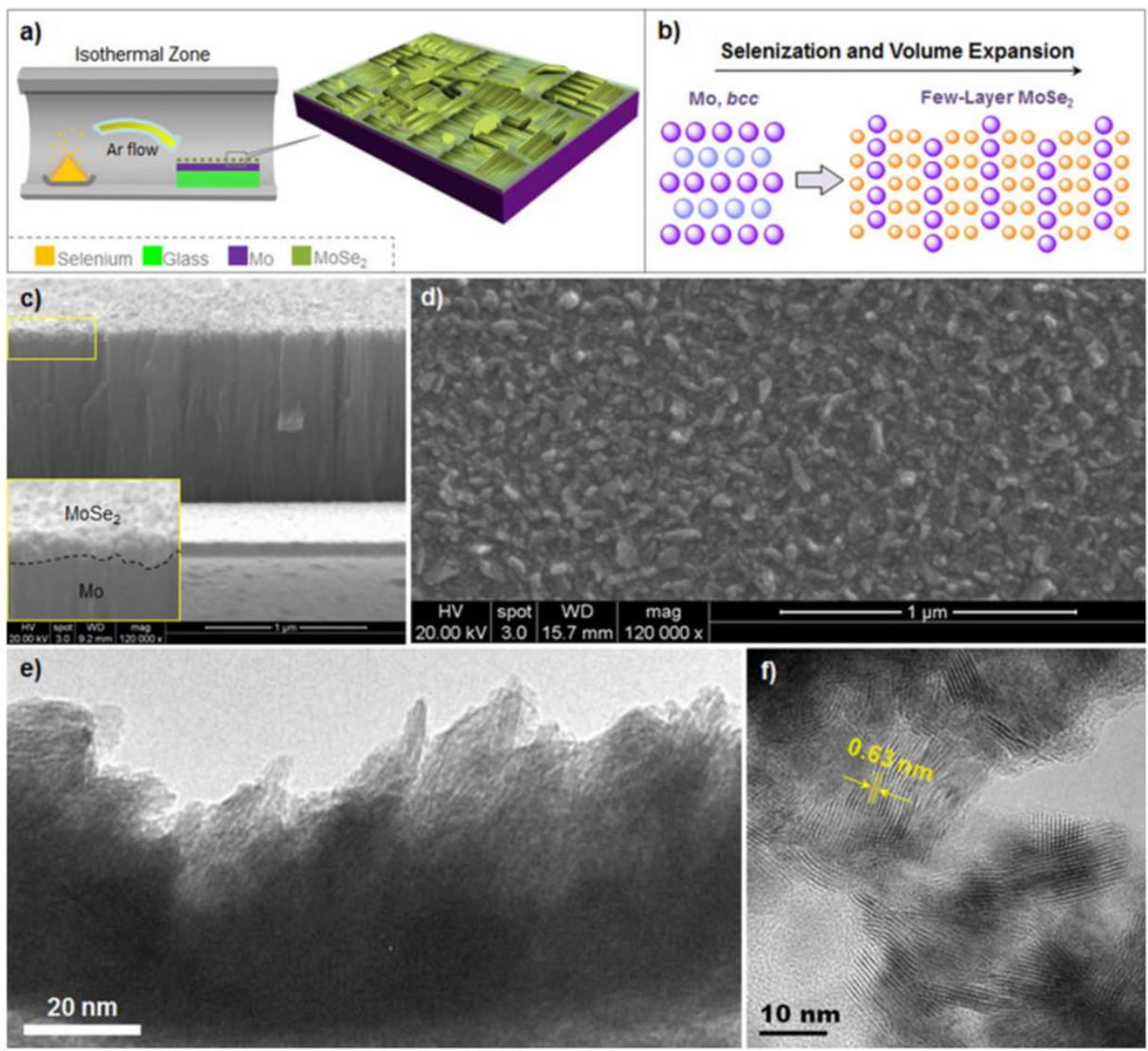

**Figure 19.** (a) Schematic representation for the selenization of Mo-coated soda-lime glass in a tube furnace for few-layer $MoSe_2$; (b) Schematic illustration showing the formation of few-layer $MoSe_2$ from body-centered cubic (*bcc*) crystal structures of Mo (c) Cross-sectional SEM image of the $MoSe_2$ on Mo surface, inset shows the borderline between $MoSe_2$ and Mo substrate; (d) SEM image of the as-synthesized $MoSe_2$ nanostructures; (e) HRTEM image of the few-layer $MoSe_2$; (f) high magnification HRTEM image of the few-layer $MoSe_2$ with interlayer spacing of $0.63-0.64$ nm. Reprinted with permission from ref. 260. Copyright 2014, Nature Publishing Group.

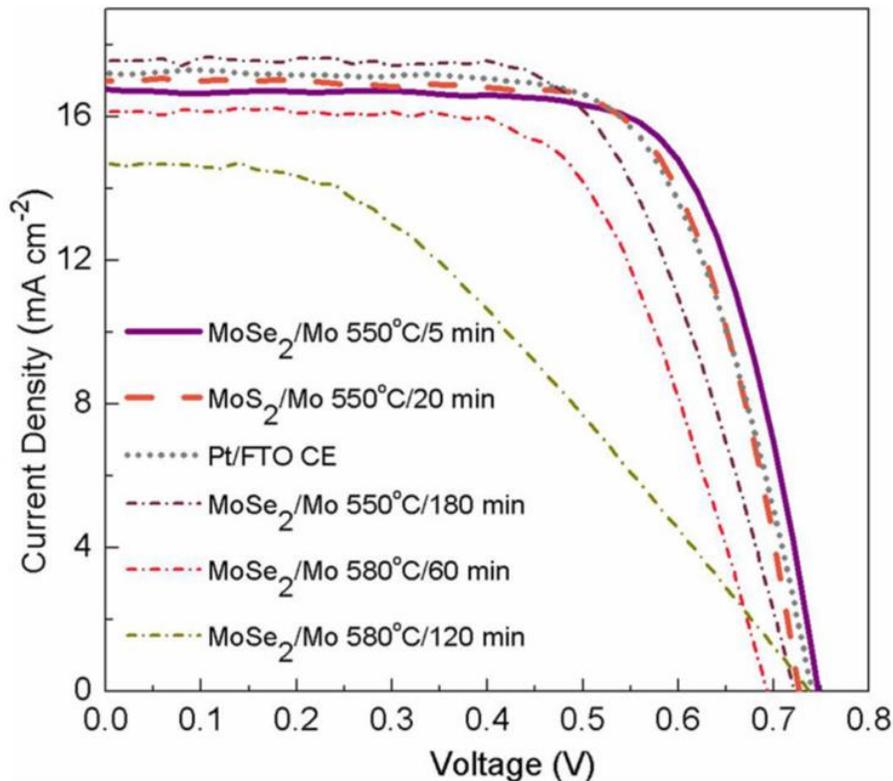

**Figure 20.** Photocurrent density-voltage (J-V) curves of DSSCs having $MoSe_2/Mo$ and $MoS_2/Mo$ counter electrodes with different temperatures and time of selenization and sulfurization and a comparison with conventional Pt/FTO counter electrodes. Reprinted with permission from ref. 260. Copyright 2014, Nature Publishing Group.

Thin films of the metal selenides $NiSe_2$, $CoSe_2$, and $MoSe_2$ were used as CEs for DSSCs for $I_3^-$ reduction by Ji et al.[262] $NiSe_2$ was found to be equally efficient to conventional Pt CEs. In comparison, $NiSe_2$ also showed a higher PCE than its sulfide analog ($NiS_2$) due to lower resistance to charge transfer. Figure 21 shows SEM images of $NiSe_2$, $CoSe_2$, and $MoSe_2$, and Tafel polarization curves and $J$–$V$ curves of the metal selenides $NiSe_2$, $CoSe_2$, $MoSe_2$, $WSe_2$, $Bi_2Se_3$, MnSe, PbSe, as well as Pt based CEs used in DSSCs. The metal selenides were deposited on FTO glass. The $NiSe_2$, $CoSe_2$, and $MoSe_2$ showed higher exchange current densities compared to other selenides. The $J$–$V$ curve measurements of the DSSCs indicated better performance of the NiSe$_2$ CE than that of the Pt CE, however the performance of DSSCs having CoSe$_2$ and MoSe$_2$ CEs was lower compared to the Pt CE.

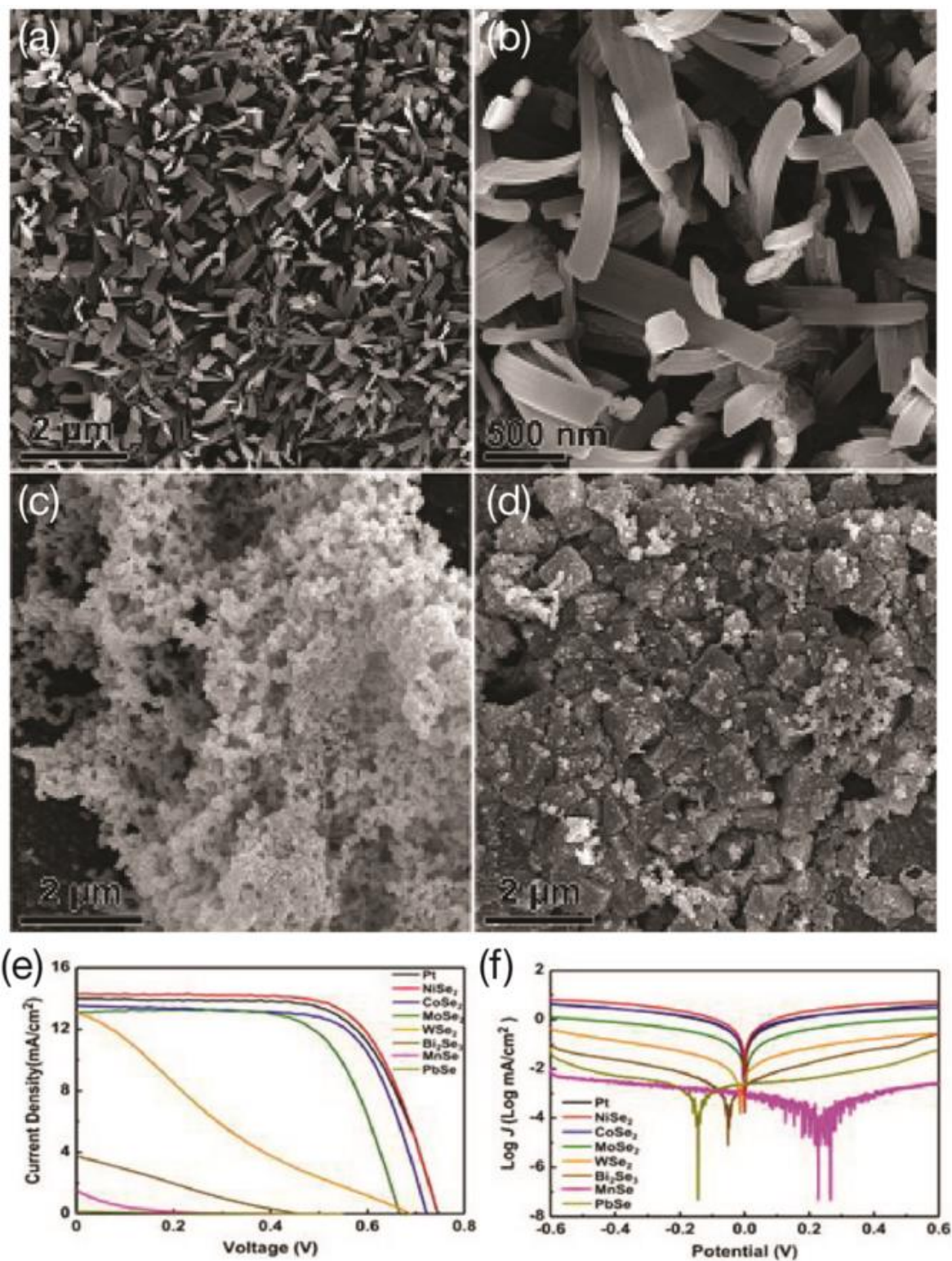

**Figure 21.** SEM images of NiSe₂ (a, b) and SEM images of CoSe₂ (c) and MoSe₂ (d). *J–V* characteristic curves (e) and Tafel polarization curves (f) of DSSCs based on CEs of made of metal selenides; NiSe$_2$, CoSe$_2$, MoSe$_2$, WSe$_2$, Bi$_2$Se$_3$, MnSe, PbSe, and Pt. Reprinted with permission from ref. 262. Copyright 2014, Elsevier.

In another study, MoSe$_2$ nanosheets were prepared using a solvothermal method which shows microsphere hierarchical architecture.[263] MoSe$_2$ nanosheets used as CE in fabricating a DSSC device showed a PCE of 9.80%, exceeding the PCE of a Pt CE based DSSC ($\eta$ = 8.17%) for the triiodide/iodide (I$_3^-$/I$^-$) redox reaction. Bi et al.[264] anchored fullerene-structured MoSe$_2$ hollow spheres on highly nitrogen-doped graphene (HNG) as a CE for a DSSC. Diethylelenetriamine (DETA) was used as a dopant for nitrogen of graphene. The hollow spheres consisted of 12-15 layers of MoSe$_2$ which formed the conductive network for facilitating rapid electron transfer in the DSCC. The MoSe$_2$ hollow spheres of 60-100 nm diameter and 8-12 nm thickness were dispersed on a HNG surface. The HNG- MoSe$_2$ hybrid has 52.4 wt.% of MoSe$_2$ content. The N content increased from 2.5% in MoSe$_2$/graphene hybrid to 12.5% in the HNG-MoSe$_2$ hybrid, which also showed high stability after 200 consecutive cycles of CV measurements. The HNG-MoSe$_2$ hybrid CE showed a PCE of 10.01%, slightly lower than a Pt CE ($\eta$ = 10.55%) under similar conditions, while the MoSe$_2$/graphene hybrid CE had a PCE of 7.34% arising from a poor fill factor of 0.60.

Composites of MoSe$_2$ nanosheets (NS) and poly(3,4 ethylenedioxythiophene):poly(styrenesulfonate) were investigated as the CE of a DSSC by Huang et al.[265] MoSe$_2$ NS acts as an electrocatalyst, while the PEDOT:PSS plays a role of a conductive binder to facilitate electron transfer between the MoSe$_2$ and the substrate. The weight ratios of MoSe$_2$ and PEDOT:PSS (MP) in the composites varied from 0.25 to 2.00. Each composite film contained 50 mg of MoSe$_2$ powder and 200 mg, 100 mg, 50 mg, and 25 mg of PEDOT: PSS, which are respectively referred to as MP-0.25, MP-0.50, MP-1.00, and MP-2.00, as per the ratios. The DSSC with a MP-1.00 composite film (equal weights of MoSe$_2$ and PEDOT:PSS) CE exhibits the highest electrocatalytic activity for the reduction of I$_3^-$. The DSSC containing the MoSe$_2$ NS/PEDOT:PSS composite film based CE shows a PCE of 7.58% , compared to the Pt CE exhibiting a PCE of 7.81% under similar experimental conditions. When MoSe$_2$ NS/PEDOT:PSS composite films coated on a titanium (Ti) foil flexible substrate was used as CE, the DSSC showed a PCE of 8.51%, compared to a PCE of 8.21% for the Pt-coated Ti foil CE. The CEs in the DSSCs show the relative order of the electrocatalytic activity as Pt > MP-1.00 > bare MoSe$_2$ > bare PEDOT:PSS, which is the same as observed by the IPCE spectra. The MP-1.00 composite film shows larger values of the heterogeneous rate constant, and the effective catalytic surface area as compared to bare MoSe$_2$ and bare PEDOT:PSS, because of MoSe$_2$ NS contents. The $R_{CT-Tafel}$ values of 3.23 $\Omega.cm^2$ for Pt, 181.46 $\Omega.cm^2$ for bare PEDOT:PSS, 3.77 $\Omega.cm^2$ for MP-1.00, and 3.11 $\Omega.cm^2$ for bare MoSe$_2$ were observed. The electrocatalytic ability of the CEs follow a trend seen in the $R_{CT}$ values measured from Tafel polarization curves and EIS. The MoSe$_2$ NS/PEDOT:PSS composite film based CEs thus show potential to replace a costly Pt electrode.

A cobalt selenide (Co$_{0.85}$Se)/MoSe$_2$/molybdenum oxide (MoO$_3$) ternary hybrid was evaluated as a CE for DSSCs.[266] Co$_{0.85}$Se/MoSe$_2$/MoO$_3$ ternary hybrids consist of nanorods, nanosheets, and nanoparticles, as confirmed by FESEM. CV showed larger current density for the Co$_{0.85}$Se/MoSe$_2$/MoO$_3$ hybrid compared with a sputtered Pt CE. The Co$_{0.85}$Se/MoSe$_2$/MoO$_3$ CE based DSSCs showed a PCE of 7.10%, much higher than that of a DSSC with a Pt CE ($\eta$ = 6.03%). For comparison, the Co$_{0.85}$Se hollow nanoparticles as a CE for DSSCs showed a PCE of 6.03%, lower compared to the Pt CE based DSSC ($\eta$ = 6.45%).[267] The transparent CEs using metal selenides alloys (M-Se; M=Co, Ni, Cu, Fe, Ru) were studied for the electrocatalytic activity for DSSCs and triiodide ($I_3^-$) reduction.[268] The DSSCs containing CEs consisting of a metal selenide alloy showed PCEs of 8.30 % for $Co_{0.85}Se$, 7.85 % for $Ni_{0.85}Se$, 6.43 % for $Cu_{0.50}Se$, 7.64 % for FeSe, and 9.22 % for $Ru_{0.33}Se$. A Pt CE based DSSC exhibited PCE of 6.18 %. Also, a nickel cobalt sulfide ($NiCo_2S_4$) nanoneedle array[269] used as a CE for a DSSC showed a PCE value of 6.9 %, which is comparable to a Pt CE ($\eta = 7.7$ %).

## 4.2 NbSe₂ Counter Electrodes

Niobium diselenide ($NbSe_2$) has attracted attention due to its superconducting properties[270-275] and can be easily processed into nanosheets, nanoflakes, nanowires and nanotubes usable for applications in field-effect transistors,[99(a)] light-emitting diodes,[276] superconductors,[277] and lithium batteries.[278]

$NbSe_2$ nanosheets, nanorods, and $NbSe_2$/C composites were used as CEs for DSSCs.[279] The morphology and structure of the $NbSe_2$ materials were characterized by SEM, TEM, and XRD while their electrochemical properties were evaluated by CV, EIS, and Tafel polarization curve measurements. The CEs based on $NbSe_2$ nanorods and $NbSe_2$ nanosheets showed lower charge transfer resistance and ionic diffusion. DSSCs having $NbSe_2$ nanosheet-based CEs achieved a PCE of 7.34%, which further increased to 7.80% for the $NbSe_2$/C composite-based CEs due to reduced series resistance, which is a PCE of 98.7% of the conventional Pt-based CEs ($\eta$ = 7.90%). Also, $NbSe_2$ nanostructures deposited via spray-coating were used to develop Pt-free CEs for DSSCs by Ibrahem et al.[280] Figure 22 shows SEM images and AFM height profiles of $NbSe_2$ nanosheets , nanorods, and nanoparticles, CV, J–V curves, IPCE spectra, and EIS (presented in Nyquist plots) of Pt and $NbSe_2$ nanostructures based CEs used in DSSCs. The morphology of the synthesized $NbSe_2$ nanostructures was analyzed by SEM and AFM techniques. SEM analysis indicated the pristine $NbSe_2$ 2D sheets were 100 μm thick. The separate $NbSe_2$ nanosheets were between 100-500 nm in lateral dimension. The length of $NbSe_2$ nanorods were up to 1.2 μm with diameters ranging between 20 to 100 nm. The average size of the $NbSe_2$ nanoparticles was between 50–100 nm. The AFM images revealed an average thickness of <8 nm for the $NbSe_2$ nanosheets, <5 nm for the $NbSe_2$ nanorods, and <3 nm for the $NbSe_2$ nanoparticles. HRTEM of the $NbSe_2$ nanosheets on their edge showed a spacing of 6.3 Å. HRTEM revealed the crystalline nature of individual $NbSe_2$ nanorods and nanoparticles. The $NbSe_2$ nanosheets, nanorods and nanparticles were studied as CEs in DSSCs as a replacement to a conventional Pt CE. The dye-absorbed $TiO_2$ electrodes were prepared by dipping electrodes into ruthenium dye 719 solution for 24 hours at room temperature. The dye solution contained 0.5 mM dye N719, [*cis*-di(thiocyanato)-N-N0-*bis*(2,20-bipyridyl-4-carboxylic acid-40-tetrabutyl-ammonium carboxylate) ruthenium(II)], and 0.5 mM chenodeoxycholic acid in a 1:1 mixture of *tert*-butanol and acetonitrile. The electrolyte solution was composed of 1-butyl-3-methylimidazolium iodide (BMII, 0.6M)), 4-*tert*-butylpyridine (0.5M), iodine (0.03M), and guanidinium thiocyanate (0.1 M) in a acetonitrile–valeronitrile mixture. The $NbSe_2$ nanosheet CEs achieved a PCE of 7.73%, compared to a PCE of 7.01% for Pt-based CEs for DSSCs. The DSSCs with $NbSe_2$ nanoparticles and nanorods based CEs show PCE values of 6.27% and 5.05%, respectively, due to low FF arising from relatively smaller surface areas, as well as low exposure on the FTO glass substrates. DSSCs having $NbSe_2$ nanosheets based CE show the best IPCE spectral response, where the peak increases from 84% for a Pt CE to 89% for a $NbSe_2$ nanosheets based CE. On the other hand, lower IPCE peak values were observed for $NbSe_2$ nanoparticles and nanorods based CEs. The charge-transfer processes at the interface of $TiO_2$/dye/electrolyte were analyzed by EIS. The $NbSe_2$ nanosheet based CEs for the DSSCs shows a middle-frequency semicircle, implying it had the highest electro-catalytic activity for the reduction of triiodide ions ($I_3^-$), and efficient generation of electrons, therefore, occurring of larger electrons at the $TiO_2$/dye/electrolyte interface.  This study suggests that $NbSe_2$ nanosheets could be used as alternative CEs to conventional Pt CEs in DSSCs because of their large surface area.

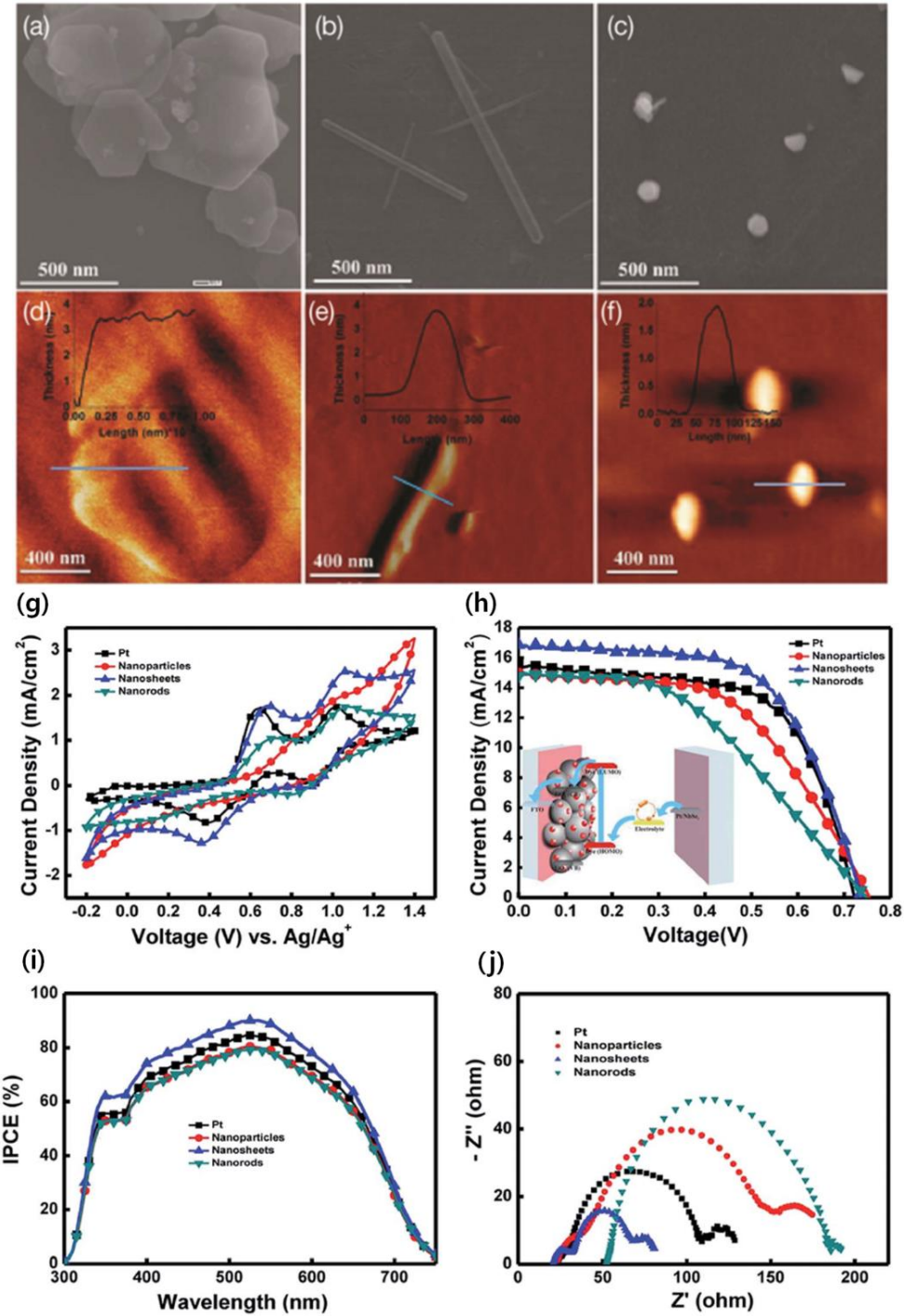

**Figure 22.** SEM images of NbSe₂ nanostructures that include nanosheets (a), nanorods (b), and nanoparticles (c), (d–f) AFM height images of NbSe₂ nanosheets (d), nanorods (e), and nanoparticles (f). The insets show height profiles measured by AFM images. (g) Cyclic voltammograms (CV) of Pt CE and NbSe₂ nanostructures based CEs. (h) J–V curves of Pt and NbSe₂ nanostructures. The inset represents DSSC structure with an energy level diagram of the materials. (i) photon-to-current conversion efficiency (IPCE) curves of Pt and NbSe₂

nanostructures, and (j) EIS Nyquist plots of Pt and NbSe$_2$ nanostructures based CEs used in DSSCs. Reprinted with permission from ref. 280. Copyright 2014, The Royal Society of Chemistry

## 4.3 TaSe$_2$ Counter Electrodes

Three transition-metal selenides (MoSe$_2$, WSe$_2$, and TaSe$_2$) were prepared by Guo et al.[281] and compared as potential CEs for fabricating DSSC devices using a solvothermal method. These transition-metal selenides show high electrocatalytic activity for the reduction of triiodide (I$_3^-$), except for MoSe$_2$, due to low adsorption and charge-transfer. Figure 23 shows SEM and TEM images of MoSe$_2$ (a and d), WSe$_2$ (b and e), and TaSe$_2$; the Nyquist plots of EIS measurements recorded at -0.75 V bias of the DSSCs based on MoSe$_2$, WSe$_2$, and Pt CEs for the triiodide/iodide (I$_3^-$/I$^-$) redox couple in the electrolyte; and their J–V curves measured under simulated sunlight illumination. The MoSe$_2$ nanosheets had thickness of about 10 nm with lateral dimensions ranging 100-150 nm. WSe$_2$ consists of interlaced nanoplates with 15 nm average thickness and a width varying between 60 and 100 nm. The SEM image of TaSe$_2$ shows fluffy nanoparticles with a wide size distribution. The HRTEM images of MoSe$_2$, WSe$_2$, and TaSe$_2$ indicated spacings of 0.282 nm, 0.283 nm, and 0.291 nm, respectively. The WSe$_2$ shows a pore size distribution (PSD) curve at about 43 nm, MoSe$_2$ at 30 nm, and TaSe$_2$ at 10 nm. The mesoporous structure of WSe$_2$ facilitates the adsorption of triiodide (I$_3^-$) and transport during the electrocatalytic activity. The BET surface areas were 95.6, 104.4 and 78.8 m$^2$/g, and the total pore volumes were 0.32 cm$^3$/g, 0.41 cm$^3$/g and 0.14 cm$^3$/g, for WSe$_2$, MoSe$_2$, TaSe$_2$, respectively. The significant difference in BET surface area and the pore size distribution led to different adsorption properties and electrocatalytic activity of the DSSCs. The R$_{CT}$ of the transition-metal selenides at the electrolyte/electrode interface indicates the level of electrocatalytic activity. The $R_{CT}$ value of 0.78 $\Omega$ cm$^2$ for the WSe$_2$ CE is smaller compared with a Pt CE (1.32 $\Omega$ cm$^2$) and a TaSe$_2$ (1.89 $\Omega$ cm$^2$) CE, while the $R_{CT}$ value of the MoSe$_2$ (2.43 $\Omega$ cm$^2$) CE was larger than that of the Pt CE. Therefore, WSe$_2$ shows better electrocatalytic activity for triiodide (I$_3^-$) reduction. The ionic $Z_N$ value of WSe$_2$ is 1.15 $\Omega$ cm$^2$, lower than that of the MoSe$_2$, and TaSe$_2$ CEs, and comparable to the Pt CE, which demonstrates a larger diffusion coefficient for triiodide (I$_3^-$) within the CEs. The conductivities measured by linear sweep voltammetry decrease in the relative order TaSe$_2$ > MoSe$_2$ > WSe$_2$. The DSSCs based on TaSe$_2$ and WSe$_2$ CEs showed larger $J_{SC}$ than that of MoSe$_2$ CE. WSe$_2$ has larger pores and high electrocatalytic activity suitable for fast regeneration and transfer of triiodide (I$_3^-$). PCE values of 7.32% for TaSe$_2$ and 7.48% for Wse$_2$ were measured, which are comparable to a sputtered Pt CE ($\eta$ = 7.91%). The lower PCE value of 6.70% for MoSe$_2$ arises from a lower $V_{OC}$ and $J_{SC}$ for the DSSC.

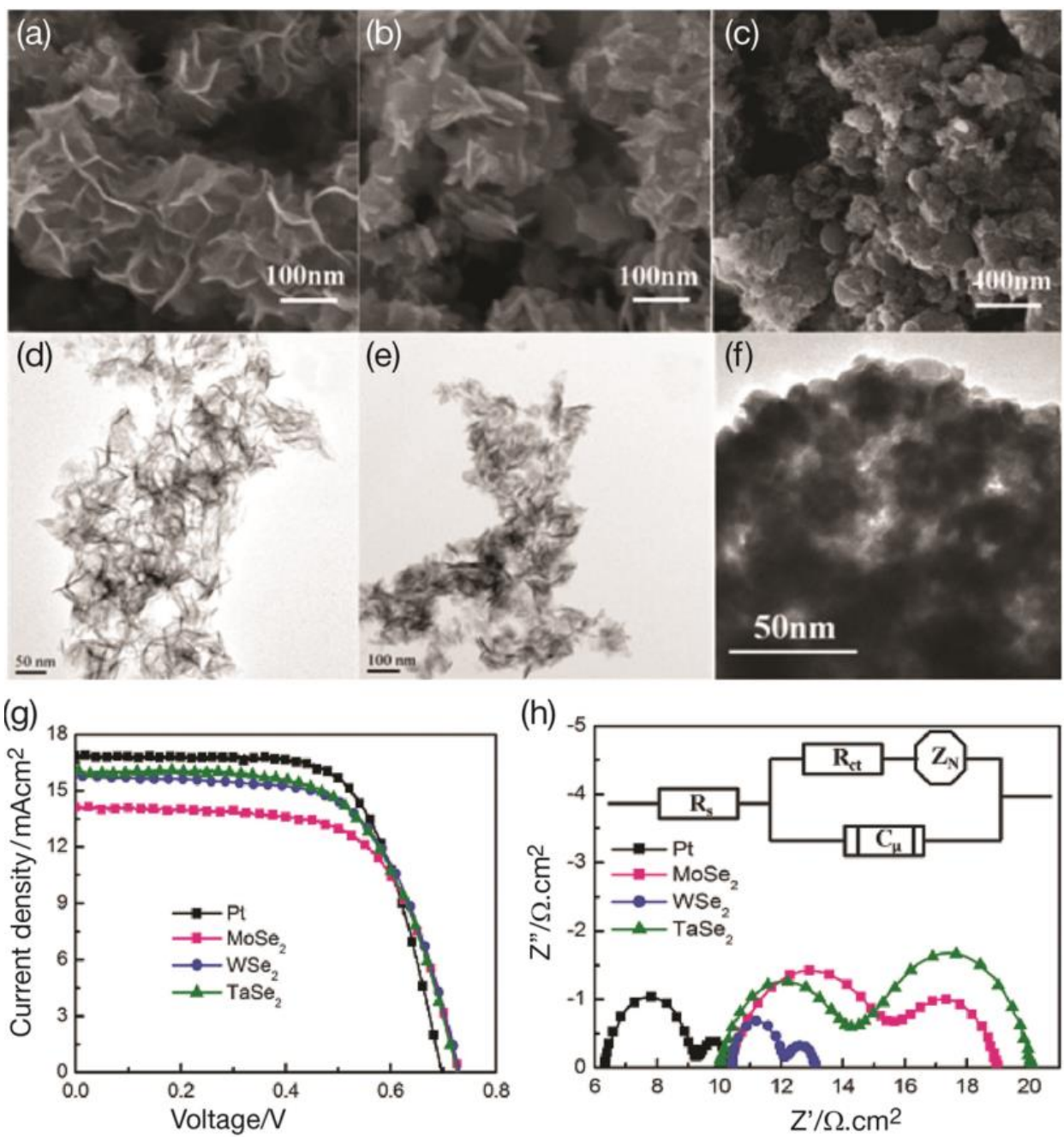

**Figure 23.** SEM and TEM images of MoSe₂ (a and d), WSe₂ (b and e), and TaSe₂ (c and f). (g) Photocurrent–voltage (J–V) curves of the DSSCs based on MoSe₂, WSe₂, and Pt CEs, measured under simulated sunlight illumination (100 mW/cm$^2$, 1.5 AM G). (h) Nyquist plots of Electrochemical impedance spectroscopy (EIS) measurements for DSSCs fabricated with two identical electrodes in the triiodide/iodide (I₃⁻/I⁻) redox couple in the electrolyte. The inset represents equivalent circuit of DSSCs for fitting Nyquist plots. Reprinted with permission from ref. 281. Copyright 2015, The Royal Society of Chemistry/Owner Societies.

## 4.4 NiSe₂ Counter Electrodes

Nickel diselenide ($NiSe_2$) has been studied as a CE of DSSCs for the reduction of $I_3^-$, which yielded higher PCE of 8.69% than that of a conventional Pt CE (8.04%) under the same experimental conditions.[282] Zhang et al.[283] developed two types of $NiSe_2$ CEs on RGO; namely, microsphere $NiSe_2$/RGO, and octahedron $NiSe_2$/RGO through a hydrothermal process. The microsphere $NiSe_2$/RGO CE exhibited higher electrocatalytic performance than a Pt CE for the reduction of triiodide ($I_3^-$) because of better carrier transfer induced by graphene nanosheets.

CEs of ternary Ni−Co compounds having different morphological structures such as nanoparticles, nanotubes, nanowires, nanoflakes, follower-like, and urchin-like have been studied for DSSCs.[284, 285] For example, the CE made of flower-like $NiCo_2S_4$/NiS microspheres[286] exhibited a PCE of 8.8%, much higher than a standard Pt CE ($\eta$ = 8.1%). A similar concept was employed by Qian et al.[287] for developing a very interesting class of CEs for DSSCs from nickel cobalt (Ni−Co) selenides having different morphological structures, due to the tuning of the Ni/Co molar ratios. The morphological structure and electrocatalytic performance of ternary Ni−Co selenides was optimized by using different Ni/Co molar ratios. Figure 24 shows the SEM images of $Co_3Se_4$, $Ni_{0.33}Co_{0.67}Se$ precursor, $Ni_{0.33}Co_{0.67}Se$, $Ni_{0.5}Co_{0.5}Se$, $Ni_{0.67}Co_{0.33}Se$, and NiSe, and also shows CV and *J-V* curves of their DSSCs. The different morphological structures were obtained by tuning the Ni/Co molar ratio, $Ni_xCo_{1-x}Se$, where, x was  0, 0.33, 0.5, 0.67, and 1.0. The 3D dandelion-like precursor of $Ni_{0.33}Co_{0.67}Se$ assembled into nanotubes having about 100 nm diameter, $Ni_{0.5}Co_{0.5}Se$ into a floccus-like microsphere structure with a diameter of 4 microns, $Ni_{0.67}Co_{0.33}Se$ microspheres compiled into nanosheets, and $Co_3Se_4$ built up rough-surface nanotubes. The specific surface areas determined from the BET method were 5.0, 10.1, 26.1, 28.9, and 35.6  $m^2$/g, for NiSe, $Co_3Se_4$, $Ni_{0.5}Co_{0.5}Se$, $Ni_{0.67}Co_{0.33}Se$ and $Ni_{0.33}Co_{0.67}Se$, respectively. Among all metal selenides, $Ni_{0.33}Co_{0.67}Se$ has the highest specific surface area, which is favorable for providing more active sites for catalysis and increasing the contact area between its CE and the electrolyte, which results in better electrochemical and photovoltaic properties of its DSSCs. The $R_{CT}$ values of CEs follows the relative order of $NiSe > Co_3Se_4 > Pt > Ni_{0.67}Co_{0.33}Se > Ni_{0.5}Co_{0.5}Se > Ni_{0.33}Co_{0.67}Se$, which implies that the electrocatalytic activity increases in a reverse order. Thus, NiSe was of the lowest activity and $Ni_{0.33}Co_{0.67}Se$ was of the highest catalytic activity for the reduction of triiodide ($I_3^-$). Therefore, higher contents of Ni in the Ni−Co selenides CEs are not favorable for electrocatalytic activity. The values of cathodic peak current density (Red-1) and the $E_{pp}$ can also help understand the electrocatalytic activities of these CEs. The authors noted the following relative order of cathodic peak current density: $NiSe$ (1.044 mA/cm$^2$) < $Co_3Se_4$ (1.849 mA/cm$^2$) < Pt (2.373 mA/cm$^2$) < $Ni_{0.67}Co_{0.33}Se$ (2.878 mA/cm$^2$) < $Ni_{0.5}Co_{0.5}Se$ (2.917 mA/cm$^2$) < $Ni_{0.33}Co_{0.67}Se$ (3.120 mA/cm$^2$). They also noted $E_{pp}$ values of $NiSe$ (602 mV) > $Co_3Se_4$ (565 mV) > Pt (460 mV) > $Ni_{0.67}Co_{0.33}Se$ (365 mV) > $Ni_{0.5}Co_{0.5}Se$ (349 mV) > $Ni_{0.33}Co_{0.67}Se$ (329 mV), which were in an agreement with EIS measurements of the CEs. The 3D dandelion-like $Ni_{0.33}S_{0.67}Se$ microspheres based CEs exhibited the highest PCE of 9.01%, exceeding that of the Pt CE ($\eta$ = 8.30%). The DSSC with a $Co_3Se_4$ CE showed a PCE of 7.95%, higher than that of NiSe CE ($\eta$ = 7.23%). These results support the notion that the ternary Ni−Co selenides possess higher electrocatalytic activities and photovoltaic properties than those of binary selenides NiSe and $Co_3Se_4$ as well as Pt CEs for the triiodide ($I_3^-$) reduction, due to their unique morphology and chemical composition.

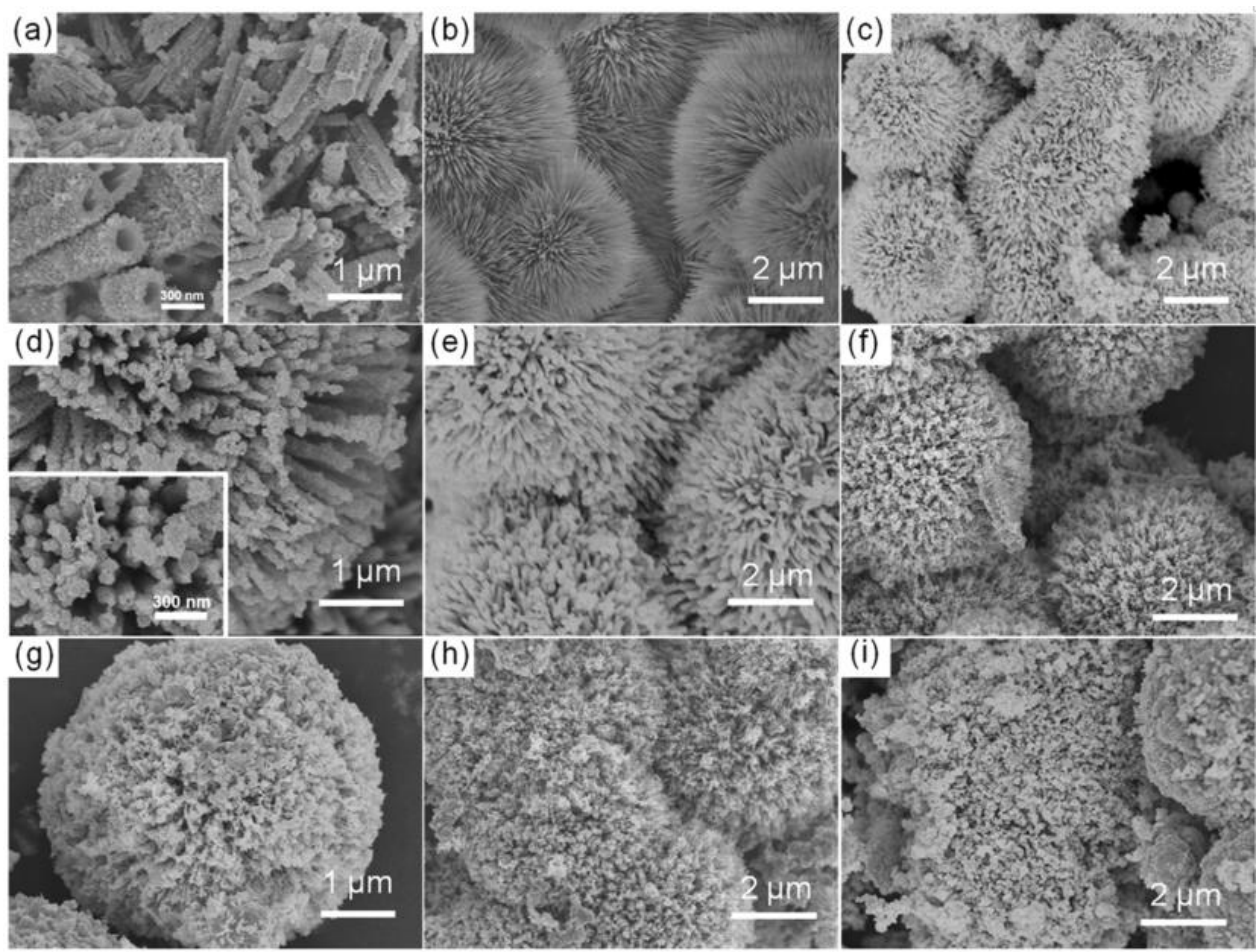

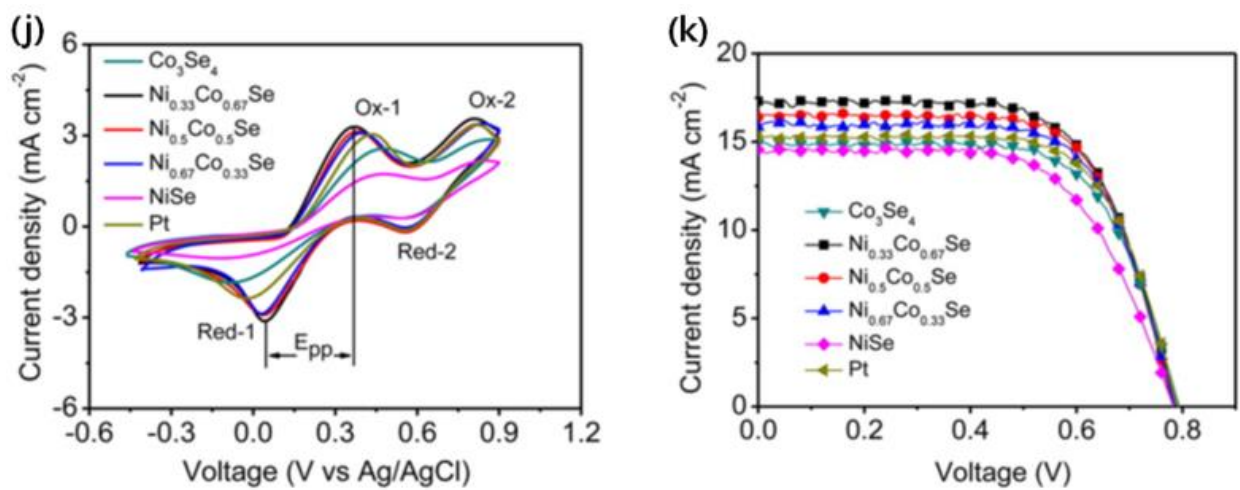

**Figure 24.** SEM images of (a) $Co_3Se_4$, (b) $Ni_{0.33}Co_{0.67}Se$ precursor, (c–e) $Ni_{0.33}Co_{0.67}Se$, (f and g) $Ni_{0.5}Co_{0.5}Se$, (h) $Ni_{0.67}Co_{0.33}Se$, and (i) NiSe. (j) CVs of DSSCs with $Co_3Se_4$, $Ni_{0.33}Co_{0.67}Se$, $Ni_{0.5}Co_{0.5}Se$, $Ni_{0.67}Co_{0.33}Se$, NiSe, and Pt-based CEs at a scan rate of 50 mV/s. (k) Photocurrent density–voltage (J-V) curves of DSSCs with $Co_3Se_4$, $Ni_{0.33}Co_{0.67}Se$, $Ni_{0.5}Co_{0.5}Se$, $Ni_{0.67}Co_{0.33}Se$,

NiSe, and Pt CEs under AM 1.5G illumination. Reprinted with permission from ref. 287. Copyright 2016, American Chemical Society.

## 4.5 FeSe$_2$ Counter Electrodes

Iron diselenide (FeSe$_2$) can be processed into nanosheets, nanocubes, flower-like structures, and nanorods, rod clusters, and microspheres,[288-294] and has applications in catalysis,[295] batteries,[296, 297] and photovoltaic devices.[298] The first-row (3d) transition metal dichalcogenides (MX$_2$) with pyrite structure (where, M = Fe, Co, Ni, and X= S, Se) also exhibit electronic, optoelectronic, and magnetic properties comparable to other TMDs.[299-302] Huang et al.[302] used 2D FeSe$_2$ nanosheets with 7 nm average thickness as a CE for a DSSC. Figure 25 shows the SEM and TEM images of the synthesized FeSe$_2$ nanosheets. The thickness of FeSe$_2$ nanosheets was found to be in the 4–7 nm range, as evaluated by TEM images. High-resolution TEM (HRTEM) indicated the crystalline nature of the nanosheets, having 0.168 nm interplanar spacing.  The FeSe$_2$ nanosheets had specific surface area of 35.02 m$^2$/g as calculated by the BET analysis. The FeSe$_2$ nanosheets exhibit a low charge-transfer resistance (R$_{CT}$) and a high electrocatalytic activity for triiodide (I$_3^-$) reduction in DSSCs. The FeSe$_2$ nanosheets CE based DSSC showed a PCE of 7.53%, comparable to a Pt CE, while the PCE of FeSe$_2$ microparticles CE based  DSSC was slightly lower ($\eta$ = 6.88%). The large surface area of the FeSe$_2$ nanosheets contributed to a higher PCE by providing more active sites for electrocatalytic activity, as well as a larger electrode/electrolyte interface that that of the FeSe$_2$ microparticles. The electrochemical stability of the CEs of FeSe$_2$ nanosheets and Pt were investigated with sequential CV scanning. The current densities of the FeSe$_2$ CE showed no change after 1000 cycles, confirming corrosion resistance to the electrolyte. The PCEs of FeSe$_2$ nanosheets under nitrogen protection (N-FeSe$_2$) and exposure to air for one week (O-FeSe$_2$) were also studied. The R$_{CT}$ of the N-FeSe$_2$ CE (0.53 $\Omega$ cm$^2$) was much lower that the O-FeSe$_2$ CE (10.13 $\Omega$ cm$^2$) and a Pt CE (1.68 $\Omega$ cm$^2$), indicating poorer electrocatalytic activity of the O-FeSe$_2$ CE compared to the N-FeSe$_2$ and Pt CEs. The Tafel polarization curves decreased in the order N-FeSe$_2$ > Pt > O-FeSe$_2$, having the same order as of exchange current density, CV and EIS measurements. The O-FeSe$_2$ nanosheets CE based DSSC has a PCE of 6.15%, much lower than that of the N-FeSe$_2$ ($\eta$ = 7.53%) CE and Pt CE ($\eta$ = 7.47%) under similar conditions.

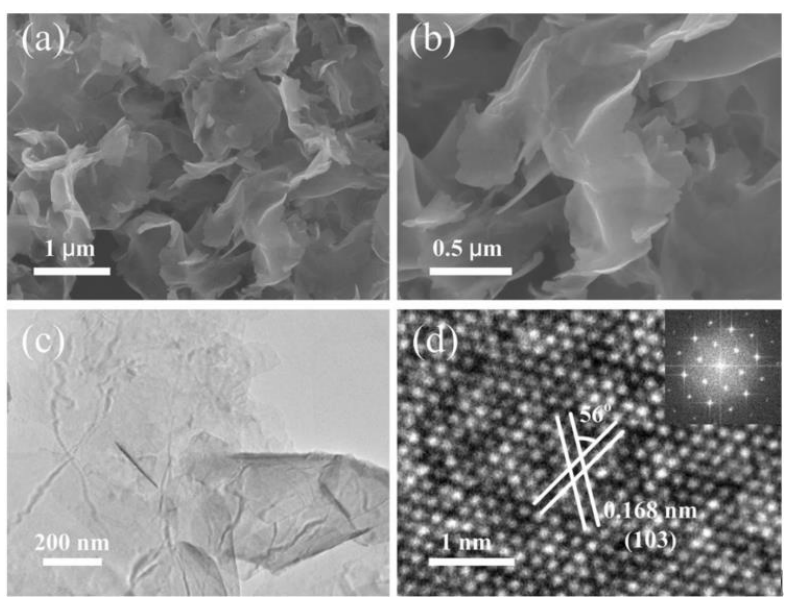

**Figure 25.** (a, b) SEM images and (c) TEM image of the as-synthesized FeSe$_2$ nanosheets, and (d) HRTEM image of a small portion of FeSe$_2$ nanosheet (inset shows SAED pattern of the FeSe$_2$ nanosheets). Reprinted with permission from ref. 302. Copyright 2015, Wiley-VCH.

3D hierarchical FeSe$_2$ microspheres using a hot-injection method were prepared and studied by Huang et al.[303] The morphologies of the FeSe$_2$ nanomaterials was controlled by the use of alkyl thiols; 1-dodecanethiol (1-DDT) or *tert*-dodecanethiol (t-DDT) and their contents were used in synthesis, which varied from irregular FeSe$_2$ micro/nanoparticles to 3D hierarchical FeSe$_2$ microspheres and consisted of ultrathin FeSe$_2$ nanosheets or urchin-like microspheres made of crystalline $FeSe_2$ nanorods having an average diameter of 650 nm. The $FeSe_2$ nanomaterials were used as CEs for DSSCs. 3D hierarchical $FeSe_2$ microspheres made of ultrathin $FeSe_2$ nanosheets showed the lowest $R_{CT}$ of 0.49 $\Omega.cm^2$ at the electrolyte/electrode interface, a lower $Z_N$ value of 0.39 $\Omega.cm^2$, and faster reaction kinetics for the reduction of $I_3^-$ to $I^-$ than that of a Pt CE ($R_{CT}$ of 1.15 $\Omega.cm^2$ and $Z_N$ value of 0.91 $\Omega.cm^2$). $R_{CT}$ values followed the relative order of $FeSe_2$ microparticles < $FeSe_2$ nanorods < Pt < $FeSe_2$ nansheets, as supported by EIS measurements. A DSSC with a $FeSe_2$ nanosheets CE exhibited a PCE of 8.39%, slightly better than that of a Pt CE (8.20%) under simulated solar illumination of 100 m/Wcm$^2$ (AM 1.5). $FeSe_2$ nanorods showed a PCE value of 8.03%, higher than that of $FeSe_2$ microparticles (7.68%). The PCE value of DSSCs is morphology dependent, where the $FeSe_2$ nanosheets CE has a high electrocatalytic activity and a larger specific surface area (30.03 m$^2$/g) than that of $FeSe_2$ nanorods (19.82 m$^2$/g). The $FeSe_2$ nanosheets CE based DSSC retained 99.5% of its initial photocurrent density, compared to 98.3% retained by the Pt CE, after simulated solar illumination of 100 m/Wcm$^2$ for 1 hour, and this indicates better stability of the $FeSe_2$ nanosheets CE than that of standard Pt CE. Also, the 3D flower-like and sphere-shaped $FeSe_2$ films were used as CEs for DSSCs.[304] The 3D flower-like $FeSe_2$-based CE exhibited a comparable PCE to a Pt CE ($\eta$ = 8.00% versus 7.87%).

## 4.6 $CoSe_2$ Counter Electrodes

Cobalt diselenide ($CoSe_2$) has been extensively studied as a catalyst for oxygen reduction reactions.[305-308] The $CoSe_2$ nanorods synthesized by a hydrothermal method was studied as CEs for DSSCs.[309] FESEM revealed a $CoSe_2$ nanorod morphology of the $CoSe_2$, which supports carrier transport from the surface of the nanorods to the redox electrolyte. The $CoSe_2$ CE also exhibited larger current density compared with a Pt CE, as measured by CV. EIS shows an $R_s$ of 8.034 $\Omega.cm^2$ and a low $R_{CT}$ of 0.097 $\Omega \cdot cm^2$ for the $CoSe_2$ CE. The DSSC with a $CoSe_2$ CE showed a PCE value of 8.38 %, higher than the DSSC based on a Pt CE ($\eta$ = 7.83%), under simulated sunlight illumination of 100 mW/cm$^2$ (AM 1.5G).

A comprehensive and detailed study was conducted by Chiu et al.[310] on composite films of $CoSe_2$/carbon ($CoSe_2$/C) deposited on FTO substrates having three different morphologies, developed using electro-deposition, followed by an annealing process at 500 $^o$C for 30 minutes in vacuum. In the first stage, three types of $CoSe_2$/carbon films containing nanowalls were deposited with an electro-deposition process employing different pH baths, while in the second stage, the morphology of the films was transformed after the annealing. The N719 dye-adsorbed $TiO_2$ film was used as a photoanode for DSSCs. $CoSe_2$/C films had three different morphologies, including nanograin (NG), nanorock (NR), and nanoclimbing-wall (NCW), which were used as CEs for DSSCs. The electrocatalytic activity of these three CEs was analyzed by CV, RDE, Tafel polarization curves, and EIS, which showed a relative order of $CoSe_2$/C-NCW > $CoSe_2$/C-NG > Pt > $CoSe_2$/C-NR as CEs in the DSSCs, a similar order as was observed for the $R_{CT}$ values obtained by EIS and Tafel measurements. The $CoSe_2$/C-NCW showed higher electrical conductivity and a large effective surface area and, therefore, the best electrocatalytic ability for triiodide ($I_3^-$) reduction. The DSSCs with $CoSe_2$/C-NG, $CoSe_2$/C-NCR, $CoSe_2$/C-NCW, and Pt CEs exhibited IPCE values between 80~95% in the 400 to 600 nm wavelength region, where the highest IPCE value of 95% was observed for the $CoSe_2$/C-NCW CE. The DSSC having a $CoSe_2$/C-NCW CE showed the highest PCE value of 8.92%, even higher than compared with the Pt CE (8.25%). The $CoSe_2$/C-NCW CEs were electro-deposited onto low-cost, flexible, and highly porous substrates, such as carbon cloth (CC, sheet resistance = 0.63 $\Omega$/sq) and nickel foam (NF, porosity = 95%, sheet resistance = 0.45 Ω/sq). Figure 26 shows photographs of the flexible nickel foam (NF) and carbon cloth (CC), SEM images of CoSe$_2$/C-NCW film on the flexible NF and CC substrates, and J–V curves of the DSSCs, with CoSe$_2$/C-NCW on NF and CC CEs measured at different light intensities (20–100 mW/cm$^2$). The CoSe$_2$/carbon-NCW CEs shell covered all the minute parts of the carbon cloth and nickel foam core shell structures. The DSSC containing the CE with CoSe$_2$/C-NCW deposited on nickel foam exhibited the highest PCE of 10.46% at 100 mW/cm$^2$ (1 Sun) and 7.90% at 20 mW/cm$^2$ (0.2 Sun). The DSSC with the CE of CoSe$_2$/C-NCW deposited on low-weight carbon cloth showed a PCE of 9.87% at 1 sun and 7.83% at 0.2 Sun. The low cost and flexible CoSe$_2$/C-NCW CEs seem to be promising materials to replace expensive Pt CEs for DSSCs for indoor, outdoor or wearable applications.

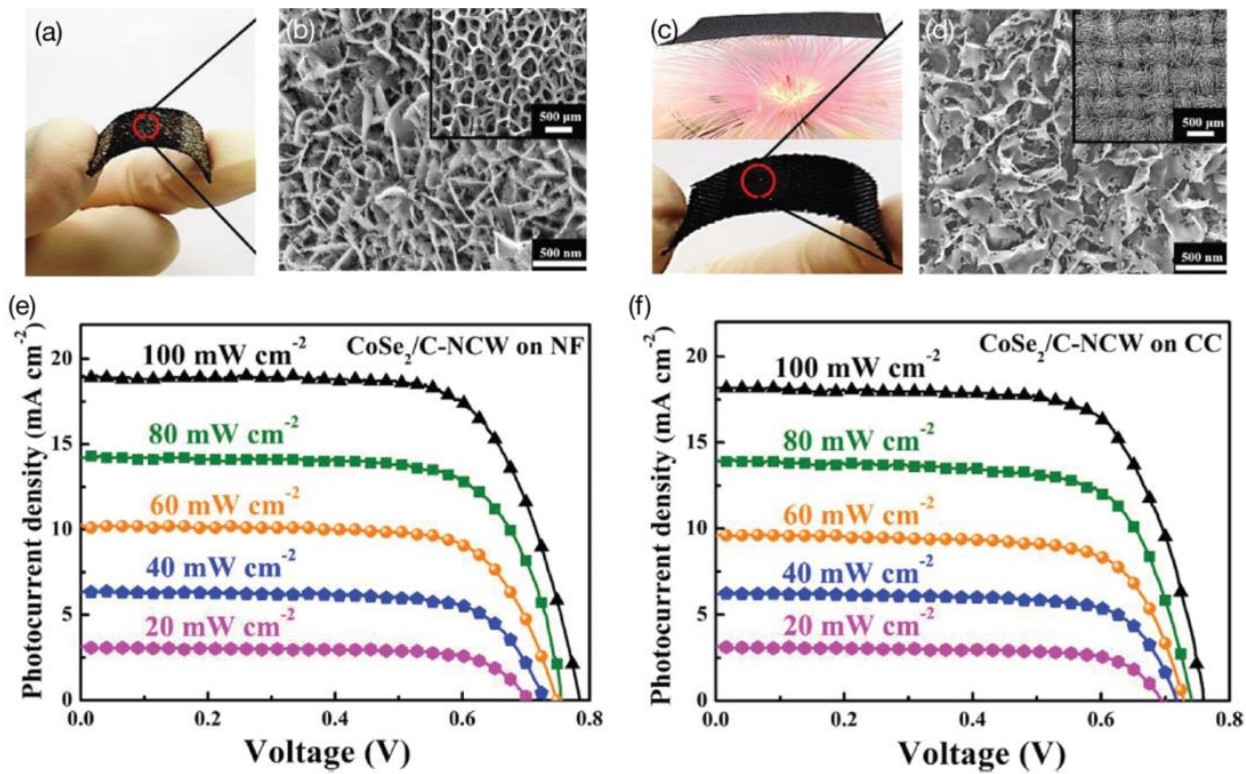

**Figure 26.** (a) Photograph of the nickel foam (NF) and (b) SEM image of CoSe$_2$/C-NCW film on the nickel foam (NF). (c) Photograph of the carbon cloth (CC) and thin CC substrate on a soft

flower having needle-like petals and (d) SEM image of $CoSe_2$/C-NCW film on carbon cloth (CC). Inset shows field-emission scanning electron microscopy (FE–SEM) images in a magnified version. (e) Photocurrent–voltage (J–V) curves of the DSSCs with $CoSe_2$/C-NCW on flexible nickel foam (NF) CE, measured at light intensities varying from 20 mW/cm² (0.2 sun) to 100 mW/cm² (1 sun). (f) Photocurrent–voltage (J–V) curves of the DSSCs with $CoSe_2$/C-NCW on flexible carbon cloth (CC) CE, under similar light intensities. Reprinted with permission from ref. 310. Copyright 2016, Elsevier.

$CoSe_2$ and RGO composites were also explored as CEs in DSSCs, which showed a PCE of 7.01 % versus a Pt CE ($\eta$ = 6.77 %).[311] $Co_{0.85}Se$ and $Ni_{0.85}Se$ was deposited on FTO glass substrate by a low-temperature hydrothermal process and were used as CEs for DSSCs by Gong et al.[312] $Co_{0.85}Se$ has a graphene-like nanostructure and possesses a large surface area, while $Ni_{0.85}Se$ is composed of aggregated particles. The graphene-like $Co_{0.85}Se$ CE showed higher electrocatalytic activity than that of the Pt CE for the reduction of triiodide ($I_3^-$). DSSCs with $Co_{0.85}Se$ CEs showed a PCE of 9.40%, significantly higher than that of a Pt CE (8.64%), under simulated solar light of 100 mW/cm² (AM 1.5G). In the case of the $Ni_{0.85}Se$ CE, the PCE of 8.32% was slightly lower than a Pt CE. Both *Jsc* and PCE values showed a relative order of $Ni_{0.85}Se$ < Pt < $Co_{0.85}Se$. The $R_{CT}$ value was found to increase in the relative order $Co_{0.85}Se$ (0.6 $\Omega$.cm²) < Pt (1.1 $\Omega$.cm²) < $Ni_{0.85}Se$ (1.8 $\Omega$.cm²), suggesting an inverse order of electrocatalytic activity of these CEs in the DSSCs. Table 4 summarizes the photovoltaic parameters of $MoSe_2$, $WSe_2$, $TaSe_2$, $NbSe_2$, $FeSe_2$, $CoSe_2$ and $Bi_2Se_3$ based CEs for DSSCs, and their comparison with a standard Pt CE.

**Table 4.** Photovoltaic parameters of $MoSe_2$, $WSe_2$, $TaSe_2$, $NbSe_2$, $FeSe_2$, $CoSe_2$ and $Bi_2Se_3$ based CEs for DSSCs. FTO glass is the common substrate used in assembling the DSSCs with different CE materials. The measurements were conducted at a simulated solar light intensity of 100 mW/cm² (AM 1.5G) unless specified. The photovoltaic parameters short-circuit photocurrent density ($J_{SC}$), open-circuit voltage ($V_{OC}$), fill factor (FF), and power conversion efficiency ($\eta$), series resistance ($R_s$), charge-transfer resistance ($R_{CT}$), electrolyte and dye used for DSSCs are summarized and compared with a standard Pt CE.

| Counter Electrodes | Redox Couple | Dye | $J_{SC}$ (mA/cm²) | $V_{OC}$ (V) | FF (%) | PCE (η, %) | $R_s$ (Ω.cm²) | $R_{CT}$ (Ω.cm²) | Ref. |
|---|---|---|---|---|---|---|---|---|---|
| MoSe$_2$/Mo (*in-situ* sulfurization) | $I^-/I_3^-$ | N719 | 16.71 | 0.746 | 72.2 | 9.00 | 1.74 | 1.39 | 260 |
| MoS$_2$/Mo (*in-situ* sulfurization) | $I^-/I_3^-$ | N719 | 16.95 | 0.726 | 70.6 | 8.69 | 1.21 | 5.25 | 260 |
| Pt reference | $I^-/I_3^-$ | N719 | 17.19 | 0.740 | 68.3 | 8.68 | 12.52 | 0.22 | 260 |
| MoSe$_2$/Mo (*in-situ* selenization) | $I^-/I_3^-$ | N719 | 15.07 | 0.805 | 67 | 8.13 | 2.64 | 0.30 | 261 |
| Pt reference | $I^-/I_3^-$ | N719 | 16.11 | 0.794 | 63 | 8.06 | 15.98 | 8.95 | 261 |
| NiSe$_2$ | $I^-/I_3^-$ | N719 | 14.3 | 0.75 | 68 | 7.3 | 20.9 | 45.0 | 262 |
| NiS$_2$ | $I^-/I_3^-$ | N719 | 14.7 | 0.72 | 52 | 5.5 | 28.4 | 50.4 | 262 |
| CoSe$_2$ | $I^-/I_3^-$ | N719 | 13.5 | 0.72 | 68 | 6.6 | 14.8 | 102.7 | 262 |
| MoSe$_2$ | $I^-/I_3^-$ | N719 | 13.0 | 0.67 | 68 | 5.9 | 16.5 | 229.8 | 262 |
| Pt reference | $I^-/I_3^-$ | N719 | 14.0 | 0.75 | 69 | 7.2 | 18.5 | 34.2 | 262 |
| MoSe$_2$ (hollow spheres) | $I^-/I_3^-$ | N749 | 16.06 | 0.704 | 38.67 | 4.46 | - | 10.24 | 264 |
| MoSe$_2$/graphene (12.5% N$_2$-doped) | $I^-/I_3^-$ | N749 | 19.73 | 0.724 | 70.07 | 10.01 | 7.18 | 3.04 | 264 |
| MoSe$_2$/graphene | $I^-/I_3^-$ | N749 | 17.12 | 0.710 | 60.41 | 7.34 | - | 8.49 | 264 |
| Graphene | $I^-/I_3^-$ | N749 | 16.67 | 0.535 | 54.12 | 4.83 | - | 16.27 | 264 |
| Pt reference | $I^-/I_3^-$ | N749 | 19.93 | 0.723 | 73.22 | 10.55 | 7.14 | 2.81 | 264 |
| MoSe$_2$/PEDOT:PSS | $I^-/I_3^-$ | N719 | 15.97 | 0.70 | 67 | 7.58 | 18.08 | 5.43 | 265 |
| Pt reference | $I^-/I_3^-$ | N719 | 16.38 | 0.74 | 65 | 7.81 | 20.19 | 4.60 | 265 |
| MoSe$_2$/PEDOT:PSS@Ti | $I^-/I_3^-$ | N719 | 16.41 | 0.75 | 69 | 8.51 | - | - | 265 |
| Pt@Ti reference | $I^-/I_3^-$ | N719 | 16.31 | 0.74 | 68 | 8.21 | - | - | 265 |
| Bare MoSe$_2$ | $I^-/I_3^-$ | N719 | 12.65 | 0.66 | 28 | 2.29 | 20.50 | 39.74 | 265 |
| Bare PEDOT:PSS | $I^-/I_3^-$ | N719 | 9.32 | 0.67 | 46 | 2.90 | 17.36 | 190.91 | 265 |
| Co$_{0.85}$Se | $I^-/I_3^-$ | N719 | 13.41 | 0.763 | 62.7 | 6.42 | - | - | 266 |
| 3Co$_{0.85}$Se/0.5MoSe$_2$/0.5MoO$_3$ | $I^-/I_3^-$ | N719 | 13.43 | 0.765 | 65.7 | 6.75 | - | - | 266 |
| 2Co$_{0.85}$Se/MoSe$_2$/MoO$_3$ | $I^-/I_3^-$ | N719 | 13.80 | 0.768 | 67.1 | 7.10 | - | - | 266 |
| Co$_{0.85}$Se/1.5MoSe$_2$/1.5MoO$_3$ | $I^-/I_3^-$ | N719 | 13.06 | 0.760 | 64.4 | 6.39 | - | - | 266 |
| MoSe$_2$/MoO$_3$ | $I^-/I_3^-$ | N719 | 12.95 | 0.754 | 62.7 | 6.12 | - | - | 266 |
| Pt reference | $I^-/I_3^-$ | N719 | 13.05 | 0.759 | 60.9 | 6.03 | - | - | 266 |
| Co$_{0.85}$Se | $I^-/I_3^-$ | N719 | 13.44 | 0.66 | 68 | 6.03 | 55.83 | 9.28 | 267 |
| Pt reference | $I^-/I_3^-$ | N719 | 14.37 | 0.67 | 67 | 6.45 | 30.64 | 13.89 | 267 |
| Co$_{0.85}$Se (front irradiation) | $I^-/I_3^-$ | N719 | 16.74 | 0.742 | 66.8 | 8.30 | - | 2.84 | 268 |
| Co$_{0.85}$Se (rear irradiation) | $I^-/I_3^-$ | N719 | 9.92 | 0.721 | 64.7 | 4.63 | - | - | 268 |
| Ni$_{0.85}$Se (front irradiation) | $I^-/I_3^-$ | N719 | 16.67 | 0.740 | 63.6 | 7.85 | - | 2.96 | 268 |
| Ni$_{0.85}$Se (rear irradiation) | $I^-/I_3^-$ | N719 | 9.26 | 0.731 | 64.6 | 4.37 | - | - | 268 |
| Cu$_{0.50}$Se (front irradiation) | $I^-/I_3^-$ | N719 | 14.55 | 0.713 | 62.0 | 6.43 | - | 5.44 | 268 |
| Cu$_{0.50}$Se (rear irradiation) | $I^-/I_3^-$ | N719 | 10.01 | 0.666 | 63.6 | 4.24 | - | - | 268 |
| FeSe (front irradiation) | $I^-/I_3^-$ | N719 | 17.10 | 0.733 | 61.0 | 7.64 | - | 4.90 | 268 |
| FeSe (rear irradiation) | $I^-/I_3^-$ | N719 | 10.49 | 0.732 | 65.8 | 5.05 | - | - | 268 |
| Ru$_{0.33}$Se (front irradiation) | I$^-$/I$_3^-$ | N719 | 18.93 | 0.715 | 68.1 | 9.22 | - | 2.77 | 268 |
| Ru$_{0.33}$Se (rear irradiation) | I$^-$/I$_3^-$ | N719 | 11.89 | 0.714 | 69.5 | 5.90 | - | - | 268 |
| Pt (front irradiation) | I$^-$/I$_3^-$ | N719 | 13.09 | 0.712 | 66.3 | 6.18 | - | 7.23 | 268 |
| Pt (rear irradiation) | I$^-$/I$_3^-$ | N719 | 9.48 | 0.652 | 57.6 | 3.56 | - | - | 268 |
| | | | | | | | | | |
| NiCo$_2$S$_4$ (hydrothermal method) | I$^-$/I$_3^-$ | N719 | 13.38 | 0.76 | 63.2 | 6.9 | - | - | 269 |
| NiCo$_2$O$_4$ (hydrothermal method) | I$^-$/I$_3^-$ | N719 | 8.2 | 0.67 | 26.7 | 1.5 | - | - | 269 |
| Pt reference | I$^-$/I$_3^-$ | N719 | 14.20 | 0.8 | 63.4 | 7.7 | - | - | 269 |
| | | | | | | | | | |
| NbSe$_2$ (nanosheets) | I$^-$/I$_3^-$ | N719 | 15.04 | 0.77 | 63 | 7.34 | 27.72 | 2.59 | 279 |
| NbSe$_2$ (nanorods) | I$^-$/I$_3^-$ | N719 | 13.94 | 0.76 | 64 | 6.78 | 19.38 | 6.21 | 279 |
| NbSe$_2$/C | I$^-$/I$_3^-$ | N719 | 15.58 | 0.77 | 65 | 7.80 | 24.07 | 3.52 | 279 |
| Pt reference | I$^-$/I$_3^-$ | N719 | 15.88 | 0.72 | 69 | 7.90 | 8.15 | 2.35 | 279 |
| | | | | | | | | | |
| NbSe$_2$ (nanosheets) | I$^-$/I$_3^-$ | N719 | 16.85 | 0.74 | 62 | 7.73 | - | - | 280 |
| NbSe$_2$ (nanorods) | I$^-$/I$_3^-$ | N719 | 14.85 | 0.74 | 46 | 5.05 | - | - | 280 |
| NbSe$_2$ (microparticles) | I$^-$/I$_3^-$ | N719 | 14.93 | 0.75 | 55 | 6.27 | - | - | 280 |
| Pt reference | I$^-$/I$_3^-$ | N719 | 15.59 | 0.72 | 62 | 7.01 | - | - | 280 |
| | | | | | | | | | |
| MoSe$_2$ (solvothermal reaction) | I$^-$/I$_3^-$ | N719 | 14.11 | 0.73 | 65 | 6.70 | 10.32 | 2.43 | 281 |
| WSe$_2$ (solvothermal reaction) | I$^-$/I$_3^-$ | N719 | 15.50 | 0.73 | 66 | 7.48 | 10.70 | 0.78 | 281 |
| TaSe$_2$ (solvothermal reaction) | I$^-$/I$_3^-$ | N719 | 15.81 | 0.73 | 64 | 7.32 | 10.00 | 1.89 | 281 |
| Pt reference | I$^-$/I$_3^-$ | N719 | 16.84 | 0.70 | 67 | 7.91 | 6.32 | 1.32 | 281 |
| | | | | | | | | | |
| NiSe$_2$ (hydrothermal reaction) | I$^-$/I$_3^-$ | N719 | 15.94 | 0.734 | 74.3 | 8.69 | 2.57 | 0.81 | 282 |
| Pt reference | I$^-$/I$_3^-$ | N719 | 15.26 | 0.731 | 72.1 | 8.04 | 2.50 | 0.97 | 282 |
| | | | | | | | | | |
| NiCo$_2$S$_4$/NiS (spin casting) | I$^-$/I$_3^-$ | N719 | 17.7 | 0.744 | 67 | 8.8 | - | - | 286 |
| NiCo$_2$S$_4$ (spin casting) | I$^-$/I$_3^-$ | N719 | 17.4 | 0.743 | 66 | 8.5 | - | - | 286 |
| Co$_9$S$_8$ (spin casting) | I$^-$/I$_3^-$ | N719 | 16.2 | 0.741 | 64 | 7.7 | - | - | 286 |
| NiS (spin casting) | I$^-$/I$_3^-$ | N719 | 14.9 | 0.735 | 63 | 6.9 | - | - | 286 |
| Pt reference | I$^-$/I$_3^-$ | N719 | 16.5 | 0.736 | 67 | 8.1 | - | - | 286 |
| | | | | | | | | | |
| Ni$_{0.33}$Co$_{0.67}$Se | I$^-$/I$_3^-$ | N719 | 17.29 | 0.789 | 67 | 9.01 | 30.40 | 1.11 | 287 |
| Ni$_{0.5}$Co$_{0.5}$Se | I$^-$/I$_3^-$ | N719 | 16.42 | 0.783 | 69 | 8.80 | 30.38 | 1.50 | 287 |
| Ni$_{0.67}$Co$_{0.33}$Se | I$^-$/I$_3^-$ | N719 | 15.89 | 0.784 | 69 | 8.59 | 30.04 | 1.96 | 287 |
| Co$_3$Se$_4$ | I$^-$/I$_3^-$ | N719 | 14.96 | 0.793 | 67 | 7.95 | 29.94 | 7.66 | 287 |
| NiSe | I$^-$/I$_3^-$ | N719 | 14.54 | 0.783 | 64 | 7.23 | 29.87 | 13.88 | 287 |
| Pt reference | I$^-$/I$_3^-$ | N719 | 15.33 | 0.791 | 69 | 8.30 | 30.30 | 2.88 | 287 |
| | | | | | | | | | |
| FeSe$_2$ (nanosheets, under N$_2$) | I$^-$/I$_3^-$ | N719 | 17.49 | 0.718 | 60 | 7.53 | 8.07 | 0.53 | 302 |
| FeSe$_2$ (nanosheets, air exposed) | I$^-$/I$_3^-$ | N719 | 15.51 | 0.708 | 56 | 6.15 | 10.13 | 6.26 | 302 |
| FeSe$_2$ (microparticles) | I$^-$/I$_3^-$ | N719 | 16.32 | 0.715 | 59 | 6.88 | 8.51 | 4.10 | 302 |
| Pt reference | I$^-$/I$_3^-$ | N719 | 17.77 | 0.725 | 58 | 7.47 | 7.94 | 1.68 | 302 |
| | | | | | | | | | |
| FeSe$_2$ microparticles (MPs) | I$^-$/I$_3^-$ | N719 | 15.63 | 0.745 | 66 | 7.68 | 9.03 | 2.32 | 303 |
| FeSe$_2$ nanosheets (NSs) | I$^-$/I$_3^-$ | N719 | 16.14 | 0.744 | 70 | 8.39 | 8.78 | 0.49 | 303 |
| FeSe$_2$ nanorods (NRs) | I$^-$/I$_3^-$ | N719 | 15.79 | 0.748 | 68 | 8.03 | 8.71 | 1.62 | 303 |
| Pt reference | I$^-$/I$_3^-$ | N719 | 15.87 | 0.750 | 69 | 8.20 | 8.62 | 1.15 | 303 |
| FeSe$_2$ (3D flower-like) | Γ$^-$/I$_3^-$ | N719 | 14.93 | 0.744 | 72.1 | 8.00 | 16.82 | 0.53 | 304 |
| FeSe$_2$ (sphere-shaped) | Γ$^-$/I$_3^-$ | N719 | 14.60 | 0.724 | 69.8 | 7.38 | 27.05 | 0.96 | 304 |
| Pt reference | Γ$^-$/I$_3^-$ | N719 | 15.13 | 0.741 | 70.2 | 7.87 | 17.01 | 0.78 | 304 |
| | | | | | | | | | |
| CoSe$_2$ (hydrothermal, 140 °C) | Γ$^-$/I$_3^-$ | N719 | 16.65 | 0.750 | 64.4 | 8.04 | 8.783 | 0.132 | 309 |
| CoSe$_2$ (hydrothermal, 160 °C) | Γ$^-$/I$_3^-$ | N719 | 17.04 | 0.743 | 66.2 | 8.38 | 8.034 | 0.097 | 309 |
| CoSe$_2$ (hydrothermal, 180 °C) | Γ$^-$/I$_3^-$ | N719 | 15.44 | 0.750 | 63.9 | 7.40 | 15.17 | 0.932 | 309 |
| Pt reference | Γ$^-$/I$_3^-$ | N719 | 16.88 | 0.743 | 62.4 | 7.83 | 12.86 | 1.923 | 309 |
| | | | | | | | | | |
| CoSe$_2$/C-NG | Γ$^-$/I$_3^-$ | N719 | 17.51 | 0.73 | 67 | 8.41 | 20.6 | 0.85 | 310 |
| CoSe$_2$/C-NR | Γ$^-$/I$_3^-$ | N719 | 15.98 | 0.73 | 67 | 7.83 | 20.6 | 1.16 | 310 |
| CoSe$_2$/C-NCW | Γ$^-$/I$_3^-$ | N719 | 18.03 | 0.73 | 67 | 8.92 | 20.6 | 0.52 | 310 |
| CoSe$_2$/C-NCW on nickel foam | Γ$^-$/I$_3^-$ | N719 | 18.86 | 0.78 | 71 | 10.46 | - | - | 310 |
| CoSe$_2$/C-NCW on carbon cloth | Γ$^-$/I$_3^-$ | N719 | 18.16 | 0.76 | 71 | 9.87 | - | - | 310 |
| Pt reference | Γ$^-$/I$_3^-$ | N719 | 16.43 | 0.74 | 67 | 8.25 | 20.6 | 1.04 | 310 |
| | | | | | | | | | |
| CoSe$_2$ | Γ$^-$/I$_3^-$ | N719 | 12.95 | 0.773 | 65 | 6.47 | - | 0.50 | 311 |
| CoSe$_2$@RGO | Γ$^-$/I$_3^-$ | N719 | 12.24 | 0.792 | 72 | 7.01 | - | 0.20 | 311 |
| Reduced graphene oxide (RGO) | Γ$^-$/I$_3^-$ | N719 | 12.11 | 0.761 | 40 | 3.66 | - | 64.75 | 311 |
| Pt reference | Γ$^-$/I$_3^-$ | N719 | 13.12 | 0.765 | 67 | 6.77 | - | 0.61 | 311 |
| | | | | | | | | | |
| Ni$_{0.85}$Se | Γ$^-$/I$_3^-$ | N719 | 15.63 | 0.739 | 72 | 8.32 | 1.8 | 1.8 | 312 |
| Co$_{0.85}$Se | Γ$^-$/I$_3^-$ | N719 | 16.98 | 0.738 | 75 | 9.40 | 2.1 | 0.6 | 312 |
| Pt reference | Γ$^-$/I$_3^-$ | N719 | 16.03 | 0.738 | 73 | 8.64 | 2.6 | 1.1 | 312 |
| | | | | | | | | | |
| Bi$_2$Se$_3$ nanoparticles | Γ$^-$/I$_3^-$ | N719 | 7.02 | 0.55 | 46 | 1.86 | - | - | 325 |
| Bi$_2$Se$_3$/graphene (40 mg) | Γ$^-$/I$_3^-$ | N719 | 15.42 | 0.78 | 50 | 6.35 | - | - | 325 |
| Bi$_2$Se$_3$/graphene (60 mg) | Γ$^-$/I$_3^-$ | N719 | 16.36 | 0.75 | 57 | 7.09 | - | - | 325 |
| Bi$_2$Se$_3$/graphene (80 mg) | Γ$^-$/I$_3^-$ | N719 | 16.01 | 0.76 | 53 | 6.66 | - | - | 325 |
| Pt reference | Γ$^-$/I$_3^-$ | N719 | 15.65 | 0.68 | 59 | 6.47 | - | - | 325 |
| | | | | | | | | | |
| ZnO (photoanode) | Γ$^-$/I$_3^-$ | N719 | 8.189 | 0.656 | 55.2 | 2.96 | 16.6 | 7.23 | 335 |
| Bi$_2$Te$_3$/ZnO (photoanode) | Γ$^-$/I$_3^-$ | N719 | 11.767 | 0.637 | 57.0 | 4.27 | 15.8 | 3.75 | 335 |

Nanograin (NG), nanorock (NR), nanoclimbing-wall (NCW), nickel foam (NF), carbon cloth (CC), rGO = reduced graphene oxide.

In the case of R$_s$ and R$_{CT}$: Some of the authors used $\Omega$ instead of $\Omega.cm^2$ for the resistances without mentioning the size of the electrode.

## 4.7 Bi$_2$Se$_3$ Counter Electrodes

Bismuth selenide (Bi$_2$Se$_3$) and bismuth telluride (Bi$_2$Te$_3$) have been previously studied as topological insulators.[313, 314] Bi$_2$Se$_3$, a semiconducting and thermoelectric material, can be processed into single layers, nanosheets, nanotubes, nanoribbons, and nanowires[315-318] and has applications in field-effect transistors,[319] sensors,[320] non-volatile memory devices,[321] photovoltaic devices,[322] and drug delivery and anti-cancer therapy.[323, 324] $Bi_2Se_3$/RGO nanocomposites were prepared by a microwave-assisted hydrothermal method as a CE for a DSSC by Zhu et al.[325] Figure 27 shows the TEM images of the graphene nanosheets, $Bi_2Se_3$ nanospheres and $Bi_2Se_3$/graphene nanocomposites containing 60 mg graphene contents. The TEM images showed that 10-15 nm $Bi_2Se_3$ nanoparticles were attached onto graphene nanosheets. SEM images showed that the graphene had a flake-like structure while the $Bi_2Se_3$ nanoparticles were large size spheres. Furthermore, the surface of the graphene oxide was found to be very smooth in comparison with graphene nanosheets doped with $Bi_2Se_3$ nanoparticles. The inclusion of $Bi_2Se_3$ nanoparticles onto graphene nanosheets was controlled to achieve high electrocatalytic activity of the CE for the reduction of triiodide ($I_3^-$). The DSSC with a $Bi_2Se_3$/graphene (60 mg) CE yielded a high PCE of 7.09%, which is comparable to a Pt CE ($\eta$ = 6.23%). The $Bi_2Se_3$/graphene nanocomposite CEs with 40 mg and 80 mg graphene content showed PCE values of 6.35% and 6.66% for the DSSCs, respectively, which was much higher than that of pure $Bi_2Se_3$ CE ($\eta$ = 1.86%), but still comparable to a Pt CE.

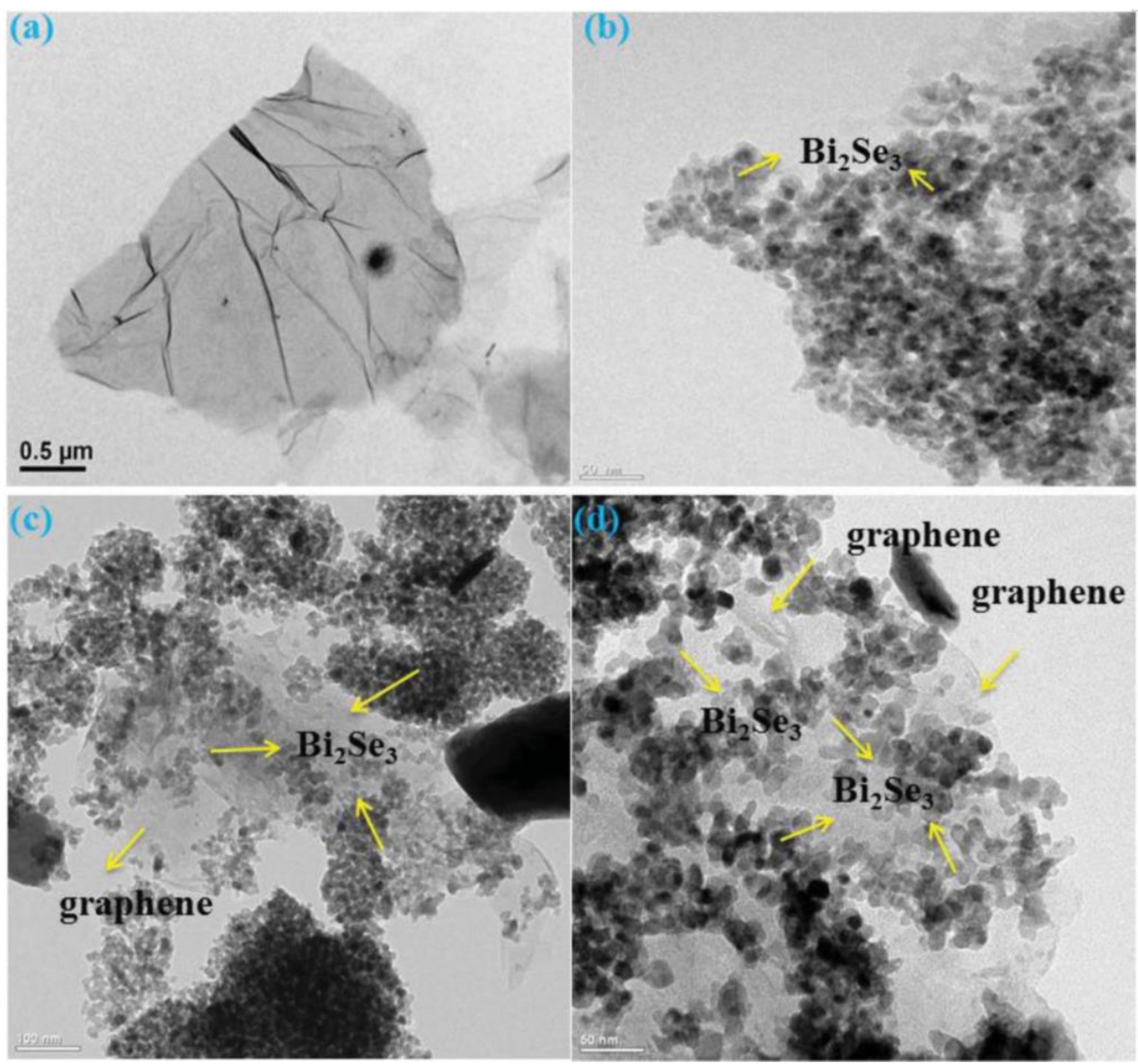

**Figure 27.** TEM images of (a) graphene nanosheet, (b) $Bi_2Se_3$ nanospheres (c) and (d) $Bi_2Se_3$/graphene nanocomposite (60 mg graphene). Reprinted with permission from ref. 325. Copyright 2016, Taylor & Francis Group.

## 5. $Bi_2Te_3$ Based Photoanodes

Bismuth telluride ($Bi_2Te_3$) has been studied for thermoelectric applications and can be processed into nanowires arrays,[326-329] nanotubes,[330] nanoplates,[331] nanosheets,[332, 333] and thin films.[334] Dou et al.[335] developed hybrid photoanodes by dispersing $Bi_2Te_3$ nanotubes into ZnO nanoparticles.

The 0.5, 1.0, 1.5, 2.0, and 2.5 wt.% of highly crystalline $Bi_2Te_3$ nanotubes were mixed with ZnO nanoparticles to fabricate photoanodes for DSSCs. Pt films deposited on FTO glass was used as a CE. The electrolyte consisted of 0.05 M $I_2$, 0.5 M LiI, and 0.1 M 4-*tert*-butylpyridine in acetonitrile-propylene carbonate (1:1) solution. In the $Bi_2Te_3$/ZnO hybrid, $Bi_2Te_3$ nanotubes provided a conduction pathway which facilitated electron transfer. The *Jsc* value increased gradually with increasing $Bi_2Te_3$ nanotubes content up to 1.5 wt.% (11.767 mA/cm$^2$) and then started decreasing as the $Bi_2Te_3$ nanotubes content exceeded 2.0 wt.% (9.957 mA/cm$^2$), due to the low dye loading on the surface of ZnO nanoparticles. A similar trend was observed for the PCE values. The DSSCs with a $Bi_2Te_3$/ZnO composite photoanode containing 1.5 wt.% $Bi_2Te_3$ nanotubes in the ZnO photoanode showed a PCE of 4.27%, which is 44.3% higher compared with a pure ZnO photoanode. The PCE values were 2.96% for the bare ZnO photoanode, and 3.74, 3.96, 4.27, 3.41 and 3.01% for the 0.5, 1.0, 1.5, 2.0 and 2.5 wt.% $Bi_2Te_3$ nanotubes content photoanodes, respectively. This study indicated that loading of thermoelectric $Bi_2Te_3$ in a ZnO photoanode improves the  photovoltaic performance of DSSCs. Wan et al.[336] prepared hexagonal $Bi_2Te_3$ nanosheets of 300-400 nm length by a hydrothermal method. The nanocomposites of the $Bi_2Te_3$ nanosheets and ZnO nanoparticles were used as photoanodes for DSSCs, where the $Bi_2Te_3$ nanosheet concentration was varied from 0 to 0.25 wt.%. The DSSCs with 0.15 wt.% of $Bi_2Te_3$ nanosheets in the $Bi_2Te_3$ nanosheet/ZnO nanoparticle composite photoanode showed a PCE of 4.10%, which was improved by 46.95% compared with the bare ZnO photoanode. The thermoelectric $Bi_2Te_3$ nanosheets in the composite photoanode helped in enhancing the electron density and reducing the temperature of the DSSCs. The $Bi_2Te_3$ nanosheet/ ZnO nanoparticles composite photoanode significantly improved the DSSC performance. A thermoelectric $Bi_2Te_3$/TiO$_2$ composite based photoanode was also used for DSSCs, where $Bi_2Te_3$ nanoplates help in converting heat into electricity and enhanced the rate of charge transfer.[337] This resulted in a 28% increase of the PCE of the DSSC. The $Bi_2Te_3$ nanoplates have also been used for doping $TiO_2$ photoanodes.[338] The effect of $Bi_2Te_3$ nanoplates size was evaluated for DSSCs, where a decrease in the size of the $Bi_2Te_3$ nanoplates led to and higher PCE. The performance of DSSCs with a $Bi_2Te_3/TiO_2$ photoanode increased by 15.3% compared to the undoped $TiO_2$ photoanode.

## 6. Long-Term Stability of TMDs Based DSSCs

The long-term stability of DSSCs is one of the most important parameters for commercial applications. The stability of DSSC devices depends upon a number of factors and the components used in their fabrication. The decrease in power conversion efficiency of a DSSC can be associated with the stability of different electrolyte components, photosensitizing dyes, aging of the $TiO_2$ photoanode, degradation of the CE (cathode), corrosion of components by electrolytes, sealant, leakage, exposure to solar irradiation, high humidity, and elevated temperature.[339-348]

A few research reports have been published on the long-term stability of TMDs based DSSCs which are briefly discussed here. Infant et al.[131] studied the stability of CVD-deposited vertically oriented $MoS_2$ thin films on an FTO surface used as a CE in a DSSC. The electrochemical stability of a $MoS_2$ CE based DSSC was analyzed by CV measurements, where electrodes were repeatedly subjected to 20 cycles at a 10 mV/s scan rate for the $I^-/I_3^-$ redox couple in the electrolyte. The $MoS_2$ based CE showed no significant change up to 20th consecutive cycle, whereas the Pt CE exhibited some changes between the cycles (Figure 28). This confirms that CVD deposited $MoS_2$ strongly adhered onto the surface of the FTO substrate.

The stability of MoS$_2$ CEs was measured under ambient conditions by storing them for 15 days, where the PCE remained at 94% of its initial efficiency value, which was much higher than that of the Pt CE.

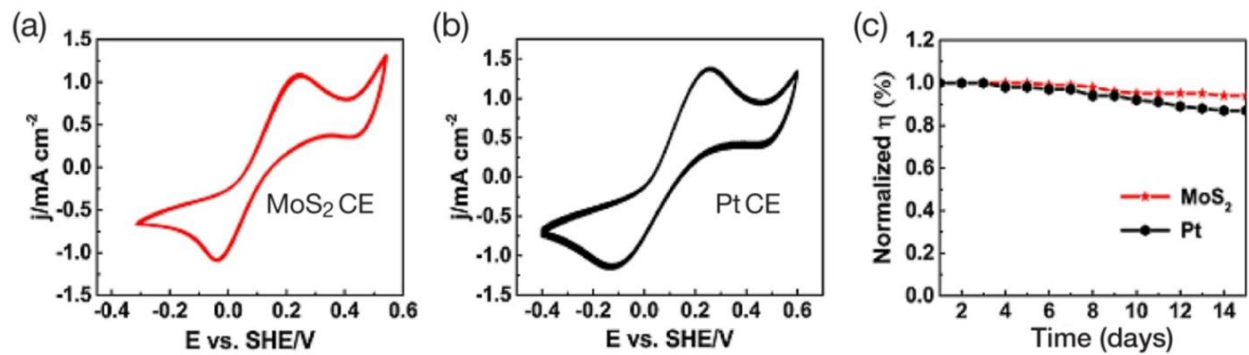

**Figure 28.** Electrochemical properties of thin films of MoS$_2$ prepared at 600 $^o$C for 15 minutes. 20 consecutive cyclic voltammogram (CV) curve of MoS$_2$ (A) and Pt (B) counter electrodes recorded at the scan rate of 10 mV/s and (C) PCE of MoS$_2$ and Pt CEs based DSSCs measured for 15 days under ambient conditions. Reprinted with permission from ref. 131. Copyright 2016, Elsevier.

For preparing Pt-free dye-sensitized solar cells, Liu et al.[350] fabricated DSSCs using MoS$_2$ and RGO composite as a CE for the reduction of triiodide (I$_3^-$) to iodide (I$^-$). AFM, XPS, and XRD confirmed the deposition of MoS$_2$ nanoparticles onto the RGO surface. The CV measurement showed a higher current density for the MoS$_2$/RGO nanocomposite based CE compared to RGO, MoS$_2$, and Pt-sputtered CEs due to an increased surface area. The MoS$_2$/RGO CE also exhibited a low R$_{CT}$ of 0.57 $\Omega$.cm$^2$ for the reduction of triiodide (I$_3^-$) to iodide (I$^-$). The MoS$_2$/RGO nanocomposite CE based DSSC showed a PCE of 6.04%, comparable to a PCE of 6.38% for the conventional Pt CE. MoS$_2$/RGO nanocomposites based CEs also have better electrochemical stability, as no degradation in current densities was observed up to 100 repeated CV tests. The stability test conducted on a DSSC having

MoS$_2$/RGO nanocomposites as CEs showed over 10% degradation in PCE over a period of 20 days, as depicted in (Figure 29a). Therefore, MoS$_2$/RGO nanocomposites based CEs were found to be stable both for environmental and consecutive electrochemical tests. Li et al.[200] prepared a composite film of TiS$_2$/PEDOT:PSS on an ITO substrate as a CE of DSSCs for the I$^-$/I$_3^-$ redox system, which exhibited a PCE as high as 7.04% and is comparable to a Pt CE. Figure 29b shows dark current density-voltage curves of DSSCs with Pt, bare TiS$_2$, bare PEDOT:PSS, and 10 wt.% TiS$_2$/PEDOT:PSS composite CEs, and also the long-term stability of a 10 wt.% TiS$_2$/PEDOT:PSS composite CE based DSSC. This another example of long-term stability of DSSCs based on TiS$_2$/PEDOT:PSS CEs .

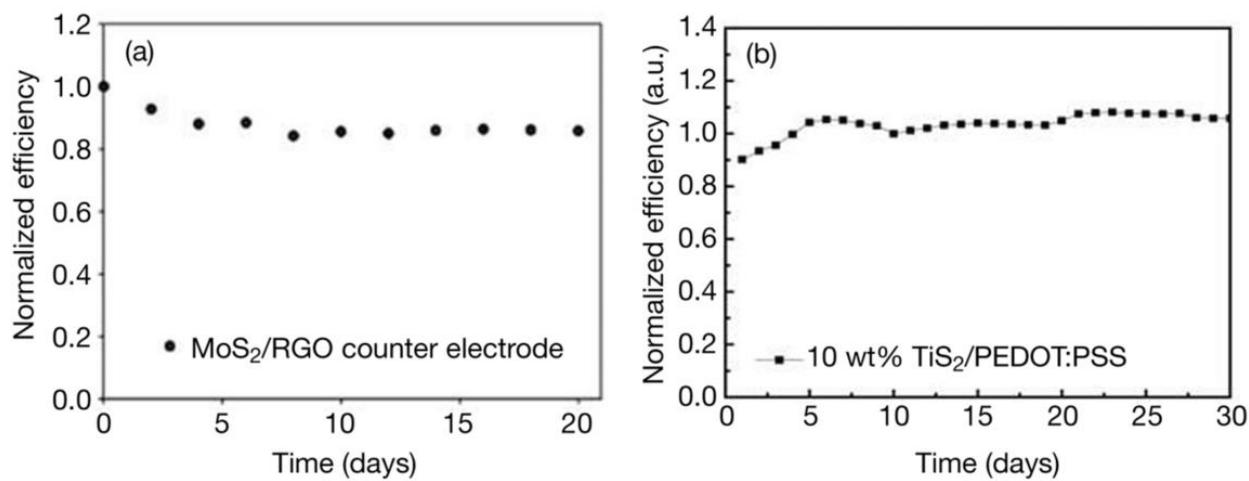

**Figure 29.** (a) Stability of a DSSC with MoS$_2$/RGO based counter electrode recorded for 20 days. Reprinted with permission from ref. 350. Copyright 2012, Royal Society of Chemistry. (b) The long-term stability of the DSSCs with 10 wt% TiS$_2$/PEDOT:PSS composite CE. Reprinted with permission from ref. 200. Copyright 2013, The Royal Society of Chemistry.

The stability of DSSCs with a TiS$_2$/graphene hybrid CE was studied by Meng et al.[199] (Figure 30). The TiS$_2$–graphene hybrid CE maintained 96% of its initial PCE value after 500 hours in air, also exhibiting higher electrochemical stability. The R$_s$ and Z$_N$ values for the TiS$_2$– graphene hybrid CE based DSSC did not change after 10 cyclic measurements. The $R_{CT}$ value of a DSSC with Pt CE significantly increased with increasing cycling number, while there was no change in the $R_{CT}$ value of the TiS$_2$–graphene hybrid after 10 cycles. This indicates better electrochemical stability of the TiS$_2$–graphene hybrid CE than the Pt CE. The stability of photovoltaic parameters of DSSCs having mesoporous CoS$_2$ nanotube arrays as CEs was recorded for 10 days by Tsai et al.[233] A slight drop in the $V_{OC}$ and FF values were observed, which resulted in a 2.2% decrease in the PCE of the CoS$_2$ nanotube array CEs. The photovoltaic parameters of the DSSC were quite stable up to 10 days, indicating good stability of DSSCs with mesoporous CoS$_2$ nanotube array CEs (Figure 31).

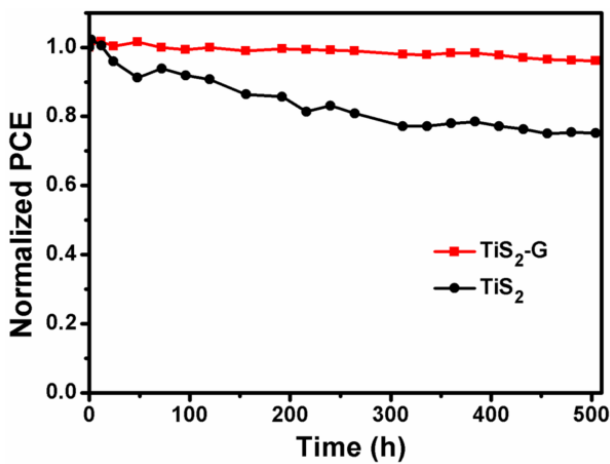

**Figure 30.** Stability of DSSCs with TiS$_2$–graphene hybrid and Pt CEs. Reprinted with permission from ref. 199. Copyright 2016, Elsevier.

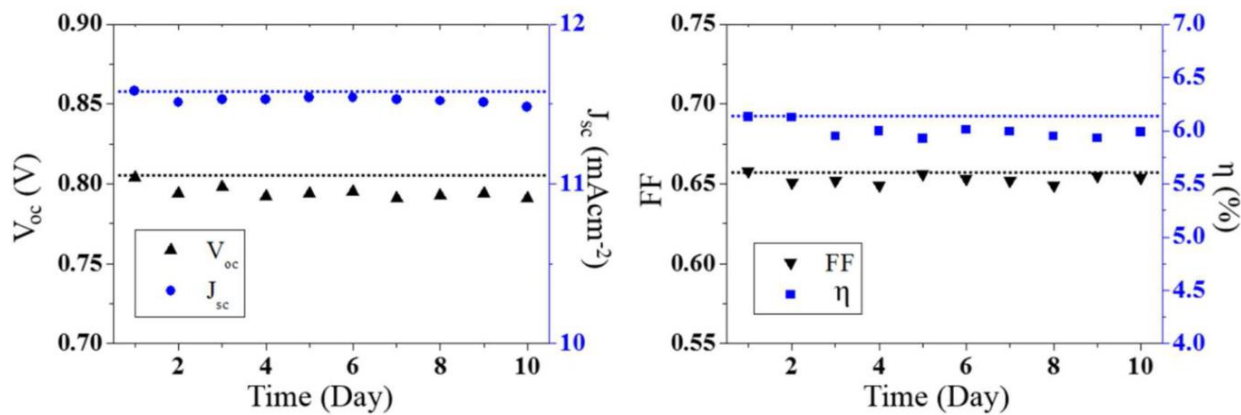

**Figure 31.** Stability of photovoltaic parameters; open-circuit voltage ($V_{OC}$), short-circuit photocurrent density ($J_{SC}$), fill factor (FF), and power conversion efficiency ($\eta$) of DSSCs with mesoporous $CoS_2$ nanotube array CE as a function of time. Reprinted with permission from ref. 233. Copyright 2015, Wiley-VCH.

The electrochemical stability of DSSCs with $FeSe_2$ nanosheets and Pt CEs in an iodine-based electrolyte was studied by Huang et al.[302] Both CEs were subjected to consecutively CV scanning. The $FeSe_2$ nanosheets-based CE showed no change in the current densities and $E_{pp}$ values up to 1000 cycles, whereas the $E_{pp}$ value for the Pt-based CE was observed to increase after 1000 consecutive cycles (Figure 32). This study confirmed a better corrosion resistance of $FeSe_2$ nanosheets CE to the iodine-based electrolyte than a Pt CE. Both CEs were also subjected to sequential EIS scanning, where negligible changes in $R_{CT}$ were noticed after 10 cycles, again indicating an excellent electrochemical stability of both CEs. Furthermore, $R_s$ and $Z_N$ values also showed no change upon repeated scanning cycles.

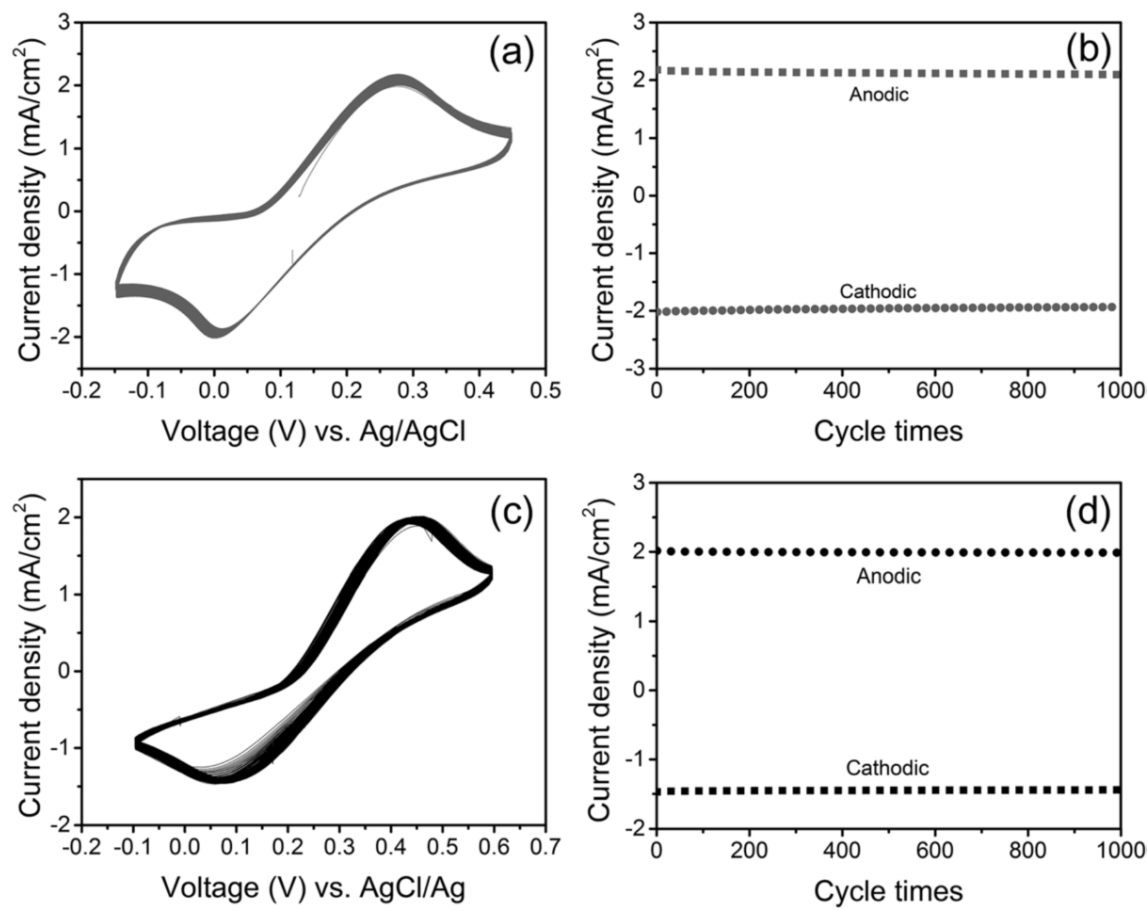

**Figure 32.** 1000 consecutive cycles of CVs of FeSe$_2$ nanosheets-based CE (a) and Pt-based CE (c) at a scan rate of 50 mV/s and the anodic as well as cathodic peak current densities up to 1000 cycles for FeSe$_2$- (b) and Pt-based CEs (d), respectively. Reprinted with permission from ref. 302. Copyright 2015, Wiley-VCH.

The stability of a FeS$_2$ nanorod based CE was also measured in an iodide (I$^-$) electrolyte up to 10 days, and CV plots showed slight change at different times of aging.[222] The hydrothermally synthesized CoSe$_2$ nanorods were used as an electrocatalyst for a DSSC for the reduction of I$_3^-$ using N719 dye by Sun et al.[351] The single crystalline CoSe$_2$ nanorods based CE showed a PCE of 10.20%, compared with a PCE of 8.17% for a Pt CE, under 1 Sun illumination. The DSSCs having CoSe$_2$ CEs were stored in daylight and their photovoltaic properties were measured every day, and showed long-term stability. These studies show an excellent electrochemical stability of TMDs-based CEs for DSSCs in iodine-based electrolyte, with no corrosion. The TMDs CEs are also quite stable when stored under ambient conditions of up to 2-3 weeks, as no significant changes were observed in the PCEs of the DSSC devices. Furthermore, TMDs based CEs should be further investigated to exhibit better electrochemical stability and environmental stability than that of standard Pt CEs, and endurance tests should be carried out to study the their stability. The DSSS devices should have a service life of at least 20 years under ambient conditions as pointed out by Grätzel.[342] The long-term stability of DSSCs is feasible by carefully selecting components and solar cell structure. The TMDs based CEs may overcome the concerns associated with scarcity, high production cost, and corrosion of Pt CEs in electrolyte solutions. Researchers in this field should address the stability for of TMDs based CEs to evaluate the performance of Pt-free DSSCs.

Some achievements and strategies to overcome the long-term stability of Pt-free DSSCs are summarized in the following section. Kato et al.[352] used Raman spectroscopy and EIS to evaluate the durability of DSSCs for 2.5 years in outdoor conditions. Both N719 dye-adsorbed $TiO_2$ CEs and carbon CEs were found to be stable. The $V_{OC}$ and FF values were slightly decreased because of increased $Z_N$ of triiodide ($I_3^-$), arising from the change in electrolyte components. Matsui et al.[353] achieved stability over 1000 hours for DSSCs at 85 ℃ and under 85% relative humidity, and recorded no degradation of the photovoltaic performance between -40 and 90 ℃ for 200 cycles. Xue et al.[354] measured thermal stability of DSSCs between -20 to 25 ℃ temperature range for 1080 hours. DSSCs with N719 dye absorbed on the $TiO_2$ photoanode retained 80% of their initial PCE values after aging for 1080 hours. The deterioration of the N719 dye was found to be the main cause for a decrease the PCE and degradation of DSSC devices. Harikisun and Desilvestro[355] evaluated photovoltaic performance of Z907-based DSSCs after continuous light-soaking at 55–60 °C for 25,600 hours, where a slight degradation was observed. The accelerated aging tests predicted a life time of 40 years for Middle European conditions while 25 years for Southern European conditions. The 10% and 20% decrease of photovoltaic performance was measured for ionic liquid and solvent based electrolytes over 1,000 hours at 180 °C, respectively.

Strategies to improve the long-term stability of DSSCs include developing new photosensitizing dyes, new non-volatile electrolytes, encapsulation, and of course new photoanodes and CEs. Like the heterojunction solar cells,[27] several strategies for improving electrochemical and thermal stability of DSSC devices have been proposed. The role of photosensitizing dyes containing π-conjugated organic systems have been studied for the stability of DSSCs. Wu et al.[356] suggested a novel concept of molecular engineering of Donor–Acceptor−π–Acceptor (D–A−π–A) based photosensitizers, not only to improve the stability of DSSCs but also to enhance the photovoltaic performance. Katoh et al.[357] compared the stability of five sensitizing dyes in DSSCs with and without π-conjugated oligothiophene moiety, which indicated that dyes with π-conjugated oligothiophene exhibit higher stability than those of without oligothiophene moiety. *Joly et al.[358] fabricated DSSCs with a new organic sensitizer (RK1) which showed a Jsc of 18.26 mA/cm$^2$, Voc of 0.76 V, and FF of 0.74, resulting in a PCE value of 10.2% under 1 Sun illumination for the triiodide/iodide ($I_3^-/I^-$) redox couple.* A similar PCE of 10.19% was achieved for the ruthenium N719 dye. When RK1 dye was used with a viscous ionic liquid electrolyte, the DSSC yielded a *Jsc* of 15.40 mA/cm$^2$, *Voc* of 0.665 V, FF of 0.69, and *a* PCE of 7.36%, with outstanding stability. The DSSC exhibited no degradation of photovoltaic performance after visible-light soaking at 65 °C for 2200 hours, but, thereafter, DSSC started degrading and retained 75% of its initial PCE at 65 °C after 5000 hours.

A new coumarin dye, namely 2-cyano-3-{5′-[1-cyano-2-(1,1,6,6-tetramethyl-10-oxo-2,3,5,6-tetrahydro-1*H*,4*H*,10*H*-11-oxa-3a-aza-benzo[*de*]anthracen-9-yl)-vinyl]-[2,2′]bithiophenyl-5-yl}-acrylic acid (NKX-2883) was developed by Wang et al.[359] to examine the stability of DSSCs in a nonvolatile electrolyte made of 0.1 M $I_2$, 0.6 M 1,2-dimethyl-3-*n*-propylimidazolium iodide (DMPImI), and 0.1 M *N*-methylbenzimidazole (NMBI) in 3-methoxypropionitrile. The NKX-2883 dye-based DSSC showed a $J_{SC}$ value of 18.8 mA/cm$^2$ and a PCE of 6.5% with 6 micron thick $TiO_2$ film. The DSSCs maintained a PCE of 6% under continuous light soaking of 100 mW/cm$^2$ (1 Sun illumination) at 50–55 °C for 1000 hours. Organic photosensitizing dyes having long alkyl chains were also proposed to improve long-term stability for both liquid and quasi-solid-state DSSCs.[360] In another study, quinoxaline based metal-free organic sensitizing dyes were utilized to introduce long-term stability in DSSCs.[361] The length of the alkyl chains on the donor unit was found to affect the performance of DSSC devices. The quasi-solid-state DSSCs with a quinoxaline-based organic dye showed a PCE of 7.14%, and maintained 100% of its initial PCE value after continuous sunlight irradiation for 1000 hours, indicating that molecular engineering of dye molecules can lead to both high PCE and long-term stability of DSSC devices.

The electrochemical stability of CEs in corrosive triiodide/iodide ($I_3^-/I^-$) electrolyte is of significant concern because it restricts commercial applications of DSSCs. To overcome this disadvantage of the $I^-/I_3^-$ redox couple, research activities have been focused on finding alternative iodine-free non-corrosive redox electrolytes.[19, 362-366] Cell-sealing conditions are also important when using liquid electrolytes. The DSSCs can be made suitable for outdoor applications by using encapsulation. The role of electrolytes has been studied in the stability of DSSCs by Sauvage et al.[367] suggesting a new electrolyte based on butyronitrile solvent with low volatility, along with thiophene-based sensitizer Na-Ru(4,40-bis(5-hexylthio)thiophen-2-yl)-2,20-bipyridine)-4-carboxylic acid-40-carboxylate-2,20-bipyridine)(thiocyanate)2, coded C106, for DSSCs which showed >95% retention of PCE value after 1000 hours at 60 $^{\circ}$C for an exposure to 100 mW/cm$^2$ light illumination. Yoon et al.[368] fabricated DSSCs with 1-propyl-3-methyl imidazolium iodide (PMII) ion-gel electrolyte with a poly(styrene-block-ethyleneoxide-block-styrene) (SEOS) triblock copolymer. The DSSC with ion-gel electrolyte retained 92% of its initial PCE up to 1440 hours, compared to 78% for the ionic liquid electrolyte. Lee et al.[369] reported long-term stability of DSSCs with organic tetrabutylammonium iodide (TBAI) or 1-methyl 3-propyl imidazolium iodide (PMII) in methoxypropionitrile-based electrolytes. The DSSCs having TBAI retained 96.9% of their initial efficiency after being stored for 1000 hours under 1 Sun light irradiation at 60 ℃. Yang et al.[370] proposed the use of poly(ethylene oxide)–poly(vinylidene fluoride) (PEO–PVDF) polymer-blend electrolytes with water and ethanol for improving stability of DSSCs. The electrical conductivity was found to increase after adding water and ethanol to the PEO–PVDF polymer-blend electrolytes. The cross-linking capability of hydroxyl-rich additives for modified electrolytes was found to have a positive impact. Chen et al.[371] demonstrated the long-time durability of DSSCs by using a succinonitrile, silica nanoparticles and 1-butyl-3-methylimidazolium tetrafluoroborate (BMI·BF$_4$) gel system which maintained 93% of its initial PCE after aging at 60 ℃ for 1000 hours. Dembele et al.[372] also demonstrated that adding 1.0 wt.% concentration of MWCNTs to TiO$_2$ photoanodes can improve both PCE and stability of a DSSC, where the PCE value increased to 4.1% compared with 3.7% for pure TiO$_2$ photoanodes. The performance of the DSSC devices was measured for 10 consecutive days under ambient light exposure. The PCE decreased about 10% for the MWCNTs/TiO$_2$ photoanodes, compared to 35% decrease in pure TiO$_2$ photoanodes.

The long-term stability of Pt-free DSSCs is of significant importance for both indoor and outdoor applications and different approaches can be examined to improve environmental stability.[356-361, 367-377] These studies show that long-term stability may be introduced in TMDs based DSSCs by similar strategies of using long alkyl chain organic dyes, ion-gel and polymer-based electrolytes, silica nanoparticles, or modifying $TiO_2$ photoanodes. Similar electrochemical and thermal stability studies should be conducted for TMD CEs for DSSC devices.

## 7. Conclusion and Perspective

The electrochemical and photovoltaic properties of DSSC devices employing CEs (CEs) of 2D layered transition metal dichalcogenides ($MoS_2$, $MoSe_2$, $WS_2$, $TiS_2$, $NbSe_2$, $TaSe_2$, $NiSe_2$, $FeSe_2$, $CoSe_2$, $SnS_2$, $Bi_2Se_3$ and their based composites) have been summarized and discussed. This data indicates that the PCEs of TMDs-based CEs surpassess conventional Pt CEs in dye-sensitized solar cell (DSSC) devices. The composites of TMDs with graphene, CNTs, carbon, carbon nanofibers (CNFs), and PEDOT:PSS also show a great potential as CEs for DSSCs. The DSSCs having CEs made of chemical vapor deposition (CVD)-grown vertically inclined $MoS_2$ films,[131] hydrothermally prepared $MoS_2$ films,[135, 141] $MoS_2$ and $MoSe_2$ thin films deposited on Mo foil,[260, 261] $MoS_2$/graphene, $MoS_2$/CNTs,[164] $MoS_2$/carbon,[171] and $MoSe_2$/PEDOT:PSS composites exhibit higher PCE values than that of Pt CEs for the reduction of triiodide ($I_3^-$) to iodide ($I^-$). Furthermore, the CEs of $TiS_2$/graphene hybrids,[199] $NiS_2$/rediuced graphene oxide (RGO) composites,[210] $FeS_2$,[221] $NbSe_2$,[280] and $FeSe_2$ nanosheets[302] also show a better photovoltaic performance than that of Pt CEs in DSSCs. The low cost and flexible $CoSe_2$/carbon-nanoclimbing-wall CE deposited on nickel foam shows the highest PCE of 10.46% versus a Pt CE ($\eta$ = 8.25%) at 1 Sun illumination (100 mW/cm$^2$ (AM 1.5G)).[310] It has been observed that the morphology of CEs also plays an important role in determining the electrocatalytic activity of DSSC devices, where CEs with large surface area nanostructures tend to show larger PCE values and a higher electrocatalytic activity of the DSSCs.[282] The DSSCs with TiS$_2$/graphene hybrids[199] and TiS$_2$/PEDOT:PSS composites[200] based CEs showed environmental stability of up to 20 and 30 days, respectively, with no degradation in the photovoltaic performance. Interestingly, thermoelectric Bi$_2$Te$_3$ nanosheet/ZnO nanoparticles composite based photoanodes[335] exhibited a 46.95% increase in PCE value compared with a bare ZnO photoanode based DSSC. Also, the DSSC with the Bi$_2$Te$_3$/TiO$_2$ composite based photoanode[336] showed a 28% increase in PCE than that of pure TiO$_2$ photoanode.

Transition metal dichalcogenides (TMDs) based materials which are analogues of 2D graphene are emerging as a great alternative to fabricating low-cost Pt-free DSSC devices. TMDs-based CEs have demonstrated better electrochemical stability than that of standard Pt CEs in iodine-based electrolyte and also under ambient conditions. In addition to long-term stability, TMDs may also cause cytotoxicity to humans, as has been observed for other nanostructured materials,[378-383] therefore, aspects of toxicity should be investigated in studies with DSSC devices.

2D TMDs based CEs are a cheap alternative to Pt CEs for DSSCs. In this review, we have summarized recent developments of TMDs used as CE materials for DSSCs which are still in their infancy. The low-cost 2D TMDs are abundantly available in nature, and can easily be processed into thin films and hybridized with other inorganic and organic materials for fabricating DSSC devices. The vast majority of 2D TMDs have yet to be studied, even as 2D graphene-based materials, but at this early stage they offer a low-cost alternative and outperform their Pt counterparts. Tremendous possibilities exist for developing new TMDs based CEs for $I_3^-/I^-$, $Co^{2+}/Co^{3+}$ and $T_2/T^-$ redox couples. The important requirements for commercial applications are ease of processing, low-cost manufacturing, high PCE value, and long-term electrochemical stability. Like graphene-based materials, more edge active sites can be created in TMDs-based CEs to facilitate more dye adsorption. Research on the use of TMDs CEs are in very early stage of developing Pt-free DSSCs. TMDs based CEs offer increased charge transfer capability and fast reaction kinetics for the reduction of triiodide ($I_3^-$) to iodide ($I^-$) in electrolyte for DSSCs and their potential can be realized in parallel to graphene. The large family of TMDs, such as $MoS_2$, $MoSe_2$, $MoTe_2$, $WS_2$, $WSe_2$, $WTe_2$, $FeSe_2$, $TaS_2$, $NbSe_2$, etc., should also be explored with other inorganic and organic materials, as the family of 2D materials is enormously large and are expected to play an important role in developing low-cost highly efficient DSSCs for commercial applications.

**Acknowledgments**

Eric Singh is thankful to Prof. Dr. G. Y. Yeom for offering him a summer research internship in his group at the School of Advanced Materials Science and Engineering, Sungkyunkwan University (SKKU), South Korea. This work was supported by the Nano•Material Technology Development Program through the National Research Foundation of Korea (NRF), funded by the Ministry of Education, Science and Technology (2016M3A7B4910429).

**Disclaimer**

The authors cannot accept liability for any kind of scientific data contained in this review article whatsoever for the accuracy of contents or any omissions or any errors or a claim of completeness.

**References**

(1) O'Regan, B. and Grätzel, M., 1991. A low-cost, high-efficiency solar cell based on dye-sensitized. *Nature*, *353*(6346), pp.737-740.

(2) Nazeeruddin, M.K., Kay, A., Rodicio, I., Humphry-Baker, R., Müller, E., Liska, P., Vlachopoulos, N. and Grätzel, M., 1993. Conversion of light to electricity by cis-$X_2$ bis(2, 2'-bipyridyl-4, 4'-dicarboxylate) ruthenium (II) charge-transfer sensitizers (X= Cl-, Br-, I-, CN-, and SCN-) on nanocrystalline titanium dioxide electrodes. *Journal of the American Chemical Society*, *115*(14), pp.6382-6390.

(3) Nazeeruddin, M.K., Pechy, P. and Grätzel, M., 1997. Efficient panchromatic sensitization of nanocrystalline $TiO_2$ films by a black dye based on a trithiocyanato–ruthenium complex. *Chemical Communications*, (18), pp.1705-1706.

(4) Mathew, S., Yella, A., Gao, P., Humphry-Baker, R., Curchod, B.F., Ashari-Astani, N., Tavernelli, I., Rothlisberger, U., Nazeeruddin, M.K. and Grätzel, M., 2014. Dye-sensitized solar cells with 13% efficiency achieved through the molecular engineering of porphyrin sensitizers. *Nature Chemistry*, *6*(3), pp.242-247.

(5) Grätzel, M., 2000. Perspectives for dye-sensitized nanocrystalline solar cells. *Progress in Photovoltaics: Research and Applications*, *8*(1), pp.171-185.

(6) Hara M. K., and Arakawa, H., 2003. Dye-sensitized solar cells, *Handbook of Photovoltaic Science and Engineering,* edited by A. Luque and S. Hegedus, Wiley, New York, pp. 663–700.

(7) Grätzel, M., 2003. Dye-sensitized solar cells. *Journal of Photochemistry and Photobiology C: Photochemistry Reviews*, *4*(2), pp.145-153.

(8)   Hardin, B.E., Snaith, H. J. and McGehee, M. D., 2012. The renaissance of dye-sensitized solar cells. *Nature Photonics*, *6*(3), pp.162-169.

(9)   Yum, J.H., Baranoff, E., Wenger, S., Nazeeruddin, M.K. and Grätzel, M., 2011. Panchromatic engineering for dye-sensitized solar cells. *Energy & Environmental Science*, *4*(3), pp.842-857.

(10)   Prakash, T., 2012. Review on nanostructured semiconductors for dye sensitized solar cells. *Electronic Materials Letters*, *8*(3), pp.231-243.

(11)   Boschloo, G. and Hagfeldt, A., 2009. Characteristics of the iodide/triiodide redox mediator in dye-sensitized solar cells. *Accounts of Chemical Research*, *42*(11), pp.1819-1826.

(12)   Hagfeldt, A., Boschloo, G., Sun, L., Kloo, L. and Pettersson, H., 2010. Dye-sensitized solar cells. *Chemical reviews*, *110*(11), pp.6595-6663.

(13)    Zhang, Q. and Cao, G., 2011. Nanostructured photoelectrodes for dye-sensitized solar cells,  *Nano Today* 6, pp.91-109

(14)   Zhang, S., Yang, X., Numata, Y. and Han, L., 2013. Highly efficient dye-sensitized solar cells: progress and future challenges. *Energy & Environmental Science*, *6*(5), pp.1443-1464.

(15)   Tétreault, N. and Grätzel, M., 2012. Novel nanostructures for next generation dye-sensitized solar cells. *Energy & Environmental Science*, *5*(9), pp.8506-8516.

(16)   Listorti, A., O'Regan, B. and Durrant, J.R., 2011. Electron transfer dynamics in dye-sensitized solar cells. *Chemistry of Materials*, *23*(15), pp.3381-3399.

(17)   Ye, M., Wen, X., Wang, M., Iocozzia, J., Zhang, N., Lin, C. and Lin, Z., 2015. Recent advances in dye-sensitized solar cells: from photoanodes, sensitizers and electrolytes to counter electrodes. *Materials Today*, *18*(3), pp.155-162.

(18)    Wu, M., Lin, X., Wang, Y., Wang, L., Guo, W., Qi, D., Peng, X., Hagfeldt, A., Grätzel, M. and Ma, T., 2012. Economical Pt-free catalysts for counter electrodes of dye-sensitized solar cells. *Journal of the American Chemical Society*, *134*(7), pp.3419-3428.

(19)    Hao, F., Dong, P., Luo, Q., Li, J., Lou, J. and Lin, H., 2013. Recent advances in alternative cathode materials for iodine-free dye-sensitized solar cells. *Energy & Environmental Science*, *6*(7), pp.2003-2019.

(20)    Wu, M. and Ma, T., 2014. Recent progress of counter electrode catalysts in dye-sensitized solar cells. *The Journal of Physical Chemistry C*, *118*(30), pp.16727-16742.

(21)    Thomas, S., Deepak, T.G., Anjusree, G.S., Arun, T.A., Nair, S.V. and Nair, A.S., 2014. A review on counter electrode materials in dye-sensitized solar cells. *Journal of Materials Chemistry A*, *2*(13), pp.4474-4490.

(22)    Nalwa, H. S. (Editor). 2004. *Encyclopedia of Nanoscience and Nanotechnology; 10-volume set.* American Scientific Publishers, Los Angeles; (b) Nalwa, H. S. (Editor). 2000. *Handbook of Nanostructured Materials and Nanotechnology; 5-volume set.* Academic Press, San Diego; (c) Nalwa, H. S. (Editor). 2009. *Nanomaterials for Energy Storage Applications;* American Scientific Publishers, Los Angeles.

(23)    Graedel, T. E., Harper, E. M., Nassar, N. T., Nuss, P., and Reck, B. K., 2015. Criticality of metals and metalloids. *Proceedings of the National Academy of Sciences of the United States of America*, *112*(14), pp. 4257-4262. DOI: 10.1073/pnas.1500415112

(24)    Minerals, Critical Minerals, and the U.S. Economy, National Research Council, National Academies Press Washington D.C., U.S. (2008); National Research Council (US) Chemical Sciences Roundtable, Assessments of Criticality, (2012).

(25)    Singh, S., and Nalwa, H. S., 2015. Graphene-based dye-sensitized solar cells: a review, *Science of Advanced Materials*, 7, 1863–1912.

(26)    Singh, E. and Nalwa, H.S. 2015, Graphene-based bulk-heterojunction solar cells: a review, *Journal of Nanoscience and Nanotechnology,* 15(9), pp. 6237-6278.

(27)    Singh, E. and Nalwa, H.S. 2015, Stability of graphene-based heterojunction solar cells, *RSC Advances,* 5(90), pp. 73575-73600.

(28)    Wilson, J.A. and Yoffe, A.D., 1969. The transition metal dichalcogenides discussion and interpretation of the observed optical, electrical and structural properties. *Advances in Physics*, *18*(73), pp.193-335.

(29)    Huang, X., Zeng, Z. and Zhang, H., 2013. Metal dichalcogenide nanosheets: preparation, properties and applications. *Chemical Society Reviews*, *42*(5), pp.1934-1946.

(30)    Chhowalla, M., Shin, H.S., Eda, G., Li, L.J., Loh, K.P. and Zhang, H., 2013. The chemistry of two-dimensional layered transition metal dichalcogenide nanosheets. *Nature Chemistry*, *5*(4), pp.263-275.

(31)    Ataca, C., Sahin, H. and Ciraci, S., 2012. Stable, single-layer $MX_2$ transition-metal oxides and dichalcogenides in a honeycomb-like structure. *The Journal of Physical Chemistry C*, *116*(16), pp.8983-8999.

(32)    Geim, A.K. and Grigorieva, I.V., 2013. Van der Waals heterostructures. *Nature*, *499*(7459), pp.419-425.

(33)    Radisavljevic, B., Radenovic, A., Brivio, J., Giacometti, V. and Kis, A. 2011. Single-layer $MoS_2$ transistors. *Nature Nanotechnology* **6**, pp.147-150.

(34)   Yun, W. S., Han, S. W., Hong, S. C., Kim, I. G. and Lee, J. D., 2012. Thickness and strain effects on electronic structures of transition metal dichalcogenides: 2H-MX$_2$ semiconductors (M= Mo, W; X= S, Se, Te). *Physical Review B*, *85*, 033305.

(35)   Eda, G. and Maier, S. A., 2013. Two-Dimensional Crystals: Managing Light for Optoelectronics, *ACS Nano,* 7, pp. 5660-5665.

(36)   Tongay, S., Suh, J., Ataca, C., Fan, W., Luce, A., Kang, J.S., Liu, J., Ko, C., Raghunathanan, R., Zhou, J. and Ogletree,; Li, J.; Grossman, J. C.; Wu, J. 2013. Defects activated photoluminescence in two-dimensional semiconductors: interplay between bound, charged, and free excitons., *Scientific Reports 3*, 2657

(37)   Tonndorf, P.; Schmidt, R.; Böttger, P.; Zhang, X.; Borner, J.; Liebig, A.; Albrecht, M.; Kloc, C.; Gorgan, O.; Zahn, D. R. T.; de Vasconcellos, S. M., and Bratschitsch R., 2013. Photoluminescence Emission and Raman Response of Monolayer MoS$_2$, MoSe$_2$, and WSe$_2$. *Optics Express* 21, pp. 4908–4916

(38)   McCreary, K.M., Hanbicki, A.T., Jernigan, G.G., Culbertson, J.C. and Jonker, B.T., 2016. Synthesis of large-area WS$_2$ monolayers with exceptional photoluminescence. *Scientific Reports*, *6*. 19159

(39)   Ueno, K., Saiki, K., Shimada, T. and Koma, A., 1990. Epitaxial growth of transition metal dichalcogenides on cleaved faces of mica. *Journal of Vacuum Science & Technology A*, *8*(1), pp.68-72.

(40)   Novoselov, K.S., Jiang, D., Schedin, F., Booth, T.J., Khotkevich, V.V., Morozov, S.V. and Geim, A.K., 2005. Two-dimensional atomic crystals. *Proceedings of the National Academy of Sciences of the United States of America*, *102*(30), pp.10451-10453.

(41)   Li, H., Lu, G., Wang, Y., Yin, Z., Cong, C., He, Q., Wang, L., Ding, F., Yu, T. and Zhang, H., 2013. Mechanical Exfoliation and Characterization of Single-and Few-Layer Nanosheets of $WSe_2$, $TaS_2$, and $TaSe_2$. *Small*, *9*(11), pp.1974-1981.

(42)   Al-Hilli, A.A. and Evans, B.L., 1972. The preparation and properties of transition metal dichalcogenide single crystals. *Journal of Crystal Growth*, *15*(2), pp.93-101.

(43)   Lee, Y.H., Yu, L., Wang, H., Fang, W., Ling, X., Shi, Y., Lin, C.T., Huang, J.K., Chang, M.T., Chang, C.S., Dresselhaus, M., Palacios, T., Li, L.-J., and Kong, J., 2013. Synthesis and transfer of single-layer transition metal disulfides on diverse surfaces. *Nano Letters*, *13*(4), pp.1852-1857.

(44)   Shi, Y., Li, H. and Li, L.J., 2015. Recent advances in controlled synthesis of two-dimensional transition metal dichalcogenides via vapour deposition techniques. *Chemical Society Reviews*, *44*(9), pp.2744-2756.

(45)   Gatensby, R., McEvoy, N., Lee, K., Hallam, T., Berner, N.C., Rezvani, E., Winters, S., O'Brien, M. and Duesberg, G.S., 2014. Controlled synthesis of transition metal dichalcogenide thin films for electronic applications. *Applied Surface Science*, *297*, pp.139-146.

(46)   Jeong, S., Yoo, D., Jang, J.T., Kim, M. and Cheon, J., 2012. Well-defined colloidal 2-D layered transition-metal chalcogenide nanocrystals via generalized synthetic protocols. *Journal of the American Chemical Society*, *134*(44), pp.18233-18236.

(47)   Whittingham, M.S. and Gamble, F.R., 1975. The lithium intercalates of the transition metal dichalcogenides. *Materials Research Bulletin*, *10*(5), pp.363-371.

(48)    Zeng, Z., Sun, T., Zhu, J., Huang, X., Yin, Z., Lu, G., Fan, Z., Yan, Q., Hng, H.H. and Zhang, H., 2012. An Effective Method for the Fabrication of Few-Layer-Thick Inorganic Nanosheets. *Angewandte Chemie International Edition*, *51*(36), pp.9052-9056.

(49)    Eng, A.Y.S., Ambrosi, A., Sofer, Z., Simek, P. and Pumera, M., 2014. Electrochemistry of transition metal dichalcogenides: strong dependence on the metal-to-chalcogen composition and exfoliation method. *ACS Nano*, *8*(12), pp.12185-12198.

(50)    Gordon, R. A., Yang, D., Crozier, E. D., Jiang, D. T. and Frindt, R. F. Structures of exfoliated single layers of $WS_2$, $MoS_2$, and $MoSe_2$ in aqueous suspension. *Physical Review B* **65**, 125407 (2002).

(51)    Cunningham, G., Lotya, M., Cucinotta, C.S., Sanvito, S., Bergin, S.D., Menzel, R., Shaffer, M.S. and Coleman, J.N., 2012. Solvent exfoliation of transition metal dichalcogenides: dispersibility of exfoliated nanosheets varies only weakly between compounds. *ACS Nano*, *6*, pp.3468-3480.

(52)    Smith, R. J., King, P.J., Lotya, M., Wirtz, C., Khan, U., De, S., O'Neill, A., Duesberg, G.S., Grunlan, J.C., Moriarty, G. and Chen, J., 2011. Large-scale exfoliation of inorganic layered compounds in aqueous surfactant solutions. *Advanced Materials*, *23*(34), pp.3944-3948.

(53)    Nicolosi, V., Chhowalla, M., Kanatzidis, M.G., Strano, M.S. and Coleman, J.N., 2013. Liquid exfoliation of layered materials. *Science*, *340*(6139), p.1226419.

(54)    Coleman, J.N., Lotya, M., O'Neill, A., Bergin, S.D., King, P.J., Khan, U., Young, K., Gaucher, A., De, S., Smith, R.J. and Shvets, I.V., Arora, S. K., Stanton, G., Kim, H. Y., Lee, K., Kim, G., Duesberg, G. S., Hallam, T.,  Boland, J. J., Wang, J. J., Donegan, J. F., Grunlan, J. C., Moriarty, G., Shmeliov, A., Nicholls, R. J., Perkins, J. M., Grieveson, E. M.,

Theuwissen, K., McComb, D. W., Nellist, P. D., and Nicolosi, V., 2011. Two-dimensional nanosheets produced by liquid exfoliation of layered materials. *Science*, *331*(6017), pp.568-571.

(55)    Niu, L., Li, K., Zhen, H., Chui, Y.S., Zhang, W., Yan, F. and Zheng, Z., 2014. Salt-Assisted High-Throughput Synthesis of Single-and Few-Layer Transition Metal Dichalcogenides and Their Application in Organic Solar Cells. *Small*, *10*(22), pp.4651-4657.

(56)    Zhang, X., Qiao, X.F., Shi, W., Wu, J.B., Jiang, D.S. and Tan, P.H., 2015. Phonon and Raman scattering of two-dimensional transition metal dichalcogenides from monolayer, multilayer to bulk material. *Chemical Society Reviews*, *44*(9), 2757-2785.

(57)    Pawbake, A.S., Pawar, M.S., Jadkar, S.R. and Late, D.J., 2016. Large area chemical vapor deposition of monolayer transition metal dichalcogenides and their temperature dependent Raman spectroscopy studies. *Nanoscale*, *8*(5), pp.3008-3018.

(58)    Verble, J.L. and Wieting, T.J., 1970. Lattice Mode Degeneracy in $MoS_2$ and Other Layer Compounds. *Physical Review Letters*, *25*(6), p.362.

(59)    G. Frey, R. Tenne, M. Matthews, M. Dresselhaus, and G. Dresselhaus, 1999. Raman and resonance Raman investigation of $MoS_2$ nanoparticles, *Physical Review B* **60**(4), pp. 2883–2892.

(60)    Castellanos-Gomez, A., Quereda, J., van der Meulen, H.P., Agraït, N. and Rubio-Bollinger, G., Spatially resolved optical absorption spectroscopy of single-and few-layer $MoS_2$ by hyperspectral imaging. *Nanotechnology*, 2016, *27*, 115705.

(61)    Balasingam, S.K., Lee, J. S. Lee and Jun, Y.; 2015. Few-layered $MoSe_2$ nanosheets as an advanced electrode material for supercapacitors, *Dalton Transactions,* 44(35), pp.15491–15498.

(62)     Zhao, W., Ghorannevis, Z., Amara, K.K., Pang, J.R., Toh, M., Zhang, X., Kloc, C., Tan, P.H. and Eda, G., 2013. Lattice dynamics in mono-and few-layer sheets of $WS_2$ and $WSe_2$. *Nanoscale*, *5*(20), pp.9677-9683.

(63)     Lee, C.; Yan, H.; Brus, L. E.; Heinz, T. F.; Hone, J.; Ryu, S., Anomalous Lattice Vibrations of Single- and Few-Layer $MoS_2$. *ACS Nano* **2010,** *4* (5), 2695-2700.

(64)     Molina-Sanchez, A.; Wirtz, L., Phonons in single-layer and few-layer $MoS_2$ and $WS_2$. *Physical Review B* **2011,** *84* (15).

(65)     Tongay, S.; Zhou, J.; Ataca, C.; Lo, K.; Matthews, T. S.; Li, J.; Grossman, J. C.; Wu, J., Thermally Driven Crossover from Indirect toward Direct Bandgap in 2D Semiconductors: $MoSe_2$ versus $MoS_2$. *Nano Letters* **2012**.

(66)     Li, H.; Zhang, Q.; Yap, C. C. R.; Tay, B. K.; Edwin, T. H. T.; Olivier, A.; Baillargeat, D., From Bulk to Monolayer MoS2: Evolution of Raman Scattering. *Advanced Functional Materials* **2012,** *22* (7), 1385-1390.

(67)     Berkdemir, A.; Gutierrez, H. R.; Botello-Mendez, A. R.; Perea-Lopez, N.; Elias, A. L.; Chia, C.-I.; Wang, B.; Crespi, V. H.; Lopez-Urias, F.; Charlier, J.-C.; Terrones, H.; Terrones, M., Identification of individual and few layers of $WS_2$ using Raman Spectroscopy, *Scientific Reports* **2013,** *doi:10.1038/srep01755*.

(68)     Sahin, H.; Tongay, S.; Horzum, S.; Fan, W.; Zhou, J.; Li, J.; Wu, J.; Peeters, F. M., 2013. Anomalous Raman spectra and thickness-dependent electronic properties of $WSe_2$. *Physical Review B*, *87* (16), 165409

(69)     Saito, R., Tatsumi, Y., Huang, S., Ling, X. and Dresselhaus, M.S., 2016. Raman spectroscopy of transition metal dichalcogenides. *Journal of Physics: Condensed Matter*, *28*(35), p.353002.

(70)    Terrones, H., Del Corro, E., Feng, S., Poumirol, J.M., Rhodes, D., Smirnov, D., Pradhan, N.R., Lin, Z., Nguyen, M.A.T., Elias, A.L. and Mallouk, T.E., 2014. New first order Raman-active modes in few layered transition metal dichalcogenides. *Scientific Reports*, *4*. 4215

(71)    Chen, S.Y., Zheng, C., Fuhrer, M.S. and Yan, J., 2015. Helicity-resolved Raman scattering of $MoS_2$, $MoSe_2$, $WS_2$, and $WSe_2$ atomic layers. *Nano Letters*, *15*(4), 2526-2532.

(72)    Fleischauer, P.D. and Bauer, R., 1988. Chemical and structural effects on the lubrication properties of sputtered $MoS_2$ films. *Tribology Transactions*, *31*(2), pp.239-250.

(73)    Spalvins, T., 1992. Lubrication with sputtered $MoS_2$ films: principles, operation, and limitations. *Journal of Materials Engineering and Performance*, *1*(3), pp.347-351.

(74)    Hilton, M.R., Bauer, R., Didziulis, S.V., Dugger, M.T., Keem, J.M. and Scholhamer, J., 1992. Structural and tribological studies of $MoS_2$ solid lubricant films having tailored metal-multilayer nanostructures. *Surface and Coatings Technology*, *53*(1), pp.13-23.

(75)    Savan, A., Pflüger, E., Voumard, P., Schröer, A. and Simmonds, M., 2000. Modern solid lubrication: recent developments and applications of $MoS_2$. *Lubrication Science*, *12*(2), pp.185-203.

(76)    Rapoport, L., Bilik, Y., Feldman, Y., Homyonfer, M., Cohen, S.R. and Tenne, R., 1997. Hollow nanoparticles of $WS_2$ as potential solid-state lubricants. *Nature*, *387*(6635), pp.791-3.

(77)    Rapoport, L., Fleischer, N. and Tenne, R., 2003. Fullerene-like $WS_2$ Nanoparticles: Superior Lubricants for Harsh Conditions. *Advanced Materials*, *15*(7-8), pp.651-655.

(78)    Liu, Y.Q., Li, C.S., Yang, J.H., Yu, Y.M. and Li, X.K., 2007. Synthesis and tribological properties of tubular $NbS_2$ and $TaS_2$ nanostructures. *Chinese Journal of Chemical Physics*, *20*(6), pp.768-772.

(79)   Polcar, T. and Cavaleiro, A., 2011. Review on self-lubricant transition metal dichalcogenide nanocomposite coatings alloyed with carbon. *Surface and Coatings Technology*, *206*(4), pp.686-695.

(80)   Yang, J.F., Parakash, B., Hardell, J. and Fang, Q.F., 2012. Tribological properties of transition metal di-chalcogenide based lubricant coatings. *Frontiers of Materials Science*, *6*(2), pp.116-127.

(81)   Krasnozhon, D., Lembke, D., Nyffeler, C., Leblebici, Y. and Kis, A., 2014. $MoS_2$ transistors operating at gigahertz frequencies. *Nano Letters*, *14*(10), pp.5905-5911.

(82)   Fang, H., Chuang, S., Chang, T.C., Takei, K., Takahashi, T. and Javey, A., 2012. High-performance single layered $WSe_2$ p-FETs with chemically doped contacts. *Nano Letters*, *12*(7), pp.3788-3792.

(83)   Wang, H., Yu, L., Lee, Y.H., Shi, Y., Hsu, A., Chin, M.L., Li, L.J., Dubey, M., Kong, J. and Palacios, T., 2012. Integrated circuits based on bilayer $MoS_2$ transistors. *Nano Letters*, *12*(9), pp.4674-4680.

(84)   Li, H., Wu, J., Yin, Z. and Zhang, H., 2014. Preparation and applications of mechanically exfoliated single-layer and multilayer $MoS_2$ and $WSe_2$ nanosheets. *Accounts of Chemical Research*, *47*(4), pp.1067-1075.

(85)   Gong, C., Zhang, H., Wang, W., Colombo, L., Wallace, R.M. and Cho, K., 2013. Band alignment of two-dimensional transition metal dichalcogenides: Application in tunnel field effect transistors. *Applied Physics Letters*, *103*(5), p.053513.

(86)   Jariwala, D., Sangwan, V.K., Lauhon, L.J., Marks, T.J. and Hersam, M.C., 2014. Emerging device applications for semiconducting two-dimensional transition metal dichalcogenides. *ACS Nano*, *8*(2), pp.1102-1120.

(87)    Schwierz, F., Pezoldt, J. and Granzner, R., 2015. Two-dimensional materials and their prospects in transistor electronics. *Nanoscale*, *7*(18), pp.8261-8283.

(88)    Chang, H.Y., Yang, S., Lee, J., Tao, L., Hwang, W.S., Jena, D., Lu, N. and Akinwande, D., 2013. High-performance, highly bendable MoS$_2$ transistors with high-k dielectrics for flexible low-power systems. *ACS Nano*, *7*(6), pp.5446-5452.

(89)    Cheng, R., Jiang, S., Chen, Y., Liu, Y., Weiss, N., Cheng, H.C., Wu, H., Huang, Y. and Duan, X., 2014. Few-layer molybdenum disulfide transistors and circuits for high-speed flexible electronics. *Nature Communications*, *5*, pp.5143.

(90)    Hwang, W.S., Remskar, M., Yan, R., Protasenko, V., Tahy, K., Chae, S.D., Zhao, P., Konar, A., Xing, H.G., Seabaugh, A. and Jena, D., 2012. Transistors with chemically synthesized layered semiconductor WS$_2$ exhibiting $10^5$ room temperature modulation and ambipolar behavior. *Applied Physics Letters*, *101*(1), p.013107.

(91)    Tosun, M., Chuang, S., Fang, H., Sachid, A.B., Hettick, M., Lin, Y., Zeng, Y. and Javey, A., 2014. High-gain inverters based on WSe$_2$ complementary field-effect transistors. *ACS Nano*, *8*(5), pp.4948-4953.

(92)    Larentis, S., Fallahazad, B. and Tutuc, E., 2012. Field-effect transistors and intrinsic mobility in ultra-thin MoSe$_2$ layers. *Applied Physics Letters*, *101*(22), p.223104.

(93) Chamlagain, B., Li, Q., Ghimire, N.J., Chuang, H.J., Perera, M.M., Tu, H., Xu, Y., Pan, M., Xaio, D., Yan, J. and Mandrus, D., 2014. Mobility improvement and temperature dependence in MoSe$_2$ field-effect transistors on parylene-C substrate. *ACS Nano*, *8*(5), pp.5079-5088.

(94) Fathipour, S., Ma, N., Hwang, W.S., Protasenko, V., Vishwanath, S., Xing, H.G., Xu, H., Jena, D., Appenzeller, J. and Seabaugh, A., 2014. Exfoliated multilayer MoTe$_2$ field-effect transistors. *Applied Physics Letters*, *105*(19), p.192101.

(95) Lin, Y.F., Xu, Y., Wang, S.T., Li, S.L., Yamamoto, M., Aparecido-Ferreira, A., Li, W., Sun, H., Nakaharai, S., Jian, W.B. and Ueno, K., 2014. Ambipolar MoTe$_2$ transistors and their applications in logic circuits. *Advanced Materials*, *26*(20), pp.3263-3269.

(96) Liu, E., Fu, Y., Wang, Y., Feng, Y., Liu, H., Wan, X., Zhou, W., Wang, B., Shao, L., Ho, C.H. and Huang, Y.S., 2015. Integrated digital inverters based on two-dimensional anisotropic ReS$_2$ field-effect transistors. *Nature Communications*, *6*, 6991

(97) Gao, D., Xue, Q., Mao, X., Wang, W., Xu, Q. and Xue, D., 2013. Ferromagnetism in ultrathin VS$_2$ nanosheets. *Journal of Materials Chemistry C*, *1*(37), pp.5909-5916.

(98) Kanazawa, T., Amemiya, T., Ishikawa, A., Upadhyaya, V., Tsuruta, K., Tanaka, T. and Miyamoto, Y., 2016. Few-layer HfS$_2$ transistors. *Scientific Reports*, *6*, 22277

(99) (a) El-Bana, M.S., Wolverson, D., Russo, S., Balakrishnan, G., Paul, D.M. and Bending, S.J., 2013. Superconductivity in two-dimensional NbSe$_2$ field effect transistors. *Superconductor Science and Technology*, *26*(12), p.125020. (b) Yang, S., Yue, Q., Cai, H., Wu, K., Jiang, C. and Tongay, S., 2016. Highly efficient gas molecule-tunable few-layer GaSe phototransistors. *Journal of Materials Chemistry C*, *4*(2), pp.248-253.

(100) Geng, X., Yu, Y., Zhou, X., Wang, C., Xu, K., Zhang, Y., Wu, C., Wang, L., Jiang, Y. and Yang, Q., 2016. Design and construction of ultra-thin MoSe$_2$ nanosheet-based heterojunction for high-speed and low-noise photodetection. *Nano Research*, *9*(9), pp.2641-2651.

(101) Zhang, E., Jin, Y., Yuan, X., Wang, W., Zhang, C., Tang, L., Liu, S., Zhou, P., Hu, W. and Xiu, F., 2015. ReS$_2$-Based Field-Effect Transistors and Photodetectors. *Advanced Functional Materials*, *25*(26), pp.4076-4082.

(102)    Yang, S., Wang, C., Ataca, C., Li, Y., Chen, H., Cai, H., Suslu, A., Grossman, J.C., Jiang, C., Liu, Q. and Tongay, S., 2016. Self-Driven Photodetector and Ambipolar Transistor in Atomically Thin GaTe-MoS$_2$ p–n vdW Heterostructure. *ACS Applied Materials & Interfaces*, *8*(4), pp.2533-2539.

(103)    Ross, J.S., Klement, P., Jones, A.M., Ghimire, N.J., Yan, J., Mandrus, D.G., Taniguchi, T., Watanabe, K., Kitamura, K., Yao, W. and Cobden, D.H., 2014. Electrically tunable excitonic light-emitting diodes based on monolayer WSe$_2$ pn junctions. *Nature Nanotechnology*, *9*(4), pp.268-272.

(104)    Yin, Z., Zhang, X., Cai, Y., Chen, J., Wong, J.I., Tay, Y.Y., Chai, J., Wu, J., Zeng, Z., Zheng, B. and Yang, H.Y., 2014. Preparation of MoS$_2$–MoO$_3$ Hybrid Nanomaterials for Light-Emitting Diodes. *Angewandte Chemie International Edition*, *53*(46), pp.12560-12565.

(105)    Lipatov, A., Sharma, P., Gruverman, A. and Sinitskii, A., 2015. Optoelectrical Molybdenum Disulfide (MoS$_2$)-Ferroelectric Memories. *ACS Nano*, *9*(8), pp.8089-8098.

(106)    Lee, J., Wang, Z., He, K., Shan, J. and Feng, P.X.L., 2013. High frequency MoS$_2$ nanomechanical resonators. *ACS Nano*, *7*(7), pp.6086-6091.

(107)    Clerici, F., Fontana, M., Bianco, S., Serrapede, M., Perrucci, F., Ferrero, S., Tresso, E. and Lamberti, A., 2016. In situ MoS$_2$ Decoration of Laser-Induced Graphene as Flexible Supercapacitor Electrodes. *ACS Applied Materials & Interfaces*, *8*(16), pp.10459-10465.

(108)    Zhou, X.; Xu, B.; Lin, Z.; Shu, D.; Ma, L. 2014, Hydrothermal Synthesis of Flower-Like MoS$_2$ Nanospheres for Electrochemical Supercapacitors. *Journal of Nanoscience and Nanotechnology*, 14, 7250−7254.

(109)    Du, G.; Guo, Z.; Wang, S.; Zeng, R.; Chen, Z.; Liu, H. Superior Stability and High Capacity of Restacked Molybdenum Disulfide as Anode Material for Lithium Ion Batteries. *Chemical Communications* 2010, 46, 1106−1108.

(110)    Zhang, Z., Shi, X., Yang, X., Fu, Y., Zhang, K., Lai, Y. and Li, J., 2016. Nanooctahedra particles assembled $FeSe_2$ microspheres embedded into sulfur-doped reduced graphene oxide sheets as a promising anode for sodium ion batteries. *ACS Applied Materials & Interfaces*. 8(22), pp.13849-13856

(111)    Xu, X., Rout, C.S., Yang, J., Cao, R., Oh, P., Shin, H.S. and Cho, J., 2013. Freeze-dried $WS_2$ composites with low content of graphene as high-rate lithium storage materials. *Journal of Materials Chemistry A*, *1*(46), pp.14548-14554.

(112)    Lopez-Sanchez, O., Alarcon Llado, E., Koman, V., Fontcuberta i Morral, A., Radenovic, A. and Kis, A., 2014. Light Generation and Harvesting in a van der Waals Heterostructure. *ACS Nano*, *8*(3), pp.3042-3048.

(113)    Perkins, F.K., Friedman, A.L., Cobas, E., Campbell, P.M., Jernigan, G.G. and Jonker, B.T., 2013. Chemical vapor sensing with monolayer $MoS_2$. *Nano Letters*, *13*(2), pp.668-673.

(114)    Late, D. J., Doneux, T. and Bougouma, M., 2014. Single-layer $MoSe_2$ based $NH_3$ gas sensor. *Applied Physics Letters*, *105*(23), p.233103.

(115)    Wang, T., Zhu, R., Zhuo, J., Zhu, Z., Shao, Y. and Li, M., 2014. Direct detection of DNA below ppb level based on thionin-functionalized layered $MoS_2$ electrochemical sensors. *Analytical Chemistry*, *86*(24), pp.12064-12069.

(116)    Pumera, M. and Loo, A.H., 2014. Layered transition-metal dichalcogenides ($MoS_2$ and $WS_2$) for sensing and biosensing. *TrAC Trends in Analytical Chemistry*, *61*, pp.49-53.

(117)    Li, Z. and Wong, S. L., 2016. Functionalization of 2D transition metal dichalcogenides for biomedical applications. *Materials Science and Engineering: C*., 70, pp.1095-1106

(118)    Singh, E., Kim, K. S., Yeom, G. Y., and Nalwa, H. S., 2017, Atomically thin-layered molybdenum disulfide ($MoS_2$) for bulk-heterojunction solar cells, *ACS Applied Materials & Interfaces*, DOI:10.1021/acsami.6b13582

(119)    Wang, Q.H., Kalantar-Zadeh, K., Kis, A., Coleman, J.N. and Strano, M.S., 2012. Electronics and optoelectronics of two-dimensional transition metal dichalcogenides. *Nature Nanotechnology*, *7*(11), 699-712.

(120)    Mak, K.F. and Shan, J., 2016. Photonics and optoelectronics of 2D semiconductor transition metal dichalcogenides. *Nature Photonics*, *10*(4), pp.216-226.

(121)    Lee, J.U., Kim, K., Han, S., Ryu, G.H., Lee, Z. and Cheong, H., 2016. Raman Signatures of Polytypism in Molybdenum Disulfide. *ACS Nano*, **2016**, *10*, 1948–1953

(122)    Beal, A.R., Knights, J.C. and Liang, W.Y., 1972. Transmission spectra of some transition metal dichalcogenides. II. Group VIA: trigonal prismatic coordination. *Journal of Physics C: Solid State Physics*, *5*(24), p.3540.

(123)    Woollam, J.A. and Somoano, R.B., 1976. Superconducting critical fields of alkali and alkaline-earth intercalates of $MoS_2$. *Physical Review B*, *13*(9), p.3843.

(124)    Wypych, F. and Schöllhorn, R., 1992. 1T-$MoS_2$, a new metallic modification of molybdenum disulfide. *Journal of the Chemical Society, Chemical Communications*, (19), pp.1386-1388.

(125)    Wang, L., Xu, Z., Wang, W. and Bai, X., 2014. Atomic mechanism of dynamic electrochemical lithiation processes of $MoS_2$ nanosheets. *Journal of the American Chemical Society*, *136*(18), pp.6693-6697.

(126)    Shirodkar, S.N. and Waghmare, U.V., 2014. Emergence of Ferroelectricity at a Metal-Semiconductor Transition in a 1 T Monolayer of $MoS_2$. *Physical Review Letters*, *112*(15), p.157601.

(127)    Acerce, M., Voiry, D. and Chhowalla, M., 2015. Metallic 1T phase $MoS_2$ nanosheets as supercapacitor electrode materials. *Nature Nanotechnology*, *10*(4), pp.313-318.

(128)    Cheng, P., Sun, K. and Hu, Y.H., 2015. Memristive Behavior and Ideal Memristor of 1T Phase $MoS_2$ Nanosheets. *Nano Letters*, *16*(1), pp.572-576.

(129)    Kappera, R., Voiry, D., Yalcin, S.E., Branch, B., Gupta, G., Mohite, A.D. and Chhowalla, M., 2014. Phase-engineered low-resistance contacts for ultrathin $MoS_2$ transistors. *Nature Materials*, *13*(12), pp.1128-1134.

(130)    Wei, W., Sun, K. and Hu, Y.H., 2016. An efficient counter electrode material for dye-sensitized solar cells—flower-structured 1T metallic phase $MoS_2$. *Journal of Materials Chemistry A*, *4*(32), pp.12398-12401.

(131)    Infant, R. S., Xu, X., Yang, W., Yang, F., Hou, L. and Li, Y., 2016. Highly active and reflective $MoS_2$ counter electrode for enhancement of photovoltaic efficiency of dye sensitized solar cells. *Electrochimica Acta*, *212*, pp.614-620.

(132)    Jiang, S., Yin, X., Zhang, J., Zhu, X., Li, J. and He, M., 2015. Vertical ultrathin $MoS_2$ nanosheets on a flexible substrate as an efficient counter electrode for dye-sensitized solar cells. *Nanoscale*, *7*(23), pp.10459-10464.

(133)    Lei, B., Li, G.R. and Gao, X.P., 2014. Morphology dependence of molybdenum disulfide transparent counter electrode in dye-sensitized solar cells. *Journal of Materials Chemistry A*, *2*(11), pp.3919-3925.

(134) Zhang, J., Najmaei, S., Lin, H. and Lou, J., 2014. $MoS_2$ atomic layers with artificial active edge sites as transparent counter electrodes for improved performance of dye-sensitized solar cells. *Nanoscale*, *6*(10), pp.5279-5283.

(135) Al-Mamun, M., Zhang, H., Liu, P., Wang, Y., Cao, J. and Zhao, H., 2014. Directly hydrothermal growth of ultrathin $MoS_2$ nanostructured films as high performance counter electrodes for dye-sensitised solar cells. *RSC Advances*, *4*(41), pp.21277-21283.

(136) Cheng, C.K. and Hsieh, C.K., 2015. Electrochemical deposition of molybdenum sulfide thin films on conductive plastic substrates as platinum-free flexible counter electrodes for dye-sensitized solar cells. *Thin Solid Films*, *584*, pp.52-60.

(137) Lin, C.H., Tsai, C.H., Tseng, F.G., Yu, Y.Y., Wu, H.C. and Hsieh, C.K., 2015. Low-Temperature Thermally Reduced Molybdenum Disulfide as a Pt-Free Counter Electrode for Dye-Sensitized Solar Cells. *Nanoscale Research Letters*, *10*, pp.1-10.

(138) Patil, S.A., Kalode, P.Y., Mane, R.S., Shinde, D.V., Doyoung, A., Keumnam, C., Sung, M.M., Ambade, S.B. and Han, S.H., 2014. Highly efficient and stable DSSCs of wet-chemically synthesized $MoS_2$ counter electrode. *Dalton Transactions*, *43*(14), pp.5256-5259.

(139) Jhang, W.H. and Lin, Y.J., 2015. Overpotential modification at the $MoS_2$ counter electrode/electrolyte interfaces by thermal annealing resulting improvement in photovoltaic performance of dye-sensitized solar cells. *Journal of Materials Science: Materials in Electronics*, *26*(6), pp.3739-3743.

(140) Antonelou, A., Syrrokostas, G., Sygellou, L., Leftheriotis, G., Dracopoulos, V. and Yannopoulos, S.N., 2015. Facile, substrate-scale growth of mono-and few-layer homogeneous $MoS_2$ films on Mo foils with enhanced catalytic activity as counter electrodes in DSSCs. *Nanotechnology*, *27*(4), p.045404.

(141)    Wu, M., Wang, Y., Lin, X., Yu, N., Wang, L., Wang, L., Hagfeldt, A. and Ma, T., 2011. Economical and effective sulfide catalysts for dye-sensitized solar cells as counter electrodes. *Physical Chemistry Chemical Physics*, *13*(43), pp.19298-19301.

(142)    Zhang, H., Choi, J., Ramani, A., Voiry, D., Natoli, S.N., Chhowalla, M., McMillin, D.R. and Choi, J.H., 2016. Engineering Chemically Exfoliated Large-Area Two-Dimensional $MoS_2$ Nanolayers with Porphyrins for Improved Light Harvesting. *ChemPhysChem*. DOI: 10.1002/cphc.201600511

(143)    Liu, W., He, S., Yang, T., Feng, Y., Qian, G., Xu, J. and Miao, S., 2014. TEOS-assisted synthesis of porous $MoS_2$ with ultra-small exfoliated sheets and applications in dye-sensitized solar cells. *Applied Surface Science*, *313*, pp.498-503.

(144)    Kim, S.S., Lee, J.W., Yun, J.M. and Na, S.I., 2015. 2-Dimensional $MoS_2$ nanosheets as transparent and highly electrocatalytic counter electrode in dye-sensitized solar cells: Effect of thermal treatments. *Journal of Industrial and Engineering Chemistry*, *29*, pp.71-77.

(145)    Hussain, S., Shaikh, S.F., Vikraman, D., Mane, R.S., Joo, O.S., Naushad, M. and Jung, J., 2015. High-Performance Platinum-Free Dye-Sensitized Solar Cells with Molybdenum Disulfide Films as Counter Electrodes. *ChemPhysChem*, *16*(18), pp.3959-3965.

(146)    Jeong, H., Kim, J.Y., Koo, B., Son, H.J., Kim, D. and Ko, M.J., 2016. Rapid sintering of $MoS_2$ counter electrode using near-infrared pulsed laser for use in highly efficient dye-sensitized solar cells. *Journal of Power Sources*, *330*, pp.104-110.

(147)    Liang, J., Li, J., Zhu, H., Han, Y., Wang, Y., Wang, C., Jin, Z., Zhang, G. and Liu, J., 2016. One-step fabrication of large-area ultrathin $MoS_2$ nanofilms with high catalytic activity for photovoltaic devices. *Nanoscale*, *8*(35), pp.16017-16025.

(148)   Tan, C., Zhao, W., Chaturvedi, A., Fei, Z., Zeng, Z., Chen, J., Huang, Y., Ercius, P., Luo, Z., Qi, X. and Chen, B., 2016. Preparation of Single-Layer $MoS_2xSe_2(1-x)$ and $Mo_xW_{1-x}S_2$ Nanosheets with High-Concentration Metallic 1T Phase. *Small*. 12 (14), pp. 1866–1874

(149)   Eksik, O., Gao, J., Shojaee, S.A., Thomas, A., Chow, P., Bartolucci, S.F., Lucca, D.A. and Koratkar, N., 2014. Epoxy nanocomposites with two-dimensional transition metal dichalcogenide additives. *ACS Nano*, *8*(5), pp.5282-5289.

(150)   Tan, C. and Zhang, H., 2015. Two-dimensional transition metal dichalcogenide nanosheet-based composites. *Chemical Society Reviews*, *44*(9), pp.2713-2731.

(151)   Lin, J.Y., Chan, C.Y. and Chou, S.W., 2013. Electrophoretic deposition of transparent $MoS_2$–graphene nanosheet composite films as counter electrodes in dye-sensitized solar cells. *Chemical Communications*, *49*(14), pp.1440-1442.

(152)   Yue, G., Lin, J.Y., Tai, S.Y., Xiao, Y. and Wu, J., 2012. A catalytic composite film of $MoS_2$/graphene flake as a counter electrode for Pt-free dye-sensitized solar cells. *Electrochimica Acta*, *85*, pp.162-168.

(153)   Yu, C., Meng, X., Song, X., Liang, S., Dong, Q., Wang, G., Hao, C., Yang, X., Ma, T., Ajayan, P.M. and Qiu, J., Graphene-mediated highly-dispersed $MoS_2$ nanosheets with enhanced triiodide reduction activity for dye-sensitized solar cells. *Carbon*, 2016, *100*, 474-483.

(154)   Lynch, P., Khan, U., Harvey, A., Ahmed, I. and Coleman, J.N., 2016. Graphene-$MoS_2$ nanosheet composites as electrodes for dye sensitised solar cells. *Materials Research Express*, *3*(3), 035007.

(155)    Li, S., Min, H., Xu, F., Tong, L., Chen, J., Zhu, C. and Sun, L., 2016. All electrochemical fabrication of $MoS_2$/graphene counter electrodes for efficient dye-sensitized solar cells. *RSC Advances*, *6*(41), pp.34546-34552.

(156)    Cheng, C.K., Lin, C.H., Wu, H.C., Ma, C.C.M., Yeh, T.K., Chou, H.Y., Tsai, C.H. and Hsieh, C.K., 2016. The Two-Dimensional Nanocomposite of Molybdenum Disulfide and Nitrogen-Doped Graphene Oxide for Efficient Counter Electrode of Dye-Sensitized Solar Cells. *Nanoscale Research Letters*, *11*(1), pp.1-9.

(157)    Fan, M.S., Lee, C.P., Li, C.T., Huang, Y.J., Vittal, R. and Ho, K.C., 2016. Nitrogen-doped graphene/molybdenum disulfide composite as the electrocatalytic film for dye-sensitized solar cells. *Electrochimica Acta*. 211, pp.164-172

(158)    Lin, J.Y., Yue, G., Tai, S.Y., Xiao, Y., Cheng, H.M., Wang, F.M. and Wu, J., 2013. Hydrothermal synthesis of graphene flake embedded nanosheet-like molybdenum sulfide hybrids as counter electrode catalysts for dye-sensitized solar cells. *Materials Chemistry and Physics*, *143*(1), 53-59.

(159)    Yue, G., Ma, X., Jiang, Q., Tan, F., Wu, J., Chen, C., Li, F. and Li, Q., 2014. PEDOT: PSS and glucose assisted preparation of molybdenum disulfide/single-wall carbon nanotubes counter electrode and served in dye-sensitized solar cells. *Electrochimica Acta*, *142*, pp.68-75.

(160)    Tai, S.Y., Liu, C.J., Chou, S.W., Chien, F.S.S., Lin, J.Y. and Lin, T.W., 2012. Few-layer $MoS_2$ nanosheets coated onto multi-walled carbon nanotubes as a low-cost and highly electrocatalytic counter electrode for dye-sensitized solar cells. *Journal of Materials Chemistry*, *22*, 24753-24759.

(161)    Lin, C.H., Tsai, C.H., Tseng, F.G., Ma, C.C.M., Wu, H.C. and Hsieh, C.K., 2016. Three-dimensional vertically aligned hybrid nanoarchitecture of two-dimensional molybdenum disulfide nanosheets anchored on directly grown one-dimensional carbon nanotubes for use as a counter electrode in dye-sensitized solar cells. *Journal of Alloys and Compounds*. http://dx.doi.org/10.1016/j.jallcom.2016.09.149

(162)    Zheng, M., Huo, J., Tu, Y., Wu, J., Hu, L. and Dai, S., 2015. Flowerlike molybdenum sulfide/multi-walled carbon nanotube hybrid as Pt-free counter electrode used in dye-sensitized solar cells. *Electrochimica Acta*, *173*, pp.252-259.

(163)    Liu, W., He, S., Wang, Y., Dou, Y., Pan, D., Feng, Y., Qian, G., Xu, J. and Miao, S., 2014. PEG-assisted synthesis of homogeneous carbon nanotubes-$MoS_2$-carbon as a counter electrode for dye-sensitized solar cells. *Electrochimica Acta*, *144*, pp.119-126.

(164)    Yue, G., Zhang, W., Wu, J. and Jiang, Q., 2013. Glucose aided synthesis of molybdenum sulfide/carbon nanotubes composites as counter electrode for high performance dye-sensitized solar cells. *Electrochimica Acta*, *112*, pp.655-662.

(165)    Lin, J.Y., Su, A.L., Chang, C.Y., Hung, K.C. and Lin, T.W., 2015. Molybdenum Disulfide/Reduced Graphene Oxide–Carbon Nanotube Hybrids as Efficient Catalytic Materials in Dye-Sensitized Solar Cells. *ChemElectroChem*, *2*(5), pp.720-725.

(166)    (a) Wang, S., Jiang, X., Zheng, H., Wu, H., Kim, S.J. and Feng, C., 2012. Solvothermal synthesis of $MoS_2$/carbon nanotube composites with improved electrochemical performance for lithium ion batteries. *Nanoscience and Nanotechnology Letters*, *4*(4), pp.378-383. (b) Wang, J.Z., Lu, L., Lotya, M., Coleman, J.N., Chou, S.L., Liu, H.K., Minett, A.I. and Chen, J., 2013. Development of $MoS_2$–CNT composite thin film from layered $MoS_2$ for lithium batteries. *Advanced Energy Materials*, *3*(6), pp.798-805.

(167) Du, T., Wang, N., Chen, H., He, H., Lin, H. and Liu, K., 2015. $TiO_2$-based solar cells sensitized by chemical-bath-deposited few-layer $MoS_2$. *Journal of Power Sources*, *275*, pp.943-949.

(168) Jhang, W.H. and Lin, Y.J., 2015. Interface modification of $MoS_2$ counter electrode/electrolyte in dye-sensitized solar cells by incorporating $TiO_2$ nanoparticles. *Current Applied Physics*, *15*(8), pp.906-909.

(169) Hung, H.C., Lin, Y.J. and Ke, Z.Y., 2016. Interface modification of $MoS_2$:$TiO_2$ counter electrode/electrolyte in dye-sensitized solar cells by doping with different Co contents. *Journal of Materials Science: Materials in Electronics*, *27*(5), pp.5059-5063.

(170) He, Z., Que, W., Xing, Y. and Liu, X., 2016. Reporting performance in $MoS_2$–$TiO_2$ bilayer and heterojunction films based dye-sensitized photovoltaic devices. *Journal of Alloys and Compounds*, *672*, pp.481-488.

(171) Yue, G., Wu, J., Xiao, Y., Huang, M., Lin, J. and Lin, J.Y., 2013. High performance platinum-free counter electrode of molybdenum sulfide–carbon used in dye-sensitized solar cells. *Journal of Materials Chemistry A*, *1*(4), pp.1495-1501.

(172) Li, W.B., Sun, M.X., He, J., Sun, S.F., Zhang, Q. and Shi, Y.Y., 2015. Preparation of $MoS_2$/Carbon Fiber Counter Electrodes and Its Application in DSSCs. *Journal of Materials Science and Engineering*, *3*, p.020.

(173) Theerthagiri, J., Senthil, R.A., Arunachalam, P., Madhavan, J., Buraidah, M.H., Santhanam, A. and Arof, A.K., 2016. Synthesis of various carbon incorporated flower-like $MoS_2$ microspheres as counter electrode for dye-sensitized solar cells. *Journal of Solid State Electrochemistry*, pp.1-10.

(174)    D. Song, M. Li, Y. Jiang, Z. Chen, F. Bai, Y. Li, B. Jiang, Facile fabrication of MoS$_2$/PEDOT-PSS composites as low-cost and efficient counter electrodes for dye-sensitized solar cells, *Journal of Photochemistry and Photobiology A,* 279 (2014) 47-51

(175)    Liang, X., Zheng, H.W., Li, X.J., Yu, Y.H., Yue, G.T., Zhang, W., Tian, J.J. and Li, T.F., 2016. Nanocomposites of Bi$_5$FeTi$_3$O$_{15}$ with MoS$_2$ as Novel Pt-free Counter Electrode in Dye-Sensitized Solar Cells, *Ceramic International* 42, pp.12888-12893

(176)    Song, J.G., Park, J., Lee, W., Choi, T., Jung, H., Lee, C.W., Hwang, S.H., Myoung, J.M., Jung, J.H., Kim, S.H. and Lansalot-Matras, C., 2013. Layer-controlled, wafer-scale, and conformal synthesis of tungsten disulfide nanosheets using atomic layer deposition. *ACS Nano*, *7*(12), pp.11333-11340.

(177)    Elias, A.L., Perea-López, N., Castro-Beltrán, A., Berkdemir, A., Lv, R., Feng, S., Long, A.D., Hayashi, T., Kim, Y.A., Endo, M. and Gutiérrez, H.R., 2013. Controlled synthesis and transfer of large-area WS$_2$ sheets: from single layer to few layers. *ACS Nano*, *7*(6), pp.5235-5242.

(178)    Nikitenko, S.I., Koltypin, Y., Mastai, Y., Koltypin, M. and Gedanken, A., 2002. Sonochemical synthesis of tungsten sulfide nanorods. *Journal of Materials Chemistry*, *12*(5), pp.1450-1452.

(179)    Zhu, Y.Q., Hsu, W.K., Terrones, H., Grobert, N., Chang, B.H., Terrones, M., Wei, B.Q., Kroto, H.W., Walton, D.R.M., Boothroyd, C.B. and Kinloch, I., 2000. Morphology, structure and growth of WS$_2$ nanotubes. *Journal of Materials Chemistry*, *10*(11), pp.2570-2577.

(180)    Rosentsveig, R., Margolin, A., Feldman, Y., Popovitz-Biro, R. and Tenne, R., 2002. WS$_2$ nanotube bundles and foils. *Chemistry of Materials*, *14*(2), pp.471-473.

(181)    Wang, Y., Li, S., Bai, Y., Chen, Z., Jiang, Q., Li, T. and Zhang, W., 2013. Dye-sensitized solar cells based on low cost carbon-coated tungsten disulphide counter electrodes. *Electrochimica Acta*, *114*, pp.30-34.

(182)    Hussain, S., Shaikh, S.F., Vikraman, D., Mane, R.S., Joo, O.S., Naushad, M. and Jung, J., 2015. Sputtering and sulfurization-combined synthesis of a transparent $WS_2$ counter electrode and its application to dye-sensitized solar cells. *RSC Advances*, *5*(125), pp.103567-103572.

(183)    Ahn, S.H. and Manthiram, A., 2016. Edge-Oriented Tungsten Disulfide Catalyst Produced from Mesoporous WO3 for Highly Efficient Dye-Sensitized Solar Cells. *Advanced Energy Materials*, *6*(3). DOI: 10.1002/aenm.201501814.

(184)    Li, S., Chen, Z. and Zhang, W., 2012. Dye-sensitized solar cells based on $WS_2$ counter electrodes. *Materials Letters*, *72*, pp.22-24.

(185)    Yue, G., Wu, J., Lin, J.Y., Xiao, Y., Tai, S.Y., Lin, J., Huang, M. and Lan, Z., 2013. A counter electrode of multi-wall carbon nanotubes decorated with tungsten sulfide used in dye-sensitized solar cells. *Carbon*, *55*, pp.1-9.

(186)    Wu, J., Yue, G., Xiao, Y., Huang, M., Lin, J., Fan, L., Lan, Z. and Lin, J.Y., 2012. Glucose aided preparation of tungsten sulfide/multi-wall carbon nanotube hybrid and use as counter electrode in dye-sensitized solar cells. *ACS Applied Materials & Interfaces*, *4*(12), pp.6530-6536.

(187)    Wilson, J.A., 1977. Concerning the semimetallic characters of $TiS_2$ and $TiSe_2$. *Solid State Communications*, *22*(9), pp.551-553.

(188)    Benesh, G.A., Woolley, A.M. and Umrigar, C., 1985. The pressure dependences of $TiS_2$ and $TiSe_2$ band structures. *Journal of Physics C: Solid State Physics*, *18*(8), p.1595.

(189)    Bourgès, C., Barbier, T., Guélou, G., Vaqueiro, P., Powell, A.V., Lebedev, O.I., Barrier, N., Kinemuchi, Y. and Guilmeau, E., 2016. Thermoelectric properties of $TiS_2$ mechanically alloyed compounds. *Journal of the European Ceramic Society*, *36*(5), pp.1183-1189.

(190)    Inoue, M., Hughes, H.P. and Yoffe, A.D., 1989. The electronic and magnetic properties of the 3d transition metal intercalates of $TiS_2$. *Advances in Physics*, *38*(5), pp.565-604.

(191)    Wan, C., Gu, X., Dang, F., Itoh, T., Wang, Y., Sasaki, H., Kondo, M., Koga, K., Yabuki, K., Snyder, G.J. and Yang, R., 2015. Flexible n-type thermoelectric materials by organic intercalation of layered transition metal dichalcogenide $TiS_2$. *Nature Materials*, *14*(6), pp.622-627.

(192)    Cucinotta, C.S., Dolui, K., Pettersson, H., Ramasse, Q.M., Long, E., O'Brian, S.E., Nicolosi, V. and Sanvito, S., 2015. Electronic Properties and Chemical Reactivity of $TiS_2$ Nanoflakes. *The Journal of Physical Chemistry C*, *119*(27), pp.15707-15715.

(193)    Chen, J., Li, S.L., Tao, Z.L., Shen, Y.T. and Cui, C.X., 2003. Titanium disulfide nanotubes as hydrogen-storage materials. *Journal of the American Chemical Society*, *125*(18), pp.5284-5285.

(194)    Prabakar, S., Bumby, C.W. and Tilley, R.D., 2009. Liquid-phase synthesis of flower-like and flake-like titanium disulfide nanostructures. *Chemistry of Materials*, *21*(8), pp.1725-1730.

(195)    Margolin, A., Popovitz-Biro, R., Albu-Yaron, A., Moshkovich, A., Rapoport, L. and Tenne, R., 2005. Fullerene-like nanoparticles of titanium disulfide. *Current Nanoscience*, *1*(3), pp.253-262.

(196)    Park, K.H., Choi, J., Kim, H.J., Oh, D.H., Ahn, J.R. and Son, S.U., 2008. Unstable Single-Layered Colloidal $TiS_2$ Nanodisks. *Small*, *4*(7), pp.945-950.

(197)    Sun, X., Bonnick, P. and Nazar, L.F., 2016. Layered TiS$_2$ Positive Electrode for Mg Batteries. *ACS Energy Letters*, *1*(1), pp.297-301.

(198)    Lin, C., Zhu, X., Feng, J., Wu, C., Hu, S., Peng, J., Guo, Y., Peng, L., Zhao, J., Huang, J. and Yang, J., 2013. Hydrogen-incorporated TiS$_2$ ultrathin nanosheets with ultrahigh conductivity for stamp-transferrable electrodes. *Journal of the American Chemical Society*, *135*(13), pp.5144-5151.

(199)    Meng, X., Yu, C., Lu, B., Yang, J. and Qiu, J., Dual integration system endowing two-dimensional titanium disulfide with enhanced triiodide reduction performance in dye-sensitized solar cells. *Nano Energy*, *22*, 59-69 (2016)

(200)    Li, C.T., Lee, C.P., Li, Y.Y., Yeh, M.H. and Ho, K.C., 2013. A composite film of TiS$_2$/PEDOT: PSS as the electrocatalyst for the counter electrode in dye-sensitized solar cells. *Journal of Materials Chemistry A*, *1*(47), pp.14888-14896.

(201)    Yang, S.L., Yao, H.B., Gao, M.R. and Yu, S.H., 2009. Monodisperse cubic pyrite NiS$_2$ dodecahedrons and microspheres synthesized by a solvothermal process in a mixed solvent: thermal stability and magnetic properties. *CrystEngComm*, *11*(7), pp.1383-1390.

(202)    Song, X., Shen, W., Sun, Z., Yang, C., Zhang, P. and Gao, L., 2016. Size-engineerable NiS$_2$ hollow spheres photo co-catalysts from supermolecular precursor for H$_2$ production from water splitting. *Chemical Engineering Journal*, *290*, pp.74-81.

(203)    Tang, C., Pu, Z., Liu, Q., Asiri, A.M. and Sun, X., 2015. NiS$_2$ nanosheets array grown on carbon cloth as an efficient 3D hydrogen evolution cathode. *Electrochimica Acta*, *153*, pp.508-514.

(204)    Gautier, F., Krill, G., Lapierre, M.F. and Robert, C., 1972. Influence of non-stoichiometry on the electrical and magnetic properties of $NiS_2$. *Solid State Communications*, *11*(9), pp.1201-1203.

(205)    Jamil, A., Batool, S.S., Sher, F. and Rafiq, M.A., 2016. Determination of density of states, conduction mechanisms and dielectric properties of nickel disulfide nanoparticles. *AIP Advances*, *6*(5), p.055120.

(206)    Wan, Z., Jia, C. and Wang, Y., 2015. In situ growth of hierarchical $NiS_2$ hollow microspheres as efficient counter electrode for dye-sensitized solar cell. *Nanoscale*, *7*(29), pp.12737-12742.

(207)    Zuo, X., Yan, S., Yang, B., Li, G., Wu, M., Ma, Y., Jin, S. and Zhu, K., 2016, Hollow spherical $NiS/NiS_2$. *Journal of Materials Science: Materials in Electronics*, 27, pp.7974

(208)    Zuo, X., Yan, S., Yang, B., Li, G., Zhang, H., Tang, H., Wu, M., Ma, Y., Jin, S. and Zhu, K., 2016. Template-free synthesis of nickel sulfides hollow spheres and their application in dye-sensitized solar cells. *Solar Energy*, *132*, pp.503-510.

(209)    Zheng, J., Zhou, W., Ma, Y., Cao, W., Wang, C. and Guo, L., 2015. Facet-dependent $NiS_2$ polyhedrons on counter electrodes for dye-sensitized solar cells. *Chemical Communications*, *51*(64), pp.12863-12866.

(210)    Li, Z., Gong, F., Zhou, G. and Wang, Z.S., 2013. $NiS_2$/reduced graphene oxide nanocomposites for efficient dye-sensitized solar cells. *The Journal of Physical Chemistry C*, *117*(13), pp.6561-6566.

(211)    Zhu, L., Richardson, B.J. and Yu, Q., 2015. Anisotropic growth of iron pyrite $FeS_2$ nanocrystals via oriented attachment. *Chemistry of Materials*, *27*(9), pp.3516-3525.

(212)    Puthussery, J., Seefeld, S., Berry, N., Gibbs, M. and Law, M., 2010. Colloidal iron pyrite (FeS$_2$) nanocrystal inks for thin-film photovoltaics. *Journal of the American Chemical Society*, *133*(4), pp.716-719.

(213)    Bi, Y., Yuan, Y., Exstrom, C.L., Darveau, S.A. and Huang, J., 2011. Air stable, photosensitive, phase pure iron pyrite nanocrystal thin films for photovoltaic application. *Nano Letters*, *11*(11), pp.4953-4957.

(214)    Cabán-Acevedo, M., Faber, M.S., Tan, Y., Hamers, R.J. and Jin, S., 2012. Synthesis and properties of semiconducting iron pyrite (FeS$_2$) nanowires. *Nano Letters*, *12*(4), pp.1977-1982.

(215)    Shukla, S., Xing, G., Ge, H., Prabhakar, R.R., Mathew, S., Su, Z., Nalla, V., Venkatesan, T., Mathews, N., Sritharan, T. and Sum, T.C., 2016. Origin of Photocarrier Losses in Iron Pyrite (FeS$_2$) Nanocubes. *ACS Nano*, *10*(4), pp.4431-4440.

(216)    Barawi, M., Ferrer, I.J., Flores, E., Yoda, S., Ares, J.R. and Sánchez, C., 2016. Hydrogen Photoassisted Generation by Visible Light and an Earth Abundant Photocatalyst: Pyrite (FeS$_2$). *The Journal of Physical Chemistry C*, *120*(18), pp.9547-9552.

(217)    Orletskii, I.G., Mar'yanchuk, P.D., Maistruk, E.V., Solovan, M.N. and Brus, V.V., 2016. Low-temperature spray-pyrolysis of FeS$_2$ films and their electrical and optical properties. *Physics of the Solid State*, *58*(1), pp.37-41.

(218)    Jiang, F., Peckler, L.T. and Muscat, A.J., 2015. Phase Pure Pyrite FeS$_2$ Nanocubes Synthesized Using Oleylamine as Ligand, Solvent, and Reductant. *Crystal Growth & Design*, *15*(8), pp.3565-3572.

(219)    Zhang, X., Scott, T., Socha, T., Nielsen, D., Manno, M., Johnson, M., Yan, Y., Losovyj, Y., Dowben, P., Aydil, E.S. and Leighton, C., 2015. Phase Stability and Stoichiometry in Thin Film Iron Pyrite: Impact on Electronic Transport Properties. *ACS Applied Materials & Interfaces*, *7*(25), pp.14130-14139.

(220)    Xu, L., Hu, Y., Zhang, H., Jiang, H. and Li, C., 2016. Confined Synthesis of $FeS_2$ Nanoparticles Encapsulated in Carbon Nanotube Hybrids for Ultrastable Lithium-Ion Batteries. *ACS Sustainable Chemistry & Engineering*, *4*(8), pp.4251-4255.

(221)    Shukla, S., Loc, N.H., Boix, P.P., Koh, T.M., Prabhakar, R.R., Mulmudi, H.K., Zhang, J., Chen, S., Ng, C.F., Huan, C.H.A. and Mathews, N., 2014. Iron Pyrite Thin Film Counter Electrodes for Dye-Sensitized Solar Cells: High Efficiency for Iodine and Cobalt Redox Electrolyte Cells. *ACS Nano*, *8*(10), pp.10597-10605.

(222)    Huang, Q.H., Ling, T., Qiao, S.Z. and Du, X.W., 2013. Pyrite nanorod arrays as an efficient counter electrode for dye-sensitized solar cells. *Journal of Materials Chemistry A*, *1*(38), pp.11828-11833.

(223)    Song, C., Wang, S., Dong, W., Fang, X., Shao, J., Zhu, J. and Pan, X., 2016. Hydrothermal synthesis of iron pyrite ($FeS_2$) as efficient counter electrodes for dye-sensitized solar cells. *Solar Energy*, *133*, pp.429-436.

(224)    Wang, Y.C., Wang, D.Y., Jiang, Y.T., Chen, H.A., Chen, C.C., Ho, K.C., Chou, H.L. and Chen, C.W., 2013. $FeS_2$ Nanocrystal Ink as a Catalytic Electrode for Dye-Sensitized Solar Cells. *Angewandte Chemie International Edition*, *52*(26), pp.6694-6698.

(225)    Kilic, B., Turkdogan, S., Astam, A., Ozer, O.C., Asgin, M., Cebeci, H., Urk, D. and Mucur, S.P., 2016. Preparation of Carbon Nanotube/$TiO_2$ Mesoporous Hybrid Photoanode with Iron Pyrite ($FeS_2$) Thin Films Counter Electrodes for Dye-Sensitized Solar Cell. *Scientific Reports*, *6*, p.27052.

(226)   Miyahara, S. and Teranishi, T., 1968. Magnetic properties of $FeS_2$ and $CoS_2$. *Journal of Applied Physics*, *39*(2), pp.896-897.

(227)   Wu, N., Losovyj, Y.B., Wisbey, D., Belashchenko, K., Manno, M., Wang, L., Leighton, C. and Dowben, P.A., 2007. The electronic band structure of $CoS_2$. *Journal of Physics: Condensed Matter*, *19*(15), p.156224.

(228)   Fang, W., Liu, D., Lu, Q., Sun, X. and Asiri, A.M., 2016. Nickel promoted cobalt disulfide nanowire array supported on carbon cloth: An efficient and stable bifunctional electrocatalyst for full water splitting. *Electrochemistry Communications*, *63*, pp.60-64.

(229)   Wang, Q., Jiao, L., Han, Y., Du, H., Peng, W., Huan, Q., Song, D., Si, Y., Wang, Y. and Yuan, H., 2011. $CoS_2$ hollow spheres: fabrication and their application in lithium-ion batteries. *The Journal of Physical Chemistry C*, *115*(16), pp.8300-8304.

(230)   Yu, L., Yang, J.F. and Lou, X.W.D., 2016. Formation of $CoS_2$ Nanobubble Hollow Prisms for Highly Reversible Lithium Storage. *Angewandte Chemie International Edition*, *55*(43), pp.13422-13426.

(231)   Zhang, D., Liu, H., Zhang, J., Wang, X., Zhang, R., Zhou, J., Zhong, J. and Yuan, B., 2016. Synthesis of Novel $CoS_2$ Nanodendrites with High Performance Supercapacitors. *International Journal of Electrochemical Science*, *11*, pp.6791-6798.

(232)   Jin, J., Zhang, X. and He, T., 2014. Self-Assembled $CoS_2$ Nanocrystal Film as an Efficient Counter Electrode for Dye-Sensitized Solar Cells. *The Journal of Physical Chemistry C*, *118*(43), pp.24877-24883.

(233)   Tsai, J.C., Hon, M.H. and Leu, I.C., 2015. Fabrication of Mesoporous $CoS_2$ Nanotube Arrays as the Counter Electrodes of Dye-Sensitized Solar Cells. *Chemistry–An Asian Journal*, *10*(9), pp.1932-1939.

(234)    Tsai, J.C., Hon, M.H. and Leu, C., 2015. Preparation of $CoS_2$ nanoflake arrays through ion exchange reaction of $Co(OH)_2$ and their application as counter electrodes for dye-sensitized solar cells. *RSC Advances*, *5*(6), pp.4328-4333.

(235)    Congiu, M., Albano, L.G.S., Decker, F. and Graeff, C.F.O., 2015. Single precursor route to efficient cobalt sulphide counter electrodes for dye sensitized solar cells. *Electrochimica Acta*, *151*, pp.517-524.

(236)    Rao, S.S., Gopi, C.V., Kim, S.K., Son, M.K., Jeong, M.S., Savariraj, A.D., Prabakar, K. and Kim, H.J., 2014. Cobalt sulfide thin film as an efficient counter electrode for dye-sensitized solar cells. *Electrochimica Acta*, *133*, pp.174-179.

(237)    Cui, X., Xie, Z. and Wang, Y., 2016. Novel $CoS_2$ embedded carbon nanocages by direct sulfurizing metal–organic frameworks for dye-sensitized solar cells. *Nanoscale*, *8*(23), pp.11984-11992.

(238)    Kim, H.J., Kim, C.W., Punnoose, D., Gopi, C.V., Kim, S.K., Prabakar, K. and Rao, S.S., 2015. Nickel doped cobalt sulfide as a high performance counter electrode for dye-sensitized solar cells. *Applied Surface Science*, *328*, pp.78-85.

(239)    Duan, X., Gao, Z., Chang, J., Wu, D., Ma, P., He, J., Xu, F., Gao, S. and Jiang, K., 2013. $CoS_2$–graphene composite as efficient catalytic counter electrode for dye-sensitized solar cell. *Electrochimica Acta*, *114*, pp.173-179.

(240)    Sun, L., Bai, Y., Zhang, N. and Sun, K., 2015. The facile preparation of a cobalt disulfide–reduced graphene oxide composite film as an efficient counter electrode for dye-sensitized solar cells. *Chemical Communications*, *51*(10), pp.1846-1849.

(241)    Hai, B., Tang, K., Wang, C., An, C., Yang, Q., Shen, G. and Qian, Y., 2001. Synthesis of $SnS_2$ nanocrystals via a solvothermal process. *Journal of Crystal Growth*, *225*(1), pp.92-95.

(242)    Zhai, C., Du, N. and Yang, H.Z.D., 2011. Large-scale synthesis of ultrathin hexagonal tin disulfide nanosheets with highly reversible lithium storage. *Chemical Communications*, *47*(4), pp.1270-1272.

(243)    Lin, Y.T., Shi, J.B., Chen, Y.C., Chen, C.J. and Wu, P.F., 2009. Synthesis and characterization of tin disulfide ($SnS_2$) nanowires. *Nanoscale Research Letters*, *4*(7), p.694.

(244)    Wang, J., Liu, J., Xu, H., Ji, S., Wang, J., Zhou, Y., Hodgson, P. and Li, Y., 2013. Gram-scale and template-free synthesis of ultralong tin disulfide nanobelts and their lithium ion storage performances. *Journal of Materials Chemistry A*, *1*(4), pp.1117-1122.

(245)    Huang, Y., Zang, H., Chen, J.S., Sutter, E.A., Sutter, P.W., Nam, C.Y. and Cotlet, M., 2016. Hybrid quantum dot-tin disulfide field-effect transistors with improved photocurrent and spectral responsivity. *Applied Physics Letters*, *108*(12), p.123502.

(246)    Giberti, A., Gaiardo, A., Fabbri, B., Gherardi, S., Guidi, V., Malagù, C., Bellutti, P., Zonta, G., Casotti, D. and Cruciani, G., 2016. Tin (IV) sulfide nanorods as a new gas sensing material. *Sensors and Actuators B: Chemical*, *223*, pp.827-833.

(247)    Du, W., Deng, D., Han, Z., Xiao, W., Bian, C. and Qian, X., 2011. Hexagonal tin disulfide nanoplatelets: A new photocatalyst driven by solar light. *CrystEngComm*, *13*(6), pp.2071-2076.

(248)    Truong, N.T.N. and Park, C., 2016. Synthesis and characterization of tin disulfide nanocrystals for hybrid bulk hetero-junction solar cell applications. *Electronic Materials Letters*, *12*(2), pp.308-314

(249)    Bai , Y., Zong , X., Yu, H., Chen, Z.-G. and Wang, L. (2014), Scalable Low-Cost $SnS_2$ Nanosheets as Counter Electrode Building Blocks for Dye-Sensitized Solar Cells. *Chemistry–A European Journal*, 20: 8670–8676. doi:10.1002/chem.201402657

(250)   Yang, B., Zuo, X., Chen, P., Zhou, L., Yang, X., Zhang, H., Li, G., Wu, M., Ma, Y., Jin, S. and Chen, X., 2015. Nanocomposite of tin sulfide nanoparticles with reduced graphene oxide in high-efficiency dye-sensitized solar cells. *ACS Applied Materials & Interfaces*, *7*(1), pp.137-143.

(251)   Sugai, S. and Ueda, T., 1982. High-pressure Raman spectroscopy in the layered materials 2H-MoS$_2$, 2HMoTe$_2$, and 2H-MoTe$_2$. *Physical Review B*, *26*(12), p.6554.

(252)   Böker, T., Severin, R., Müller, A., Janowitz, C., Manzke, R., Voß, D., Krüger, P., Mazur, A. and Pollmann, J., 2001. Band structure of MoS$_2$, MoSe$_2$, and α−MoTe$_2$: Angle-resolved photoelectron spectroscopy and ab initio calculations. *Physical Review B*, *64*(23), p.235305.

(253)   Heda, N.L., Dashora, A., Marwal, A., Sharma, Y., Srivastava, S.K., Ahmed, G., Jain, R. and Ahuja, B.L., 2010. Electronic properties and Compton profiles of molybdenum dichalcogenides. *Journal of Physics and Chemistry of Solids*, *71*(3), pp.187-193.

(254)   Rhim, S.H., Kim, Y.S. and Freeman, A.J., 2015. Strain-induced giant second-harmonic generation in monolayered 2H-MoX$_2$ (X= S, Se, Te). *Applied Physics Letters*, *107*(24), p.241908.

(255)   Caramazza, S., Marini, C., Simonelli, L., Dore, P. and Postorino, P., 2016. Temperature dependent EXAFS study on transition metal dichalcogenides MoX$_2$ (X= S, Se, Te). *Journal of Physics: Condensed Matter*, *28*(32), p.325401.

(256)   Jiang, H., 2012. Electronic band structures of molybdenum and tungsten dichalcogenides by the GW approach. *The Journal of Physical Chemistry C*, *116*(14), pp.7664-7671.

(257)   Ahuja, U., Joshi, R., Kothari, D.C., Tiwari, H. and Venugopalan, K., 2016. Optical Response of Mixed Molybdenum Dichalcogenides for Solar Cell Applications Using the Modified Becke–Johnson Potential. *Zeitschrift für Naturforschung A*, *71*(3), pp.213-223.

(258)    Abderrahmane, A., Ko, P.J., Thu, T.V., Ishizawa, S., Takamura, T. and Sandhu, A., 2014. High photosensitivity few-layered MoSe$_2$ back-gated field-effect phototransistors. *Nanotechnology*, *25*(36), p.365202.

(259)    Chuang, H.J., Chamlagain, B., Koehler, M., Perera, M.M., Yan, J., Mandrus, D., Tománek, D. and Zhou, Z., 2016. Low-resistance 2D/2D ohmic contacts: A universal approach to high-performance WSe$_2$, MoS$_2$, and MoSe$_2$ transistors. *Nano Letters*, *16*(3), pp.1896-1902.

(260)    Lee, L.T.L., He, J., Wang, B., Ma, Y., Wong, K.Y., Li, Q., Xiao, X. and Chen, T., 2014. Few-layer MoSe$_2$ possessing high catalytic activity towards iodide/tri-iodide redox shuttles. *Scientific Reports*, *4*, 4063-4069.

(261)    Chen, H., Xie, Y., Cui, H., Zhao, W., Zhu, X., Wang, Y., Lü, X. and Huang, F., 2014. In situ growth of a MoSe$_2$/Mo counter electrode for high efficiency dye-sensitized solar cells. *Chemical Communications*, *50*(34), pp.4475-4477.

(262)    Ji, I.A., Choi, H.M. and Bang, J.H., 2014. Metal selenide films as the counter electrode in dye-sensitized solar cell. *Materials Letters*, *123*, pp.51-54.

(263)    Jia, J., Wu, J., Dong, J., Tu, Y., Lan, Z., Fan, L. and Wei, Y., 2016. High-Performance Molybdenum Diselenide Electrodes Used in Dye-Sensitized Solar Cells and Supercapacitors. *IEEE Journal of Photovoltaics*, *6*(5), pp.1196-1202.

(264)    Bi, E., Chen, H., Yang, X., Ye, F., Yin, M. and Han, L., 2015. Fullerene-Structured MoSe$_2$ Hollow Spheres Anchored on Highly Nitrogen-Doped Graphene as a Conductive Catalyst for Photovoltaic Applications. *Scientific Reports*, *5*, 13214

(265)    Huang, Y.J., Fan, M.S., Li, C.T., Lee, C.P., Chen, T.Y., Vittal, R. and Ho, K.C., 2016. MoSe$_2$ nanosheet/poly(3, 4-ethylenedioxythiophene):poly (styrenesulfonate) composite film as a Pt-free counter electrode for dye-sensitized solar cells. *Electrochimica Acta*, *211*, pp.794-803.

(266)    Dong, J., Wu, J., Jia, J., Hu, L. and Dai, S., 2015. Cobalt/molybdenum ternary hybrid with hierarchical architecture used as high efficient counter electrode for dye-sensitized solar cells. *Solar Energy*, *122*, pp.326-333.

(267)    Q. Jiang, G. Hu, 2015. Co0.85Se hollow nanoparticles as Pt-free counter electrode materials for dye-sensitized solar cells, *Material Letters,* 153 pp. 114-117.

(268)    Duan, Y., Tang, Q., Liu, J., He, B. and Yu, L., 2014. Transparent Metal Selenide Alloy Counter Electrodes for High-Efficiency Bifacial Dye-Sensitized Solar Cells. *Angewandte Chemie International Edition*, *53*(52), pp.14569-14574.

(269)    Banerjee, A.; Upadhyay, K. K.; Bhatnagar, S.; Tathavadekar, M.; Bansode, U.; Agarkar, S.; Ogale, S. B., 2014. Nickel Cobalt Sulfide Nanoneedle Array as an Effective Alternative to Pt as a Counter Electrode in Dye Sensitized Solar Cells. *RSC Advances* 4, pp. 8289−8294.

(270)    Jerome, D., Grant, A.J. and Yoffe, A.D., 1971. Pressure enhanced superconductivity in $NbSe_2$. *Solid State Communications*, *9*(24), pp.2183-2185.

(271)    Galvan, D.H., Kim, J.H., Maple, M.B., Avalos-Borja, M. and Adem, E., 2000. Formation of $NbSe_2$ nanotubes by electron irradiation. *Fullerene Science and Technology*, *8*(3), pp.143-151.

(272)    Staley, N.E., Wu, J., Eklund, P., Liu, Y., Li, L. and Xu, Z., 2009. Electric field effect on superconductivity in atomically thin flakes of $NbSe_2$. *Physical Review B*, *80*(18), p.184505.

(273)    Ivanovskaya, V.V., Enyashin, A.N., Medvedeva, N.I. and Ivanovskii, A.L., 2003. Electronic properties of superconducting $NbSe_2$ nanotubes. *Physica Sstatus Solidi (b)*, *238*(3), pp.R1-R4.

(274)    Tsuneta, T., Toshima, T., Inagaki, K., Shibayama, T., Tanda, S., Uji, S., Ahlskog, M., Hakonen, P. and Paalanen, M., 2003. Formation of metallic NbSe$_2$ nanotubes and nanofibers. *Current Applied Physics*, *3*(6), pp.473-476.

(275)    Hitz, E., Wan, J., Patel, A., Xu, Y., Meshi, L., Dai, J., Chen, Y., Lu, A., Davydov, A.V. and Hu, L., 2016. Electrochemical intercalation of lithium ions into NbSe$_2$ nanosheets. *ACS Applied Materials & Interfaces*, *8*(18), pp.11390-11395.

(276)    Reynolds, K.J., Frey, G.L. and Friend, R.H., 2003. Solution-processed niobium diselenide as conductor and anode for polymer light-emitting diodes. *Applied Physics Letters*, *82*(7), pp.1123-1125.

(277)    (a)Yoshida, M., Ye, J., Nishizaki, T., Kobayashi, N. and Iwasa, Y., 2016. Electrostatic and electrochemical tuning of superconductivity in two-dimensional NbSe$_2$ crystals. *Applied Physics Letters*, *108*(20), p.202602. (b) Zhu, X., Guo, Y., Cheng, H., Dai, J., An, X., Zhao, J., Tian, K., Wei, S., Zeng, X.C., Wu, C. and Xie, Y., 2016. Signature of coexistence of superconductivity and ferromagnetism in two-dimensional NbSe$_2$ triggered by surface molecular adsorption. *Nature Communications*, *7*. p.11210

(278)    Murphy, D.W., Trumbore, F.A. and Carides, J.N., 1977. A new niobium selenide cathode for nonaqueous lithium batteries. *Journal of the Electrochemical Society*, *124*(3), pp.325-329.

(279)    Guo, J., Shi, Y., Zhu, C., Wang, L., Wang, N. and Ma, T., 2013. Cost-effective and morphology-controllable niobium diselenides for highly efficient counter electrodes of dye-sensitized solar cells. *Journal of Materials Chemistry A*, *1*(38), pp.11874-11879.

(280)    Ibrahem, M.A., Huang, W.C., Lan, T.W., Boopathi, K.M., Hsiao, Y.C., Chen, C.H., Budiawan, W., Chen, Y.Y., Chang, C.S., Li, L.J., Tsai, C.H., Chu, C. C., 2014. Controlled mechanical cleavage of bulk niobium diselenide to nanoscaled sheet, rod, and particle structures for Pt-free dye-sensitized solar cells. *Journal of Materials Chemistry A*, *2*(29), pp.11382-11390.

(281)    Guo, J., Liang, S., Shi, Y., Hao, C., Wang, X. and Ma, T., 2015. Transition metal selenides as efficient counter-electrode materials for dye-sensitized solar cells. *Physical Chemistry Chemical Physics*, *17*(43), pp.28985-28992.

(282)    Gong, F., Xu, X., Li, Z., Zhou, G. and Wang, Z.S., 2013. $NiSe_2$ as an efficient electrocatalyst for a Pt-free counter electrode of dye-sensitized solar cells. *Chemical Communications*, *49*(14), pp.1437-1439.

(283)    Zhang, X., Jing, T.Z., Guo, S.Q., Gao, G.D. and Liu, L., 2014. Synthesis of $NiSe_2$/reduced graphene oxide crystalline materials and their efficient electrocatalytic activity in dye-sensitized solar cells. *RSC Advances*, *4*(92), pp.50312-50317.

(284)    Xiao, J.; Wan, L.; Yang, S.; Xiao, F.; Wang, S. Design Hierarchical Electrodes with Highly Conductive $NiCo_2S_4$ Nanotube Arrays Grown on Carbon Fiber Paper for High-Performance Pseudocapacitors. *Nano Letters* 2014, 14, 831−838.

(285)    Wang, Q.; Liu, B.; Wang, X.; Ran, S.; Wang, L.; Chen, D.; Shen, G., 2012. Morphology Evolution of Urchin-Like $NiCo_2O_4$ Nanostructures and Their Applications as Psuedocapacitors and Photoelectrochemical Cells. *Journal of Materials Chemistry* 22, pp. 21647−21653.

(286)    Huo, J.; Wu, J.; Zheng, M.; Tu, Y.; Lan, Z., 2016. Flower-Like Nickel Cobalt Sulfide Microspheres Modified with Nickel Sulfide as Pt-Free Counter Electrode for Dye-Sensitized Solar Cells. *Journal of Power Sources*, 304, pp.266−272.

(287)    Qian, X., Li, H., Shao, L., Jiang, X. and Hou, L., 2016. Morphology-Tuned Synthesis of Nickel Cobalt Selenides as Highly Efficient Pt-Free Counter Electrode Catalysts for Dye-Sensitized Solar Cells. *ACS Applied Materials & Interfaces*, *8*(43), pp.29486-29495. DOI: 10.1021/acsami.6b09966

(288)   Coleman, R.G., 1959. New occurrences of ferroselite (FeSe$_2$). *Geochimica et Cosmochimica Acta*, *16*(4), pp.296-301.

(289)   Lutz, H.D. and Müller, B., 1991. Lattice vibration spectra. LXVIII. Single-crystal Raman spectra of marcasite-type iron chalcogenides and pnictides, FeX$_2$ (X= S, Se, Te; P, As, Sb). *Physics and Chemistry of Minerals*, *18*(4), pp.265-268.

(290)   Bastola, E., Bhandari, K.P., Matthews, A.J., Shrestha, N. and Ellingson, R.J., 2016. Elemental anion thermal injection synthesis of nanocrystalline marcasite iron dichalcogenide FeSe$_2$ and FeTe$_2$. *RSC Advances*, *6*(74), pp.69708-69714

(291)   Mao, X., Kim, J.G., Han, J., Jung, H.S., Lee, S.G., Kotov, N.A. and Lee, J., 2014. Phase-Pure FeSe$_x$ (x= 1, 2) Nanoparticles with One-and Two-Photon Luminescence. *Journal of the American Chemical Society*, *136*(20), pp.7189-7192.

(292)   Yuan, B., Luan, W. and Tu, S.T., 2012. One-step synthesis of cubic FeS$_2$ and flower-like FeSe$_2$ particles by a solvothermal reduction process. *Dalton Transactions*, *41*(3), pp.772-776.

(293)   Xu, J., Jang, K., Lee, J., Kim, H.J., Jeong, J., Park, J.G. and Son, S.U., 2011. Phase-selective growth of assembled FeSe$_2$ nanorods from organometallic polymers and their surface magnetism. *Crystal Growth & Design*, *11*(7), pp.2707-2710.

(294)   Shi, W., Zhang, X., Che, G., Fan, W. and Liu, C., 2013. Controlled hydrothermal synthesis and magnetic properties of three-dimensional FeSe$_2$ rod clusters and microspheres. *Chemical Engineering Journal*, *215*, pp.508-516.

(295)    Chang, X., Jian, J., Cai, G., Wu, R. and Li, J., 2016. Three-dimensional $FeSe_2$ microflowers assembled by nanosheets: Synthesis, optical properties, and catalytic activity for the hydrogen evolution reaction. *Electronic Materials Letters*, *12*(2), pp.237-242.

(296)    Park, G.D., Kim, J.H. and Kang, Y.C., 2016. Large-scale production of spherical $FeSe_2$-amorphous carbon composite powders as anode materials for sodium-ion batteries. *Materials Characterization*, *120*, pp.349-356.

(297)    Ganga, B.G., Ganeshraj, C., Krishna, A.G. and Santhosh, P.N., 2013. Electronic and optical properties of $FeSe_2$ polymorphs: solar cell absorber. *arXiv preprint arXiv:1303.1381*.

(298)    Jin, R., Zhao, K., Pu, X., Zhang, M., Cai, F., Yang, X., Kim, H. and Zhao, Y., 2016. Structural and photovoltaic properties of $FeSe_2$ films prepared by radio frequency magnetron sputtering. *Materials Letters*, *179*, pp.179-181.

(299)    Kong, D., Cha, J.J., Wang, H., Lee, H.R. and Cui, Y., 2013. First-row transition metal dichalcogenide catalysts for hydrogen evolution reaction. *Energy & Environmental Science*, *6*(12), pp.3553-3558.

(300)    Yang, J., Cheng, G.H., Zeng, J.H., Yu, S.H., Liu, X.M. and Qian, Y.T., 2001. Shape control and characterization of transition metal diselenides $MSe_2$ (M= Ni, Co, Fe) prepared by a solvothermal-reduction process. *Chemistry of Materials*, *13*(3), pp.848-853.

(301)    Chen, X. and Fan, R., 2001. Low-temperature hydrothermal synthesis of transition metal dichalcogenides. *Chemistry of Materials*, *13*(3), pp.802-805.

(302)    Huang, S., He, Q., Chen, W., Qiao, Q., Zai, J. and Qian, X., 2015. Ultrathin $FeSe_2$ Nanosheets: Controlled Synthesis and Application as a Heterogeneous Catalyst in Dye-Sensitized Solar Cells. *Chemistry–A European Journal*, *21*(10), pp.4085-4091

(303)    Huang, S., He, Q., Chen, W., Zai, J., Qiao, Q. and Qian, X., 2015. 3D hierarchical FeSe$_2$ microspheres: controlled synthesis and applications in dye-sensitized solar cells. *Nano Energy*, *15*, pp.205-215.

(304)    Wang, W., Pan, X., Liu, W., Zhang, B., Chen, H., Fang, X., Yao, J. and Dai, S., 2014. FeSe$_2$ films with controllable morphologies as efficient counter electrodes for dye-sensitized solar cells. *Chemical Communications*, *50*(20), pp.2618-2620.

(305)    Kong, D., Wang, H., Lu, Z. and Cui, Y., 2014. CoSe$_2$ nanoparticles grown on carbon fiber paper: an efficient and stable electrocatalyst for hydrogen evolution reaction. *Journal of the American Chemical Society*, *136*(13), pp.4897-4900.

(306)    Liu, Y., Cheng, H., Lyu, M., Fan, S., Liu, Q., Zhang, W., Zhi, Y., Wang, C., Xiao, C., Wei, S. and Ye, B., 2014. Low overpotential in vacancy-rich ultrathin CoSe$_2$ nanosheets for water oxidation. *Journal of the American Chemical Society*, *136*(44), pp.15670-15675.

(307)    Wu, R., Xue, Y., Liu, B., Zhou, K., Wei, J. and Chan, S.H., 2016. Cobalt diselenide nanoparticles embedded within porous carbon polyhedra as advanced electrocatalyst for oxygen reduction reaction. *Journal of Power Sources*, *330*, pp.132-139.

(308)    McCarthy, C.L., Downes, C.A., Schueller, E.C., Abuyen, K. and Brutchey, R.L., 2016. Method for the Solution Deposition of Phase-Pure CoSe$_2$ as an Efficient Hydrogen Evolution Reaction Electrocatalyst. *ACS Energy Letters*, *1*(3), pp.607-611.

(309)    Dong, J., Wu, J., Jia, J., Wu, S., Zhou, P., Tu, Y. and Lan, Z., 2015. Cobalt selenide nanorods used as a high efficient counter electrode for dye-sensitized solar cells. *Electrochimica Acta*, *168*, pp.69-75.

(310)    Chiu, I.T., Li, C.T., Lee, C.P., Chen, P.Y., Tseng, Y.H., Vittal, R. and Ho, K.C., 2016. Nanoclimbing-wall-like CoSe$_2$/carbon composite film for the counter electrode of a highly efficient dye-sensitized solar cell: A study on the morphology control. *Nano Energy*, *22*, pp.594-606.

(311)   Choi, W.Y., 2015. Synthesis of $CoSe_2$/RGO Composites and Their Application in a Counter Electrode of Dye-sensitized Solar Cells. Graduate school of Ulsan National Institute of Science and Technology (UNIST), South Korea http://scholarworks.unist.ac.kr/handle/201301/10604

(312)   Gong, F., Wang, H., Xu, X., Zhou, G. and Wang, Z.S., 2012. In situ growth of $Co_{0.85}Se$ and $Ni_{0.85}Se$ on conductive substrates as high-performance counter electrodes for dye-sensitized solar cells. *Journal of the American Chemical Society*, *134*(26), pp.10953-10958.

(313)   Zhang, H., Liu, C.X., Qi, X.L., Dai, X., Fang, Z. and Zhang, S.C., 2009. Topological insulators in $Bi_2Se_3$, $Bi_2Te_3$ and $Sb_2Te_3$ with a single Dirac cone on the surface. *Nature Physics*, *5*(6), pp.438-442.

(314)   Sharma, P.A., Sharma, A.L., Hekmaty, M., Hattar, K., Stavila, V., Goeke, R., Erickson, K., Medlin, D.L., Brahlek, M., Koirala, N. and Oh, S., 2014. Ion beam modification of topological insulator bismuth selenide. *Applied Physics Letters*, *105*(24), p.242106.

(315)   Sun, Y., Cheng, H., Gao, S., Liu, Q., Sun, Z., Xiao, C., Wu, C., Wei, S. and Xie, Y., 2012. Atomically thick bismuth selenide freestanding single layers achieving enhanced thermoelectric energy harvesting. *Journal of the American Chemical Society*, *134*(50), pp.20294-20297.

(316)   Batabyal, S.K., Basu, C., Das, A.R. and Sanyal, G.S., 2006. Solvothermal synthesis of bismuth selenide nanotubes. *Materials Letters*, *60*(21), pp.2582-2585

(317)   Tang, H., Wang, X., Xiong, Y., Zhao, Y., Zhang, Y., Zhang, Y., Yang, J. and Xu, D., 2015. Thermoelectric characterization of individual bismuth selenide topological insulator nanoribbons. *Nanoscale*, *7*(15), pp.6683-6690.

(318)   Zou, Y., Chen, Z.G., Huang, Y., Yang, L., Drennan, J. and Zou, J., 2014. Anisotropic electrical properties from vapor–solid–solid grown $Bi_2Se_3$ nanoribbons and nanowires. *The Journal of Physical Chemistry C*, *118*(35), pp. 20620-20626.

(319)   Zhu, H., Zhao, E., Richter, C.A. and Li, Q., 2014. Topological Insulator $Bi_2Se_3$ Nanowire Field Effect Transistors. *ECS Transactions*, *64*(17), pp.51-59.

(320)   Fan, Y., Hao, G., Luo, S., Qi, X. and Zhong, J., 2015. Photoresponse Properties of $Bi_2Se_3$ Nanoplates. *Science of Advanced Materials*, *7*(8), pp.1589-1593.

(321)   Zhang, X., Wen, F., Xiang, J., Wang, X., Wang, L., Hu, W. and Liu, Z., 2015. Wearable non-volatile memory devices based on topological insulator $Bi_2Se_3$/Pt fibers. *Applied Physics Letters*, *107*(10), p.103109.

(322)   Yuan, Z., Wu, Z., Bai, S., Cui, W., Liu, J., Song, T. and Sun, B., 2015. Layered bismuth selenide utilized as hole transporting layer for highly stable organic photovoltaics. *Organic Electronics*, *26*, pp.327-333.

(323)   Li, Z., Hu, Y., Howard, K.A., Jiang, T., Fan, X., Miao, Z., Sun, Y., Besenbacher, F. and Yu, M., 2015. Multifunctional Bismuth Selenide Nanocomposites for Antitumor Thermo-Chemotherapy and Imaging. *ACS Nano*, *10*(1), pp.984-997.

(324)   Li, Z., Liu, J., Hu, Y., Howard, K.A., Li, Z., Fan, X., Chang, M., Sun, Y., Besenbacher, F., Chen, C. and Yu, M., 2016. Multimodal Imaging-Guided Antitumor Photothermal Therapy and Drug Delivery Using Bismuth Selenide Spherical-Sponge. *ACS Nano*,*10*(10), pp 9646–9658

(325)   *Zhu, L., Cho, K. Y., and Oh, W. C.,* 2016, Microwave-assisted synthesis of $Bi_2Se_3$/reduced graphene oxide nanocomposite as efficient catalytic counter electrode for dye-sensitized solar cell, *Fullerenes, Nanotubes and Carbon Nanostructures*, 24(10), pp. 622-629

(326)   Prieto, A.L., Sander, M.S., Martín-González, M.S., Gronsky, R., Sands, T. and Stacy, A.M., 2001. Electrodeposition of ordered $Bi_2Te_3$ nanowire arrays. *Journal of the American Chemical Society*, *123*(29), pp.7160-7161.

(327)   Ma, L., Zhang, Q., Zhao, Q., Li, Z., Ji, C. and Xu, X.J., 2016. Synthesis and Characterization of $Bi_2Te_3$/Te Superlattice Nanowire Arrays. *Journal of Nanoscience and Nanotechnology*, *16*(1), pp.1207-1210.

(328)   Lee, J., Kim, J., Moon, W., Berger, A. and Lee, J., 2012. Enhanced Seebeck coefficients of thermoelectric $Bi_2Te_3$ nanowires as a result of an optimized annealing process. *The Journal of Physical Chemistry C*, *116*(36), pp.19512-19516.

(329)   Gooth, J., Zierold, R., Sergelius, P., Hamdou, B., Garcia, J., Damm, C., Rellinghaus, B., Pettersson, H.J., Pertsova, A., Canali, C. and Borg, M., 2016. Local Magnetic Suppression of Topological Surface States in $Bi_2Te_3$ Nanowires. *ACS Nano*, *10*(7), pp.7180-7188.

(330)   Du, R., Hsu, H.C., Balram, A.C., Yin, Y., Dong, S., Dai, W., Zhao, W., Kim, D., Yu, S.Y., Wang, J. and Li, X., 2016. Robustness of topological surface states against strong disorder observed in $Bi_2Te_3$ nanotubes. *Physical Review B*, *93*(19), p.195402.

(331)   Son, J.S., Choi, M.K., Han, M.K., Park, K., Kim, J.Y., Lim, S.J., Oh, M., Kuk, Y., Park, C., Kim, S.J. and Hyeon, T., 2012. n-Type nanostructured thermoelectric materials prepared from chemically synthesized ultrathin $Bi_2Te_3$ nanoplates. *Nano Letters*, *12*(2), pp.640-647.

(332)    Tsai, H.W., Wang, T.H., Chan, T.C., Chen, P.J., Chung, C.C., Yaghoubi, A., Liao, C.N., Diau, E.W.G. and Chueh, Y.L., 2014. Fabrication of large-scale single-crystal bismuth telluride ($Bi_2Te_3$) nanosheet arrays by a single-step electrolysis process. *Nanoscale*, *6*(14), pp.7780-7785.

(333)    Miao, L., Yi, J., Wang, Q., Feng, D., He, H., Lu, S., Zhao, C., Zhang, H. and Wen, S., 2016. Broadband third order nonlinear optical responses of bismuth telluride nanosheets. *Optical Materials Express*, *6*(7), pp.2244-2251.

(334)    Park, N.W., Ahn, J.Y., Umar, A., Yoon, S.G. and Lee, S.K., 2016. Thermoelectric Properties of n-Type Bismuth Telluride ($Bi_2Te_3$) Thin Films Prepared by RF Sputtering. *Science of Advanced Materials*, *8*(6), pp.1172-1176.

(335)    Dou, Y., Wu, F., Fang, L., Liu, G., Mao, C., Wan, K. and Zhou, M., 2016. Enhanced performance of dye-sensitized solar cell using $Bi_2Te_3$ nanotube/ZnO nanoparticle composite photoanode by the synergistic effect of photovoltaic and thermoelectric conversion. *Journal of Power Sources*, *307*, pp.181-189.

(336)    Wan, K., Wu, F., Dou, Y., Fang, L. and Mao, C., 2016. Enhance the performance of dye-sensitized solar cells by $Bi_2Te_3$ nanosheet/ZnO nanoparticle composite photoanode. *Journal of Alloys and Compounds*, *680*, pp.373-380.

(337)    Chen, T., Guai, G.H., Gong, C., Hu, W., Zhu, J., Yang, H., Yan, Q. and Li, C.M., 2012. Thermoelectric $Bi_2Te_3$-improved charge collection for high-performance dye-sensitized solar cells. *Energy & Environmental Science*, *5*(4), pp.6294-6298.

(338)    Hu, L., Fang, L., Wu, F. and Wang, J., 2014. Effect of the Size of $Bi_2Te_3$ nanoplates on the Performance of $TiO_2$-based Novel Hybrid DSSC Combined PV and TE Effect. In *MRS Proceedings* (Vol. 1640, pp. mrsf13-1640). Cambridge University Press.

(339)    Hinsch, A., Kroon, J.M., Kern, R., Uhlendorf, I., Holzbock, J., Meyer, A. and Ferber, J., 2001. Long-term stability of dye-sensitised solar cells. *Progress in Photovoltaics: Research and Applications*, *9*(6), pp.425-438.

(340)    Pettersson, H. and Gruszecki, T., 2001. Long-term stability of low-power dye-sensitised solar cells prepared by industrial methods. *Solar Energy Materials and Solar Cells*, *70*(2), pp.203-212.

(341)    Sommeling, P.M., Späth, M., Smit, H.J.P., Bakker, N.J. and Kroon, J.M., 2004. Long-term stability testing of dye-sensitized solar cells. *Journal of Photochemistry and Photobiology A: Chemistry*, *164*(1), pp.137-144.

(342)    Grätzel, M., 2006. Photovoltaic performance and long-term stability of dye-sensitized meosocopic solar cells. *Comptes Rendus Chimie*, *9*(5), pp.578-583.

(343)    Dai, S., Weng, J., Sui, Y., Chen, S., Xiao, S., Huang, Y., Kong, F., Pan, X., Hu, L., Zhang, C. and Wang, K., 2008. The design and outdoor application of dye-sensitized solar cells. *Inorganica Chimica Acta*, *361*(3), pp.786-791.

(344)    Takeda, Y., Kato, N., Higuchi, K., Takeichi, A., Motohiro, T., Fukumoto, S., Sano, T. and Toyoda, T., 2009. Monolithically series-interconnected transparent modules of dye-sensitized solar cells. *Solar Energy Materials and Solar Cells*, *93*(6), pp.808-811.

(345)    Asghar, M.I., Miettunen, K., Halme, J., Vahermaa, P., Toivola, M., Aitola, K. and Lund, P., 2010. Review of stability for advanced dye solar cells. *Energy & Environmental Science*, *3*(4), pp.418-426.

(346)    Kontos, A.G., Stergiopoulos, T., Likodimos, V., Milliken, D., Desilvesto, H., Tulloch, G. and Falaras, P., 2013. Long-term thermal stability of liquid dye solar cells. *The Journal of Physical Chemistry C*, *117*(17), pp.8636-8646.

(347)    Jiang, R., Anderson, A., Barnes, P.R., Xiaoe, L., Law, C. and O'Regan, B.C., 2014. 2000 hours photostability testing of dye sensitised solar cells using a cobalt bipyridine electrolyte. *Journal of Materials Chemistry A*, *2*(13), pp.4751-4757.

(348)    Sauvage, F., 2014. A review on current status of stability and knowledge on liquid electrolyte-based dye-sensitized solar cells. *Advances in Chemistry*, *2014*, Article ID 939525, http://dx.doi.org/10.1155/2014/939525.

(349)    Yun, S., Lund, P.D. and Hinsch, A., 2015. Stability assessment of alternative platinum free counter electrodes for dye-sensitized solar cells. *Energy & Environmental Science*, *8*(12), pp.3495-3514.

(350)    Liu, C.J., Tai, S.Y., Chou, S.W., Yu, Y.C., Chang, K.D., Wang, S., Chien, F.S.S., Lin, J.Y. and Lin, T.W., 2012. Facile synthesis of $MoS_2$/graphene nanocomposite with high catalytic activity toward triiodide reduction in dye-sensitized solar cells. *Journal of Materials Chemistry*, *22*, 21057-21064.

(351)    Sun, H., Zhang, L. and Wang, Z. S., 2014. Single-crystal $CoSe_2$ nanorods as an efficient electrocatalyst for dye-sensitized solar cells, *Journal of Materials Chemistry A*, 2 pp. 16023-16029.

(352)    Kato, N., Takeda, Y., Higuchi, K., Takeichi, A., Sudo, E., Tanaka, H., Motohiro, T., Sano, T. and Toyoda, T., 2009. Degradation analysis of dye-sensitized solar cell module after long-term stability test under outdoor working condition. *Solar Energy Materials and Solar Cells*, *93*(6), pp.893-897.

(353)    Matsui, H., Okada, K., Kitamura, T. and Tanabe, N., 2009. Thermal stability of dye-sensitized solar cells with current collecting grid. *Solar Energy Materials and Solar Cells*, *93*(6), pp.1110-1115.

(354)   Xue, G., Guo, Y., Yu, T., Guan, J., Yu, X., Zhang, J., Liu, J. and Zou, Z., 2012. Degradation mechanisms investigation for long-term thermal stability of dye-sensitized solar cells. *International Journal of Electrochemical Science*, *7*, pp.1496-1511.

(355)   Harikisun, R. and Desilvestro, H., 2011. Long-term stability of dye solar cells. *Solar Energy*, *85*(6), pp.1179-1188.

(356)   Wu, Y., Zhu, W.H., Zakeeruddin, S.M. and Grätzel, M., 2015. Insight into D–A− π–A Structured Sensitizers: A Promising Route to Highly Efficient and Stable Dye-Sensitized Solar Cells. *ACS Applied Materials & Interfaces*, *7*(18), pp.9307-9318.

(357)   Katoh, R.; Furube, A.; Mori, S.; Miyashita, M.; Sunahara, K.; Koumura, N.; Hara, K., 2009. Highly Stable Sensitizer Dyes for Dye-Sensitized Solar Cells: Role of the Oligothiophene Moiety. *Energy & Environmental Science*, 2, pp. 542−546.

(358)   *Joly, D., Pellejà, L., Narbey, S., Oswald, F., Chiron, J., Clifford, J.N., Palomares, E. and Demadrille, R., 2014. A robust organic dye for dye sensitized solar cells based on iodine/iodide electrolytes combining high efficiency and outstanding stability. Scientific Reports, 4, p.4033.*

(359)   Wang, Z.S., Cui, Y., Hara, K., Dan-oh, Y., Kasada, C. and Shinpo, A., 2007. A High-Light-Harvesting-Efficiency Coumarin Dye for Stable Dye-Sensitized Solar Cells. *Advanced Materials*, *19*(8), pp.1138-1141.

(360)   Kim, S., Kim, D., Choi, H., Kang, M.S., Song, K., Kang, S.O. and Ko, J., 2008. Enhanced photovoltaic performance and long-term stability of quasi-solid-state dye-sensitized solar cells via molecular engineering. *Chemical Communications*, (40), pp.4951-4953.

(361)    Lu, X., Feng, Q., Lan, T., Zhou, G. and Wang, Z.S., 2012. Molecular engineering of quinoxaline-based organic sensitizers for highly efficient and stable dye-sensitized solar cells. *Chemistry of Materials*, *24*(16), pp.3179-3187.

(362)    Yanagida, S., Yu, Y. and Manseki, K., 2009. Iodine/iodide-free dye-sensitized solar cells. *Accounts of Chemical Research*, *42*(11), pp.1827-1838.

(363)    Tian, H. and Sun, L., 2011. Iodine-free redox couples for dye-sensitized solar cells. *Journal of Materials Chemistry*, *21*(29), pp.10592-10601.

(364)    Wang, M., Grätzel, C., Zakeeruddin, S.M. and Grätzel, M., 2012. Recent developments in redox electrolytes for dye-sensitized solar cells. *Energy & Environmental Science*, *5*(11), pp.9394-9405.

(365)    Ahmad, S., Bessho, T., Kessler, F., Baranoff, E., Frey, J., Yi, C., Grätzel, M. and Nazeeruddin, M.K., 2012. A new generation of platinum and iodine free efficient dye-sensitized solar cells. *Physical Chemistry Chemical Physics*, *14*(30), pp.10631-10639.

(366)    Cong, J., Yang, X., Kloo, L. and Sun, L., 2012. Iodine/iodide-free redox shuttles for liquid electrolyte-based dye-sensitized solar cells. *Energy & Environmental Science*, *5*(11), pp.9180-9194.

(367)    Sauvage, F., Chhor, S., Marchioro, A., Moser, J.E. and Grätzel, M., 2011. Butyronitrile-based electrolyte for dye-sensitized solar cells. *Journal of the American Chemical Society*, *133*(33), pp.13103-13109.

(368)    Yoon, J., Kang, D., Won, J., Park, J.Y. and Kang, Y.S., 2012. Dye-sensitized solar cells using ion-gel electrolytes for long-term stability. *Journal of Power Sources*, *201*, pp.395-401.

(369)    Lee, K.M., Chiu, W.H., Lu, M.D. and Hsieh, W.F., 2011. Improvement on the long-term stability of flexible plastic dye-sensitized solar cells. *Journal of Power Sources*, *196*(20), pp.8897-8903.

(370)    Yang, Y., Zhou, C.H., Xu, S., Hu, H., Chen, B.L., Zhang, J., Wu, S.J., Liu, W. and Zhao, X.Z., 2008. Improved stability of quasi-solid-state dye-sensitized solar cell based on poly (ethylene oxide)–poly(vinylidene fluoride) polymer-blend electrolytes. *Journal of Power Sources*, *185*(2), pp.1492-1498.

(371)    Chen, Z., Yang, H., Li, X., Li, F., Yi, T. and Huang, C., 2007. Thermostable succinonitrile-based gel electrolyte for efficient, long-life dye-sensitized solar cells. *Journal of Materials Chemistry*, *17*(16), pp.1602-1607.

(372)    Dembele, K.T., Nechache, R., Nikolova, L., Vomiero, A., Santato, C., Licoccia, S. and Rosei, F., 2013. Effect of multi-walled carbon nanotubes on the stability of dye sensitized solar cells. *Journal of Power sources*, *233*, pp.93-97.

(373)    Kim, S.R., Parvez, M.K., In, I., Lee, H.Y. and Park, J.M., 2009. Novel photo-crosslinkable polymeric electrolyte system based on poly (ethylene glycol) and trimethylolpropane triacrylate for dye-sensitized solar cell with long-term stability. *Electrochimica Acta*, *54*(26), pp.6306-6311.

(374)    Sauvage, F., Fischer, M.K., Mishra, A., Zakeeruddin, S.M., Nazeeruddin, M.K., Bäuerle, P. and Grätzel, M., 2009. A Dendritic Oligothiophene Ruthenium Sensitizer for Stable Dye-Sensitized Solar Cells. *ChemSusChem*, *2*(8), pp.761-768.

(375)    Parvez, M.K., In, I., Park, J.M., Lee, S.H. and Kim, S.R., 2011. Long-term stable dye-sensitized solar cells based on UV photo-crosslinkable poly (ethylene glycol) and poly (ethylene glycol) diacrylate based electrolytes. *Solar Energy Materials and Solar Cells*, *95*(1), pp.318-322.

(376)   Bella, F., Pugliese, D., Nair, J.R., Sacco, A., Bianco, S., Gerbaldi, C., Barolo, C. and Bongiovanni, R., 2013. A UV-crosslinked polymer electrolyte membrane for quasi-solid dye-sensitized solar cells with excellent efficiency and durability. *Physical Chemistry Chemical Physics*, *15*(11), pp.3706-3711.

(377)   Bella, F., Griffini, G., Gerosa, M., Turri, S. and Bongiovanni, R., 2015. Performance and stability improvements for dye-sensitized solar cells in the presence of luminescent coatings. *Journal of Power Sources*, *283*, pp.195-203.

(378)   Singh, S. and Nalwa, H. S., 2007. Nanotechnology and health safety–toxicity and risk assessments of nanostructured materials on human health, *Journal of Nanoscience and Nanotechnolology* 7, pp.3048-3070

(379)   Zhao, Y. L., and Nalwa, H. S. (Editors.), 2007. *Nanotoxicology-Interactions of Nanomaterials with Biological Systems,* American Scientific Publishers, Los Angeles.

(380)   Hu, X. and Zhou, Q., 2013.Health and Ecosystem Risks of Graphene, *Chemical Reviews*, 113, pp.3815–3835

(381)   Chng, E.L.K. and Pumera, M., 2015. Toxicity of graphene related materials and transition metal dichalcogenides. *RSC Advances*, *5*(4), pp.3074-3080.

(382)   Hao, J., Song, G., Liu, T., Yi, X., Yang, K., Cheng, L. and Liu, Z., 2016. In Vivo Long-Term Biodistribution, Excretion, and Toxicology of PEGylated Transition-Metal Dichalcogenides $MS_2$ (M= Mo, W, Ti) Nanosheets. *Advanced Science*. DOI: 10.1002/advs.201600160

(383)   Teo, W.Z., Chng, E.L.K., Sofer, Z. and Pumera, M., 2014. Cytotoxicity of Exfoliated Transition-Metal Dichalcogenides ($MoS_2$, $WS_2$, and $WSe_2$) is Lower Than That of Graphene and its Analogues. *Chemistry–A European Journal*, *20*(31), pp.9627-9632.